\documentclass[final,12pt,3p,longtitle]{elsarticle}

\biboptions{sort&compress}
\journal{Physics Reports}

\usepackage{latexsym}
\usepackage{graphicx}
\usepackage{amsfonts}
\usepackage{amssymb,amscd}
\usepackage[dvips]{epsfig}                   
\usepackage{array}
\usepackage{wrapfig}
\usepackage{graphics}

\newcommand*{\La}{\cal{L}}
\newcommand*{\no}{\noindent}
\newcommand*{\bea}{\begin{eqnarray}}
\newcommand*{\eea}{\end{eqnarray}}
\newcommand*{\be}{\begin{equation}}
\newcommand*{\ee}{\end{equation}}
\newcommand*{\pd}{\partial}
\newcommand*{\pdm}{\pd_{\mu}}
\newcommand*{\pdn}{\pd_{\nu}}

\newcommand*{\pref}[1]{(\ref{#1})}

\newcommand*{\mn}{{\mu\nu}}

\newcommand*{\prefr}[2]{(\ref{#1}-\ref{#2})} 
\newcommand*{\nn}{\nonumber}

\newcommand*{\tr}{\mathrm{tr}}
\newcommand*{\atanh}{\mathrm{atanh}}

\newcommand*{\indexsep}{,}
\newcommand*{\tl}{\mathrm{tl}}

\begin{document}

\begin{frontmatter}
 
\title{Gauge bosons at zero and finite temperature}

\author{Axel Maas}

\address{Institute of Theoretical Physics, Friedrich-Schiller-University Jena, Max-Wien-Platz 1, D-07743 Jena, Germany}

\begin{abstract}

Gauge theories of the Yang-Mills type are the single most important building block of the standard model of particle physics and beyond. They are an integral part of the strong and weak interactions, and in their Abelian version of electromagnetism. Since Yang-Mills theories are gauge theories their elementary particles, the gauge bosons, cannot be described without fixing a gauge. Therefore, to obtain their properties a quantized and gauge-fixed setting is necessary.

Beyond perturbation theory, gauge-fixing in non-Abelian gauge theories is obstructed by the Gribov-Singer ambiguity, which requires the introduction of non-local constraints. The construction and implementation of a method-independent gauge-fixing prescription to resolve this ambiguity is the single most important first step to describe gauge bosons beyond perturbation theory. Proposals for such a procedure, generalizing the perturbative Landau gauge, are described here. Their implementation are discussed for two example methods, lattice gauge theory and the quantum equations of motion.

After gauge-fixing, it is possible to study gauge bosons in detail. The most direct access is provided by their correlation functions. The corresponding two- and three-point correlation functions are presented at all energy scales. These give access to the properties of the gauge bosons, like their absence from the asymptotic physical state space, particle-like properties at high energies, and the running coupling. Furthermore, auxiliary degrees of freedom are introduced during gauge-fixing, and their properties are discussed as well. These results are presented for two, three, and four dimensions, and for various gauge algebras.

Finally, the modifications of the properties of gauge bosons at finite temperature are presented. Evidence is provided that these reflect the phase structure of Yang-Mills theory. However, it is found that the phase transition is not deconfining the gauge bosons, although the bulk thermodynamical behavior is of a Stefan-Boltzmann type. The resolution of this apparent contradiction is also presented. In addition, this resolution provides an explicit and constructive solution to the Linde problem.

Thus, the technical and conceptual framework presented here can be taken as a basis how to determine correlation functions in Yang-Mills theory, therefore opening up the avenue to investigate theories of direct practical relevance. The status of this effort will be briefly described, alongside with connections to other approaches to Yang-Mills theory beyond perturbation theory.

\end{abstract}


\end{frontmatter}

\newpage
\setcounter{secnumdepth}{4}
\setcounter{tocdepth}{4}
\tableofcontents
\newpage

\section{Introduction}\label{sintro}

The best theoretical description of particle physics is currently arguably the standard-model of particle physics \cite{pdg,Bohm:2001yx}, despite indirect as well as conceptual evidence for its incompleteness \cite{Bohm:2001yx,Morrissey:2009tf}. The most significant of the latter is the missing gravitational interactions. Otherwise, the standard model contains all known interactions: The strong and weak nuclear force, electromagnetism, as well as the, at time of this writing still hypothetical \cite{Flacher:2008zq,Baak:2011ze} Higgs interaction.

The basic building block of the standard model, as well as of a multitude of extensions of the standard model \cite{Bohm:2001yx,Morrissey:2009tf,Sannino:2009za}, are gauge theories \cite{Bohm:2001yx}. These consists of a gauge sector, described by an Abelian or non-Abelian Yang-Mills theory, and a number of matter fields coupled to these gauge fields. The only exception to this rule is the Higgs interaction, which takes the form of a Yukawa theory \cite{Bohm:2001yx}.

Thus, the entire standard model of particle physics rests on the description of gauge fields using Yang-Mills theories. It is this central element of particle physics with which the following will be concerned, dropping almost always all matter fields. The aim is to describe the elementary degrees of freedom of Yang-Mills theory, the gauge bosons, as well as their interaction. In this context the fact that Yang-Mills theory is a gauge theory will play a central role, as that makes the description of the elementary excitations necessarily gauge-dependent. This subtlety will be discussed in detail in the following.

The starting point of any such discussion is the classical Lagrangian of Yang-Mills theory \cite{Yang:1954ek,Bohm:2001yx}
\bea
\La&=&-\frac{1}{4}F_\mn^a F^\mn_a\label{intro:la}\\
F_\mn^a&=&\pdm A_\nu^a-\pdn A_\mu^a+g f^a_{bc} A_\mu^b A_\nu^c\label{intro:fst}.
\eea
\no Herein the fields $A_\mu^a$ are the gauge fields describing the gauge bosons, called here for simplicity gluons. Of course, these could easily be selected to be the weak isospin bosons or the hypercharge gauge bosons. Furthermore, the parameters of the Lagrangian are the coupling constant $g$, and the numbers $f^{abc}$, which  are the structure constants of the associated gauge algebra ${\cal A}$. This gauge algebra can, in principle, be any semi-simple Lie algebra. In case of the standard model it is ${\cal A}=$su$(3)\times$su$(2)\times$u$(1)$ \cite{O'Raifeartaigh:1986vq}. The factor su$(3)$ generates the strong interactions, and the factor su$(2)\times$u$(1)$ yields, after mixing, the weak and electromagnetic interactions. Herein, the discussion will be restricted to non-Abelian gauge algebras, as without matter fields Abelian gauge theories are trivial theories of non-interacting gauge bosons, as they miss any kind of interactions between the gauge bosons \cite{Bohm:2001yx}. 

For a semi-simple algebra the interactions of Yang-Mills theory factorize into the dynamics of Yang-Mills theories with independent simple algebras. They would only be linked by the presence of matter fields, which in the standard model are the leptons, the quarks, and, possibly, the Higgs field. In the standard model, these matter fields couple only minimally to the gauge fields, leading in case of the electroweak sector also to a hiding of the symmetry by the Higgs mechanism \cite{Bohm:2001yx}. It thus suffices to consider here simple gauge algebras.

However, even without the presence of the matter fields, non-Abelian Yang-Mills theories are far from trivial. As long as the gauge algebra is non-Abelian the structure constants $f^{abc}$ are non-vanishing. Then, due to the quadratic part of the field-strength tensor \pref{intro:fst}, interactions between the gauge fields are introduced. These interactions are found to obey asymptotic freedom, i.\ e., they become weaker with increasing energy \cite{Bohm:2001yx}. On the other hand, with decreasing energy they become stronger. As a consequence, perturbation theory can only capture the leading behavior at large energies, but fails completely when the energy of a process reaches the typical scale of the theory, which is denoted as $\Lambda_{\mathrm{YM}}$ in the following. E.\ g., in the strong interactions, this scale is of the order of 1 GeV. With reaching this scale, all types of genuine non-perturbative effects are present, in particular the existence of bound states like glueballs, or the confinement of gluons \cite{Montvay:1994cy}.

Since these effects are qualitatively present even in the absence of matter fields, this is the reason to first investigate here the simpler case of neglecting matter fields, and concentrate on Yang-Mills theories with a simple Lie algebra. Understanding this system non-perturbatively then provides a firm basis for going back to the standard model. It is one possible version of this necessary foundation of understanding gauge theories with which the following will be concerned.

At the heart of many obstacles to be encountered in this process is the fact that in Yang-Mills theory redundant degrees of freedom are introduced for the sake of having a local quantum field theory \cite{Haag:1992hx}. As a consequence, Yang-Mills theory is a gauge theory. One of the most striking features of the Lagrangian \pref{intro:la} is therefore its invariance under local gauge transformations, which take the infinitesimal form
\bea
A_\mu^a&\to& A_\mu^a+D_\mu^{ab}\phi_b\label{intro:igt}\\
D_\mu^{ab}&=&\delta^{ab}\pdm+g f^{ab}_c A_\mu^c\nn
\eea
\no where $\phi^a$ are some arbitrary functions. Therefore, the gauge fields themselves cannot be entities of the physical reality, as any observations should be independent of the chosen gauge\footnote{For the theory to be consistent it is required that no anomalies are present \cite{Henneaux:1992ig}. Whether this is the case for Yang-Mills theories beyond perturbation theory has not yet been proven, but no evidence exists which suggests the presence of anomalies, and their absence will therefore be assumed here.}. Consequently, any particle-like excitations associated with the gauge fields, also cannot represent physical observable particles. In fact, the particle-field duality \cite{Haag:1992hx} turns out to be only a high-energy property \cite{Bohm:2001yx}, and at low energies the fields lack a structure which can be easily interpreted as a particle-like excitation. The theory must therefore in some way remove them from the physical spectrum, an effect which cannot be captured by perturbation theory \cite{Bohm:2001yx}: They are said to be confined. Indeed, there appears to also exist no mechanism, which could turn the gauge-dependent gluons into gauge-independent ones \cite{Haag:1992hx}, since otherwise it should be possible to experimentally detect isolated gluon-like states. Within experimental uncertainties, this is not the case \cite{pdg}. These properties will be discussed extensively in section \ref{quant:schwinger}.

Since the confinement of gluons is not described by perturbation theory, which has asymptotically observable gluon states \cite{Bohm:2001yx}, it is necessarily non-perturbative. Furthermore, the behind-the-moon-problem illustrates that the mechanism is also nonlocal \cite{Kugo:1979gm}. This confinement of gluons is not yet fully understood. Here, possible scenarios will be described in section \ref{quant:confinement}, though important connections still remain to be made, see in particular section \ref{sreltop}. In addition, the underlying mechanism could very well be gauge-dependent, but even an understanding in one gauge would be substantial progress. However, the gauge-invariant statement of gluon confinement must be valid in any gauge.

With this in mind, in the following the discussion will be essentially restricted to the Landau gauge, which will be introduced below\footnote{Reviews, which also cover some parts of the following, though with partially different emphasis, are \cite{Alkofer:2000wg,Fischer:2006ub,Binosi:2009qm,Boucaud:2011ug,Vandersickel:2012tg}.}. However, the statement of what Landau gauge means is non-trivial beyond perturbation theory, as a consequence of the Gribov-Singer ambiguity \cite{Gribov:1977wm,Singer:1978dk}. This ambiguity and its consequence will be discussed in great detail in sections \ref{snpgf} and \ref{quant:resgrisin}. One possibility is to split the perturbatively unique Landau gauge in a family of gauges beyond perturbation theory. This will be discussed in more detail in section \ref{quant:resgrisin}. Occasionally, comments on the state of affairs in other gauges will also be made, but this is not the main focus here. Of course, eventually the aim can only be to understand the same mechanism in at least a sizable number of different gauges, similar to the concept of gauge-parameter independence of perturbation theory \cite{Bohm:2001yx}.

Irrespective of the particular choice, once a gauge is unambiguously fixed, the concept of gluons is well-defined, and questions relating to their properties can be posed. This will be done in section \ref{szerot}. The methods used for this purpose are described beforehand in section \ref{smethods}, and are the equation of motions and lattice gauge theory. These will be used jointly to determine the correlation functions, a concept introduced in section \ref{sscorfunc}, of gluons. From the two-point functions, the propagators, it is possible to infer properties of gluons. By determining the corresponding three-point functions it is possible to determine their relative interaction strength, the running gauge coupling, but also estimates of decay rates and fusion processes. Ultimately, this will give access to much more complicated quantities like scattering amplitudes in the future.

These investigations can be extended to finite temperature, which will be done in chapter \ref{sfinitet}. This setting is of quite some importance when considering the early universe, but also for its laboratory-based recreation in heavy-ion collisions \cite{Jacobs:2007dw,Andronic:2009gj,Leupold:2011zz,Friman:2011zz}. Here, the change of the properties of gluons will be read off from the temperature-dependent propagators, which will be determined once more by the quantum equations of motion and lattice gauge theory simultaneously. It is found that gluons are not deconfined at any temperature, though they generate a thermodynamic potential which at high temperature coincides with that of a gas of non-interacting gluons. This apparent contradiction will be resolved, showing that long-range physics is sub-leading to hard effects for bulk thermodynamics. As a by-product, this will yield a constructive resolution of the Linde problem \cite{Linde:1980ts}.

Finally, the state of the art, the required steps to make this a full formal procedure, and possible pitfalls will be summarized in section \ref{ssum}. Also the results will be set into perspective to other investigations and other theories, in particular QCD, over the course of this text, in particular in section \ref{sbym}. The final result of the presented set of methods and concepts is a framework for the description of gauge theories beyond perturbation theory at all energy scales. Though by far not yet a simple out-of-the-box solution given the complexity of the subject, it represents a versatile toolbox to describe the gauge-invariant, observable physics of gauge theories using explicitly the elementary degrees of freedom, the gauge bosons. It thus makes explicit contact between the basic structure of the theory and the consequences of it, which can be measured in experiment, irrespective of the involved scale, and without the addition of further parameters to the original theory, as is necessary in model or effective theory abstractions. The continued development of this framework and its application to various problems is therefore a very active field of research. Especially, it has already been applied to a wide range of theories, from Yang-Mills theory itself, as described here, over QCD \cite{Alkofer:2000wg,Fischer:2006ub}, Higgs physics \cite{Maas:2010nc}, Technicolor extensions of the standard model \cite{Maas:2011jf,Sannino:2009za,Doff:2009kq}, to quantum gravity \cite{Eichhorn:2011pc,Eichhorn:2010tb,Litim:2003vp}.

The starting point of all of these investigations is, of course, quantized Yang-Mills theory.

\section{Quantizing Yang-Mills theory in Landau gauge}\label{squant}

\subsection{The structure of gauge orbits}

As noted, the Lagrangian \pref{intro:la} is invariant under the infinitesimal gauge transformation \pref{intro:igt}. In fact, it is also invariant under the finite gauge transformation
\bea
A_\mu^{(h)}&=& hA_\mu h^{-1}+h\pdm h^{-1}\nn\\
A_\mu&=&\tau_a A_\mu^a\nn\\
h&=&\tau_a \phi^a\nn
\eea
\no where $\tau^a$ are the generators of the gauge algebra in the fundamental representation. As a consequence, the set of fields connected by gauge transformations
\be
{\cal G}(A_\mu)=\{A_\mu^{(h)}\;\forall h\}\nn,
\ee
\no are all equivalent representations of a given, fixed space-time history $A_\mu$ of the gluon field. Such a space-time history will be called a configuration in the following. The set ${\cal G}$ depending on such a configuration is called its associated gauge orbit. The value of all objects which do not change under a gauge-transformation and thus are equal for all members of the gauge orbit, e.\ g.\ the action, are called gauge-invariant. On the other hand, objects which are changing under gauge-transformations will be called gauge-dependent. In general, they will depend on which element of the set ${\cal G}$ is selected to calculate them. The chosen element of ${\cal G}$ is called the representative of the gauge orbit. However, this is not necessary. Gauge-dependent quantities may still be invariant under the choice of elements from a subset of ${\cal G}$, or may change only between different subsets of ${\cal G}$.

To determine the expectation value $<{\cal Q}>$ of a quantity ${\cal Q}(A_\mu)$, depending on the field variables $A_\mu$, the path-integral formalism can be used\footnote{From here on everything will be given in Euclidean space-time in natural units. The reconstruction of any Minkowski quantity can be done either by explicit Wick rotation \cite{Peskin:1995ev} or by usage of the Schwinger function reconstruction \cite{Haag:1992hx}. In a lattice formulation it can be proven that this is possible for all gauge-invariant quantities \cite{Seiler:1982pw}. A proof for gauge-dependent quantities, necessary to make the theory well-defined in Euclidean space-time, is still lacking, but henceforth it will be assumed to be possible. See \cite{Alkofer:2003jj} for more details.} \cite{Rivers:1987hi}. The value of $<{\cal Q}>$ is then given by
\be
<{\cal Q}>=\int {\cal D} A_\mu {\cal Q}(A_\mu) e^{-\int d^4x \La}\label{quant:unfixedpi},
\ee
\no where the measure ${\cal D} A_\mu$ is normalized such that $<1>=1$. The functional integral is over all gauge orbits of all configurations. It is then necessary to ensure that the integral is only sampling each configuration with the same weight using adequate normalization to obtain a well-defined result for ${\cal Q}$. For gauge-invariant quantities, this is sufficient \cite{Montvay:1994cy,Seiler:1982pw}. For a gauge-dependent quantity, like the gauge-fields, this is not sufficient. The integration over all representatives of a gauge-orbit yields that all such expectation values vanish \cite{Elitzur:1975im}. To obtain a non-zero value, it is necessary to define how to select a representative, or a weighted subset of representatives, of each gauge orbit, which is equivalent to choosing a coordinate system in terms of the field variables \cite{Bohm:2001yx}. This has to be done by introducing a gauge-fixing function into the path integral weight, which cuts the domain of integration. Only when performing the integral over such a reduced gauge orbit, non-zero values for gauge-dependent observables can be obtained.

How to perform this is not only interesting as a conceptual question in its own right, nor just because it is not possible to discuss gluon properties without fixing a gauge, but also for practical reasons. As will be discussed in section \ref{quant:giobs}, it is very convenient to use gauge-dependent quantities in intermediate stages to finally obtain gauge-invariant, observable quantities.

\subsection{Perturbative gauge-fixing}

The simplest way to implement a selection criterion is obtained in perturbation theory. Furthermore, the implementation in perturbation theory is also a viable starting point to generalize to the full theory.

In perturbation theory, it is sufficient to implement a local condition to select a unique representative on the gauge orbit \cite{Bohm:2001yx,Fuster:2005eg}. One possibility to do so is by requiring
\be
\pdm A_\mu^a=0\label{quant:lg},
\ee
\no the Landau gauge condition. This gauge condition will be used exclusively here, i.\ e., whatever other conditions a field configuration satisfies in the following, it will always satisfy \pref{quant:lg} as well.

The Landau gauge is a limiting case of the class of covariant gauges \cite{Bohm:2001yx}, which are well suited for perturbative purposes. Many other gauge conditions, like Coulomb gauge, maximal Abelian gauges, interpolating gauges, axial gauges, Curci-Ferrari gauges, background-gauges, and others have also been used. However, Landau gauge has a number of distinct advantages. One is that it is manifestly covariant, and none of the technical complications associated with non-manifest covariance appear \cite{Burnel:2008zz}. Furthermore, renormalization is most simple in Landau gauge, as the degree of the divergence is the lowest of all possible, i.\ e., just logarithmically divergent and not quadratically divergent in four dimensions. As a consequence, all correlation functions in lower than four dimensions are finite. Finally, in Landau gauge, transverse tensor structures and longitudinal tensor structures of correlation functions, with respect to gluon momenta, are as decoupled as possible \cite{Fischer:2008uz}. In particular, the minimum number of tensor structures is required when treating non-amputated correlation functions. All these properties make Landau gauge technically convenient. As a consequence, it is by all means the best-studied gauge beyond perturbation theory so far.

The condition \pref{quant:lg} is sufficient to perturbatively single out one representative of each gauge orbit, up to global gauge transformations. This follows from the fact that no transformation functions $\phi^a$ exist, which takes a gauge copy satisfying the Landau gauge by an infinitesimal transformation \pref{intro:igt} into a gauge copy which also satisfies the Landau gauge condition \pref{quant:lg}. When expanding a finite gauge transformation in powers of the coupling constant, this also holds.

The Landau gauge condition \pref{quant:lg} can be introduced as a restriction of the space of all gauge orbits to a hypersurface of representatives satisfying the Landau gauge condition by either use of the Faddeev-Popov procedure \cite{Faddeev:1967fc,Bohm:2001yx}, or by the more general anti-field formalism \cite{Henneaux:1992ig,Fuster:2005eg}. Both equivalent formalisms introduce two additional auxiliary anti-commuting, scalar fields $c^a$ and $\bar{c}^a$, the ghost and anti-ghost fields belonging to the adjoint representation of the gauge algebra. This permits to construct a local formulation of the restriction of the gauge orbit to a hypersurface. As a result, the path integral \pref{quant:unfixedpi} takes the form
\bea
<{\cal Q}>&=&\lim_{\xi\to 0}\int{\cal D}A_\mu{\cal D}c{\cal D}\bar{c} {\cal Q}(A_\mu,c,\bar{c})e^{-\int d^dx {\La}_g}\label{quant:gfpi}\\
{\La}_g&=&{\La}+\frac{1}{2\xi}(\pdm A_\mu^a)^2+\bar{c}_a\pdm D_\mu^{ab} c_b\nn,
\eea
\no in $d$ dimensions. The introduction of a gauge parameter $\xi$ is necessary in an intermediate step to make all possible inversions of operators well-defined. In implementations using lattice gauge theory, see section \ref{zerot:lat}, it is actually not necessary to make explicit references to it, and also in calculations using functional methods, see section \ref{zerot:dse}, it is possible to remove this parameter rather early on. It will therefore be mostly left implicit hereafter. It should be noted that for all such expressions implicitly necessary weight-factors have been included in the measure to ensure that still $<1>=1$ is valid.

Of course, given the presence of auxiliary fields it is possible to also construct quantities ${\cal Q}$ including them. From the appearance of the covariant derivative it follows that the ghosts interact in non-Abelian gauge theories with the gauge fields. Thus it is mandatory to include them in any calculations, which makes explicit use of the gauge-fixed path integral in the form \pref{quant:gfpi}. However, they do not represent physical particles, and as such will not appear in observable physical states, but only in intermediate states. Nonetheless, it is possible to form bound states from them, called ghost balls, or as gluon-ghost hybrids \cite{Kugo:1979gm}. Such bound-states are still unphysical, and thus still not be observable, even if they are color singlets.

After having obtained a perturbatively gauge-fixed setting, it is possible to determine correlation functions in perturbation theory, using standard methods \cite{Rivers:1987hi}. For going beyond perturbation theory, it is quite useful to first introduce the concept of correlation functions in more detail, as they will play a role when attempting to gauge-fix beyond perturbation theory in the way proposed here.

\subsection{Correlation functions}

\subsubsection{Propagators and vertices}\label{sscorfunc}

The basic entities, which determine the complete partition function, are the Green's or correlation functions \cite{Rivers:1987hi}. In their simplest version, these are polynomials of $n$ fields
\be
<A_{a_1}(x_1)...A_{a_n}(x_n)>=\int{\cal D} A_i A_{a_1}(x_1)...A_{a_n}(x_n) e^{-\int d^4x\La}\label{quant:fullgreen}
\ee
\no with $a_i$ a multi-index encompassing the type of the field as well as color and Lorentz indices. The number $n$ of fields involved will henceforth be called the order of the correlation function. From these the connected correlation functions can be derived, which are defined for the two-point functions as
\be
<A_1 A_2>_c=<A_1 A_2>-<A_1><A_2>\nn,
\ee
\no and correspondingly for higher $n$-point functions \cite{Bohm:2001yx}. Furthermore, the amputated correlation functions, or vertex functions, $\Gamma^i$ fulfill the relation \cite{Rivers:1987hi}
\be
<A_{a_1}(x_1)...A_{a_n}(x_n)>_{c}=\int dy_1...dy_n\frac{1}{\Gamma^{A_{a_1}A_{b_1}}(x_1-y_1)...\Gamma^{A_{a_n}A_{b_n}}(x_n-y_n)}\Gamma^{A_{b_1}...A_{b_n}}(y_1,...,y_n)\nn.
\ee
\no In particular, for the two-point functions, the propagators\footnote{The inverse appears essential due to the amputation of the two-point correlation functions.}, the relation 
\be
<A_a(x)A_b(y)>_c=\Gamma^{A_aA_b-1}(x-y)\equiv D^{A_aA_b}(x-y)\nn,
\ee
\no holds. It has been assumed that the system is translationally invariant, which will be the situation encountered throughout. It should be noted that for Yang-Mills theory the one-point functions in all covariant gauges, and in particular in Landau gauge, always vanish, as otherwise Lorentz symmetry would be broken. Therefore, connected and full Green's functions coincide for two- and three-point correlation functions, and a difference is only encountered for higher correlation functions.

The relation \pref{quant:fullgreen} permits the definition of the generating functional for the various correlation functions \cite{Alkofer:2000wg}. The equations for the full correlation functions are obtained from
\bea
&&\left(-\frac{\delta}{\delta A_a}\left[\frac{\delta}{\delta j}\right]+j_a\right) Z[j]=0\label{quant:genfcf}\\
&&Z[j]=\int{\cal D} A_a e^{-S+\int d^d x j^a A_a}\label{quant:sourcedpi},
\eea
\no where the $j_a$ are sources to be set to zero at the end of the derivation. The derivative in the argument of the expression of the action implies that all appearances of the fields have to be replaced by a derivative with respect to the sources, inheriting their indices. This equation yields the one-point correlation functions. Higher correlation functions are obtained by taking further functional derivatives of this equation with respect to the fields.

The connected correlation functions are obtained from
\bea
&&-\frac{\delta S}{\delta A_a}\left[\frac{\delta W}{\delta j}+\frac{\delta}{\delta j}\right]+j_a=0\label{quant:gencon}\\
&&W=\ln Z\nn,
\eea
\no where $W$ is also called the free energy. The vertex functions are determined from
\bea
&&\frac{\delta\Gamma}{\delta A_a}-\frac{\delta S}{\delta A_a}\left[A_a+\frac{\delta^2 W}{\delta j_a\delta j_b}\frac{\delta}{\delta j_b}\right]=0\label{quant:gendse}\\
&&\Gamma[\phi]=-W+\int d^dx \frac{\delta W}{\delta j_a} j_a\nn.
\eea
\no Thus, the Legendre transform $\Gamma$ of $W$, the effective action, is the generating functional of vertex functions. In particular \cite{Bohm:2001yx}
\bea
\Gamma^{A_{a_1}...A_{a_n}}(x_1,...,x_n)&=&\frac{\delta^n\Gamma}{\delta A_{a_1}(x_1)...\delta A_{a_n}(x_n)}\label{quant:vertexdef}\\
\Gamma^{A_a A_b}(x,y)&=&D^{A_aA_b-1}(x-y)=\frac{\delta^2\Gamma}{\delta A_a(x)\delta A_b(y)}=\left(\frac{\delta^2W}{\delta j_a(x)\delta j_b(y)}\right)^{-1}\nn.
\eea
\no The order of the field indices $A_i$ is relevant not only because of assignment of the arguments, but also if anti-commuting fields appear.

From \pref{quant:vertexdef} follows the reconstruction of the original path-integral \pref{quant:sourcedpi} as \cite{Bohm:2001yx}
\be
Z[j]=\sum_{n=0}^\infty\int d^dx_1...d^dx_n <A_{a_1}(x_1)...A_{a_n}(x_n)>j_{a_1}(x_1)...j_{a_n}(x_n).\label{quant:recon}
\ee
\no Similar reconstructions can be performed for $W$ and $\Gamma$ in terms of the connected correlation functions and the vertex functions, respectively. The equations for the vertex functions generated by derivations of \pref{quant:gendse} with respect to the fields are known as Dyson-Schwinger equations \cite{Rivers:1987hi,Alkofer:2000wg}, and are described in more detail in section \ref{zerot:dse}.

As a general convention, in the following the function, or functions, which modify a propagator or vertex away from its tree-level value will be called a dressing function. E.\ g., for a propagator $\Gamma$, its dressing function $\gamma$ is defined as
\be
\Gamma_{ab}=\Gamma^0_{ab}\frac{1}{\gamma}\nn,
\ee
\no where the inversion is used to connect to the literature \cite{Alkofer:2000wg}\footnote{Note that in the context of (functional) renormalization-group equations often the inverse of $\gamma$ is called a dressing function \cite{Pawlowski:2005xe}.}, while $\Gamma^0_{ab}$ is the tree-level propagator. In general, for both propagators and vertices, additional tensor structures both in Lorentz and color indices can appear beyond tree-level. In this case more than one dressing function will be needed. They will be defined as dimensionless factor functions, and the new tensor structures are taken to not contain any information beyond those needed to specify its transformation properties and engineering dimensions. E.\ g., if the gluon 2-point vertex were transverse at tree-level, but acquired a non-trivial longitudinal dressing beyond tree-level, this would read
\be
\Gamma_\mn^{A^2 ab}=\delta^{ab}\left(\delta_\mn-\frac{p_\mu p_\nu}{p^2}\right)\frac{p^2}{Z(p^2)}+\delta^{ab}\frac{p_\mu p_\nu}{p^2}\frac{p^2}{L(p^2)}\nn.
\ee
\no Of course, this is not the case in Landau gauge \cite{Bohm:2001yx,Cucchieri:2008zx}. In particular, the transverse gluon propagator dressing function will be denoted by $Z$ throughout\footnote{Note that this function corresponds to the trace in both color and Lorentz indices of the full gluon propagator, multiplied by $p^2$, thus being $p^2\langle A_\mu^a(-p)A_\mu^a(p)\rangle/((d-1)N_A)$ with the dimension $d$ of space-time and $N_A$ the size of the adjoint representation of the gauge algebra. It is thus always positive semi-definite.}, while the ghost propagator dressing function will be denoted by $-G$. As a general shorthand, $D_\mn$ will be used for the gluon propagator $\Gamma_\mn^{AA-1}$, $D$ for $Z(p)/p^2$, and $D_G$ for the ghost propagator $\Gamma^{\bar{c}c-1}$, and color indices are kept explicit if not a factor of $\delta^{ab}$ has been factored out, i.\ e., $D^{ab}(p)=\delta^{ab}D(p)$.

\subsubsection{The construction of gauge-invariant observables}\label{quant:giobs}

By expanding the equation \pref{quant:gendse} and its functional derivatives in the coupling (or another parameter), a perturbative set of equations for the correlation functions is obtained \cite{Rivers:1987hi}. This permits straightforward computations of the perturbative behavior of the correlation functions. This process can be simplified by the use of Feynman rules \cite{Bohm:2001yx}.

In particular, the matrix elements of two-body decays are described by the three-point correlation functions, while two-to-two scattering cross sections matrix elements can be derived from the four-point correlation functions \cite{Bohm:2001yx,Peskin:1995ev}. The latter are the lowest correlation functions which can be combined to yield a momentum-dependent gauge-invariant quantity in Yang-Mills theory. Their computation is therefore the minimal requirement to obtain gauge-invariant information in general. In perturbation theory, their calculation can often be reduced to simpler calculations by, e.\ g., Cutosky rules \cite{Peskin:1995ev}. Any such gauge-invariant result will necessarily always have all color indices in some way contracted, as on any non-contracted index a gauge-transformation could act, and thus would modify the result.

Note that gauge-invariant physics also influences gauge-dependent correlation functions, for any order. E.\ g., a gluon can emit a virtual glueball, and reabsorb it later. Turning the argument around, there must exist indirect ways to infer gauge-invariant information from the gauge-dependent correlation functions. Therefore, in principle, it should be possible to obtain this information also from lower-order correlation functions. An example for this will be the determination of the phase transition temperature and order parameters from the two-point correlation functions in section \ref{sfinitet:props}.

It should be noted that, though the final result is gauge-invariant, it is often possible to choose a gauge which is especially suited to calculate a particular gauge-invariant quantity. This makes gauge-fixed calculations, though requiring more degrees of freedom, for many purposes quite attractive. E.\ g., almost all of perturbation theory is formulated in this way \cite{Bohm:2001yx,Peskin:1995ev}, though it is not necessary when only calculating gauge-invariant quantities \cite{Montvay:1994cy,Arnone:2005fb,Rosten:2010vm}. The calculation of the hadronic spectrum in lattice gauge theory utilizes also gauge-fixed intermediate stages for a more efficient treatment \cite{Gattringer:2010zz}.

Thus, the complete stage for perturbative calculations is now set. Unfortunately, many interesting effects, like bound-states, phase transitions, hiding or breaking of symmetries, confinement, etc.\ cannot be obtained in perturbation theory \cite{Bohm:2001yx}, and a treatment beyond perturbation theory is necessary. This is the main topic of this work.

\subsection{Non-perturbative gauge-fixing and Gribov copies}\label{snpgf}

To calculate gauge-dependent quantities non-perturbatively, in particular correlation functions, it is necessary to obtain a gauge-fixing which is valid beyond perturbation theory. However, beyond perturbation theory, gauge conditions like the Landau-gauge condition \pref{quant:lg} have no longer a unique solution for a given configuration. There are several explicit examples illustrating this fact \cite{Gribov:1977wm,Maas:2005qt,Sobreiro:2005ec,Esposito:2004zn,Capri:2012ev}. Such independent solutions are called Gribov copies, and the associated ambiguity of the gauge condition is termed the Gribov-Singer ambiguity \cite{Gribov:1977wm,Singer:1978dk}. This is a property which pertains to any non-Abelian gauge theory of Yang-Mills type\footnote{In curved space-times, the properties of Gribov copies can change substantially, see e.\ g.\ \cite{Canfora:2010zf,Anabalon:2010um,Canfora:2011xd}. Furthermore, Gribov copies also appear in gravity \cite{Das:1978gk,Esposito:2004zn,Baulieu:2012ay}. Both findings are beyond the scope of this review.}.

Unfortunately, it turns out that it is impossible to construct any kind of purely local gauge condition to single out a unique representative for a configuration \cite{Singer:1978dk}\footnote{This is only proven for covariant gauges \cite{Singer:1978dk}. However, it has not yet been possible to find any local gauge condition which resolves the problem, and at best it can be traded in for a different problem of similar complexity, as is done, e.\ g, in direct Laplacian gauges \cite{Vink:1992ys,Durr:2002jc,vanBaal:1994ai}. For recent approaches to circumvent this problem, see e.\ g.\ \cite{Quadri:2010vt}.}. This is a rather general result, which is based on the structure of the gauge orbits in non-Abelian gauge theories. As a consequence, it is only possible to specify a unique representative of a gauge-orbit when this specification is done non-locally.

\subsection{Proposals for resolving the Gribov-Singer ambiguity}\label{quant:resgrisin}

\subsubsection{Gribov regions}

In principle, it would be possible to average, in a well-defined way, over the Gribov copies to obtain also non-perturbatively a well-defined gauge, similar to, e.\ g., covariant gauges in perturbation theory. However, such Hirschfeld gauges \cite{Hirschfeld:1978yq,Fujikawa:1978fu} induce significant cancellations, and no practical implementation has been constructed so far, but only conceptually developed \cite{Neuberger:1986xz,vonSmekal:2008en,vonSmekal:2007ns,vonSmekal:2008ws}. Thus, the alternative is to select by some prescription a single representative or a smaller subset for each gauge orbit, which therefore satisfies further\footnote{The global gauge freedom will be left unfixed. Fixing it would lead, e.\ g., to Polyakov gauges \cite{Marhauser:2008fz} or aligned gauges \cite{Maas:2012ct}.} constraints in addition to \pref{quant:lg}.

Thus, further (non-local) constraints are required. To implement the constraints, a sequence of further conditions can be applied. In all cases, the first step taken is always to reduce the perturbative gauge freedom by implementing a local gauge condition, here the Landau gauge \pref{quant:lg}.

The remaining set of Gribov copies is called the residual gauge orbit \cite{Maas:2008ri}. Since the condition is perturbatively unique, only finite gauge transformations connect two different elements of the residual gauge orbit \cite{Dudal:2008sp}. This is trivially so, since any infinitesimal gauge transformation will move along the gauge orbit automatically out of the gauge-fixing hypersurface implemented by the perturbative gauge-fixing by construction.

The first restriction of the residual gauge orbit taken here is to constrain it to the so-called first Gribov region \cite{Gribov:1977wm}. This Gribov region is defined by the requirement that the Faddeev-Popov operator $M$
\be
M^{ab}=-\pdm D^{ab}_\mu\label{quant:fpo},
\ee
\no is strictly positive semi-definite, i.\ e., all of its eigenvalues are zero or positive. This region can be shown to be bounded and convex \cite{Zwanziger:2003cf}, and the Faddeev-Popov operator has zero eigenvalues only on the boundary of this region\footnote{Apart from trivial zero modes due to constant eigenmodes, which will always be implicitly factored out.}, the so-called Gribov horizon. It can be shown that all gauge orbits pass at least once through the first Gribov region \cite{Dell'Antonio:1991xt}, and hence no physical information is lost by restricting to it. The boundedness is a remarkable fact, as it implies that when calculating physical observables no arbitrarily large field fluctuations have to be taken into account. It contains the origin of field-space, and thus perturbation theory, as well. This follows from the fact that in the vacuum case \pref{quant:fpo} reduces to the positive semi-definite Laplacian \cite{Maas:2005qt}. Thus by restricting to the first Gribov region, ordinary perturbation theory is always included.

Besides this first Gribov region, the remainder of the residual gauge orbit is a set of further Gribov regions. These are separated by further concentric Gribov horizons, each having more and more negative eigenvalues. The number of negative eigenvalues increases by one by passing the boundaries of these regions, but stays constant inside \cite{Gribov:1977wm,Sobreiro:2005ec}. It is expected that every residual gauge orbit passes through every Gribov region, though there is not yet an explicit proof of this.

This restriction can be implemented using a $\theta$-function \cite{Gribov:1977wm,Zwanziger:1992qr} in the perturbative gauge-fixed path integral \pref{quant:gfpi}
\bea
\langle{\cal Q}\rangle&=&\lim_{\xi\to 0}\int{\cal D}A_\mu{\cal D}c{\cal D}\bar{c} {\cal Q}\theta\left(-\pdm D_\mu^{ab}\right)(A_\mu,c,\bar{c})e^{-\int d^4x {\La}_g}\label{quant:fgr}\\
\theta\left(-\pdm D_\mu^{ab}\right)&=&\mathop{\Pi}_{i}\theta(\lambda_i)\nn,
\eea
\no where $\lambda_i$ is the $i$th eigenvalue of the Faddeev-Popov operator \pref{quant:fpo}. Thus, only if all eigenvalues are positive the $\theta$-function contributes. To make contact to the implementation of this restriction in lattice gauge theory below in section \ref{smethods:lgf}, the definition $\theta(0)=1$ has to be made for the step function.

Unfortunately, a unique, method-independent prescription how to effectively implement this restriction to the first Gribov region explicitly has not yet been constructed. There are, however, a number of possibilities, which have been explored.

A proposal for how to implement this restriction using additional ghost fields, and thus in a similar way as in perturbation theory, has been made by approximating the $\theta$-function by a $\delta$-function with the argument that in a high-dimensional space only the boundary contains an appreciable part of the volume \cite{Zwanziger:1992qr}. This generates the so-called Zwanziger Lagrangian. However, due to subtleties related to the definition of the step-function it is not yet proven that this is a valid procedure, though it has many interesting properties, and has been investigated in great detail, see e.\ g.\ \cite{Zwanziger:1992qr,Capri:2007ix,Dudal:2007cw,Dudal:2008sp,Dudal:2008rm,Baulieu:2008fy,Dudal:2008xd,Gracey:2009mj,Huber:2009tx,Zwanziger:2009je,Zwanziger:2010iz,Gracey:2010df,Nakajima:2009rw,Dudal:2011gd} and especially the review \cite{Vandersickel:2012tg}. Furthermore, no Gribov copy, or any gauge copy in general, is preferred compared to another \cite{Maas:2010wb}. It would thus be completely legitimate to always chose the innermost Gribov copy for each gauge orbit. If (almost) all gauge orbits have a representative away from the Gribov horizon, this would yield distinctively different results for gauge-dependent quantities, e.\ g.\ the expectation value of the lowest Faddeev-Popov eigenvalue. Thus, such a replacement is already implementing a certain selection of Gribov copies, and thus corresponds to an extended gauge-fixing procedure. This is completely correct, provided (almost) all gauge orbits have Gribov copies on the Gribov horizon. Though not proven, this appears very likely.

Another proposal \cite{Maas:2009se} how to enforce this restriction explicitly stems from empiric observations, and insights gained from two-dimensional Yang-Mills theory in Coulomb gauge \cite{Reinhardt:2008ij}. These suggest that a possible method-independent characterization of the first Gribov region would be the requirement that the ghost propagator
\be
D_G^{ab}=\langle\bar{c}^a c^b\rangle=\langle M^{ab-1}\rangle\label{quant:ghp},
\ee
\no is negative-semi-definite in position space. Though at least the former statement is clearly a necessary condition, it has not yet been possible to show that this is a sufficient condition. Therefore, this has the status of a proposal which requires further investigations. However, for the methods employed here there are some method-dependent means how to restrict to the first Gribov region. This is described in section \ref{smethods}.

After restricting to the first Gribov region, the remainders of the residual gauge orbits still possess a large number of Gribov copies \cite{Dell'Antonio:1991xt,Semenov:1982,vanBaal:1997gu,Mehta:2009zv,Hughes:2012hg}. This set will also be denoted as the residual gauge orbit in the following, to avoid the term residual of the residual gauge orbit. In fact, in an infinite volume this number is likely infinite, and in a finite volume $V$ it appears to be a rapidly rising function of $V$ \cite{Mehta:2009zv,Mehta:2011sp,Hughes:2012hg}, possibly even proportional to $\exp V$ \cite{Maas:2009se}. 

\begin{figure}
\includegraphics[width=0.5\textwidth]{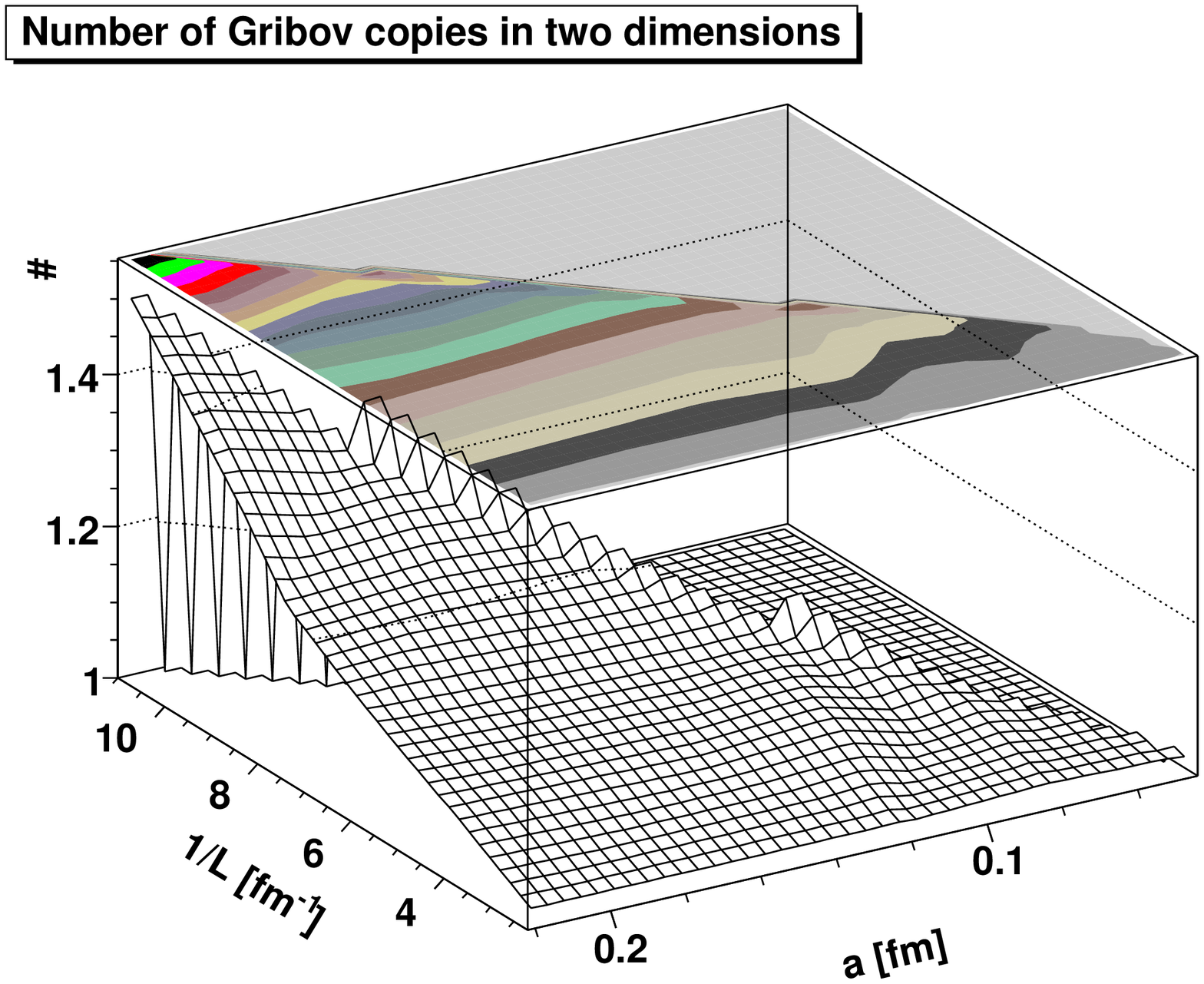}\includegraphics[width=0.5\textwidth]{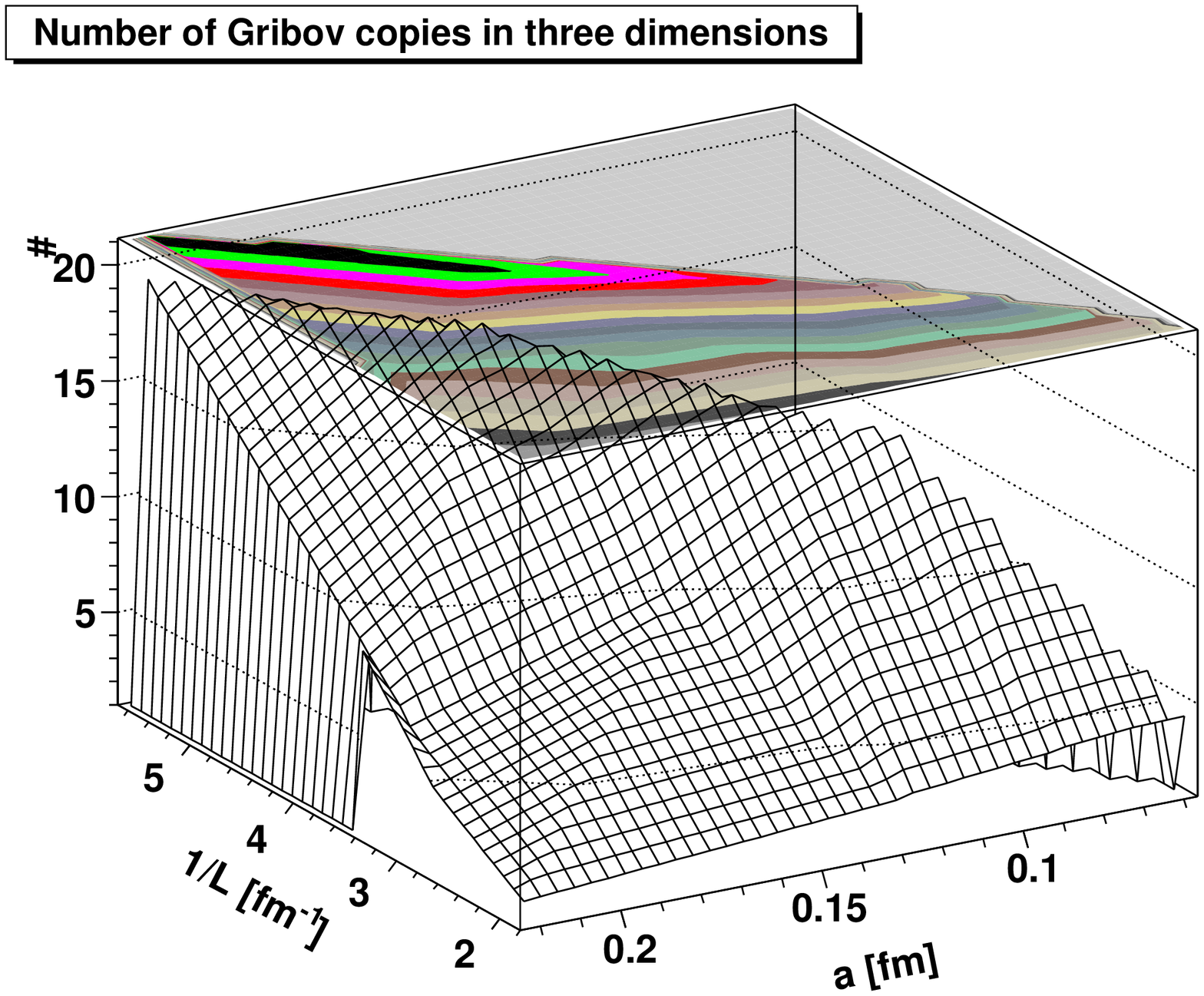}\\
\begin{minipage}[c]{0.5\linewidth}
\begin{center}\includegraphics[width=\textwidth]{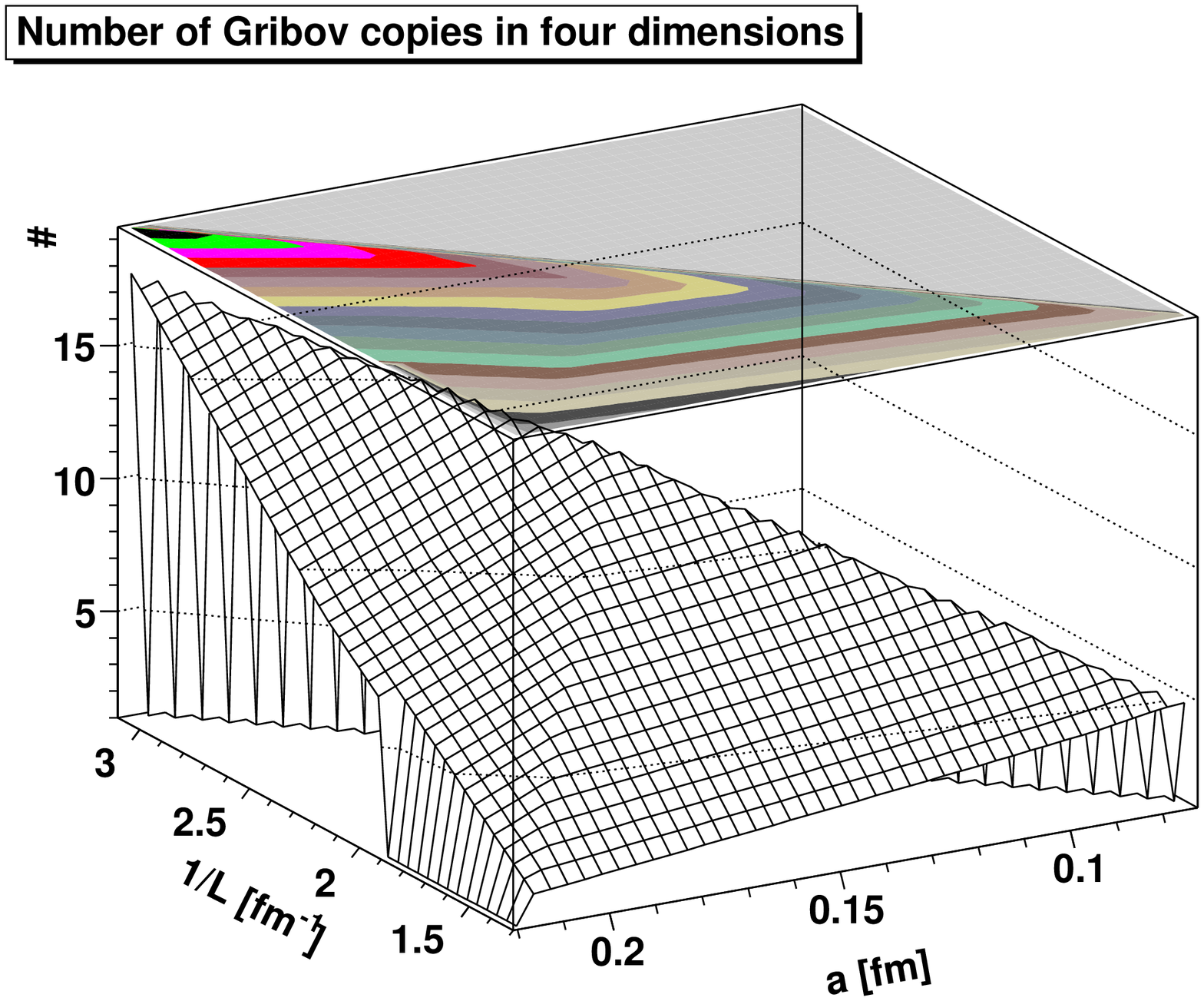}\end{center}
\end{minipage}
\begin{minipage}[c]{0.5\linewidth}
\caption{\label{fig:gc}Lower limit for the number of Gribov copies \cite{Maas:2009se,Maas:unpublished} as a function of volume $L^d$ and a lattice-regulator $1/a$ \cite{ZinnJustin:2002ru} for two (top-left panel), three (top-right panel), and four (bottom panel) dimensions.}
\end{minipage}
\end{figure}

Actually, counting Gribov copies is in practice a non-trivial problem \cite{Maas:2011ba}, since two Gribov copies are different if and only if they differ at least at one space-time point after factorizing all possible global gauge transformations and all space-time transformations. This implies that for the decision whether two representatives of a gauge orbit are identical or Gribov copies, it is required to compare their field values at every space-time point\footnote{It appears that Gribov copies differ from each other over some large domain \cite{Heinzl:2007cp}, so in practice already a coarse search can yield that two candidates are different. However, to ensure that they are the same requires a check of the whole space-time point by point.}. It is also in general non-trivial how to find all Gribov copies, so that they can be counted. Thus, except for very special systems \cite{Mehta:2009zv}, only a lower limit can be posed on the number of Gribov copies. Examples for these are shown for two, three, and four space-time dimension on a finite lattice\footnote{Details of the lattice method are given in section \ref{zerot:lat}. All units here and hereafter have been fixed by setting the string tension to (440 MeV)$^2$, see for details \cite{Cucchieri:2008qm,Maas:2007uv}.} in figure \ref{fig:gc}. It should be noted that, though two dimensions has trivial dynamics \cite{Dosch:1978jt}, gauge fixing has the same subtleties as in higher dimensions. Two-dimensional Yang-Mills theory is therefore an ideal laboratory to study these issues without the obscuring dynamics \cite{Maas:2007uv}. Remarkably, the number of Gribov copies not only increases with volume but in three and four dimensions also with increasing cutoff, while the latter seems not to be the case in two dimensions \cite{Maas:2011ba,Maas:unpublished}. This is one of many hints encountered in this work that two-dimensional Yang-Mills theory has indeed some distinct features.

Once more, it should be noted that one Gribov copy has no intrinsic difference compared to another Gribov copy, since they are physically equivalent. Thus any choice of a Gribov copy to represent the residual gauge orbit is equally acceptable. This is nicely illustrated by stochastic quantization, in which it is found that there is no stochastic force acting along a gauge orbit, and thus in the stochastic equilibration process no point on a gauge orbit is preferred \cite{Damgaard:1987rr,Baulieu:1981ec,Zwanziger:2003cf,Zwanziger:2001kw}. This, of course, is just in disguise the problems encountered when defining the path integral, which require to introduce a gauge condition in the first place.

\subsubsection{Minimal Landau gauge}

As stated, the residual gauge orbits inside the first Gribov horizon possess further Gribov copies. It is therefore necessary to constraint the choice of Gribov copies further. There are two strategies mainly in use currently for that purpose. Both were originally motivated by studies on a finite lattice \cite{Mandula:1987rh,Cucchieri:1995pn}.

The first method is essentially a stochastic approach. In this case, instead of specifying conditions for selecting a Gribov copy, a random Gribov copy is chosen for each residual gauge orbit \cite{Cucchieri:1995pn}. This prescription, termed the minimal Landau gauge, therefore averages over Gribov-copy-dependent properties when calculating correlation functions. Assuming the choice to be ergodic, unbiased, and well-behaved, this implies that this prescription is equivalent to averaging over the residual gauge orbit \cite{Maas:2011ba,Maas:unpublished}. However, a constructive prescription how to make this choice in a path integral formulation is only developing \cite{Maas:2011ba,Maas:2010wb,Maas:unpublished}. Precise definitions of this gauge therefore exist only as operational definitions in terms of algorithms in lattice gauge theory \cite{Cucchieri:1995pn}. The second approach attempts to characterize Gribov copies and make a choice based on these characteristics. Two possibilities for this characterization will be presented in the next two sections.

The central element of all operational definitions of the minimal Landau gauge is the fact that any Gribov copy in the first Gribov region maximizes the functional \cite{vanBaal:1997gu,Zwanziger:1993dh}
\bea
F[A]&=&1-\frac{1}{V}\int d^d x A_\mu^a A_\mu^a\label{quant:abslg}\\
\langle F[A]\rangle&=&1-\frac{N_g}{2^d\pi^{d/2}\Gamma\left(1+\frac{d}{2}\right)V}\int dp p^{d-1} D_{\mu\mu}^{aa}(p)\nn\\
D_\mn^{ab}&=&\langle A_\mu^a A_\nu^b\rangle\nn,
\eea
\no on each configuration, where $D_\mn^{ab}$ is again the gluon propagator. This implies that this gauge minimizes the integrated weight of the gluon propagator. That this is indeed satisfying the minimal Landau gauge follows from the fact that the first derivative of \pref{quant:abslg} is the Landau gauge condition, and the Hessian is the Faddeev-Popov operator \cite{Zwanziger:1993dh}. If any given algorithm finds one of all the maxima with equal probability, it would be a faithful representation of the distribution along the residual gauge orbit, and also be ergodic.

This gauge has a further important potential. It has been argued that all Gribov copy selection procedures in the infinite-volume and continuum limit, the so-called thermodynamic limit, should yield the same result \cite{Zwanziger:2003cf}, and thus therefore the one of the minimal Landau gauge. The basic idea behind this argument is essentially of thermodynamic nature: The configuration space is in this limit infinite-dimensional, thus almost all Gribov copies of almost all residual gauge orbits should lie at the boundary of the first Gribov region. The important assumption is then that all Gribov copies at the boundary have similar correlation functions. Then the same argument as in thermodynamics holds, and the equilibrium behavior should emerge. Since the minimal Landau gauge is already averaging, this equilibrium behavior should coincide with the one found in the minimal Landau gauge. However, this argument is only expected to hold for any finite polynomial in the gauge fields. E.\ g., the ghost propagator \pref{quant:ghp} cannot be expressed as a finite series in polynomials of the fields, and therefore the argument does not need to hold for it.

\begin{figure}
\includegraphics[width=0.5\textwidth]{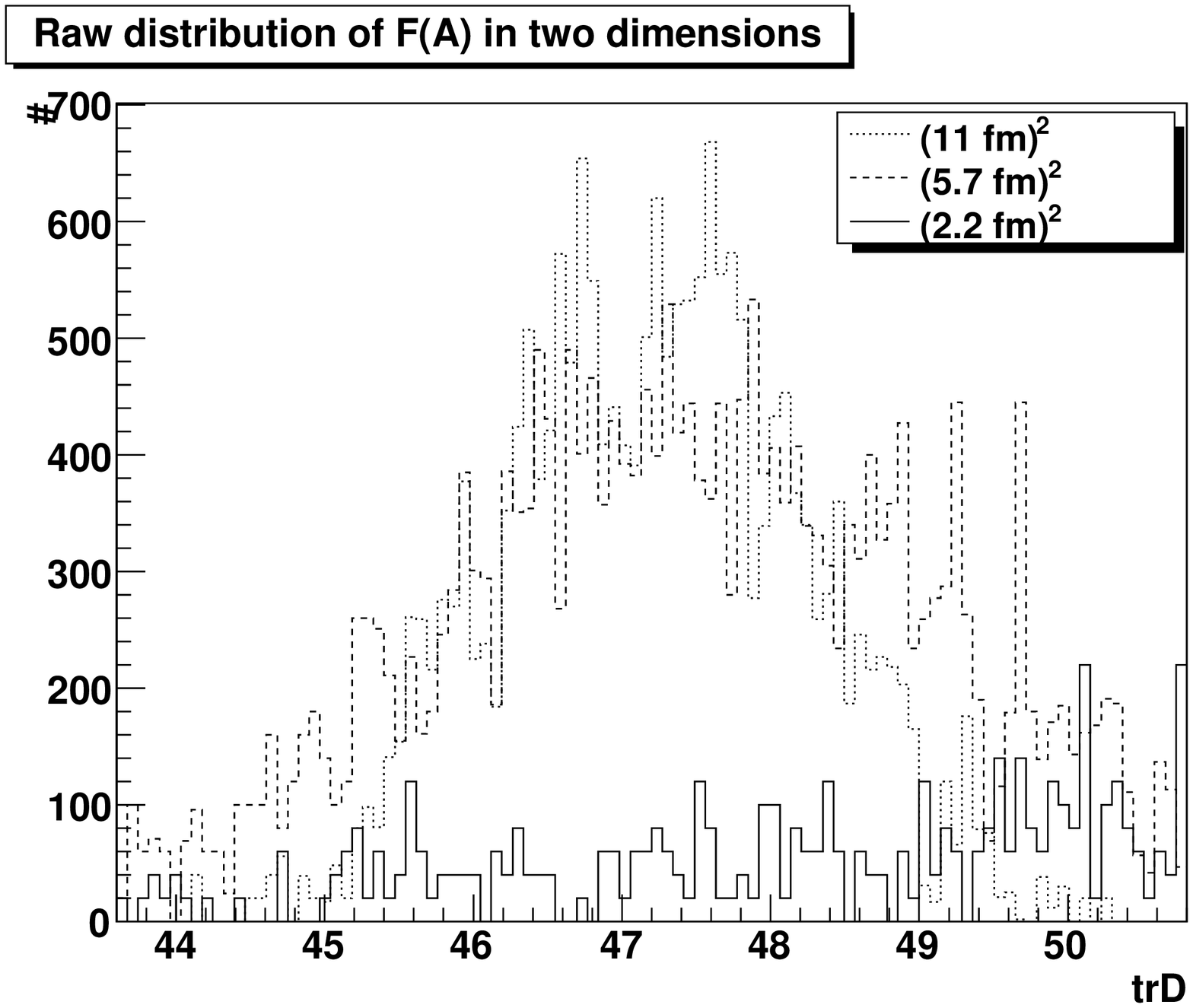}\includegraphics[width=0.5\textwidth]{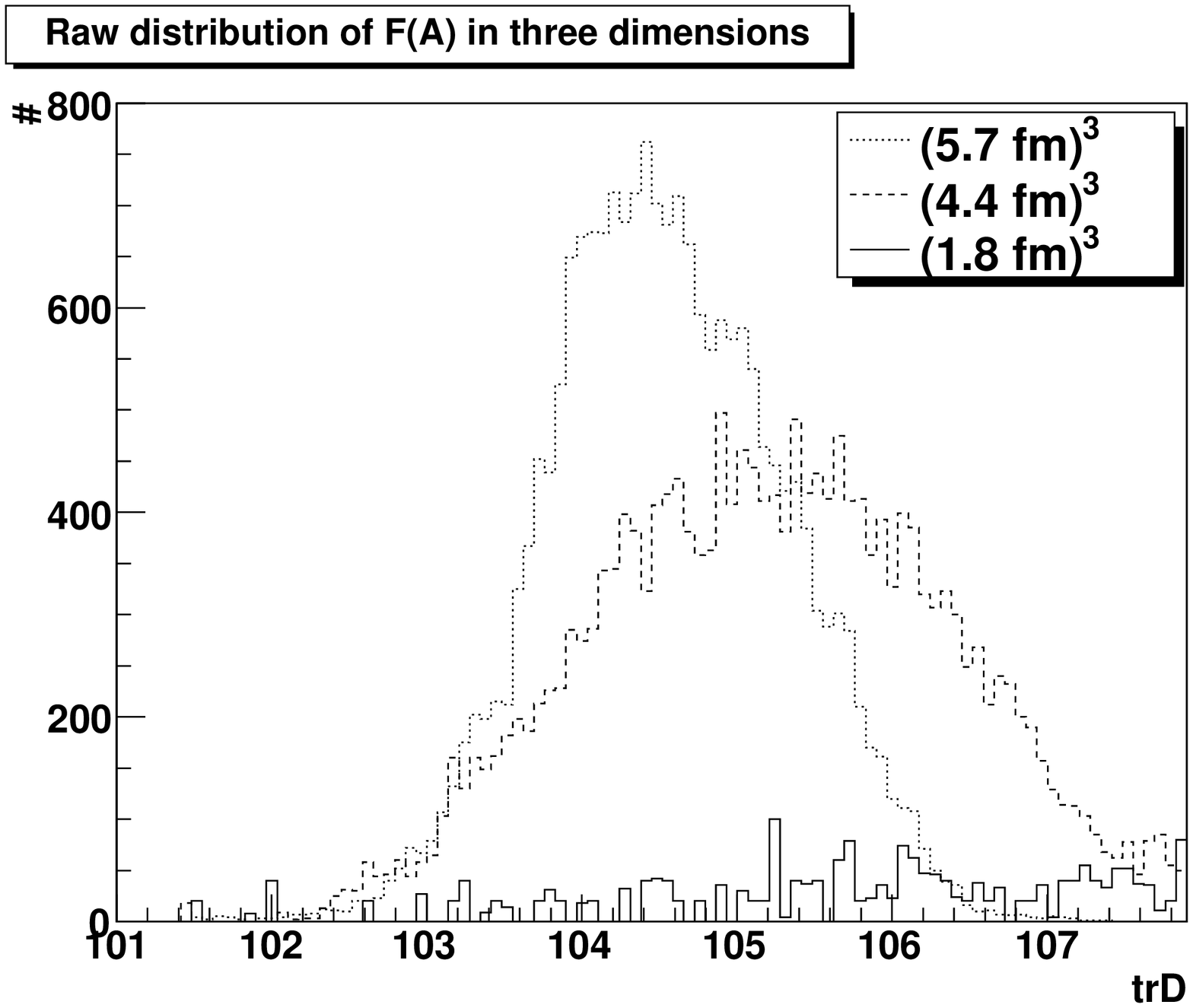}\\
\begin{minipage}[c]{0.5\linewidth}
\begin{center}\includegraphics[width=\textwidth]{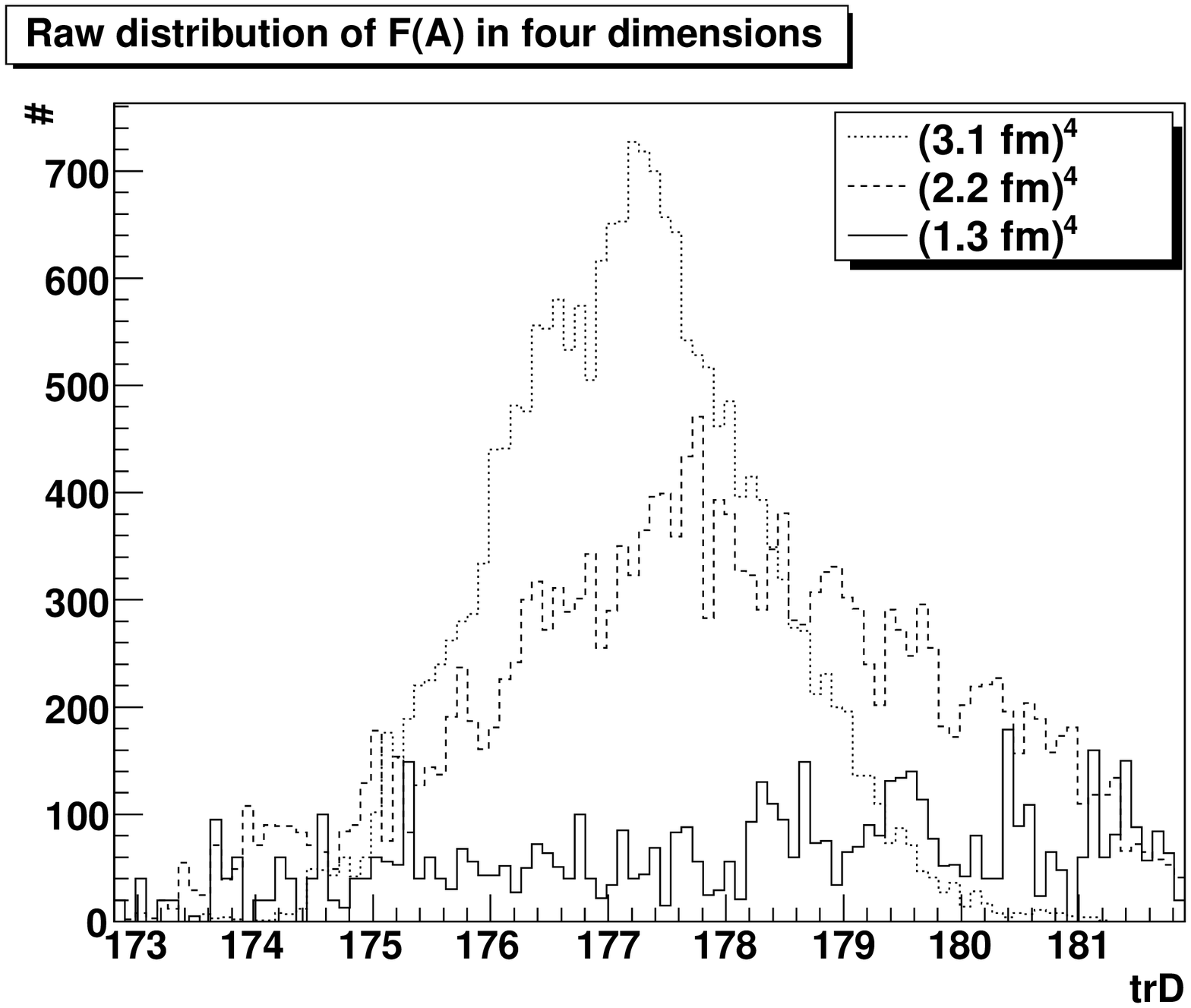}\end{center}
\end{minipage}
\begin{minipage}[c]{0.5\linewidth}
\caption{\label{fig:raw-d}The distribution of \pref{quant:abslg} for a fixed discretization of $a=0.22$ fm, after subtracting one and rescaling by a factor of -1000 \cite{Maas:unpublished}. Results are shown for two (top-left panel), three (top-right panel), and four dimensions (bottom panel). Here and hereafter always 1000 configurations with 20 checked Gribov copies each have been used.}
\end{minipage}
\end{figure}

\begin{figure}
\begin{minipage}[c]{0.5\linewidth}
\begin{center}
\includegraphics[width=\textwidth]{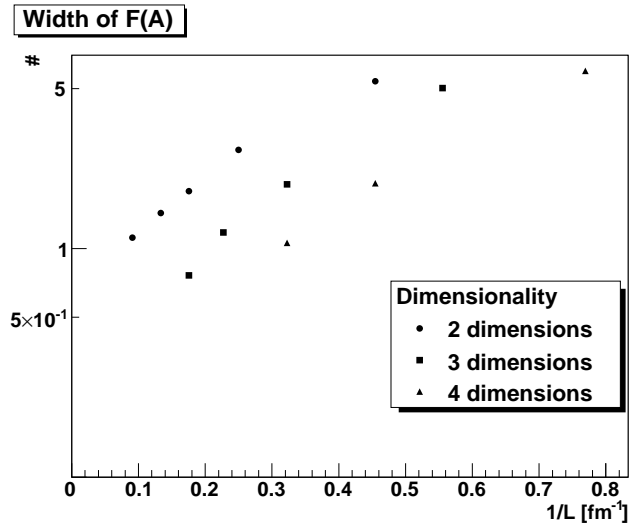}
\end{center}
\end{minipage}
\begin{minipage}[c]{0.5\linewidth}
\caption{\label{fig:width-d}Lower limit of the width of the distribution of \pref{quant:abslg} for a fixed discretization of $a=0.22$ fm on a lattice \cite{Maas:unpublished}, after subtracting one and rescaling by a factor of -1000.}
\end{minipage}
\end{figure}

Take as an example the expression \pref{quant:abslg} itself, which is a finite polynomial in the fields. Its distribution over the residual gauge orbit is shown in figure \ref{fig:raw-d}. It is clearly visible how the distribution of \pref{quant:abslg} moves towards a more peaked distribution, and is actually well described by a Gaussian, with a volume-dependent width, shown in figure \ref{fig:width-d}. Thus, it appears plausible that this quantity, and thus likely also the gluon propagator, will have the same value, irrespective of the representative chosen along the residual gauge orbit.

However, this statement does not take renormalization into account \cite{Maas:2011ba}. In particular, \pref{quant:abslg} has to be regularized, to be meaningful in the continuum limit. It is not a-priori clear how this quantum effect affects the distribution, as it involves the regularization of a composite, non-local operator. From U(1) gauge theory it is known that this may cause significant problems \cite{deForcrand:1994mz}.

\begin{figure}
\includegraphics[width=0.5\textwidth]{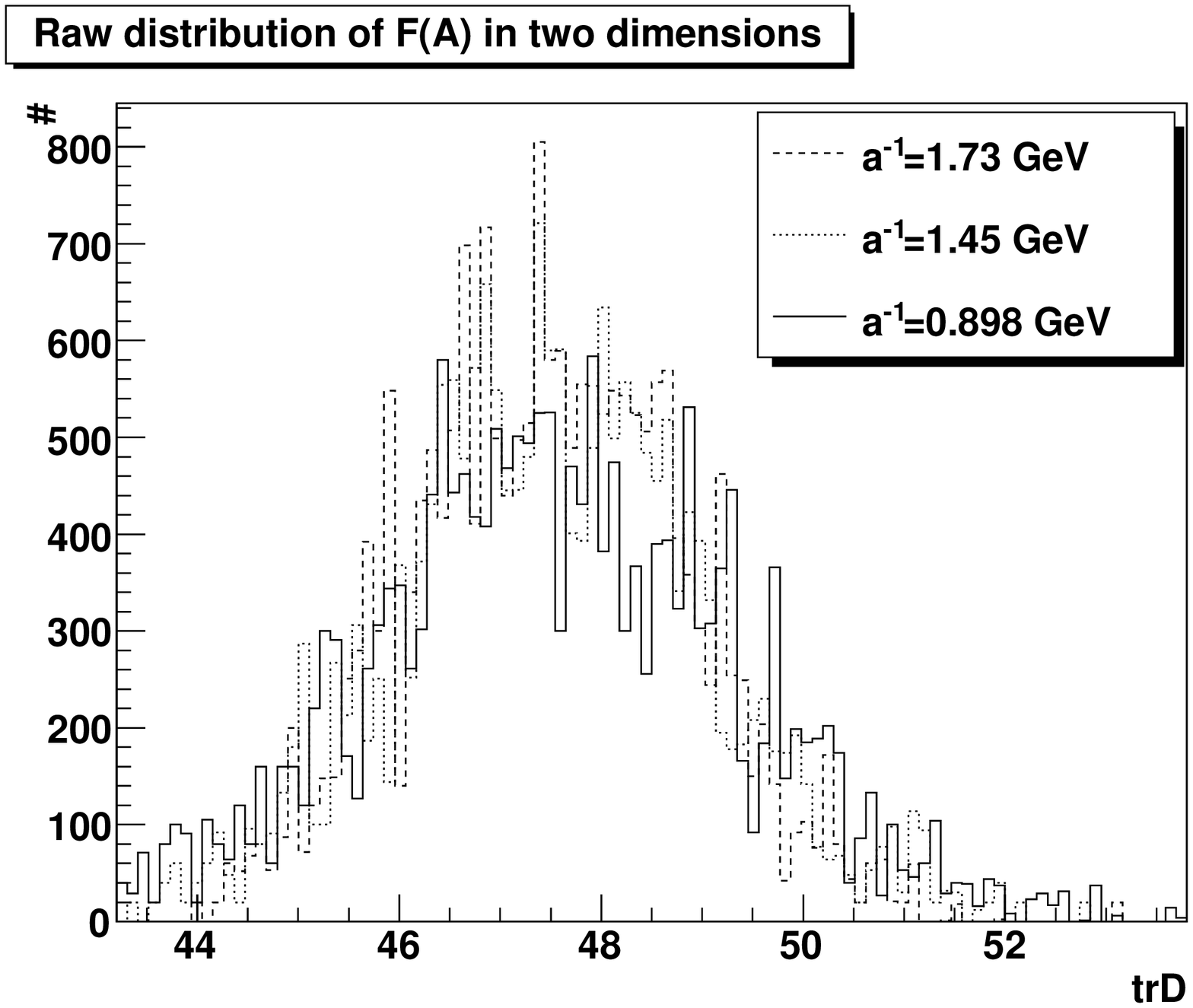}\includegraphics[width=0.5\textwidth]{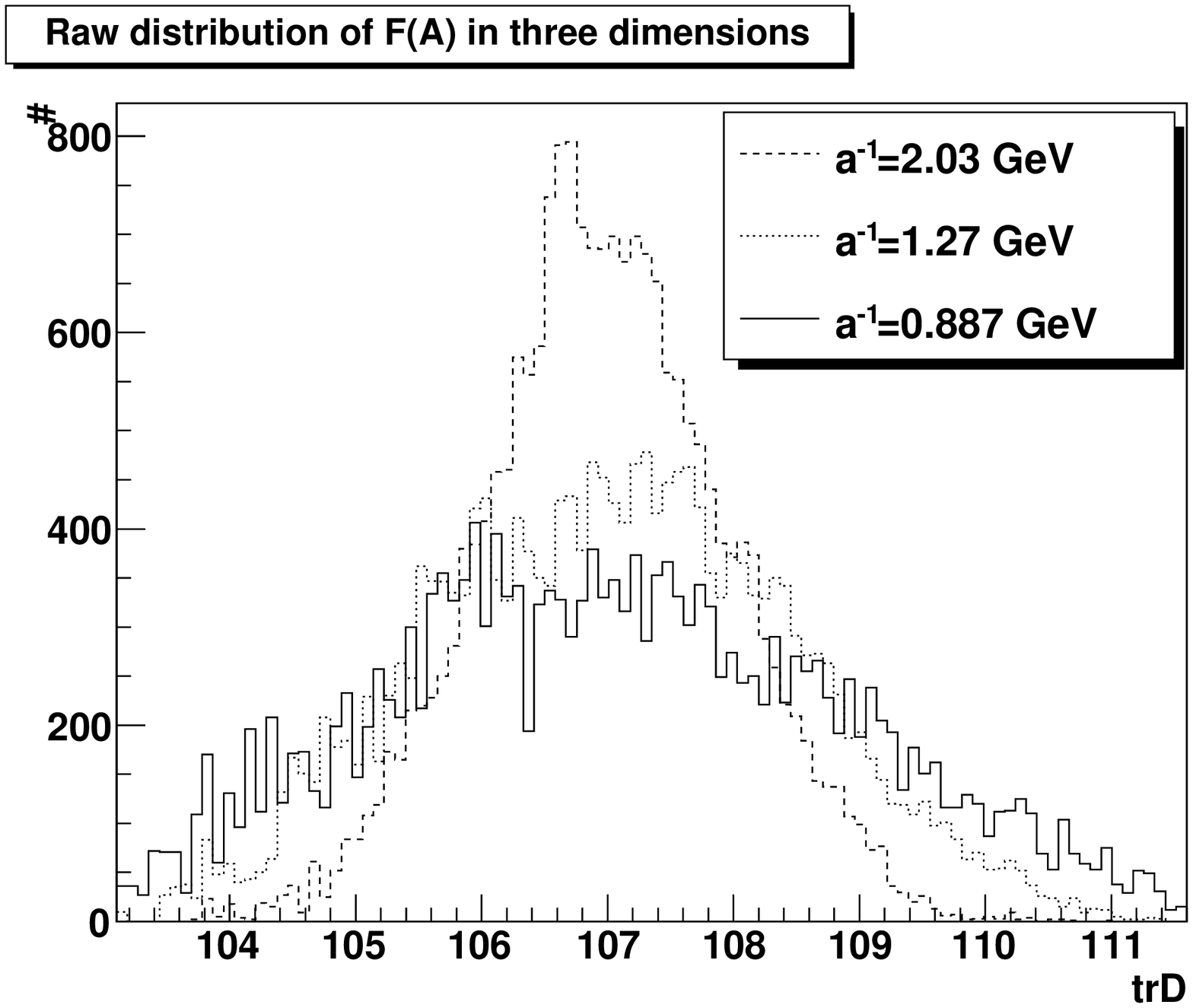}\\
\begin{minipage}[c]{0.5\linewidth}
\begin{center}\includegraphics[width=\textwidth]{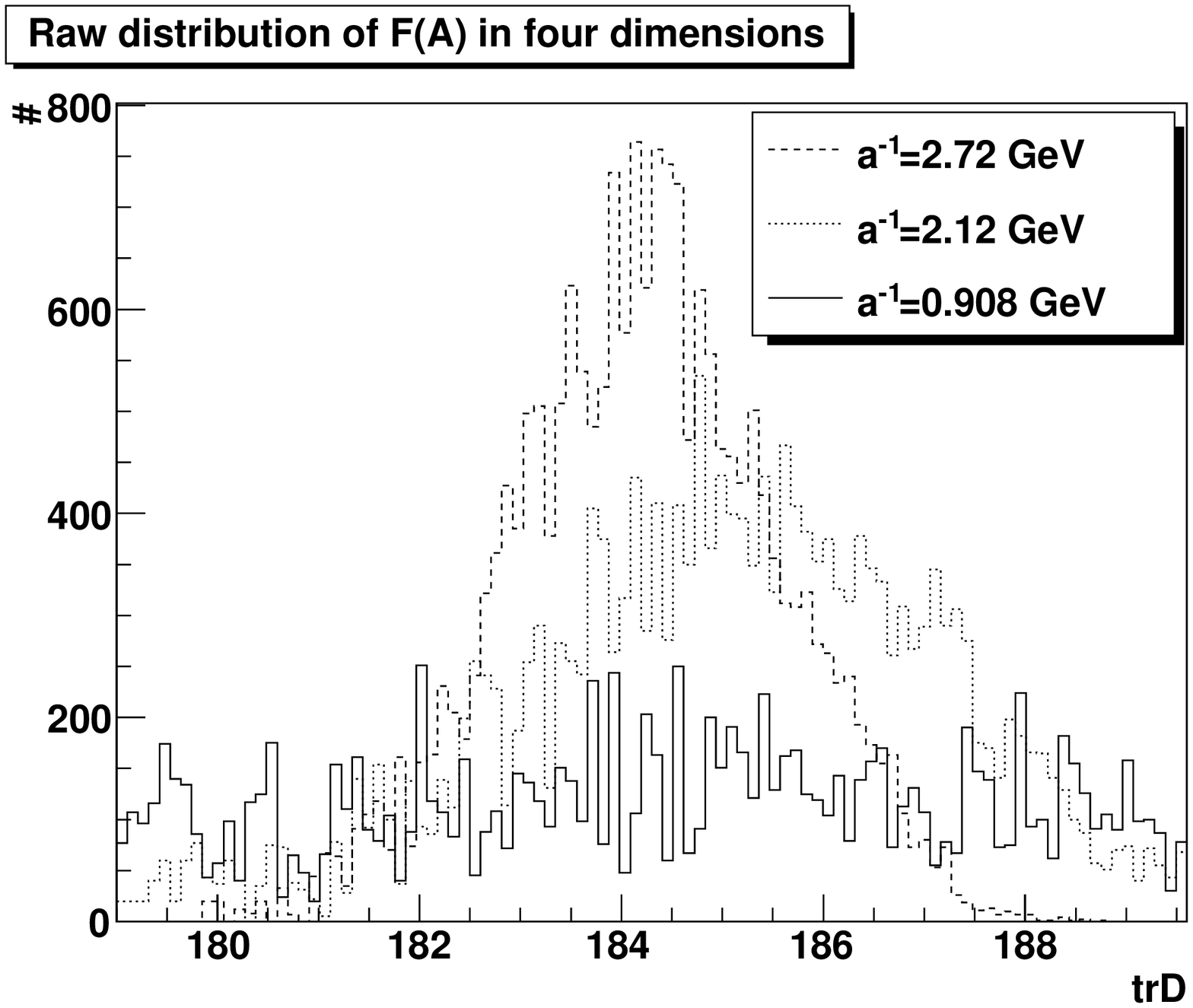}\end{center}
\end{minipage}
\begin{minipage}[c]{0.5\linewidth}
\caption{\label{fig:raw-d-ren}The change of \pref{quant:abslg} when the cut-off is increased, and only a multiplicative renormalization is performed \cite{Maas:unpublished}. The top-left panel shows a volume of (5.7 fm)$^2$ in two dimensions, the top-right panel a volume of (3.1 fm)$^3$ in three dimensions, and the bottom panel a volume of (1.3 fm)$^4$ in four dimensions. A one is subtracted before renormalization.}
\end{minipage}
\end{figure}

The consequences of changing the cut-off and performing only a multiplicative renormalization\footnote{Note that volume effects can also play a role here, and thus volumes as large as possible should be used.} are shown in figure \ref{fig:raw-d-ren}. While in two dimensions, where almost no Gribov copies are present, there is little change, the distribution appreciably changes in higher dimensions. Thus, regularization of this composite, non-local operator plays an important role in understanding the thermodynamic limit, and how stringent a finite polynomial of the fields has to take the same value irrespective of the choice of Gribov copy. It is therefore better to to investigate instead the gluon propagator, and see whether it takes a unique value in the thermodynamic limit. This will be discussed at length below in section \ref{zerot:prop}.

\subsubsection{Absolute Landau gauge}\label{sec:abslg}

\begin{figure}
\begin{center}
\includegraphics[width=0.6\textwidth]{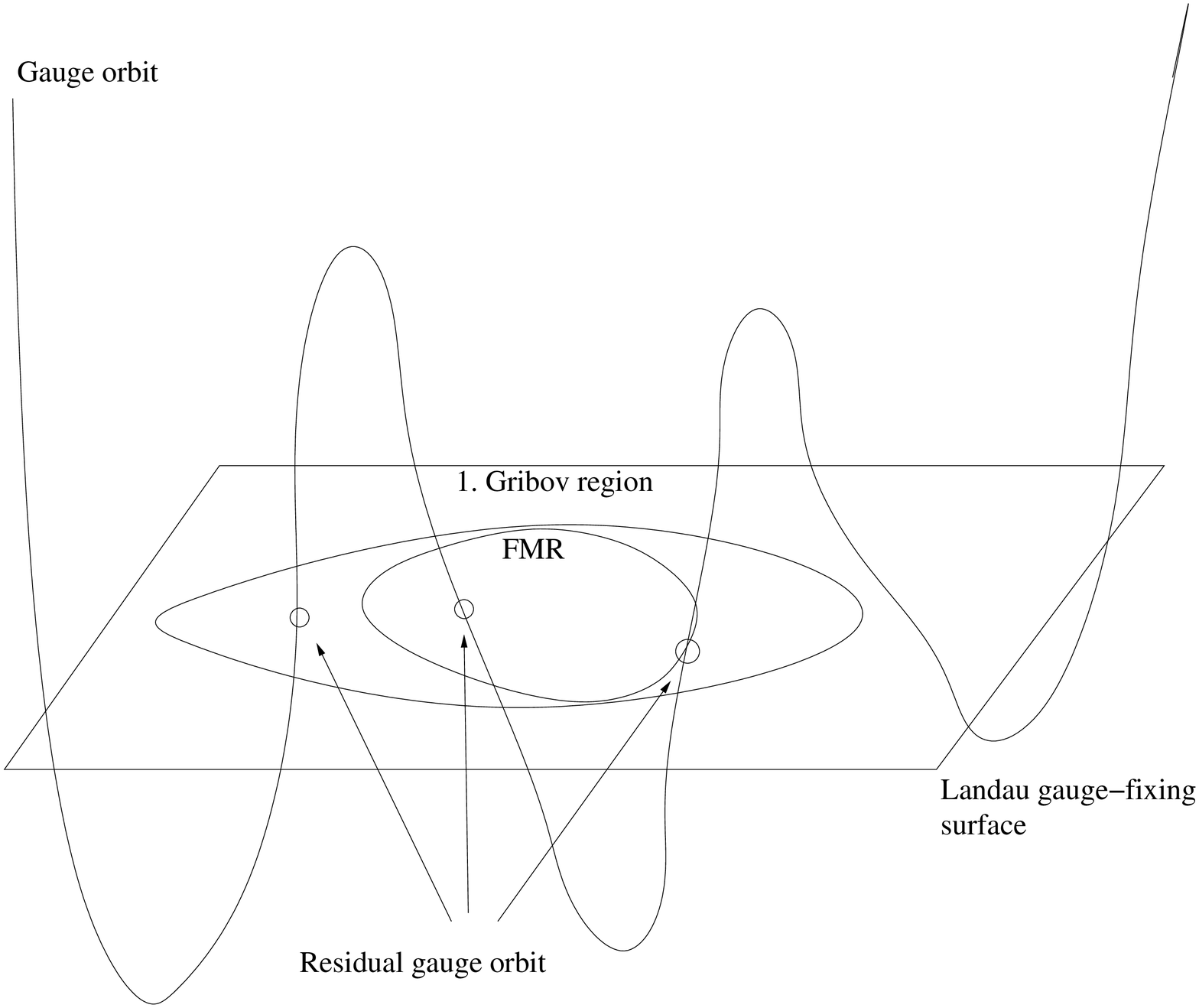}
\end{center}
\caption{\label{fig:sketch}A sketch of the geometry of field configuration space. A gauge orbit is shown, which intersects the gauge-fixing hypersurface, defined by \pref{quant:lg}, six times in the visible part. Three of these Gribov copies are inside the first Gribov region. These three form the residual gauge orbit, after the restriction to the first Gribov region has been performed. As is visible, they are not connected by infinitesimal transformations. One of the Gribov copies is (necessarily) inside the fundamental modular region (FMR), and in this case a second Gribov copy is located very close to the boundary of the fundamental modular region, which itself touches the Gribov horizon.}
\end{figure}

An alternative way to choose a representative on the residual gauge orbit is the absolute Landau gauge \cite{Zwanziger:1993dh,vanBaal:1997gu,Maas:2008ri}, which makes a very definite choice rather than a random choice. This gauge choice is derived from the following observation. The functional \pref{quant:abslg} has, up to topological identifications \cite{vanBaal:1997gu}, a unique absolute maximum\footnote{It should be noted that in some theories the relative and absolute maxima become arbitrarily close in the thermodynamic limit \cite{deForcrand:1994mz}.}. The resulting set of absolute maxima, called the fundamental modular domain or region \cite{Zwanziger:1993dh}, in analogy to conformal field theory, is by definition embedded in the first Gribov region, and includes the origin. It is less trivial to show that it is convex and bounded, and thus connected \cite{Zwanziger:2003cf}. It can furthermore be shown that part of the boundary of the fundamental modular domain coincides with the Gribov horizon in the thermodynamic limit only \cite{Zwanziger:1993dh,Zwanziger:2003cf}. All possibly remaining degenerate absolute minima are on the boundary. This boundary has actually a quite rough structure \cite{Greensite:2004ur}, including wedge singularities, and topological configurations, like e.\ g.\ instantons \cite{Maas:2005qt}, are located there. By construction, in this region the gluon propagator has its least integrated weight. A sketch of the situation is shown in figure \ref{fig:sketch}.

Based on this observation, the absolute Landau gauge is defined as selecting the Gribov copy which belongs to the fundamental modular domain \cite{Cucchieri:1997dx,Cucchieri:1997ns}. This condition can be realized by either checking the absolute minimization of \pref{quant:abslg} explicitly or by the introduction of a suitable weight factor in the path integral \cite{Zwanziger:1991gz}. In case the residual gauge orbit has more than one Gribov copy on the boundary of the fundamental modular domain, again a random choice is made \cite{Maas:2008ri}. It should be noted that if the thermodynamic arguments made before were correct, the absolute Landau gauge and the minimal Landau gauge will coincide in the thermodynamic limit, at least for correlation functions being finite polynomials of the gauge fields.

It should be noted once more that the relevant quantity \pref{quant:abslg} has to be renormalized. As has been illustrated by figure \ref{fig:raw-d-ren}, this is likely not just a multiplicative renormalization. However, an additive renormalization, which is in principle Gribov-copy-selection-dependent, can, in principle, make a local maximum a global maximum, and this is not yet fully understood. In addition, finding the global minimum along the gauge orbit is in general a hard problem, since this requires to know all Gribov copies inside the first Gribov region. Therefore, actually reaching the absolute Landau gauge is in practice very complicated, see chapter \ref{smethods}.

\subsubsection{Landau-$B$ gauges}\label{squant:dse}

A further proposal are the so-called Landau-$B$ gauges. They are motivated by two observations.

One is that all information on a theory is contained in its correlation functions \cite{Rivers:1987hi}. As a consequence, if two gauges differ, at least one correlation function has to differ at at least one momentum configuration\footnote{To the extent that the correlation functions have to be analytic functions, this implies that they then differ for almost all momenta.}. E.\ g., the perturbative Landau gauge condition \pref{quant:lg} can be cast into a condition on the gluon propagator as \cite{Cucchieri:2008zx,Maas:2009se}
\be
p_\mu p_\nu D_\mn^{ab}=\langle p_\mu A_\mu^a(-p) p_\nu A_\nu^b(p)\rangle=0\nn,
\ee
\no for all momenta equally, and thus locally, and for each configuration individually. Thus the gauge condition in Fourier space already implies the vanishing of the longitudinal part of the gluon propagator. Similar conditions can also be posed for covariant gauges in general \cite{Cucchieri:2008zx}. It appears therefore possible that the non-local condition to select a particular representative of the residual gauge orbit should also be specifiable by a condition on one or more correlation functions. Since the conditions need only to be non-local in position space, they can be local in momentum space.

The second observation comes from the solutions of the Dyson-Schwinger equations (DSEs), as will be discussed below. There a family of solutions is found \cite{Boucaud:2008ky,Fischer:2008uz,RodriguezQuintero:2010wy,Boucaud:2008ji,Kondo:2009wk}, which are parametrized by the value of the ghost dressing function at zero momentum \cite{Fischer:2008uz}, or, equivalently, by the infrared value of an effective ghost-gluon coupling \cite{Boucaud:2008ky,Boucaud:2008ji,RodriguezQuintero:2010wy}. It has been speculated that this should also be possible at a non-zero momentum instead \cite{Maas:2009se,Fischer:2008uz}, but this possibility has not been followed yet. Therefore, here only the zero momentum case will be discussed. Of course, in a finite volume, and due to the trivial zero-modes of the Faddeev-Popov operator \cite{Suman:1995zg}, the ghost propagator cannot be evaluated at zero momentum. Therefore, in this context always the momentum closest to zero will be used, and it is understood that the infinite-volume limit has to be taken eventually.

The idea that this quantity could be of interest is supported by the fact that in almost all cases a differing value of \pref{quant:abslg} also leads to a differing value of the ghost propagator at zero momentum\footnote{For simplicity, in the following always zero momentum is used, with the understanding to replace it by the lowest momentum accessible.} \cite{Maas:2008ri,Maas:2009se,Cucchieri:1997dx,Sternbeck:2005tk,Bogolubsky:2005wf,Bakeev:2003rr}. However, without knowledge of all Gribov copies, this cannot be made an exact statement at the current time.

This observation can now be used to construct a set of gauges, the Landau-$B$ gauges \cite{Maas:2009se}. The following construction has been found to work on small finite volumes \cite{Maas:2009se,Maas:unpublished,Maas:2010nc}, and also at strong coupling \cite{Maas:2009ph}. The situation in the thermodynamic limit will be discussed below. The basic ingredient of the Landau-$B$ gauges is a second, strictly positive, non-perturbative gauge parameter $B$, which can be selected within a  certain range to be determined below.

To implement the gauge, an auxiliary parameter $b$ is needed, which is defined as
\be
b=\lim_{p^2\to 0}\frac{G(p^2,\mu^2)}{G(\mu^2,\mu^2)}\label{quant:lbg}.
\ee
\no By definition, this quantity is renormalization-group invariant. In two and three dimensions, $\mu$ can actually be chosen to be infinite \cite{Maas:2009se}, since the correlation functions do not diverge, and then $G(\infty,\infty)=1$, see section \ref{ssuv}. The parameter $b$ is taken to be defined Gribov copy-wise, i.\ e., on each individual Gribov copy\footnote{In a finite volume, the actual momentum is not zero. Therefore, the fact that a Gribov copy is not rotationally and translationally symmetric has to be taken into account in practical calculations, see \cite{Maas:2009se,Maas:2010nc,Maas:unpublished}.}. Thus, strictly speaking, $b$ is, as $F(A)$, a functional of the configuration, and should also be written as $b(A)$ and the appearing ghost dressing functions as $G(A,p^2,\mu^2)$. This will be suppressed throughout.

The gauge is now defined \cite{Maas:2009se} by the condition that on each residual gauge orbit the Gribov copy is selected as a representative which has a value of $b$, which is closest to the gauge parameter $B$. If there exist more than one Gribov copy satisfying this condition, once more a degeneracy is encountered, and in this case, as before, a random choice should be made. The element of randomness has to be such that it selects in equal measure copies which have a positive or negative distance to $B$. This gauge construction implies that $<b>=B$, and thus determines the averaged ghost dressing function at zero momentum. If $B$ should be selected such that this cannot be realized, e.\ g.\ because $B$ is larger than $b$ for all copies and configurations, the resulting ghost propagator is as close as possible to $B$. In this sense, it is possible to define a minB and a maxB gauge \cite{Maas:2009se}, by always selecting the Gribov copy with the smallest and largest value of $b$ for each orbit, respectively.

It should be possible to construct a corresponding weight for the path integral, which implements this condition. But neither the existence nor the form of such a construction has yet been proven, there exists only proposals for it \cite{Maas:unpublished,Maas:2010wb}.

Given the path integral in Landau gauge restricted to the first Gribov horizon \pref{quant:fgr}, one proposal to do so is by modifying the path integral to
\bea
<{\cal O}>&=&\lim_{\eta\to 0}\lim_{\xi\to 0}\int{\cal D}A_\mu{\cal D}c{\cal D}\bar{c} {\cal O}(A_\mu,c,\bar{c})\theta\left(-\pdm D_\mu^{ab}\right)e^{-\int d^4x {\La}_g}\nn\\
&&\times\exp\left(-\frac{1}{\eta}\left(\left(\frac{1}{V}\int d^dxd^dy\pdm^x\bar{c}^a(x)\pdm^yc^a(y)\right)-Z_B\gamma\right)^2\right)\label{quant:bgauge}.
\eea
\no Here, the representation of the ghost dressing function at zero momentum as
\bea
G^{aa}(0)&=&\lim_{p\to 0}p^2 D_G^{aa}(p)=\lim_{p\to 0}p^2 \langle\bar{c}^a(-p) c^a(p)\rangle\nn\\
&=&\lim_{p\to 0}\frac{1}{V}\int d^dxd^dy p^2e^{ip(x-y)}\langle\bar{c}^a(y)c^a(x)\rangle\nn\\
&=&\lim_{p\to 0}\frac{1}{V}\int d^dxd^dy\pdm^x\pdm^ye^{ip(x-y)}\langle\bar{c}^a(y)c^a(x)\rangle\nn\\
&=&\lim_{p\to 0}\frac{1}{V}\int d^dxd^dye^{ip(x-y)}\langle\pdm^y\bar{c}^a(y)\pdm^xc^a(x)\rangle=\left\langle\frac{1}{V}\int d^dxd^dy\pdm^y\bar{c}^a(y)\pdm^xc^a(x)\right\rangle\nn
\eea
\no has been used, and $Z_B\gamma=B$ guarantees the correct renormalization. The volume factor $V$ remains due to the translational invariance of the zero mode. The Gaussian weight thus inserted in \pref{quant:bgauge} then enforces that in the limit the ghost dressing function has to reach the pre-determined value, satisfying \pref{quant:lbg}, for each Gribov copy. Note that this additional term explicitly breaks perturbative BRST symmetry, and is non-local. Furthermore, the order of the limits is important. The statement of this inclusion is, however, merely that the ghost dressing function should become $B$ at zero momentum, which can therefore be considered a boundary condition. This permits to implement this constraint also without explicitly taking into account this non-local term in the path integral in self-consistent calculations, as will be done in section \ref{zerot:dse}. Of course, in lattice gauge theory just the appropriate Gribov copy will be selected. It should be noted that it is likely possible, analogously to conventional covariant gauges \cite{Bohm:2001yx}, to remove the limit in $\eta$ in \pref{quant:bgauge}. This yields then a gauge averaging over non-perturbative Gribov copies, with a Gaussian weight centered at $Z_B\gamma$.

However, the expression \pref{quant:bgauge} is only appropriate if, up to a measure zero set, all configurations have at least one Gribov copy which can satisfy the condition\footnote{I am grateful to D.\ Zwanziger for a discussion on this topic.}. Otherwise gauge-invariant quantities would be affected. If a $B$ value is chosen for which this is not the case, the maxB or minB gauges are explicit examples of such a case, the expression \pref{quant:bgauge} cannot be correct. Instead, at best it can only be imposed that the condition $<b>=B$ can be satisfied, but not $b=B$. This can be implemented in the path-integral by using a Lagrange parameter\footnote{Which on a finite lattice will additionally depend on the volume and the discretization.} $\lambda(B)$, like temperature, as a gauge parameter, which gives \pref{quant:bgauge} the form \cite{Maas:2010wb,Maas:2011ba,Maas:unpublished}
\bea
<{\cal O}>&=&\lim_{\xi\to 0}\int{\cal D}A_\mu{\cal D}c{\cal D}\bar{c} {\cal O}(A_\mu,c,\bar{c})\theta\left(-\pdm D_\mu^{ab}\right)e^{-\int d^4x {\La}_g}\nn\\
&&\times\exp\left(N(B)+\lambda(B)\frac{1}{V}\int d^dxd^dy\pdm^x\bar{c}^a(x)\pdm^yc^a(y)\right)\label{quant:bgauge2},
\eea
\no where $N$ is a, possibly orbit-dependent, normalization, such that gauge-invariant quantities are not affected by the averaging. If the normalization is orbit-independent, it can be absorbed in the path integral measure. Otherwise the same statements about \pref{quant:bgauge2} hold true as for \pref{quant:bgauge}. In general, the Lagrange parameter $\lambda$ may depend on global properties of the theory, in particular the value of the zero-momentum ghost propagator for $\lambda=0$. Indeed, for at least this one value $\lambda=0$ this gauge condition is definitely valid, as it is just implies to average over the residual gauge orbit. In addition, it is trivially possible to define
\be
\lambda=Z_B^{-1} B\label{zblambda},
\ee 
\no and give the Lagrange parameter the meaning of an unrenormalized gauge parameter. Note that there are still some values which cannot be imposed, like negative values of $B$, much like negative temperatures (usually) cannot be employed. It thus still remains to identify the permitted set of $B$ values. The minimal Landau gauge appears in such a setup at $\lambda=0$, since it is equivalent to averaging over all Gribov copies with a flat weight. This implies that due to \pref{zblambda} the minimal Landau gauge is a fixed-point of the renormalization group with respect to both the perturbative gauge parameter of the covariant gauge and this non-perturbative gauge parameter.

At the present time there is neither proof, nor rock-solid evidence that this eliminates the Gribov-Singer ambiguity in the same way as the absolute Landau gauge, i.\ e., by uniquely identifying a Gribov copy for at least some value of $B$, which would imply for this/these value(s) of $B$ the admissibility of the construction \pref{quant:bgauge}. However, when averaging over the residual gauge orbit using the prescription \pref{quant:bgauge2} the Gribov-Singer ambiguity is lifted by construction: At least in the same form as the minimal Landau gauge, it provides a prescription how to treat the residual gauge orbit. This follows, since if indeed there exists a set of residual gauge orbits with non-zero measure where there are Gribov copies degenerate in this condition, then, as in minimal Landau gauge, just random choices will be made. The gauge-fixing conditions then returns to the same operative level of gauge definition as for the minimal Landau gauge. Of course, in such a case the non-perturbative gauge-dependence between different Landau-$B$ gauges will be trivially just a lower limit to the variability which would be obtained when resolving any further degeneracies, like in the absolute Landau gauge. Thus, all three gauges, minimal Landau gauge, absolute Landau gauge, and Landau-$B$ gauges, are complete resolutions of the Gribov-Singer ambiguity: In all cases it is fully specified how to treat Gribov copies.

After specifying this gauge, it is now an important question what the actual distribution of $b$ is. A particular interesting question is, whether there are holes in the distribution of $b$, whether it has a finite width in the thermodynamic limit, and so on. Results from functional continuum calculation \cite{Fischer:2008uz} suggest that the possible range is $[B_0,\infty)$, with some lower limit $B_0$ not precisely determined yet, and including the limit $B\to\infty$. The latter would be required to be implemented by another limit in \pref{quant:bgauge} to be taken after the limit $\eta\to 0$, and also in \pref{quant:bgauge2}. However, even in the opposite extreme case that the interval should shrink to a single value, the construction is not wrong, since the ghost dressing function has to take some value at zero momentum, being it finite or infinite. In this case, all what happens, is that the Landau-$B$ gauges are reduced to the minimal Landau gauge.

\begin{figure}
\includegraphics[width=0.5\textwidth]{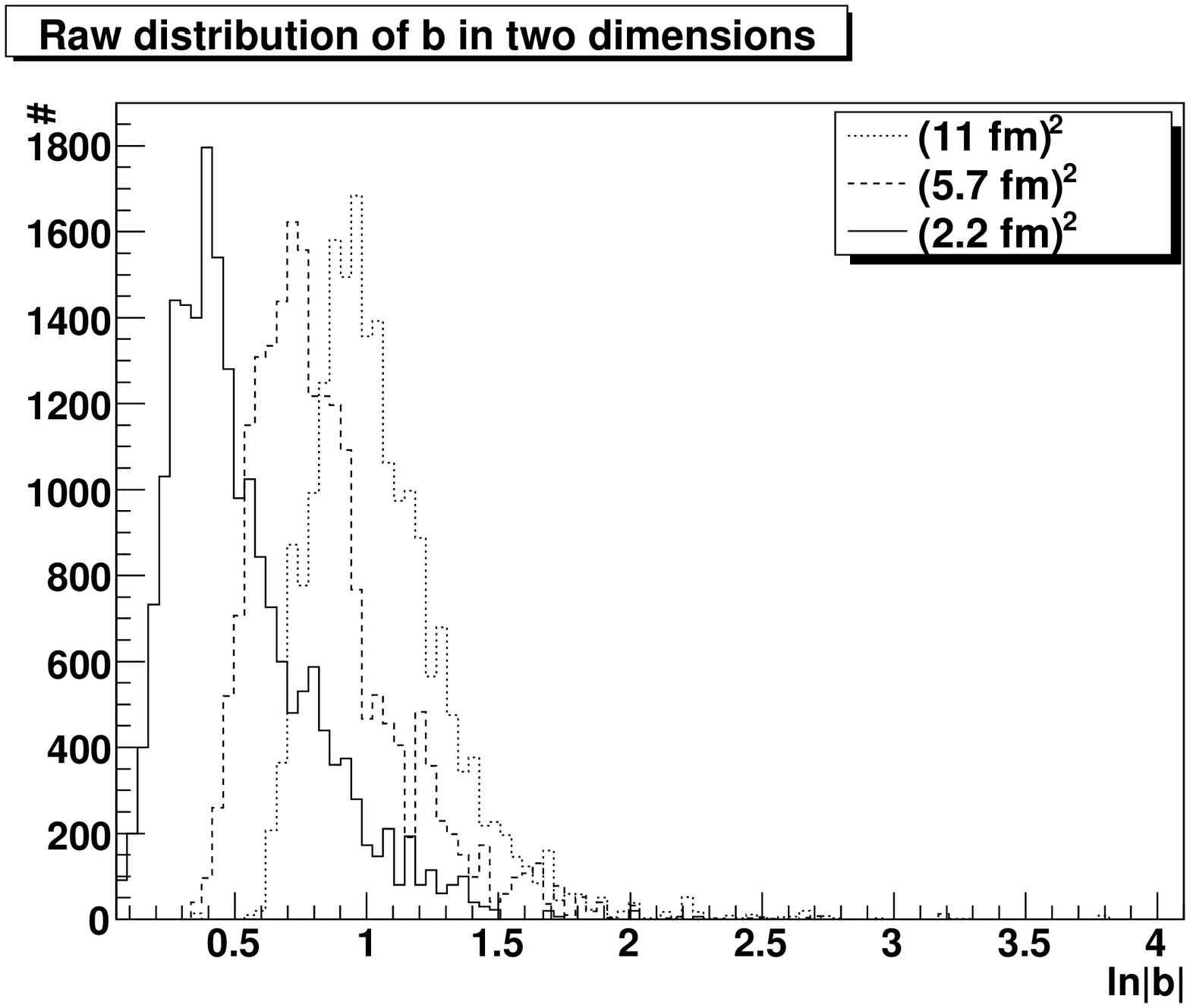}\includegraphics[width=0.5\textwidth]{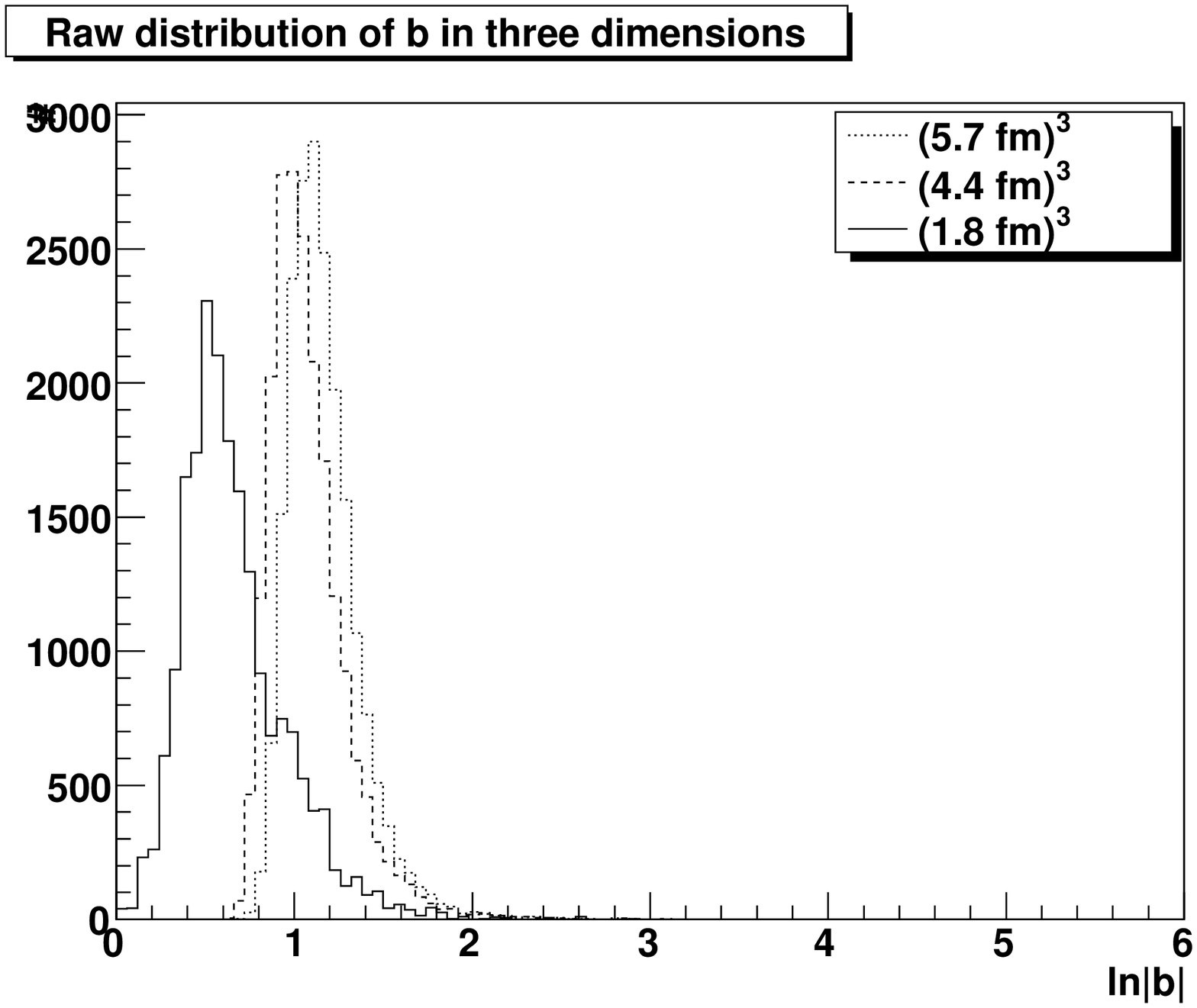}\\
\begin{minipage}[c]{0.5\linewidth}
\begin{center}\includegraphics[width=\textwidth]{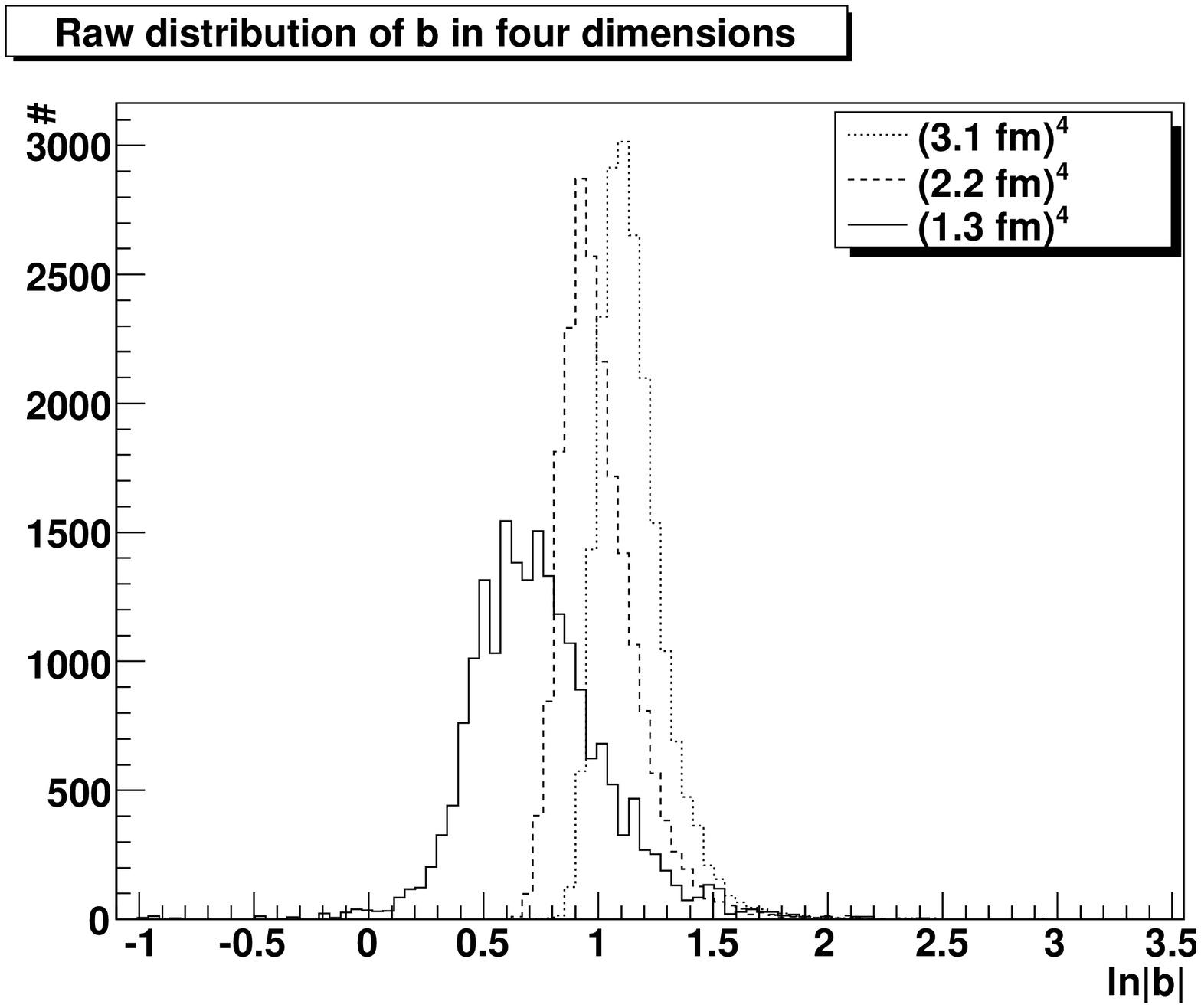}\end{center}
\end{minipage}
\begin{minipage}[c]{0.5\linewidth}
\caption{\label{fig:raw-b}The distribution of $b$ for a fixed discretization of $a=0.22$ fm \cite{Maas:unpublished}. Results are shown for two (top-left panel), three (top-right panel), and four dimensions (bottom panel). The absolute value is necessary due to some peculiarities of the method in determining the $b$ value in the algorithms employed here \cite{Cucchieri:2006tf,Boucaud:2005gg}, which are negligible when determining lattice averages, and in particular for the thermodynamic limit \cite{Maas:unpublished,Maas:2010nc}.}
\end{minipage}
\end{figure}

Thus, the permitted range of $B$ is primarily of practical importance. The distribution of $b$ as a function of volume is shown in figure \ref{fig:raw-b}. First of all, the distribution is very strongly asymmetric, with a long tail towards large values, and thus quite different from the Gaussian one found for $F(A)$. It is furthermore strongly volume-dependent. However, as can be seen from the result for two dimensions where on small volumes almost no Gribov copies exist, this is to quite some extent a 'trivial' finite volume effect, which is also seen in the minimal Landau gauge. These just stem from the fact that in a finite volume the lowest momentum is non-zero, but decreasing with volume.

\begin{figure}
\includegraphics[width=0.5\textwidth]{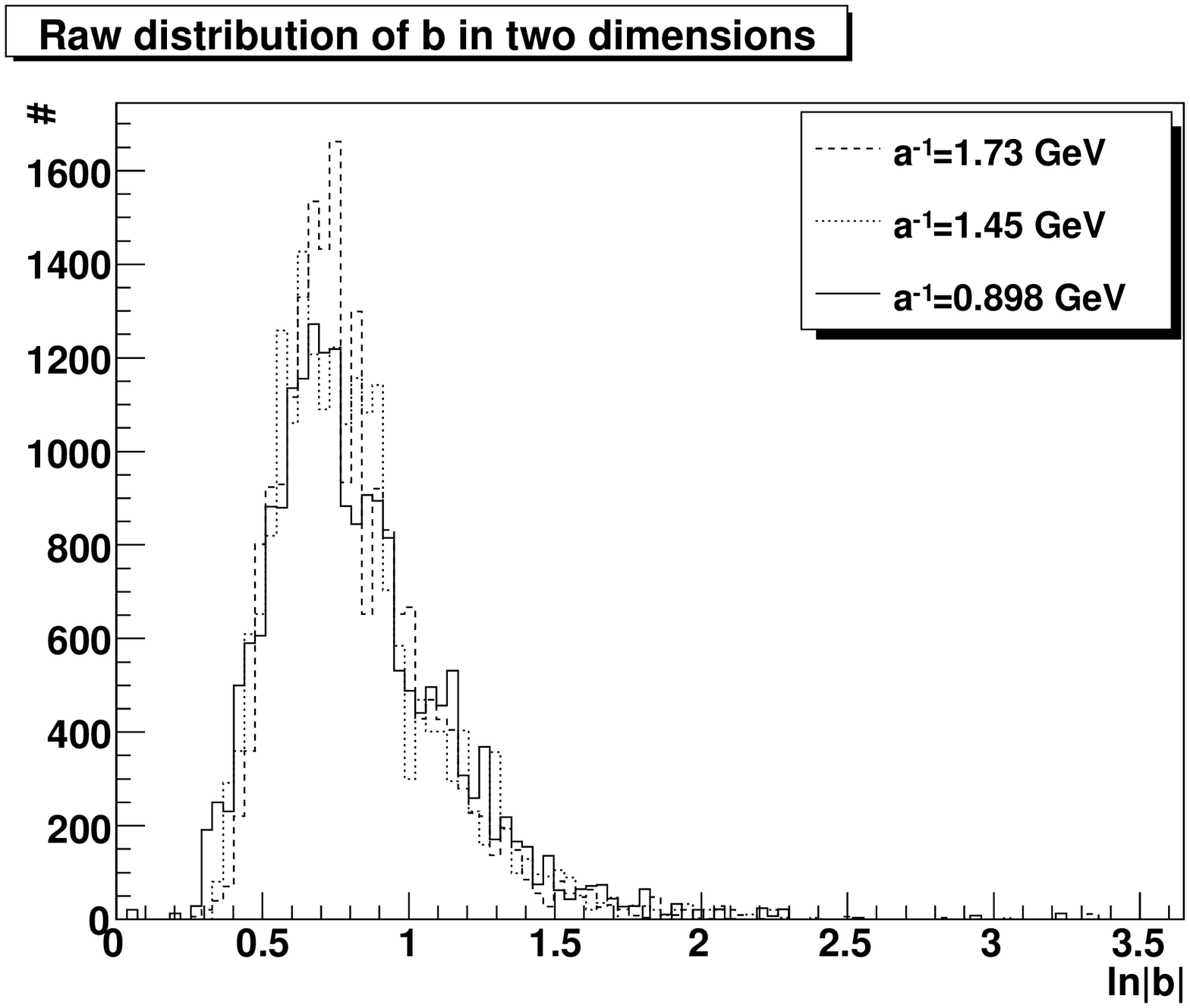}\includegraphics[width=0.5\textwidth]{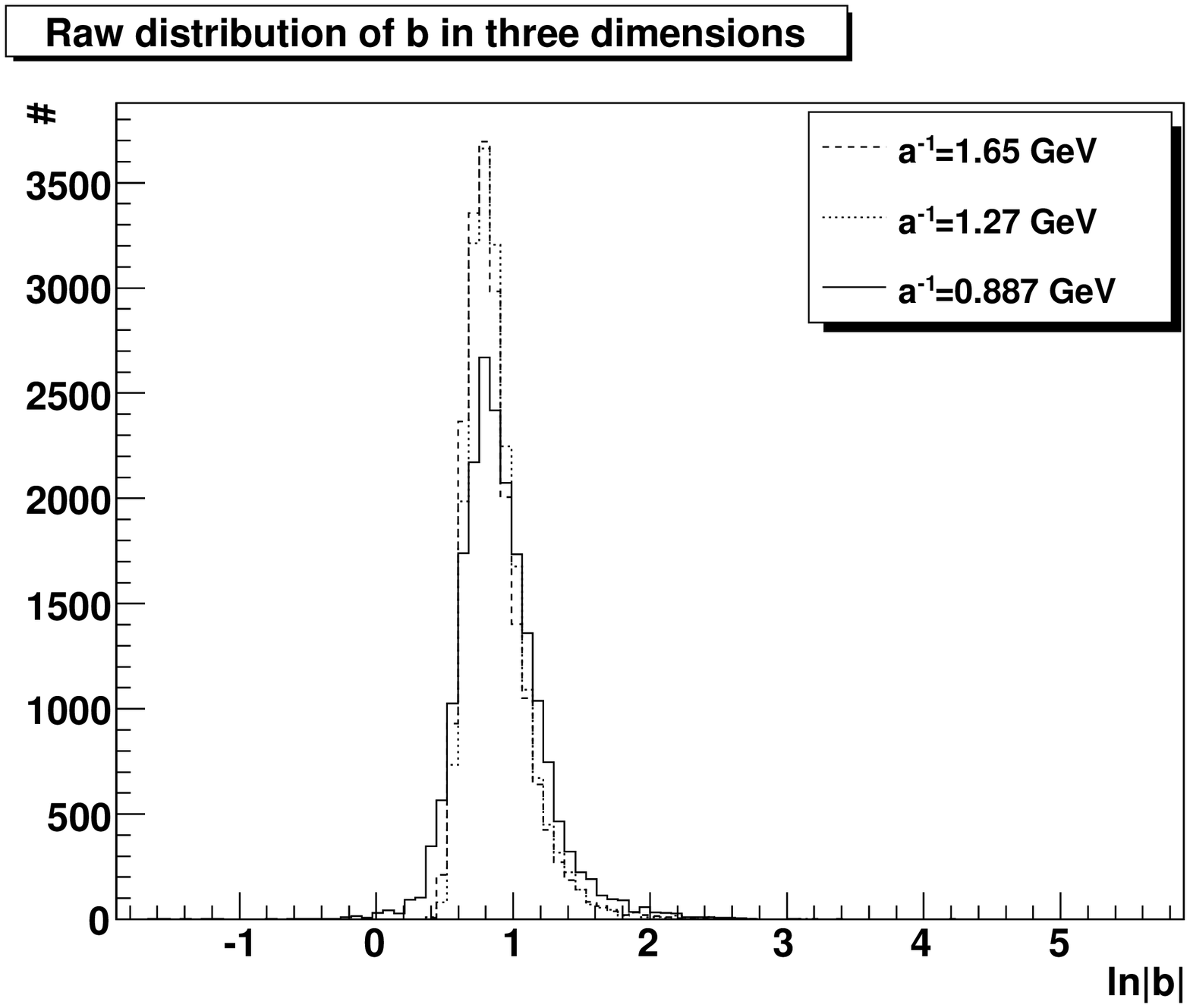}\\
\begin{minipage}[c]{0.5\linewidth}
\begin{center}\includegraphics[width=\textwidth]{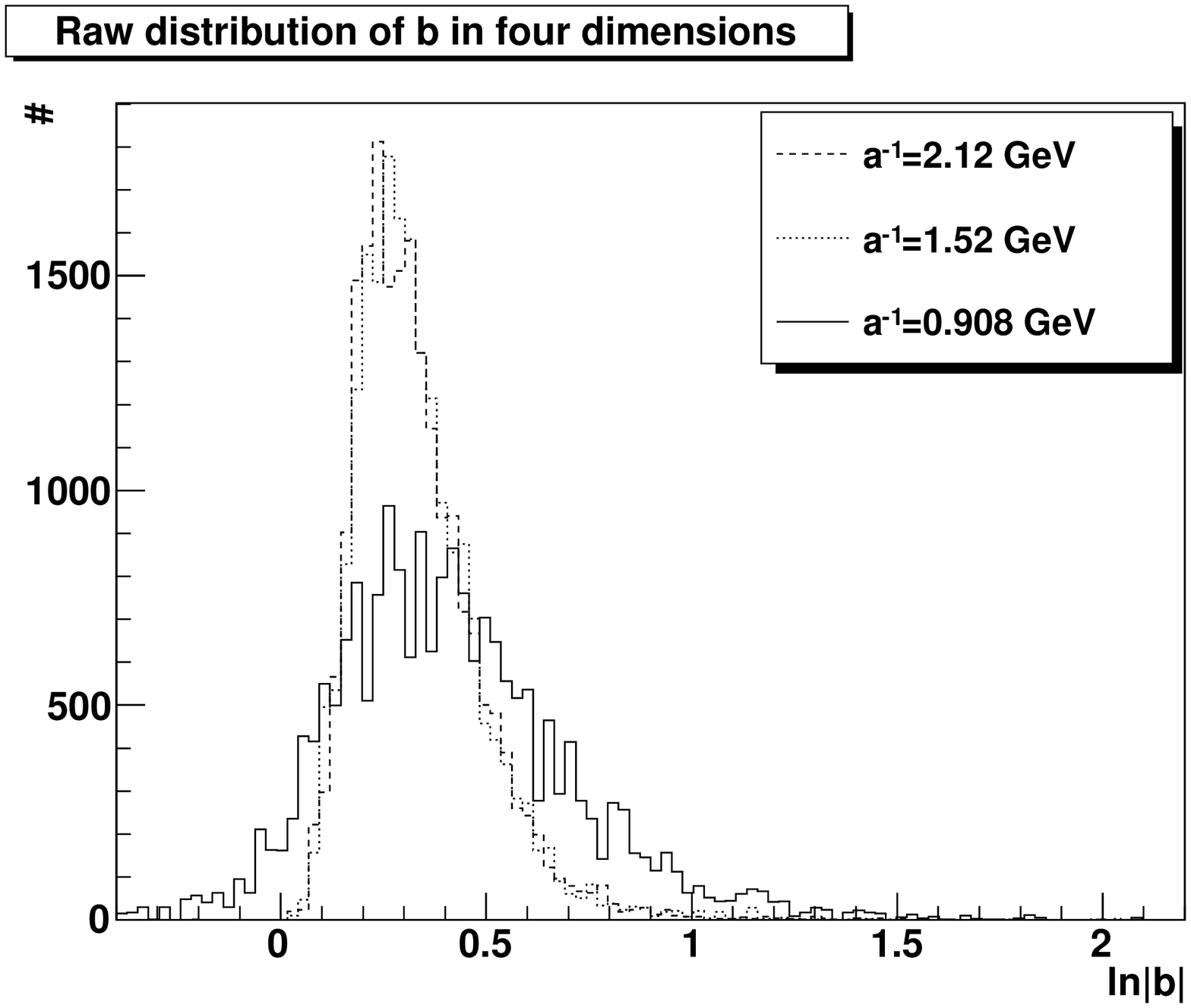}\end{center}
\end{minipage}
\begin{minipage}[c]{0.5\linewidth}
\caption{\label{fig:raw-b-ren}The change of \pref{quant:abslg} when the cut-off is increased for $\mu=\infty$ in two and three dimensions, and $\mu=2$ GeV in four dimensions. The top-left panel shows a volume of (5.7 fm)$^2$ in two dimensions, the top-right panel a volume of (3.1 fm)$^3$ in three dimensions, and the bottom panel a volume of (1.3 fm)$^4$ in four dimensions \cite{Maas:unpublished}.}
\end{minipage}
\end{figure}

On the other hand, the renormalization properties shown in figure \ref{fig:raw-b-ren} are much better than for $F(A)$. If the discretization errors are not too large, i.\ e., the number of lattice points is sufficiently large, then there are essentially no cut-off effects visible for the distribution. This is to be expected from a renormalization-group invariant quantity, which $b$ is. From available results, it appears that a cut-off larger than 1 GeV is at least necessary for reaching the appropriate behavior, and at least a lattice size of $20^d$.

\begin{figure}
\includegraphics[width=0.5\textwidth]{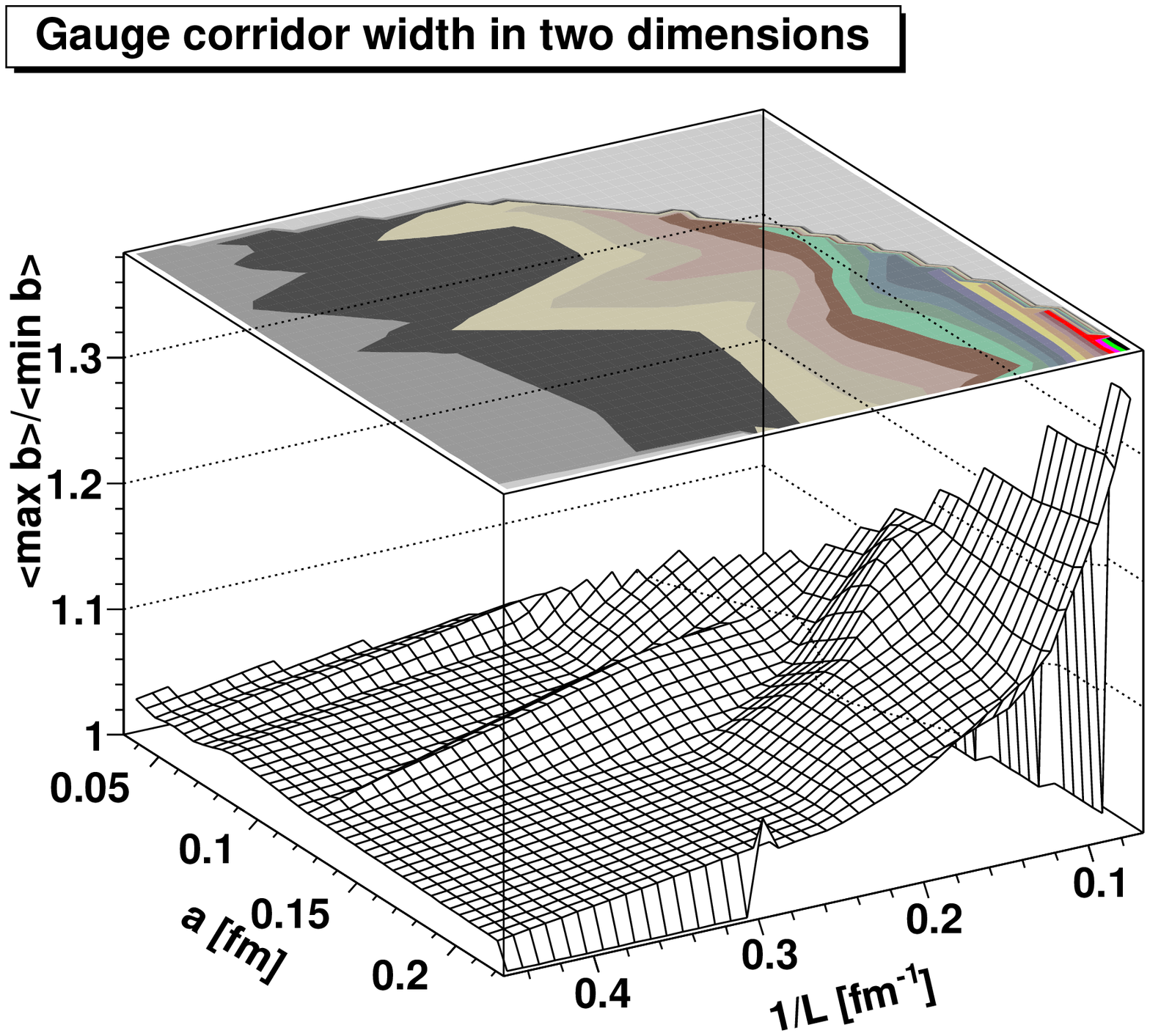}\includegraphics[width=0.5\textwidth]{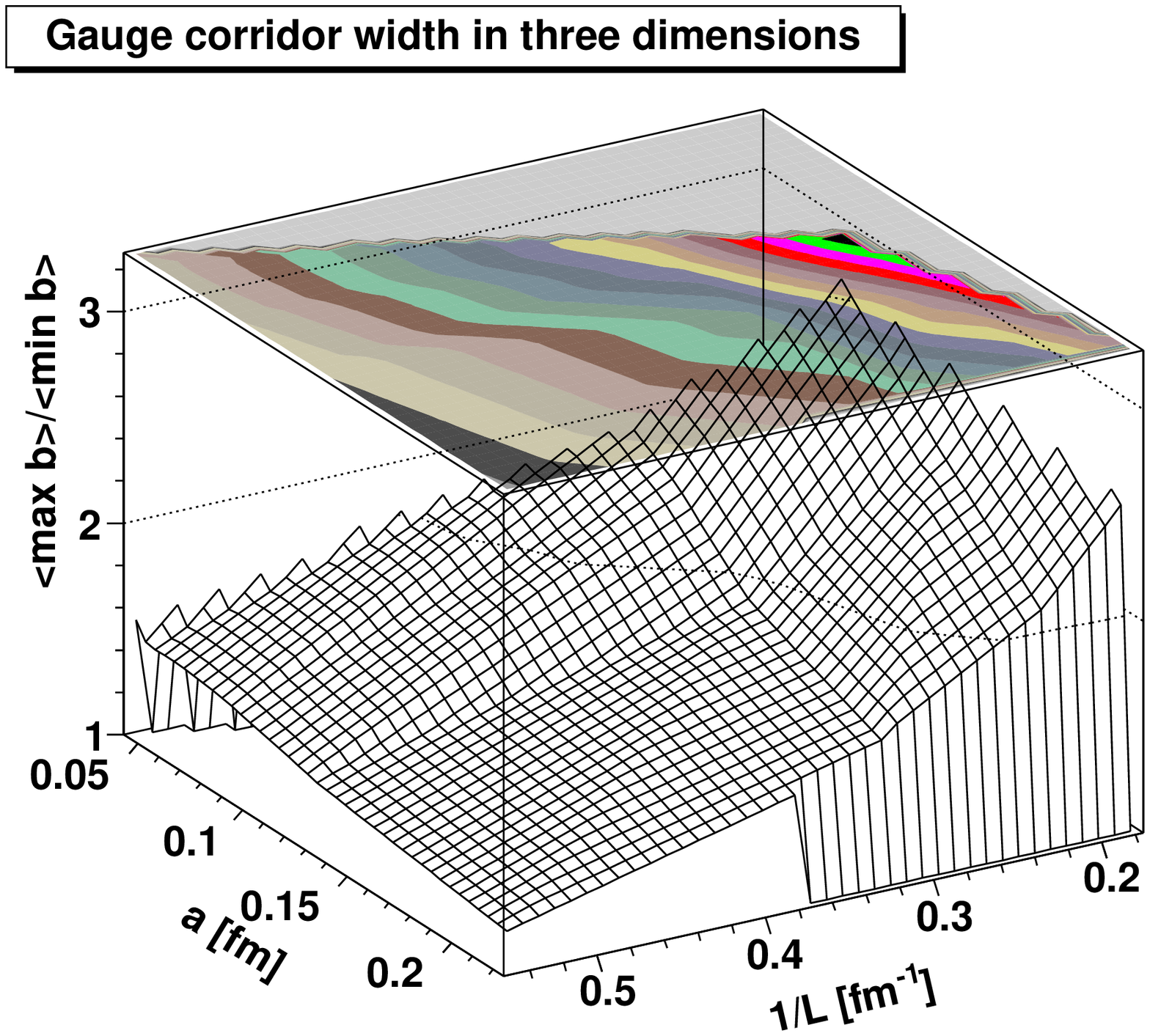}\\
\begin{minipage}[c]{0.5\linewidth}
\begin{center}\includegraphics[width=\textwidth]{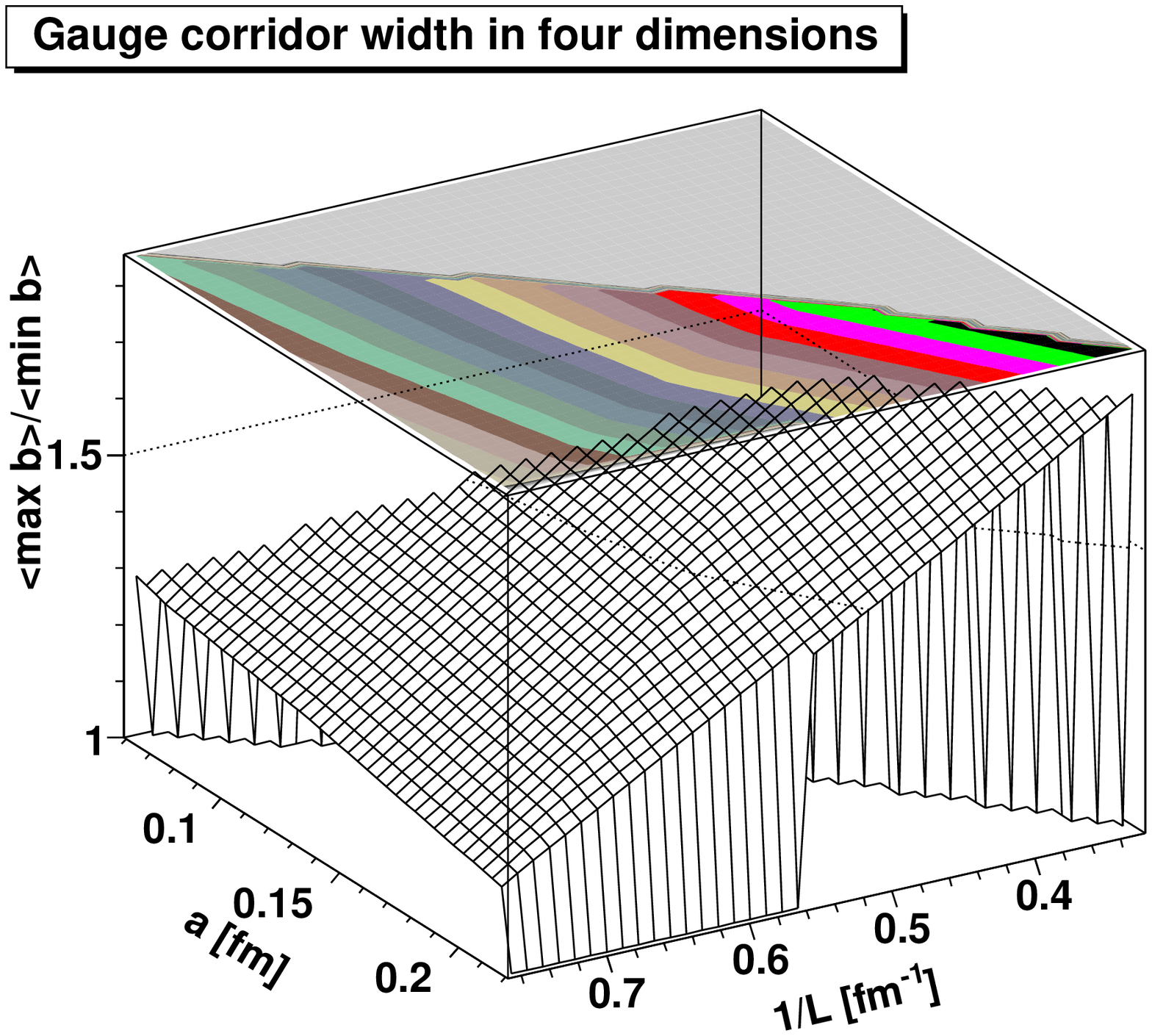}\end{center}
\end{minipage}
\begin{minipage}[c]{0.5\linewidth}
\caption{\label{fig:gcorridor}The gauge corridor width $\langle\max b\rangle/\langle\min b\rangle$ in two (top-left panel), three (top-right panel), and four (bottom panel) dimensions \cite{Maas:2009se,Maas:unpublished}. Note that these results are lower and upper bounds to this ratio.}
\end{minipage}
\end{figure}

\begin{figure}
\begin{center}\includegraphics[width=\textwidth]{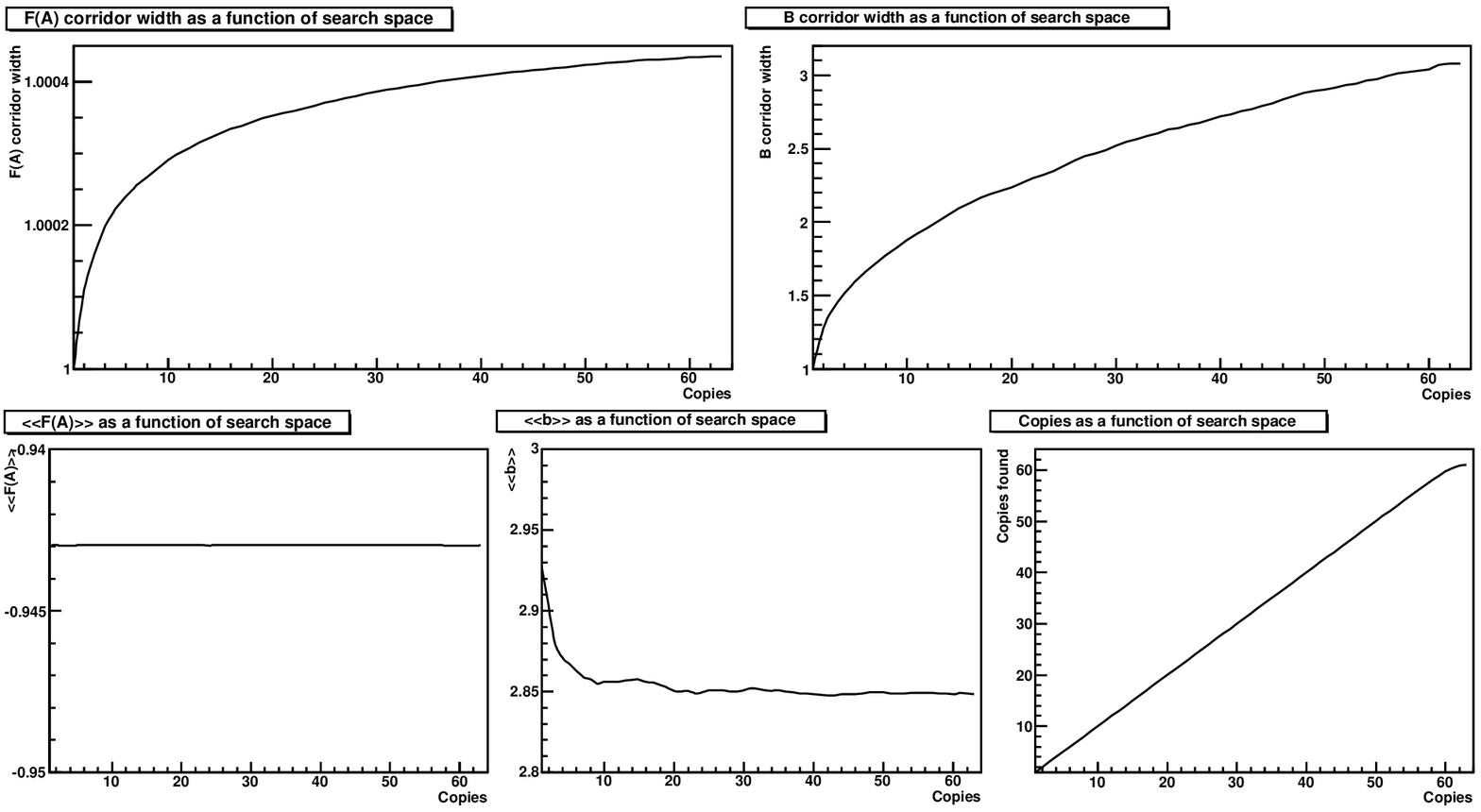}\end{center}
\caption{\label{fig:copydep}Various observables \cite{Maas:unpublished} as a function of the size of the search space \cite{Cucchieri:1997dx,Maas:2009se} for a $36^3$ lattice with lattice regulator $a=(1.6  $ GeV$)^{-1}$. Note that also the number of configurations (here 1632) taken into account can modify this result. Lower left panel: Double gauge orbit and configuration averaged value of $F(A)$. Lower middle panel: Double gauge orbit and configuration averaged value of $b$. Lower right panel: Number of Gribov copies found. Upper left panel: Width of the $F(A)$ corridor. Upper right panel: Width of the $B$ corridor.}
\end{figure}

Hence, it remains to get a better view on the development of the permitted interval for $B$. A useful possibility is to determine the width of the gauge corridor $[\langle\min b\rangle,\langle\max b\rangle]$, given by $\langle\max b\rangle/\langle\min b\rangle$. Taking the ratio removes any over-all rescaling effects due to trivial renormalization and volume effects. This ratio is shown in figure \ref{fig:gcorridor}. The results shown are only approximations in two respects. Because of the inherent problem to find all Gribov copies with any given method, the results shown give only a lower limit to the size of the interval. That is an issue which has to be taken into account at any rate for any results displayed in this work: In almost all cases only lower limits of the possible gauge variations are available, and though an extrapolation in the number of Gribov copies is possible in principle \cite{Maas:2010nc,Silva:2004bv}, there is no guarantee. The effect of taking different search spaces into account is illustrated in figure \ref{fig:copydep}. A detailed discussion can be found in \cite{Maas:2011ba,Maas:unpublished}.

Irrespective of this, the corridor in all dimensions, even in the case of two dimensions with its small number of Gribov copies, quickly opens with increasing volume. The increase of the ratio of upper to lower bound of the corridor is actually not affected by trivial finite volume effects, but can be affected by non-trivial finite volume effects. Still, this is the strongest impact of the choice of Gribov copies for the investigated volumes which has been found so far. Though it is tempting to conclude from these results that the interval of permitted $B$ values is indeed larger than a point, and has an infinite upper limit, experience has shown that this cannot so easily be decided without investigating the volume dependence over a much wider range \cite{Sternbeck:2007ug,Cucchieri:2008fc,Cucchieri:2007rg,Bogolubsky:2009dc}. In particular, convincing evidence has been found that at least in three and four dimensions the lower limit is necessarily finite \cite{Sternbeck:2007ug,Cucchieri:2008fc,Bogolubsky:2009dc}, and thus the value of $\langle\min b\rangle$ has to flatten out. This effect is not seen yet. Furthermore, the dependence is on the discretization is non-trivial. While in two and three dimensions the corridor width increases with finer discretization, this is not the case in two dimensions. This reminds of the case with the number of Gribov copies, shown in figure \ref{fig:gc}, where two dimensions also behaves differently than three and four dimensions.

Note that the construction so far has been entirely based on selection of Gribov copies, without returning to the findings in functional calculations. Whether indeed, as in functional calculations, the upper limit of the $B$ corridor is infinite in the thermodynamic limit cannot be decided yet. The evidence in favor or disfavor of this possibility will be discussed at length in section \ref{zerot:prop}. Some rather general comments on whether this is possible at all will also be given in section \ref{sec:kugo}. However, the discussion here shows that the expectation value of the ghost dressing function at zero momentum can be brought into contact with Gribov copies. If either \pref{quant:bgauge} or \pref{quant:bgauge2} can furthermore be shown to be correct in general, this would imply that selecting the zero-momentum value of the ghost propagator in functional calculations is indeed equivalent to treating Gribov copies.

The fact that giving a definite prescription how to deal with Gribov copies can be based on a desired behavior of the ghost propagator at zero momentum is already showing that the Gribov-Singer ambiguity can be resolved in this way in a well-defined manner, in the same sense as the minimal Landau gauge resolves it. It thus remains to formally understand how this is resembled in the approach starting with the continuum theory. Of course, this is by no means the only possibility to resolve the ambiguity, and any definite prescription how to treat Gribov copies serves equally well. See e.\ g.\ \cite{Serreau:2012cg} for a recent alternative proposal.

\subsection{Gauge relations}

\begin{figure}
\includegraphics[width=\textwidth]{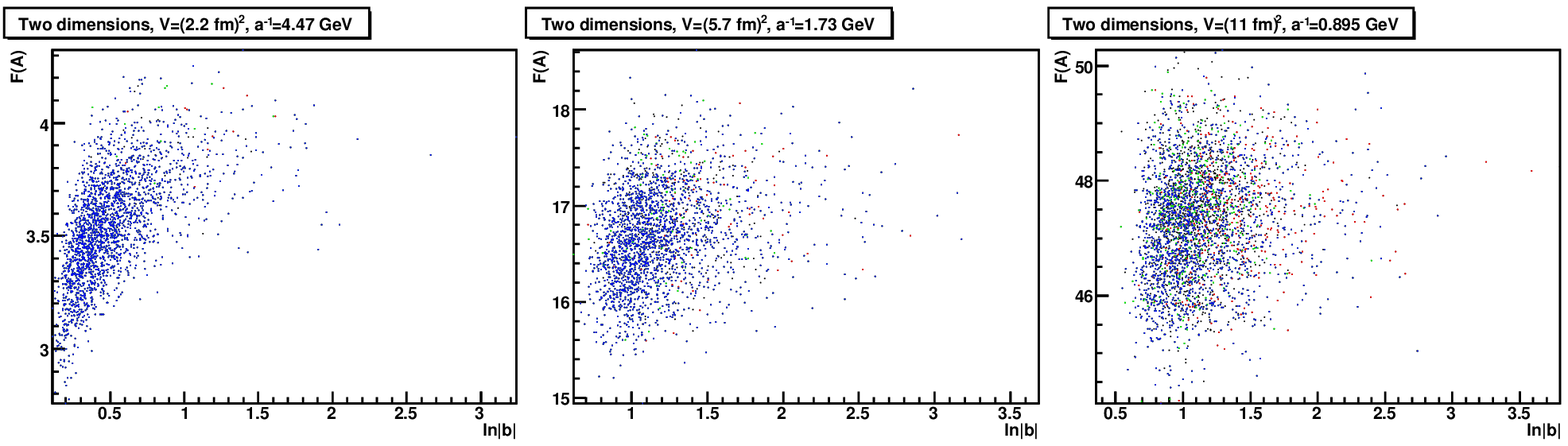}\\
\includegraphics[width=\textwidth]{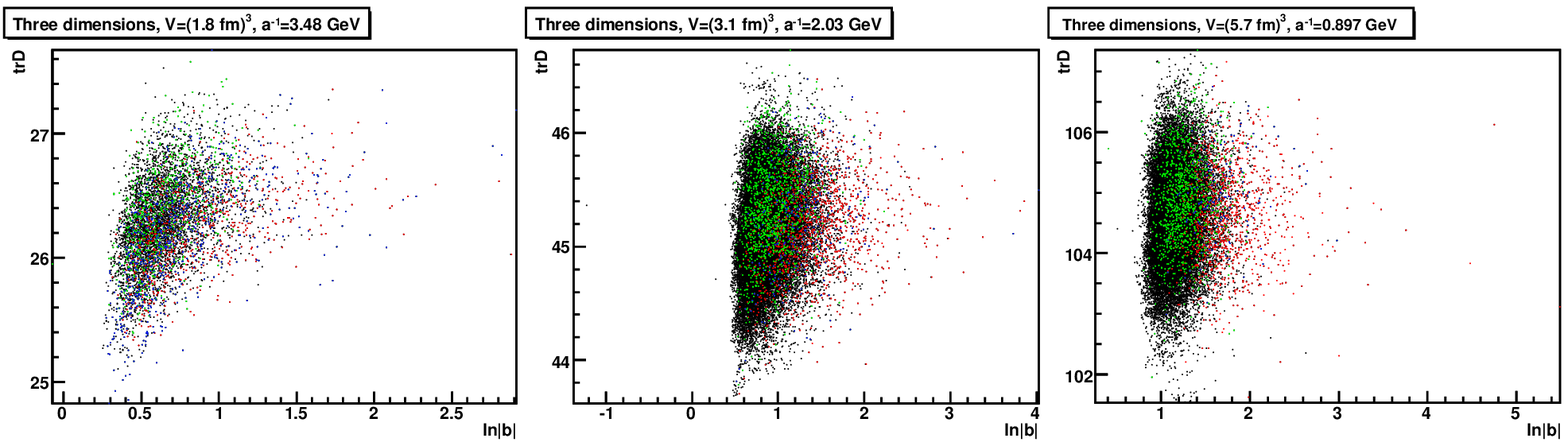}\\
\includegraphics[width=\textwidth]{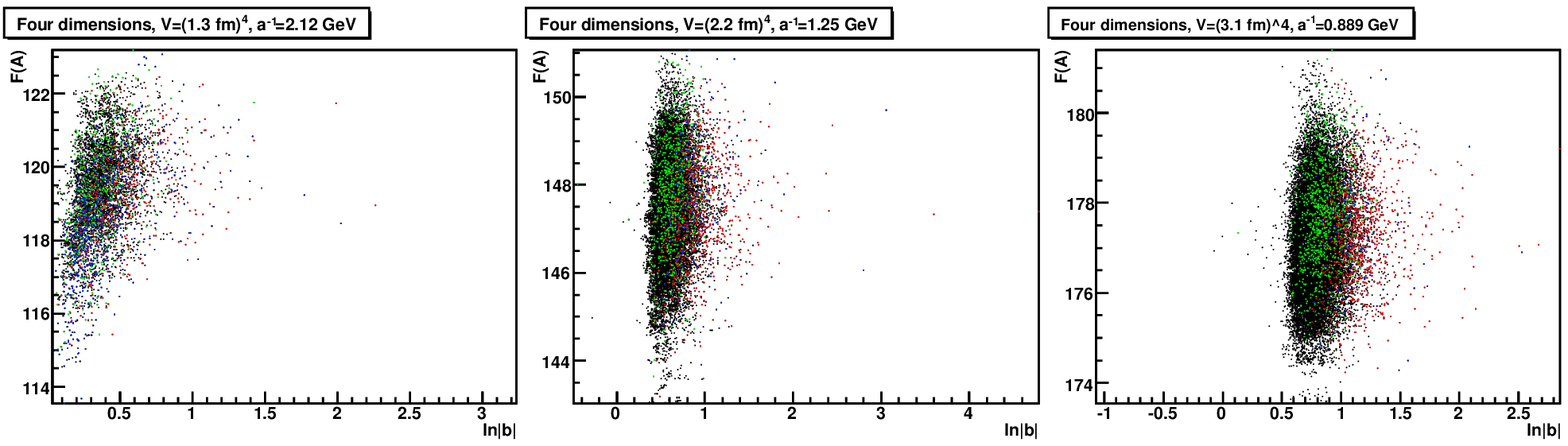}
\caption{\label{fig:fgr}The first Gribov region projected to the coordinates $F(A)$ (after subtraction of one and multiplication by -1000) and $b$. Top, middle, and bottom panel show results for two, three, and four dimensions, respectively. Always three different volumes are shown, with increasing volume from left to right. Red points denote copies on the would-be horizon (see text), green points are in the fundamental modular domain, blue points have both these properties, and black points neither. More details can be found in \cite{Maas:unpublished}.}
\end{figure}

Before turning to the more practical matter of calculating the correlation functions, it provides further insight to investigate the relation between the two quantities $F(A)$ and $b$. A first insight is given by the projection of the first Gribov region to these two coordinates. This is shown in figure \ref{fig:fgr}. It is visible that, for the largest volumes studied here, there is essentially no correlation between the two coordinates. Thus, it can be expected that when using one of the parameters to fix the gauge, the results for the other parameter will be essentially its average value. Thus, it will have a similar value as in the minimal Landau gauge.

The next observation comes from classifying the different Gribov copies. One possibility is to select the fundamental modular region copies, at least as good as the algorithm can identify them. The second possibility is to select the largest value of $b$. In minimal Landau gauge, for sufficiently large volumes, this quantity is dominated by the lowest eigenvalue of the Faddeev-Popov operator \cite{Maas:2010nc,Cucchieri:2008fc,Sternbeck:2005vs}, and thus this would be in one-to-one correspondence with the Gribov horizon. If this were to hold in general, $b$ can also be used to characterize the Gribov horizon.

It is then seen in figure \ref{fig:fgr} that the set of copies belonging to both categories shrinks with increasing volume and cut-off. Thus, less and less gauge orbits have a Gribov copy belonging to both categories. Hence, it appears that though such a common boundary exists \cite{Zwanziger:1993dh,Zwanziger:2003cf}, it is only a small part of the total boundary of the Gribov region. But this may also be an artifact: The number of copies being found satisfying both criteria is found to increase when the search space increases \cite{Maas:unpublished}, so there is the chance that this particular Gribov copy of each orbit is often just not found among the large number of copies. However, this would also imply that the standard algorithms used for minimal Landau gauge will not sample this type of copies very well.

Because both coordinates appear to be quite unrelated, it is a natural possibility to select combined gauge choices. E.\ g., the Gribov copy for a configuration could be selected which maximizes
\be
xF(A)+yb\nn,
\ee
\no for some value of $x$ and $y$, or any other function of the two coordinates. However, it does not appear to be possible to require the satisfaction of two independent exact conditions for $F(A)$ and $b$. Another possible choice would be just minimizing instead of maximizing $F(A)$, which could be called an inverse Landau gauge \cite{Silva:2004bv,Maas:2009se}. Here, these possibilities will not be pursued further, and only the set of Landau-$B$ gauges together with the absolute and minimal Landau gauge will be investigated.

\section{Methods}\label{smethods}

It is actually a quite non-trivial problem to determine the correlation functions beyond perturbation theory. Various methods have been employed for this. However, much more successful than any individual method was the combination of two or more of them to compensate the respective disadvantages. For the task of describing gluons, the combination of lattice gauge theory and functional methods has been most useful, and will be employed here. Alternative approaches employed are in particular methods based on effective actions \cite{Capri:2007ix,Dudal:2007cw,Dudal:2008sp,Dudal:2008rm,Baulieu:2008fy,Dudal:2008xd,Gracey:2009mj,Zwanziger:2009je,Zwanziger:2010iz,Tissier:2010ts,Frasca:2007uz}, but also stochastic quantization approaches \cite{Zwanziger:2001kw,Zwanziger:2003cf,Zwanziger:2002ia}, both of which will not be detailed further.

The functional methods can be split further, in the quantum equations of motion on the one hand, and functional renormalization group methods on the other. Though the quantum equations of motion will be mainly used here, the related approach using functional renormalization group equations will be shortly introduced in section \ref{zerot:erg}.

An enormous advantage of all of these approaches, in contrast to models, is that perturbation theory is always included. Because of asymptotic freedom, it becomes manifest for Yang-Mills theory at large momenta. Conversely, this implies that use can be made of the wealth of perturbative results in all of these methods.

\subsection{Lattice gauge theory}\label{zerot:lat}

One of the most successful methods to investigate gauge theories in the non-perturbative domain is lattice gauge theory \cite{Montvay:1994cy,Gattringer:2010zz}. In this approach, space-time is discretized and only a finite volume is investigated, usually with the geometry of a hypercube having $N$ sites in each direction space in intervals of length $a$, and having thus a physical volume $a^dN^d=L^d=V$. Effectively, thus, quantum field theory is approximated by a finite system and thus quantum mechanics. Only by taking the appropriate thermodynamic limit in the end, the original quantum field theory is recovered. However, the existence of this limit is still an unproven assumption. The results shown in figures \ref{fig:gc}-\ref{fig:raw-d-ren} and \ref{fig:raw-b}-\ref{fig:fgr} in the previous chapter were all obtained with this method.

This method is particularly useful to determine gauge-invariant quantities \cite{Montvay:1994cy,Gattringer:2010zz}. Furthermore, in this case it is known how to establish the connection to Minkowski space-time \cite{Seiler:1982pw}. The most dramatic successes of lattice gauge theory have been obtained by making use of the fact that it is very well approachable by numerical methods \cite{Gattringer:2010zz}, permitting to evaluate path integrals directly. Though this process just delivers final numbers, it permits insight into the hadronic spectrum and many other results, and is essentially only limited by the possibilities of the numerical algorithms. Though these made enormous progress over the years, there are still substantial challenges. These include very light dynamical fermions, isospin breaking, finite density, disparate scales, supersymmetry, critical slowing down, and other problems.

Here, however, the focus is different and is on the possibility to determine correlation functions directly by numerical evaluation. With the numerical approach to lattice gauge theory, it is only possible to evaluate full correlation functions, defined by \pref{quant:fullgreen}. To obtain vertex functions, the connected contributions have to be determined, and these have to be amputated. The latter has quite some implications in Landau gauge, as will be discussed in more detail in section \ref{zerot:sti}.

\subsubsection{The action}

The detailed approach to obtain the correlation functions is a multi-step process \cite{Montvay:1994cy,Gattringer:2010zz}. Usually, the formulation of Yang-Mills theory using lattice gauge theory is in terms of the group elements rather than of the algebra-valued gauge fields themselves. In principle, any group representation of the theory is equally valid, however, some turn out to be more amendable to numerical calculations than others. A dramatic example is the case for the gauge algebra su($N$). In this case a formulation in terms of the group SU($N$) is much more efficient with the available algorithms  \cite{Gattringer:2010zz} than for the gauge group SU($N$)/Z$_N$ \cite{Mendes:1996ze,Burgio:2006xj,Barresi:2006gq}, which would be more appropriate for the standard model \cite{O'Raifeartaigh:1986vq}. Since the gluonic correlation functions are invariant under group transformations which are mapped to the identity in the algebra, e.\ g.\ center transformations for SU($N$), the notion of algebra and group will be used synonymously in the following.

In general, the gauge fields $A_\mu^a(x)$ are thus represented by the corresponding group-valued link variables $U_\mu(x)=\exp(iaA_\mu^a\tau_a)$ for each lattice site $x$ and direction $\mu$. Of course, from these the original algebra-valued gauge fields can always be reconstructed, though this in general produces errors of the size ${\cal O}(a^2)$, where $a$ is the lattice spacing. E.\ g., for SU(2) the relation is given by\footnote{Care has to be taken if the center symmetry of the gauge group is broken \cite{Karsch:1994xh}. In this case, \pref{lat:gfields} can at best be used for configurations with a positive, real Polyakov loop.}
\be
A_\mu^a=\frac{\sqrt{\beta}}{4ia}\tr \tau^a (U_\mu-U_\mu^+)+{\cal O}(a^2)\label{lat:gfields}
\ee
\no where $\beta=4N_F/g^2$ is related to the bare coupling $g$, which has dimension $a^{-\frac{4-d}{2}}$, and $N_F$ is the dimension of the fundamental representation of the gauge algebra. Similar relations hold for other gauge algebras.

The discretized lattice action $S_L$ is then given in terms of these link variables. The form used here is the so-called Wilson action \cite{Montvay:1994cy}
\bea
S_L&=&\beta\sum_{x,\nu>\mu}\left(1-\frac{1}{N_F}\tr U_\mn\right)\nn\\
U_\mn&=&U_\mu(x)U_\nu(x+e_\mu)U_\mu(x+e_\nu)^+U_\nu(x)^+\nn,
\eea
\no with the unit vector in the hypercubic lattice $e_\mu$ in the direction $\mu$ and the (gauge-invariant) plaquette $U_\mn$. The size of the hypercubic lattice is $N^d$. This lattice action agrees with the continuum action $S$ up to corrections of order $a^2$, and thus coincides with it in the limit $a\to 0$. This behavior can be improved \cite{Gattringer:2010zz}, but this will not be necessary here. Studies of correlation functions using improved actions can be found, e.\ g., in \cite{Gong:2008td}.

Using a Monte Carlo simulation, a set of (independent) gauge orbits $\{{\cal G}\}$, being a set ${\cal G}=\{U_\mu\}$ of $dN^d$ link variables $U_\mu$ each, is generated. For this purpose, various methods are available, which will not be detailed here, as they are not specific for the purpose of calculating correlation functions. See \cite{Gattringer:2010zz} for a comprehensive introduction, and for the particular algorithm used for most of the results presented here \cite{Cucchieri:2006tf}.

It should be noted that there are quite a number of technical problems still associated with the generation of the configurations, not the least to determine the value of the cut-off, and thus of $a$ as a function of $\beta$. For the present purpose, standard methods are sufficiently well known \cite{Montvay:1994cy,Gattringer:2010zz}, and will not be detailed here. The important technical point is that in all cases here the scale is set by giving the intermediate distance string tension the value (440 MeV)$^2$, see for details e.\ g.\ \cite{Maas:2007uv,Cucchieri:2008qm,Maas:2007af}.

\subsubsection{Fixing the gauge}\label{smethods:lgf}

A typical Monte-Carlo algorithm generates a set of $N_C$ independent field configurations $\{U_\mu\}$ using the weight $\exp(-S_L)$, i.\ e., each configuration contains link variables for all directions and lattice sites. Each of these link configurations are in a random gauge, and have to be transformed into the desired gauge first before the gauge-fixed correlation functions can be evaluated. Otherwise, the gauge-dependent correlation functions average to correlation functions of Gaussian random variables, according to Elitzur's theorem \cite{Elitzur:1975im}.

The gauge-fixing to satisfy the the Landau gauge condition \pref{quant:lg} and to restrict to copies inside the first Gribov region can be performed by minimizing the expression
\be
\sum_{x,\mu} \tr U_{\mu}(x)\sim 1+\mathrm{const.}\sum_{x,\mu,a} A_\mu^a A_\mu^a+{\cal O}(a^2)\label{quant:lgtgf},
\ee
\no which agrees with \pref{quant:abslg} in the continuum limit up to prefactors \cite{Cucchieri:1995pn}. For this purpose, a gauge transformation $g(x)$ on the lattice is searched for, such that the gauge fields obtained from the gauge transformed links $u_\mu(x)=g(x)U_\mu(x)g(x+e_\mu)^+$ satisfy the Landau gauge condition. To check whether this condition is indeed fulfilled, it is possible to calculate, e.\ g., the gauge-fields by \pref{lat:gfields} and check whether they are transverse \cite{Cucchieri:1995pn}. In fact, despite that these fields can be transverse up to the numerical precision, due to the discretization errors the corresponding continuum fields will only be transverse up to order ${\cal O}(a^2)$.

Actually, this direct local evaluation is not very sensitive to long-wavelength fluctuations. Therefore, a more useful quantity to evaluate is \cite{Cucchieri:1995pn}
\bea
&&\frac{1}{d}\sum_\mu\frac{1}{N_\mu}\sum_c\frac{1}{[\tr(Q_\mu \tau_c)]^2}\times\sum_{x_\mu}(\tr\{[q_\mu(x_\mu)-Q_\mu]\tau_c\})^2\label{lat:e6}\\
q_\mu(x_\mu)&=&\frac{1}{2i}\sum_{x_\nu,\nu\neq\mu}\big[g(x)U_\mu(x)g(x+e_\mu)^+-g(x+e_\mu)U_\mu(x)^+ g(x)^+ \big] \nn \\
Q_\mu&=&\frac{1}{N_\mu}\sum_{x_\mu}q_\mu(x_\mu)\nn,
\eea
\no where $N_\mu$ is the size of the hypercubic lattice in direction $\mu$. A typical good numerical condition on this quantity is to require it to be below $10^{-12}$ \cite{Cucchieri:1995pn,Cucchieri:2006tf}.

To determine the gauge transformation $g(x)$, several algorithms have been developed, see \cite{Cucchieri:1995pn} for a detailed discussion. A particularly useful example of such an algorithm for contemporary lattice sizes is the stochastic overrelaxation algorithm. Essentially, it updates $g(x)$ at each lattice site in turn\footnote{This can be accelerated by using a checker-board update instead of a lexicographical update.} by either choosing the new $g(x)$ such that it minimizes \pref{quant:lgtgf} locally or by only rotating it such that \pref{quant:lgtgf} is left unchanged. Which of both possibilities is performed is chosen randomly. The probabilities assigned for both possibilities significantly influence the performance of the algorithm, and depend on the volume, discretization, and dimensionality of the lattice investigated \cite{Maas:2007uv,Cucchieri:1995pn}. It is most convenient to permit the algorithm to adapt these probabilities during run-time \cite{Cucchieri:2006tf}. However, the best value for the probability is also very much configuration-dependent \cite{Maas:2008ri}. Therefore, no optimal choice exists, and the best choice is one which optimizes the algorithm on the average.

The gauge fields obtained so minimize \pref{quant:lgtgf} and therefore fulfill the perturbative Landau gauge condition \pref{quant:lg}. Furthermore, it can be shown that they are in the first Gribov region \cite{Zwanziger:1993dh}, by noting that at the minima of \pref{quant:lgtgf} the Hessian of its second derivatives has to be positive (semi-)definite, and that this Hessian coincides with the Faddeev-Popov operator \pref{quant:fpo}. Actually, getting out of the first Gribov region using lattice calculations is not entirely trivial. This can be achieved using, e.\ g., stochastic quantization \cite{Pawlowski:2009iv} or gauge-fixed Monte-Carlo simulations \cite{vonSmekal:2008ws,vonSmekal:2008es}.

However, without further specification no minimum is preferred and thus a random Gribov copy is obtained within the first Gribov region. Choosing this copy for proceeding is therefore equivalent to implementing the minimal Landau gauge. To obtain other gauges, it is necessary to specify this copy further. Since for this purpose knowledge of more copies is necessary, this requires access to further Gribov copies. Up to now, no systematic way of generating new Gribov copies is known. Therefore, a stochastic approach is used \cite{Cucchieri:1997dx,Silva:2004bv,Maas:2009se}. For this it is important to note that any kind of local algorithm which attempts to minimize \pref{quant:lgtgf} will find the closest minimum to its starting point, or even with some amount of randomness in its search, a minimum very close to its starting point. Therefore, restarting the gauge-fixing algorithm with a different initial guess for $g(x)$ will in general produce another Gribov copy, if there exists more than one copy, though with an unlucky guess the previous one will be found\footnote{This assumes perfect numerical accuracy. On finite machines, numerical artifact copies will be found, but can be identified with knowledge on the precision of the algorithm \cite{Maas:2008ri}. Further artificial Gribov copies exist due to the lattice regularization \cite{Giusti:2001xf}, but these become irrelevant when taking the continuum limit.}. By repeating this sufficiently often, a set of Gribov copies is generated. In this set of copies then the copy which matches the non-perturbative supplemental gauge condition, e.\ g.\ absolute Landau gauge or a Landau-$B$ gauge, as good as possible is selected to represent the gauge orbit. Since this is a high-dimensional optimization problem, it is in general not possible to guarantee finding all copies \cite{Maas:2008ri}. Therefore, gauges like the absolute Landau gauge or the maxB gauge can never be reached with certainty, though sophisticated algorithms can be devised to make the search at least somewhat more efficient \cite{Maas:2008ri,Silva:2004bv,Oliveira:2003wa,Yamaguchi:1999hq,Bogolubsky:2005wf,Bogolubsky:2007bw}. For gauges which do not strive to find an extreme value, like a Landau-$B$ gauge with some non-extremal value of $B$, this problem only manifests itself in the generation of more stochastic noise.

Any of these methods finally selects one particular Gribov copy with particular, gauge-fixed links $u_\mu$ for each configuration. This yields the set of gauge-fixed configurations $\{\{u_\mu\}\}$. These represent the gauge orbits\footnote{If the aim is to implement gauges averaging with a specific prescription over gauge copies, like general covariant gauges or Hirschfeld-like gauges, this requires to average over the set of Gribov copies found with the required weight functions and over all Gribov regions. Such gauges will not be considered further here. See, e.\ g., \cite{Cucchieri:2008zx,vonSmekal:2008ws,Cucchieri:2009kk,vonSmekal:2008es,Maas:2011ba,Maas:unpublished}.}. Note that no ghost fields need to be introduced. Gauge-fixing on the lattice, in the original spirit of gauge-fixing, is due to selecting a representative for each gauge orbit.

\subsubsection{Correlation functions}

Once such a set of Gribov copies representing the gauge orbits has been specified, correlation functions can be determined. This is done by determining the correlation function on each gauge orbit's representative and then average over all gauge orbits. Thus, for some correlation function $\Omega^{A_1...A_n}$ being the expectation value of the product of $n$ fields $A_n$, the lattice result for the correlation function from $N_C$ gauge orbits is given by
\be
\Omega^{A_1...A_n}=\langle A_1...A_n\rangle=\frac{1}{N_C}\sum_i A^i_1...A_n^i\label{smethod:latcor},
\ee
\no where the index $i$ labels the representatives of the $N_C$ gauge orbits. In the limit of $N_C\to\infty$ the function obtained is indeed the correlation function \cite{Montvay:1994cy}. At finite $N_C$, statistical fluctuations will be present, and a stochastic error can be assigned to the correlation functions at each point \cite{Montvay:1994cy,sokal:errors}. Furthermore, on any finite lattice, there will be systematic errors due to the finite volume and discretization and the presence of the non-trivial geometrical structure introduced by choosing a hypercube. Only by taking the thermodynamic limit (or extrapolating to this limit) the continuum correlation functions in an infinite volume can be obtained. These lattice artifacts will be discussed in more detail in section \ref{methods:latart}.

Here, the correlation functions in momentum space are in most cases more interesting than those in position space. Therefore, instead of evaluating the correlation functions first in position space and then Fourier-transforming it is more convenient to transform the gluon fields directly, which is done by \cite{Rothe:2005nw}
\be
A_\mu^a(p)=e^{- \frac{i\pi P_{\mu}}{N_\mu}} \sum_X e^{2\pi i \sum_\mu \frac{P_\mu X_\mu}{N_\mu}} A_\mu^a(x) ,\label{eq:Aofp}
\ee
\no where on a finite lattice the components $P_\mu$ of $P$ have the integer values $-N_\mu/2 + 1\, , \ldots , \, N_\mu/2 \, $ and the components $X_\mu$ of $X$ are the (integer) coordinates in the lattice ranging from $0$ to $N_\mu-1$. Due to the periodic boundary conditions in position space, the fields are also periodic in momentum space, and thus only the momenta $0...N_\mu/2$ are independent.

The physical momenta $p$ to which the integer lattice ones $P$ correspond are
\be
p_\mu=\frac{2}{a}\sin\frac{P_\mu\pi}{N_\mu}\nn.
\ee
\no The prefactor in \pref{eq:Aofp} comes from the implicit mid-point definition of the gauge-fields when using the links, and has to be taken into account. The gluon propagator, e.\ g., on a single configuration is then given by
\be
D_\mn^{ab}(p)=\frac{1}{V}A_\mu^a(p)A_\nu^b(-p)\nn,
\ee
\no where use has been made of the reality of the gluon field in position space and $V$ is the total volume of the lattice, stemming from the normalization of the Fourier sum. The three-gluon correlation function is then given by
\be
\frac{1}{V}A_\mu^a(p)A_\nu^b(q)A_\rho^c(k),
\ee
\no with $p+q+k=0$, and so on.

A bit more complicated is to determine quantities involving the ghost fields. In the lattice calculations a gauge copy is selected directly, and the Jacobian is never required, and thus no ghost fields appear. Therefore, the only possibility to obtain correlation functions involving ghost fields is by returning to the path integral \pref{quant:gfpi}. Integrating out the ghost fields gives expressions for correlation functions involving the ghost fields in terms of the Faddeev-Popov operator \pref{quant:fpo}. 

To make use of these relations requires to determine the Faddeev-Popov operator on the lattice. It is defined in terms of its action on an arbitrary function $\omega^a$ as a function of the gauge-fixed link variables as \cite{Zwanziger:1993dh}
\bea
M(y,x)^{ab}\omega_b(x)&=&c\left(\sum_x\big(G^{ab}(x)\omega_b(x)+\sum_\mu A_\mu^{ab}(x)\omega_b(x+e_\mu)+B_\mu^{ab}(x)\omega_b(x-e_\mu)\big)\right)\nn\\
G^{ab}(x)&=&\sum_\mu \tr(\{\tau^a,\tau^b\}(u_\mu(x)+u_\mu(x-e_\mu)))\nn\\
A_\mu^{ab}(x)&=&-2\tr(\tau^a \tau^bu_\mu(x))\nn\\
B_\mu^{ab}(x)&=&-2\tr(\tau^a \tau^bu_\mu^{+}(x-e_\mu))\nn,
\eea
\no where $c$ is a normalization parameter, depending on the normalization of the generators  $\tau^a$ of the gauge algebra.

E.\ g., integrating out the ghost fields for the correlation functions of a ghost and an anti-ghost, the ghost propagator, then yields its representation in terms of the Faddeev-Popov operator as
\be
<\bar{c}^a(x)c^b(y)>=<M^{ab-1}(x,y)>\nn.
\ee
\no Therefore, to determine the ghost propagator, and actually all correlation functions involving a ghost and an anti-ghost field, requires to invert the Faddeev-Popov operator. This needs to be done numerically in lattice simulations. Similar to the case of fermion fields \cite{Gattringer:2010zz}, there are two numerical possibilities widely used, both based on a (bi-)conjugate gradient algorithm \cite{Press:1997nr,Meurant:2006}, exploiting the positive-semi-definiteness of the Faddeev-Popov operator in the first Gribov region. They differ by the source vector, point source \cite{Boucaud:2005gg,Cucchieri:2006tf,Maas:2010nc} or plain-wave source \cite{Suman:1995zg,Cucchieri:1997dx,Cucchieri:2006tf}. The advantage of a point source is that the inversion has to be done only once to get the full Fourier spectrum of the ghost propagator after Fourier transformation, while this has to be done separately for each mode with a plain-wave source. The drawback of the point source is that it suffers from significantly larger statistical fluctuations, limiting its use essentially to only calculating the propagator. A quantitative comparison of both methods for the ghost propagator and the ghost-gluon vertex can be found in \cite{Cucchieri:2006tf}.

The ghost number is a conserved quantum number in Landau gauge\footnote{There is a second symmetry of the ghost fields in Landau gauge which relates ghosts and anti-ghosts \cite{Alkofer:2000wg}. In principle, this symmetry could be broken \cite{Dudal:2003dp,Kondo:2001tm}, but this does not seem to be the case \cite{Cucchieri:2005yr}.}. Hence, knowledge of the inverse Faddeev-Popov operator is then already sufficient to construct all higher order ghost correlation functions. E.\ g., the ghost-gluon vertex is given by \cite{Cucchieri:2004sq}
\be
<A_\mu^a(x)\bar{c}^b(y)c^c(z)>=<A_\mu^a(x) M^{bc-1}(y,z)>\nn.
\ee
\no Once more, this can be proven by integrating out the ghost fields.

Thus, in general, the construction of vertices is a straight-forward generalization of the calculation of the propagators \cite{Cucchieri:2006tf}. However, the relation \pref{smethod:latcor} already shows that the correlation functions obtained are the full correlation functions. To compare it to the vertex functions, lattice results always have to be reduced to their connected and amputated part. The connected part is not a problem for correlation functions up to order three, as any one-point correlation function vanishes in Landau gauge, but can become very cumbersome at higher orders \cite{Cucchieri:2006tf,Maas:2007uv}. The fact that the correlation functions are non-amputated implies that only the tensor structures transverse in all gluon momenta can be obtained in lattice calculations in Landau gauge, but not the longitudinal ones. To isolate then the individual transverse tensor structures requires projection of the connected correlation functions with appropriate tensors \cite{Ball:1980ax}, and then amputation by division with appropriate products of propagators. Details of this procedure can be found in \cite{Cucchieri:2004sq} for the ghost-gluon vertex and in \cite{Cucchieri:2006tf} for the three-gluon vertex. Higher vertices have not yet been determined in lattice calculations, mainly due to the statistical noise which increases quickly with the number of fields involved.

It should be noted that the correlation functions of an individual configuration will not exhibit translational nor rotational symmetry, and thus a propagator will depend on both $x$ and $y$ separately, instead of only the difference $x-y$. Only after performing the average \pref{smethod:latcor} these symmetries will become manifest in the limit $N_C\to\infty$, within discretization errors and hypercubic artifacts. Thus, the averaged propagators will indeed depend only on $x-y$.

\subsubsection{Lattice artifacts}\label{methods:latart}

The previous description produces the correlation functions for a finite set of momenta. This set ranges from a lower cut-off, which is essentially given by $1/L$, where $L$ is the extension of the lattice in the direction of the momentum, up to a largest momentum of size $1/a$, where $a$ is the lattice spacing. At the infrared cutoff, the system becomes aware of its extension, and the fact that it is a cube, while at large momenta it becomes aware of its hypercubic symmetry, and that there is a lattice. Thus, in principle, only for momenta fulfilling the relation \cite{Fischer:2007pf}
\be
\frac{1}{L}\ll p\ll\frac{1}{a}\label{lat:limrange},
\ee
\no the lattice can be expected to give reliable results. Thus the strength of the lattice formulation lies in an intermediate momentum window. An example for such artifacts in the case of finite volume, i.\ e.\ finite $L$, will be given in section \ref{zerot:prop} in figure \ref{fig:volume} below. 

Though not directly comparable, an example of how misleading such artifacts could be comes from the treatment of full QCD. There, for several hadronic observables the results are still far from the experimentally known values because the quarks are heavy, despite that the pion mass is already quite close to the physical one (see, e.\ g., \cite{Engel:2010my} for a more detailed discussion and further references). Also, expected states still have not been seen, like pion scattering states. In this case also physical effects, like isospin breaking, may play a role. Thus, this should serve only as a warning, but should not be treated as a general permission to distrust the lattice results.

A less obvious source of artifacts appears when $n$-point vertices, instead of propagators, are considered: The $n-1$ independent momenta cannot be positioned arbitrarily with respect to each other, but only in a way compatible with the geometry of the lattice. E.\ g., on a two-dimensional lattice, there are very few possibilities to construct two momenta such that they have the same size as their sum, as the 60$^\circ$ angle between them can only be realized for very particular sizes in lattice units \cite{Maas:2007uv}. Thus, it is not always possible to easily reconstruct momentum configurations from the continuum, and this becomes worse the larger the number of momentum vectors involved is compared to the number of dimensions. Of course, the finer the lattice is, the weaker this problem becomes, but at lattice sizes currently accessible, this can be a serious restriction\footnote{This is connected to the problem of assigning spin and orbital angular momentum to states, since the rotation group is the discrete hypercubic group H$_d$, rather than the continuous group SO($d$) \cite{Gattringer:2010zz}. However, this is of little relevance for the topic here.}.

There are further problems. One is that, since the lattice is discrete, any function is actually only given by a set of numbers. Thus, it is a-priori impossible to predict how well the results will approximate the continuum correlation functions in the complex plane, as a general statement of function theory \cite{Alkofer:2003jj}. This is of particular seriousness when considering the continuation to Minkowski space. A second problem is that there is still no proof that lattice gauge theory has as limit the desired continuum theory \cite{Montvay:1994cy}, though it appears unlikely that this should not be the case\footnote{In fact, at times the position is taken that the lattice is the only well-defined version of Yang-Mills theory. However, given that the universe is finite, and gravity plays a role at some point, this is more a philosophical question. Even if it were the case, the continuum field theory could then still be seen as a possible approximation of the lattice theory.}.

A particular problem for the gauge-fixed setting followed here is that in addition to the conventional Gribov copies there are also additional Gribov copies which are lattice artifacts \cite{Giusti:2001xf}. These will disappear when taking the limit to the continuum field theory in an infinite volume, but may contribute otherwise. However, it appears that this does not provide a new qualitative dimension to the problem of Gribov copies \cite{Maas:2008ri}, and in the present treatment lattice Gribov copies are therefore not distinguished from other finite volume and discretization artifacts.

On top of all these principal problems, there is of course a practical problem. Most non-perturbative insights, including those presented here, come from numerical calculations. Thus, computational resources are a very practical limitation, especially when considering accessible volumes and discretizations \cite{Fischer:2007pf,Cucchieri:2008fc,Cucchieri:2007rg,Maas:2007uv,Bogolubsky:2009dc}. Thus, the investigation of disparate energy scales is strongly restricted by computation time. A possibility is to use extrapolations, but since neither finite volume nor finite discretization are approximations which are under analytical control, extrapolations cannot be considered to be ultimately reliable. It is known from solid-state physics \cite{Landau:2005mc} and lower-dimensional QED \cite{Goecke:2008zh} that these can be easily misleading.

Hence, the lattice approach is alone not sufficient, at least at the current time, to treat the problems of interest, e.\ g., in the standard model and beyond. To complement it, in the following continuum functional methods will be paired with this approach. Their particular strengths are exactly at the extremities of scales: in the far infrared and ultraviolet and for light and heavy states, as well as combinations of all this. Thus, they are best where the lattice artifacts are strongest, and both approaches naturally compensate their respective weaknesses and combine their strengths.

\subsection{Quantum equations of motion}\label{zerot:dse}

\subsubsection{Formulating the equations}

The quantum equations of motion, or Dyson-Schwinger equations (DSEs), can be obtained for full, connected, or vertex functions. The starting point are the respective equations for the generating functionals \pref{quant:genfcf}, \pref{quant:gencon}, and \pref{quant:gendse}. Here, the most useful one will be the one for vertex functions \pref{quant:gendse}, though \pref{quant:gencon} will be valuable in section \ref{zerot:sti}.

The equation \pref{quant:gendse} yields an equation for the one-point vertex function for the selected field. Repeated derivations with respect to other fields then generate a hierarchy of equations for higher-$n$-point correlation functions by virtue of \pref{quant:vertexdef}. This derivation is essentially an algorithmic problem, and can be formalized and thus automatized \cite{Alkofer:2008nt,Huber:2011qr}. It should be noted that the equations may differ in form depending on the order of derivatives \cite{Alkofer:2008nt}, though of course this only corresponds to rearrangements due to identities relating different vertex functions.

Before being able to obtain a set of equations it is first necessary to perform the gauge-fixing. To enforce the Landau-gauge condition the usual Faddeev-Popov formulation \pref{quant:gfpi} can be used \cite{Bohm:2001yx}. Thus, there will be not only gluon fields but also ghost and anti-ghost fields, and both types of fields have to be included in the construction of correlation functions. This fixes the perturbative part of the gauge.

The further restrictions are less simple to implement. First of all, it has been shown that the form of the equations is invariant whether they are derived on the whole of field-space, or within a fixed Gribov region \cite{Zwanziger:2001kw,Maas:2011ba}. The reason is that the Faddeev-Popov determinant, which appears after integrating out the ghost fields, is by definition zero on the boundaries of these regions, effectively cutting off the integration.

However, this is not sufficient to fix which of the Gribov regions is used. One possibility is to assume that to replace the $\theta$ function in \pref{quant:fgr} by a $\delta$ function \cite{Zwanziger:1992qr} is a valid procedure. This possibility will not be pursued here, see \cite{Zwanziger:1992qr,Capri:2007ix,Dudal:2007cw,Dudal:2008sp,Dudal:2008rm,Baulieu:2008fy,Dudal:2008xd,Gracey:2009mj,Huber:2009tx,Zwanziger:2009je,Zwanziger:2010iz,Gracey:2010df,Nakajima:2009rw,Dudal:2011gd} for investigations based on this assumption.

An alternative is to enforce the selection of the Gribov region by selection conditions on the set of vertex functions $\{\Gamma^1,...,\Gamma^n,...\}$, which satisfy the various non-local gauge conditions. As addressed in section \ref{quant:resgrisin}, there is not yet a sufficient condition known to restrict the set of solutions to those belonging to the first Gribov region. Here, the assumption will be made that it is sufficient to require that the ghost two-point function $\Gamma^{\bar{c}c}$ is negative semi-definite in both position and momentum space. The remaining solution manifold can then be reduced further. E.\ g., implementing the absolute Landau gauge \pref{quant:abslg} corresponds to selecting the solution of the DSEs which has the smallest integrated diagonal gluon propagator. Implementing a Landau-$B$ gauge can be done by implementing \pref{quant:lbg} as a restriction for the ghost propagator, i.\ e., requiring it to coincide with the desired value of $B$,
\be
B=\lim_{p\to 0}\frac{p^2 D_G(p^2,\mu^2)}{\mu^2 D_G(\mu^2,\mu^2)}\label{zerot:lbgdse}.
\ee
\no Thus, also $B$ appears here explicitly. Note that this prescription has the same effect as the Landau-$B$ gauge condition, in particular in contrast to the condition \pref{quant:lbg}, this is a condition on the averaged ghost propagator and not on an individual configuration. Hence, at this level this cannot be a precise statement about non-perturbative gauge-fixing in a functional approach, despite that the same results are obtained. Of course, if either of the conjectured restrictions \pref{quant:bgauge} or \pref{quant:bgauge2} were correct, and DSEs could be derived despite the non-locality involved, this would be an alternative. This will not be done here, but is should be noted that a modification of type \pref{quant:bgauge2}, because it is bilinear in the ghost fields, will only appear in the DSE of the ghost propagator, and will only contribute to the position-space averaged ghost propagator \cite{Maas:unpublished}.

It remains to formulate the case of the minimal Landau gauge. The only proposal of constructing an equivalent gauge choice with continuum methods is so far based on \pref{quant:bgauge2} with the Lagrange parameter set to zero. Assuming this to be to correct and furthermore to be equivalent to the above ansatz in equation \pref{zerot:lbgdse}, there is then an indirect way: In this case, in fact a Landau-$B$ gauge is chosen, but the value of $B$ is the one obtained in the minimal Landau gauge in lattice calculations. This, of course, requires an outside source like lattice gauge theory for this information. Since the minimal Landau gauge just produces the average value of the ghost propagator, it will yield the average value of $B$, thus giving an indirect handle on fixing the minimal Landau gauge in DSEs. This prescription is correct, even if the ghost propagator coincided for all Landau-$B$ gauges in the thermodynamic limit. If, however, the Landau-$B$ gauges did not yet completely exhaust all permitted constraints to select a Landau gauge, this would imply that further constraints on further correlation functions can be implemented as well, and would have to be specified if they were averaged over as well. This procedure leads to results in agreement between lattice and functional calculations, within the approximations made, as shown in section \ref{zerot:prop}.

A practical implementation of the Landau-$B$ gauge, and thus the minimal Landau gauge, can then be obtained using the DSE for the ghost propagator. Since it is also the simplest DSE, it will now serve as an example for deriving DSEs \cite{Alkofer:2000wg,Rivers:1987hi}.

Neglect for a moment the problem of Gribov copies. The ghost sector of the action is then given by
\be
S_{gh}=\int d^dz\bar c^c(z)\pd_\rho(\delta^{cd}\pdm+gf^{cde}A^e_\rho(z))c^d(z)\label{ghostaction}.
\ee
\noindent Inserting this into \pref{quant:gendse} and differentiating\footnote{All derivatives used are left-derivatives.} with respect to $\bar c^a(x)$ yields
\be
\left(\tilde{Z}_3\pd^{2x}c^a(x)-gf^{ade}\pdm^xA^e_\mu(x)c^d(x)+\frac{\delta\Gamma}{\delta\bar c^a(x)}\right)e^W=0\nn,
\ee
\noindent where the $x$-index on a $\pd$ indicates the variable with respect to which to derive, and the wave-function renormalization of the ghost with the wave-function renormalization constant $\tilde{Z}_3$ has been taken explicitly into account. The renormalization constants of the interactions are absorbed into the derivatives of $\Gamma$, and will only reappear if these are written in terms of vertex functions, and the coupling constant is already the renormalized one. Replacing the fields by their respective derivatives and dividing, after taking the derivatives, by $\exp(W)$, yields
\be
\tilde{Z}_3\pd^{2x}c^a(x)-gf^{ade}\pdm^x\left(\frac{\delta W}{\delta j_\mu^e(x)}\frac{\delta W}{\delta\bar\eta^d(x)}+\frac{\delta^2 W}{\delta j_\mu^e(x)\delta\bar\eta^d(x)}\right)+\frac{\delta\Gamma}{\delta\bar c^a(x)}=0\nn,
\ee
\no where $j$ and $\eta(\bar{\eta})$ are the sources for the gluon and the (anti-)ghost, respectively. As a general feature of such derivations, products of single derivatives of $W$ appear. When deriving such terms again only once with respect to the fields, always at least one single derivative remains, which can be replaced by a classical field. When setting the classical sources to zero at the end, the classical fields are also set to zero and therefore these terms always vanish. Hence they can already be neglected at this stage of the calculation, and will not appear furthermore. Note that this is not necessarily true if more than one further derivative would be applied.

So the remaining expression is
\be
\tilde{Z}_3\pd^{2x}c^a(x)-gf^{ade}\pdm^x\frac{\delta^2 W}{\delta j_\mu^e(x)\delta\bar\eta^d(x)}+\frac{\delta\Gamma}{\delta\bar c^a(x)}=0\nn.
\ee
\no To obtain the equation for the ghost propagator, this equation is derived once more with respect to $c^b(y)$ which leads to
\be
\tilde{Z}_3\pd^{2x}\delta^{ab}\delta(x-y)-gf^{ade}\pdm^x\frac{\delta^3 W}{\delta c^b(y)\delta j_\mu^e(x)\delta\bar\eta^d(x)}+\frac{\delta^2\Gamma}{\delta c^b(y)\delta\bar c^a(x)}=0\label{methods:gdse1}.
\ee
\no The last term will finally lead to the appearance of the ghost propagator, since $W$ is the Legendre transform of $\Gamma$. The second term of \pref{methods:gdse1} yields the interaction part. Using
\be
\frac{\delta^2 W}{\delta j^e_\mu(x)\delta\bar\eta^d(x)}=-\int d^dzd^dw\frac{\delta^2 W}{\delta j_\nu^f(z)\delta j^e_\mu(x)}\frac{\delta^2\Gamma}{\delta\bar c^g(w)\delta A_\nu^f(z)}\frac{\delta^2 W}{\delta\eta^g(w)\delta\bar\eta^d(x)}\nn,
\ee
\no where the minus arises due to the anti-commuting derivatives. Since all mixed two-point functions vanish in Landau gauge, it is possible to write down the result in position space
\bea
D^{ab-1}_G(x-y)&=&\tilde{Z}_3\pd^{2x}\delta(x-y)\nn\\
&&+gf^{ade}\pdm^x\int d^dzd^dw D^{ef}_{\mu\nu}(x-z)D^{dg}_G(x-w)\Gamma^{c\bar cA;bgf}_\nu(y,w,z).\nn
\eea
\no Herein the gluon propagator and the ghost-gluon vertex appear.

\begin{figure}
\includegraphics[width=\textwidth]{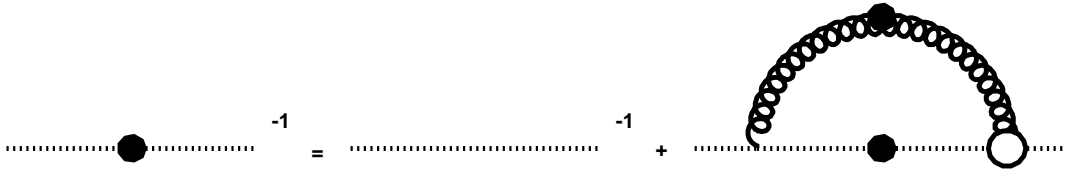}
\caption{The DSE \pref{zerot:ghe} for the ghost. Dotted lines are ghosts, curly lines are gluons. Lines with a dot are fully dressed propagators, white circles are full vertices, and all other vertices are bare.}
\label{fig:gheq}
\end{figure}

Replacing all expressions with their Fourier-transformed\footnote{All momenta are always defined incoming and momentum conservation at the vertices is taken into account. Hence in principle one of the arguments of the vertices could be dropped, but since this depends on conventions, all are kept.} and afterwards dropping\\ $\int d^dp/(2\pi)^d \exp(-ip(x-y))$ produces the result in momentum space as
\bea
D^{ab-1}_G(p)&=&-\tilde{Z}_3\delta^{ab}p^2\nonumber\\
&&+\int\frac{d^dq}{(2\pi)^d}\Gamma_{0\mu}^{c\bar{c}A\indexsep dae}(q,p,-q-p)D^{ef}_{\mu\nu}(p+q)D^{dg}_G(q)\Gamma^{c\bar cA\indexsep bgf}_\nu(p,q,-p-q)\label{zerot:ghe}\nn,
\eea
\no where the tree-level ghost-gluon vertex
\be
\Gamma_{0\mu}^{c\bar{c}A;abc}(p,q,k)=igf^{abc}q_\mu\label{smethod:tlcca}
\ee
\no has been made explicit. The resulting equation can be depicted graphically, similarly to Feynman graphs. For the case of the ghost equation, this is shown in figure \ref{fig:gheq}.

The DSE for the ghost propagator \pref{zerot:ghe} already shows the generic features of a DSE for a primitively divergent vertex function. These differ from the ones for non-primitively divergent vertex functions only by the appearance of the tree-level term \cite{Rivers:1987hi,Huber:2009wh}, in this case the bare ghost propagator. The further terms are self-energy contributions, which have the form of loop integrals like in perturbation theory, but differ from them by the appearance of the full vertex functions. Thus, these equations are self-consistency equations and take the form of non-linear integral equations \cite{Rivers:1987hi}.

Another property is the generic structure of the equations, which is
\be
\Gamma^n(p_1,...,p_{n-1})=\Gamma^n_0(p_1,...,p_{n-1})+\Pi(p_1,...,p_{n-1},\Gamma^2,\Gamma^3,...,\Gamma^n,\Gamma^{n+1},\Gamma^{n+2}),\nn
\ee
\no where it has been used that $\Gamma^1=0$ in Yang-Mills theory and $\Gamma^n_0$ denotes the tree-level value of $\Gamma^n$. The important property is that $\Gamma^n$ does not only depend on all vertex functions with lower $n$, but in general also on those with\footnote{In a theory with higher than quartic terms in the fields even higher $n$-point correlation functions can appear.} $n+1$ and $n+2$. Thus, this set of equations, being the analogue of the classical equations of motion, is coupled. In fact, it forms an infinite hierarchy of equations for the correlation functions.

\begin{figure}
\includegraphics[width=\textwidth]{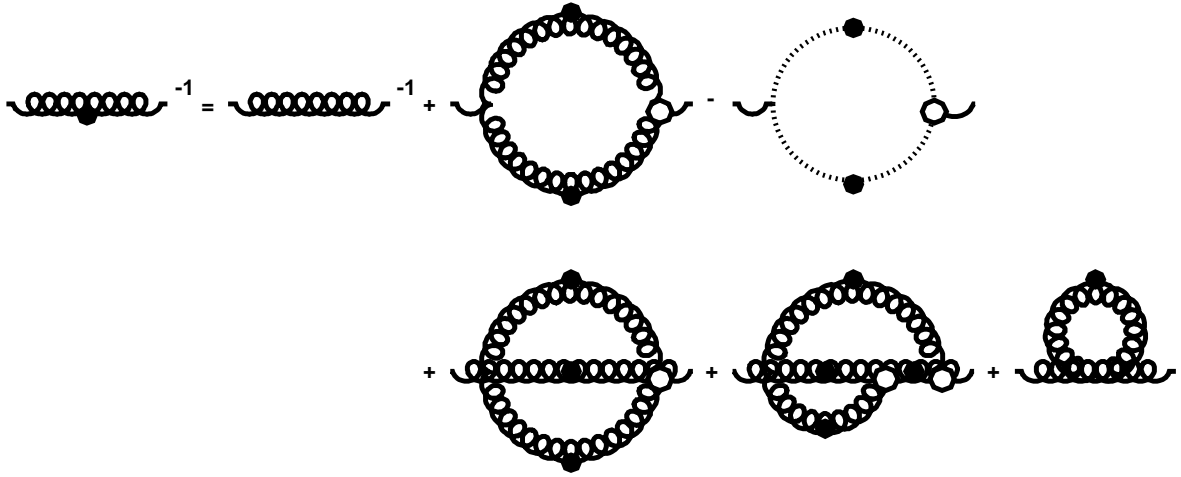}
\caption{\label{fig:gleq}The DSE for the gluon propagator.}
\end{figure}

Furthermore, the number of self-energy diagrams proliferates quickly with the number of legs \cite{Rivers:1987hi,Alkofer:2008nt}, but at most two-loop diagrams will be obtained. In addition, in contrast to perturbation theory \cite{Rivers:1987hi,Negele:1988vy}, the self-consistency of the equations guarantees that the equations of motion are still well-defined, though their practical treatment becomes increasingly complicated. Already the gluon equation, shown in figure \ref{fig:gleq}, gives an impression of the complexity. An important aside is here that the sign of the ghost-loop for tree-level quantities is opposite from the one of the gluon loop at one-loop order. Thus, the gluons provide an anti-screening behavior at large momenta, while the ghosts provide a screening behavior. Adding both up, this will finally yield the asymptotic freedom of Yang-Mills theory, as the gluons dominate at large momenta \cite{Peskin:1995ev}.

\begin{figure}
\includegraphics[width=\textwidth]{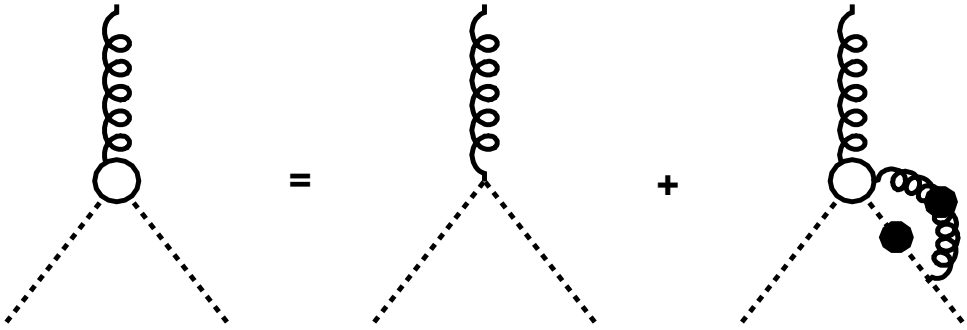}
\caption{\label{fig:ghglv}The DSE for the ghost-gluon vertex.}
\end{figure}

The simplest vertex equation is the one for the ghost-gluon vertex, shown in figure \ref{fig:ghglv}. The particular structure of this vertex will play a great role in the following sections. Various further examples for DSEs for vertices can be found in \cite{Alkofer:2008nt,Schleifenbaum:2004id,Alkofer:2008dt,Kellermann:2008iw,Huber:2011qr}.

Returning now to the problem of Gribov copies, the form of the ghost propagator DSE suggests a way to implement the Landau-$B$ gauges \cite{Fischer:2008uz,Maas:2009se}. Multiplying the equation \pref{zerot:ghe} by $p^2$ and neglecting for now the color structure\footnote{This will be justified by the results presented in section \ref{szerot}.}, it becomes an equation for the ghost dressing function. By subtracting the ghost equation from itself at zero momentum, it takes the form
\be
\frac{1}{G(p)}-\frac{1}{G(0)}=\Pi^{\bar{c}c}(p)-\Pi^{\bar{c}c}(0)\nn,
\ee
\no where $\Pi^{\bar{c}c}$ is the ghost self-energy. Multiplying then by the ghost dressing function at $\mu$ yields
\be
\frac{G(\mu)}{G(p)}-\frac{1}{B}=G(\mu)\left(\Pi^{\bar{c}c}(p)-\Pi^{\bar{c}c}(\mu)\right)\nn.
\ee
\no Since the value of $G(\mu)$ is fixed by the renormalization condition, this completely specifies the equation, and introduces the $B$ parameter directly. As noted above, this method can also be used to solve the equations in the minimal Landau gauge. Only the absolute Landau gauge requires a different approach. In this case, the full solution manifold of the equations has to be constructed, and then the one solution has to be selected which absolutely minimizes the (renormalized) expression \pref{quant:abslg}.

The remaining problem is then to find a solution to the DSEs with a positive semi-definite ghost propagator. For such non-linear integral equations this is quite a formidable task, and rarely possible exactly. In particular, it is in general necessary to truncate the DSEs, which usually implies to neglect most of the equations for the higher correlation functions. This will be detailed below in section \ref{sec:trunc}. Before turning to this task, it is useful to first introduce a related formalism, the (functional) renormalization group equations.

Note that with this the non-perturbative part of the gauge-fixing, if correct, has been reformulated as a boundary condition to the DSEs. This is also directly visible from the fact that in \pref{quant:bgauge2} 
the added term to the action is a pure boundary-term, i.\ e.\ it includes only the ghost fields at infinity when using the Gauss theorem .Such boundary conditions can play indeed a significant role in the treatment of DSEs, even when discussing genuine physical, instead of mere gauge-fixing, features \cite{Garcia:1996np,Guralnik:2007rx,Bender:2010hf}. The boundary conditions are therefore possibly not only permissible, but perhaps even a necessary ingredient for DSEs.

\subsubsection{Renormalization group methods}\label{zerot:erg}

The functional renormalization group equations (FRG(E)s) \cite{Pawlowski:2005xe,Rosten:2010vm,Kopietz:2010zz} are very similar in nature to the quantum equations of motion. Without going into further details, they are, essentially, functional derivatives of the quantum equations of motion.

The most important element in their construction is to introduce regulator functions. Since physical observables in a renormalized theory are independent of such regulators, this independence can then be used to derive equations for the correlation functions. A solution is then obtained in the limit of removing the regulator at the end. If the system of equations is truncated, the choice of the regulator can have a non-trivial impact \cite{Pawlowski:2003hq,Litim:2001up}, though for an exact solution its choice is irrelevant.

Thus, Instead of deriving the functional renormalization group equations from the quantum equations of motion, a better starting point is the effective action $\Gamma$ where a regulator has been inserted. This then yields a starting equation, similar to \pref{quant:gendse} for DSEs, as a starting point. The FRGEs can then be derived directly from this equation,
\be
\frac{\delta \Gamma_k}{\delta A}=\frac{\delta S}{\delta A}\left[\int d^d x\frac{1}{\Gamma^2_k(x,y)+R_k(x,y)}\frac{\delta}{\delta A}+A\right]\nn,
\ee
\no where $R$ is the regulator function and $k$ is the regulating scale, which has to be sent to zero to obtain the full result for the correlation functions. From this generating functional, similar as for the quantum equations of motion, a hierarchy of coupled non-linear integral equations for the vertex functions can be obtained.

However, the structure of these equations is significantly different from the ones of the equations of motion. One obvious difference is the appearance of the regulator function. More importantly, the appearing loop diagrams are always only one loop. The resulting drawback is that only full vertex functions appear, instead of also bare ones as in the equations of motion \cite{Pawlowski:2005xe}.

The two systems of equations are thus sufficiently different that finding a solution to both of them is a non-trivial consistency check \cite{Fischer:2006vf,Fischer:2009tn}. It has been shown that indeed the one-parameter family of solutions, which is obtained from the DSEs by selecting the $B$ parameter, can also be obtained with FRGEs \cite{Fischer:2008uz,Weber:2011aa}, in particular the one for $B=\infty$ \cite{Fischer:2008uz,Pawlowski:2003hq}. On the other hand, the fact that indeed both systems are solved by the same solution is sufficiently restrictive to show that the infrared structure of the vertex functions has to be qualitatively unique in the $B=\infty$ case \cite{Fischer:2006vf,Fischer:2009tn}, under certain weak assumptions.

Furthermore, the freedom in choosing the regulator can be used to compensate for the effects of truncations \cite{Pawlowski:2005xe,Litim:2001up}. As a consequence, the results from functional studies for propagators most closely resembling the ones from lattice calculations up to now have been obtained using FRGEs \cite{Fischer:2008uz,Pawlowski:unpublished}. Nonetheless, for the solution of functional renormalization group equations otherwise the same applies concerning the necessity of a truncation as in the case of the quantum equations of motion.

In total, both sets of equations have, from a practical point of view, their individual strengths and weaknesses. Exploiting the combination of both of them has so far been most useful \cite{Fischer:2006vf,Fischer:2009tn,Huber:2009wh}, and promises to be so in the future as well. Given that both systems of equations are very similar nevertheless, for simplicity here only the case of DSEs will be treated, with only given some additional results from FRGEs where they provide particular useful additional insights.

\subsubsection{Ultraviolet}\label{ssuv}

At large momenta, the leading behavior of the correlation functions can be obtained using perturbation theory in the coupling constant, due to asymptotic freedom \cite{Bohm:2001yx,Peskin:1995ev}. The corresponding equations are obtained by expanding all correlation functions in the DSEs in the coupling constant \cite{Rivers:1987hi}. As a consequence, the contributions from correlation functions of order $n+m$ in equations for correlation functions of order $n$ are then subleading by $m$ powers of the coupling constant, and can thus be neglected if the expansion order is less than $m$. Therefore, at any given order of the perturbative expansion the number of equations is finite, and identical to the ones constructed with conventional Feynman rules.

Since the results in two, three, and four dimensions are interesting and have turned out to be important to the understanding of Yang-Mills theory, it is worthwhile to also analyze the perturbative expansion in all cases. However, in two dimensions the perturbative integrals are ill-defined, and require an infrared cut-off, not leading to a viable perturbative expansion \cite{Dudal:2008xd}\footnote{Strictly speaking, this also applies to three dimensions \cite{Jackiw:1980kv}, but there the breakdown of perturbation theory only occurs at higher order, and is thus not relevant for the purpose here. In particular, the Linde problem at finite temperature \cite{Linde:1980ts}, to be discussed in chapter \ref{sfinitet}, is also connected to this.}. Thus, only the case of three and four dimensions remain, which exhibit a qualitatively different behavior. Therefore, it is worthwhile to consider the situation in four and three dimensions separately.

The characteristic properties of perturbation theory are already obtained from the leading corrections to tree-level. At this order, all vertices become their tree-level counter-parts in the propagator equations. Corrections to the vertices are then of higher order in the coupling constant. The much more complex case of higher-order corrections including higher-order correlation functions is in principle analogous, though much more challenging in detail \cite{Bohm:2001yx,Collins:1984xc}.

The generic properties of correlation functions at large momenta can already be obtained from a dimensional analysis. This can be illustrated most easily for the case of the propagators. For this analysis it is important that Yang-Mills theory is renormalizable in four dimensions, and in Landau gauge the highest appearing divergence is logarithmic \cite{Collins:1984xc}.

As a consequence, in four dimensions, the generic propagator behaves as \cite{Bohm:2001yx}
\be
p^2D(p)\to \zeta\left(1+\omega\ln\frac{p^2}{\Lambda_\mathrm{YM}^2}\right)^\alpha+{\cal O}(g^4)\nn,
\ee
\no where $\zeta$ is a normalization factor and $\Lambda_\mathrm{YM}$ is a dynamically generated scale. The appearance of it, called dimensional transmutation \cite{Peskin:1995ev,Alkofer:2000wg}, is a pure quantum effect. The classical theory is scale invariant, and thus the propagator would have to scale like the canonical dimension of the fields, and thus $p^2 D(p)$ would be constant. The quantum effects break this scale (or dilatation) symmetry of the classical theory, and give the theory a scale, $\Lambda_\mathrm{YM}$. Since this is the only scale, $p^2 D(p)$ can only depend on $p^2/\Lambda_\mathrm{YM}^2$. The logarithmic behavior is then typical for such renormalizable field theories like Yang-Mills theory.

In contrast, in three dimensions the propagators asymptotically take the form
\be
p^2 D(p)\to 1+\frac{a g^2}{p}+{\cal O}\left(\frac{g^4}{p^2}\right)\label{uv:3dexp}
\ee
\no with some constant $a$. This is the simplest, asymptotically free possibility just by power-counting, since $g^2$ has dimension of mass. The classical theory is therefore not scale-invariant, in contrast to the four-dimensional case.

Though, as stated, in the two-dimensional case perturbation theory is not applicable, the same arguments, and the fact that $g$ now has the dimension of mass, suggest that the asymptotic behavior will be
\be
p^2 D(p)\to 1+\frac{a g^2}{p^2}+{\cal O}\left(\frac{g^4}{p^4}\right)\label{uv:2dexp},
\ee
\no which turns out to be approximately correct \cite{Maas:2007uv,Maas:2010qw}. It is thus similar to the behavior in three dimensions. It should be noted that the possibility for the expansions \pref{uv:3dexp} and \pref{uv:2dexp} are solely based on a dimensional analysis, and therefore remain valid even if perturbation theory is not applicable.

Higher order correlation functions show, of course, a more complicated structure, as the number of external momenta increases \cite{Bohm:2001yx}. Still, the characteristic perturbative behavior is logarithmic in four dimensions, and polynomial corrections to the tree-level value in lower dimensions. Note that in lower dimensions, operator-product expansion corrections \cite{Peskin:1995ev,Collins:1984xc} could mix with the perturbative contributions, in contrast to four dimensions, where they are sub-leading.\\

\no{\bf Three dimensions}\\

At leading order in the perturbative expansion the two-loop diagrams in the gluon equation drop out, since they are of order $g^4$, compared to the order $g^2$ one-loop diagrams. A second important point to this order is resummation, i.\ e.\ the inclusion of all one-particle reducible contributions for a self-consistent solution of the DSEs to a given order \cite{Alkofer:2000wg,Rivers:1987hi,Bohm:2001yx}. In four dimensions, resummed and ordinary perturbation theory do not coincide. However, in three and two dimensions resummation only produces effects from order $g^4$ on. This follows from dimensional arguments alone, since only tree-level expressions can contribute at order $g^2$ in the loop diagrams, such that a polynomial correction of type $g^2/p$ appears. The next higher orders are of $g^4$, and thus require a factor $1/p^2$. However, there resummation sets in, as factors $(g^2/p)^2$ appear in next-to-leading-order from the leading order. Therefore, resummation effects do not occur at leading order, but only from next-to-leading order onwards.

The next problem is that each of the loop integrals in the propagator equations is possibly linearly divergent from superficial power-counting, and only their combined result is finite. The latter can be deduced from the fact that Yang-Mills theory has only a logarithmic divergence in Landau gauge in four dimensions \cite{Collins:1984xc}. To cope with these divergences in perturbation theory it is convenient to use dimensional regularization\footnote{Which in odd dimensions implicitly already fixes the renormalization scheme \cite{Maas:2004se,Collins:1984xc,Bohm:2001yx}.} \cite{Peskin:1995ev,Bohm:2001yx} to perform the integrals. However, dimensional regularization requires an analytic continuation to non-integer dimensions. It is a-priori not clear whether such an analytic continuation exists for the non-perturbative correlation functions \cite{ZinnJustin:2002ru}. Even if it exists, it is generally not known beforehand, and thus dimensional regularization can only be applied if an assumption on this structure is made \cite{Kizilersu:2000dk}. The alternative is to use a different regularization scheme.

It is thus more instructive to use instead of dimensional regularization the same regularization scheme used later for the non-perturbative calculations. This is even more important as, besides the lattice, no non-perturbative regularization scheme is known, which preserves manifest gauge covariance \cite{ZinnJustin:2002ru}. This is not in itself a problem, as this merely implies that the counter-terms will also not be gauge-covariant\footnote{Note that such a procedure modifies the corresponding Slavnov-Taylor identities in a well-defined way \cite{Fischer:2008uz,Pawlowski:2005xe,Ellwanger:1994iz}.} \cite{Collins:1984xc}. Only the renormalized correlation functions will then be again gauge-covariant. Thus, studying this effect already in perturbation theory, where it is possible to compare to a gauge-covariant regularization scheme, provides insights into how to treat these problems in the non-perturbative calculations.

This does not matter for the ghost equation, which due to the non-renormalization of the ghost-gluon vertex\footnote{An explicit leading-order confirmation can be found in \cite{Schleifenbaum:2004di}.} \cite{Taylor:1971ff,Fischer:2005qe,Watson:2000ph} is actually fully finite in three dimensions. Performing the loop integral yields the result for the ghost dressing function as \cite{Maas:2004se}
\be
\frac{\delta^{ab}}{G(p)}=\delta^{ab}-\frac{g^2C_A\delta^{ab}}{16p}\label{ghp:3dpert},
\ee
\no where $C_A$ is the adjoint Casimir of the gauge group. Thus, the perturbative correction is polynomial in the momentum. Hence, at very large momenta just the tree-level behavior remains, in a manifest display of asymptotic freedom. On the other hand, at a small, finite momentum the dressing function diverges, a manifestation of the Landau pole, and the explicit breakdown of perturbation theory.

The equation for the gluon propagator is different. The cutoff-regularization produces spurious divergences in this case. First, it is useful to note the result obtained using dimensional regularization, yielding \cite{Maas:2004se}
\bea
\frac{\delta^{ab}\left(\delta_\mn-\frac{p_\mu p_\nu}{p^2}\right)}{Z(p)}&=&\delta^{ab}\left(\delta_\mn-\frac{p_\mu p_\nu}{p^2}\right)-\delta^{ab}\frac{g^2C_A}{64p}\left(\delta_\mn+\frac{p_\mu p_\nu}{p^2}\right)-\delta^{ab}\frac{g^2C_A}{32p}\left(5\delta_\mn-6\frac{p_\mu p_\nu}{p^2}\right)\nn\\
&=&\left(\delta_\mn-\frac{p_\mu p_\nu}{p^2}\right)\delta^{ab}\left(1-\frac{11g^2C_A}{64p}\right)\nn,
\eea
\no where in the first line the first term is the tree-level contribution, the second term comes from the ghost loop, the third term from the gluon loop, and the tadpole vanishes upon dimensional regularization. The individual diagrams are not individually transverse, but only their sum reproduces this property \cite{Peskin:1995ev}. This cancellation of the longitudinal components is actually guaranteed by the gauge condition \cite{Cucchieri:2008zx} and by the Slavnov-Taylor identities (STIs), as discussed in section \ref{zerot:sti}.

This should be compared to the result obtained using a cut-off regulator,
\bea
\frac{\delta^{ab}\left(\delta_\mn-\frac{p_\mu p_\nu}{p^2}\right)}{Z(p)}&=&\delta^{ab}\left(\delta_\mn-\frac{p_\mu p_\nu}{p^2}\right)\nn\\
&&-\delta^{ab}\frac{g^2C_A}{64p}\left(\delta_\mn+\frac{p_\mu p_\nu}{p^2}\right)-\frac{g^2 C_A\Lambda}{6\pi^2p^2}\delta^{ab}\delta_\mn\nn\\
&&-\delta^{ab}\frac{g^2C_A}{32p}\left(5\delta_\mn-6\frac{p_\mu p_\nu}{p^2}\right)-\frac{4g^2 C_A\Lambda}{6\pi^2p^2}\delta^{ab}\delta_\mn\nn\\
&&+\frac{2g^2 C_A\Lambda}{3\pi^2p^2}\delta^{ab}\delta_\mn\nn,
\eea
\no where the first line is the tree-level part, the second line comes from the ghost loop, the third from the gluon loop, the last from the tadpole, and $\Lambda$ is the cut-off. The finite part agrees with the previous result. However, there remains a longitudinal part proportional to the cut-off, which is spurious, and behaves like a gluon mass. It emerges since a cut-off regulator does not respect the gauge symmetry \cite{Peskin:1995ev,Collins:1984xc,ZinnJustin:2002ru}. As a consequence, the gluon propagator is no longer transverse, as it ought to be in Landau gauge.

However, this is not a problem. Such spurious divergences or other artifacts of a non-gauge-invariant regulator can always be compensated for by non-gauge-invariant counter-terms \cite{Collins:1984xc}. In this case, the counter-term Lagrangian to be added takes the form
\bea
{\La}_c&=&A_\mu^a(x) A_\mu^a(x)\delta m^2\nn\\
\delta m^2&=&-\frac{g^2 C_A\Lambda}{6\pi^2}\nn.
\eea
\no which has the dimension of a mass squared since the coupling constant squared in three dimensions has the dimension of a mass. This counter-term makes the gluon self-energy once more finite and transverse, and in agreement with the result using dimensional regularization. It should be noted that these counter-terms are fully fixed by the requirements of gauge invariance, as in the present case, and thus no additional arbitrary renormalization constants appear.

Similarly, any other kind of such artifacts of a non-gauge invariant regularization procedure can be absorbed in corresponding counter-terms. Thus, the one-loop perturbative result for the gluon dressing function becomes
\be
Z(p)=\left(1-\frac{11g^2C_A}{64p}\right)^{-1}\label{gp:3dpert}.
\ee
\no This already shows that in the non-perturbative treatment, where dimensional regularization is not possible without knowledge of the analytic structure, such counter terms will naturally appear, and have to be taken into account.\\

\no{\bf Four dimensions}\\

The situation in four dimensions is different for two reasons. One is that the dressing functions are not finite, but actually divergent, and therefore have to be renormalized even when using dimensional regularization. The second is that resummation already sets in at one-loop order. This is due to the fact that corrections are logarithmic instead of polynomial, and two logarithms at different order can be combined into one, as the momentum dependence is the same. Thus, to obtain a self-consistent solution of the DSEs at a given order requires to perform resummed perturbation theory in four dimensions. Since the DSEs are inherently self-consistent, only such a solution makes sense as a perturbative approximation to the full solution. The resummed solutions can be obtained with a variety of methods from ordinary perturbation theory \cite{Bohm:2001yx,Peskin:1995ev}, but here simply an appropriate ansatz will be made.

The form of the perturbative dressing functions in four dimensions is \cite{Bohm:2001yx}
\bea
G(p)&=&G(s)\left[\omega\log\left(\frac{p^2}{s^2}\right)+1\right]^\delta\label{ghp:4dpert}\\
Z(p)&=&Z(s)\left[\omega\log\left(\frac{p^2}{s^2}\right)+1\right]^\gamma\label{gp:4dpert},
\eea
\no with the anomalous dimensions $\delta$ and $\gamma$, and the parameter $\omega$, which all three will be determined from self-consistency, the subtraction point $s$, and the values $Z(s)$ and $G(s)$ of the dressing functions at the subtraction point.

The perturbative domain is reached for momenta large compared to the scale $\Lambda_{\mathrm{YM}}$. It is furthermore sufficient to only inspect the large momentum tail of the integrals, cutting off the loops in the infrared at $p$, the scale of the external momenta \cite{Fischer:2002hna,Fischer:2003zc,vonSmekal:1997vx}.

Since the external momentum is small, it is justified to approximate the internal dressing functions by $G(p+q)\approx G(q)$ and then performing the angular integrals explicitly. The explicit form of the ghost equation is in this limit given by
\be
\frac{1}{G(p^2)}=\tilde Z_3-\frac{3g^2C_A}{64\pi^2}\int_{p^2}^{\Lambda^2}dy\frac{G(y)Z(y)}{y} \nn,
\ee
\no with $y=q^2$ and the wave-function renormalization constant $\tilde Z_3$, and the upper cut-off just indicates the use of a regulator. As in three dimensions, there are no spurious divergences, even when using a cut-off regulator. The (logarithmically) divergent upper boundary of the integral can then be absorbed by the wave-function renormalization constant. Putting in the ans\"atze \pref{ghp:4dpert} and \pref{gp:4dpert} yields that the solutions must fulfill
\bea
\omega&=&-\frac{3g^2C_A}{64\pi^2\delta}\nn\\
\delta&=&-\frac{1}{2}(\gamma+1)\nn
\eea
\no to be self-consistent solutions, reproducing the known relations \cite{Bohm:2001yx} between these perturbative parameters.

For the gluon equation, a new problem appears. In fact, there is no self-consistent solution to it and the ghost equation simultaneously, as long as the three-gluon vertex retains its tree-level value \cite{Fischer:2002hna,vonSmekal:1997vx}. The reason is that, due to resummation, high-order effects of the vertices also contribute to generate a self-consistent solution. This does not affect the ghost equation, due to the non-renormalization of the ghost-gluon vertex. Therefore, it would be necessary to solve the corresponding equations of propagators and vertices simultaneously. This can be addressed better with the renormalization-group improved perturbation theory, but this is a rather lengthy calculation \cite{Peskin:1995ev,Bohm:2001yx}. Here, it is sufficient to note that this yields the missing quantity $\delta$ as
\be
\delta=-\frac{9}{44}\nn.
\ee
\no In fact, the knowledge of \pref{ghp:4dpert} and \pref{gp:4dpert} can then be used in turn to limit the structure of the three-gluon vertex \cite{Kizilersu:2009kg}. This is important for truncations to be discussed below.

Again, asymptotic freedom yields that these are indeed the leading ultraviolet contributions to the propagators, and at high-energies the non-perturbative contributions will only be  sub-leading corrections to it, falling off as inverse powers of the momenta \cite{Peskin:1995ev}.

As a final remark it should be noted that the gauge algebra does not enter the ultraviolet behavior in a qualitative manner, but only alters some of the coefficients, and in four dimensions the value of the anomalous exponent at higher orders, but without changing its sign.

\subsubsection{Infrared}

In analogy to finding the (resummed) self-consistent solutions at large momenta, the aim of a non-perturbative infrared analysis is to find self-consistent solutions to the equations at asymptotically small momenta. Two qualitatively different types of solutions have been found over the time, the so-called scaling and decoupling solutions\footnote{Originally, as a third possibility a singular solution has been proposed, embodying the concept of infrared slavery \cite{Mandelstam:1979xd,Buttner:1995hg,Alkofer:2000wg}.  This possibility is at the current time not a consistent solution of functional studies anymore, and there are no indications using lattice gauge theory that it could exist, see chapter \ref{szerot}. In fact, there is a proof \cite{Zwanziger:2012se} that such an enhancement is not possible.}. Here, the latter name is chosen for the characteristic momentum-behavior of its running coupling in a certain scheme, which is infrared vanishing. It is also referred to as a massive solution \cite{Binosi:2009qm} because the gluon exhibits a finite screening mass. However, as discussed below this occurs also in the scaling case with a finite or infinite screening mass, the later being an overscreening behavior. Thus, this distinction is not entirely satisfactory. The most elementary difference is found in the ghost dressing function, which is for the scaling case infrared divergent, while it is infrared finite, like for a photon, in the decoupling case. Thus it is possible to refer to the decoupling case as the finite-ghost case, which will be done here.

For a discussion of the infrared behavior, it is quite useful to discuss finite-ghost-like and scaling-like behavior separately, which turn out to be quite distinct. Though the scaling solution will be formally obtained in the limit of $B\to\infty$, it is not settled which is the relation of the scaling case to this gauge-fixing condition, in particular if it turns out that only finite $B$ values or even only a single value of $B$ is admissible in the first Gribov region. This will be discussed further in section \ref{sec:kugo}. Until then, the results for the scaling solution will be just displayed alongside with those of the finite-ghost type, which are obtained for finite values of $B$.

In any case, a momentum is infrared if it is very small compared to all other scales. At zero temperature the only scale of Yang-Mills theory is $\Lambda_\mathrm{YM}$, and thus an infrared momentum is a momentum satisfying \cite{vonSmekal:1997vx,Fischer:2007pf}
\be
p\ll\Lambda_\mathrm{YM},
\ee
\no while at finite temperature infrared momenta satisfy
\be
p\ll\Lambda_\mathrm{YM},\;T\label{meth:irscale},
\ee
\no irrespective of the relative size of $T$ and $\Lambda_\mathrm{YM}$. If Yang-Mills theory is coupled to matter, then the momenta also have to be much smaller than any mass or other dimensionful parameters introduced by this coupling.

The quantification of much smaller than all other scales, and thus the value of the energy scale beyond which an asymptotic behavior is expected, is a highly non-trivial question. Investigations of string-breaking in four dimensions indicate that the typical length scales where dynamical effects can still be encountered in Yang-Mills theory coupled to matter is of the order of 50 MeV, or about 10 fm \cite{Bali:2005fu}. In the context of Wilson lines, this domain is known as the $N$-ality domain, i.\ e., the domain where only asymptotic properties are relevant \cite{Greensite:2003bk}. Another indication is that the typical scale obtained in lattice calculation for $\Lambda_\mathrm{YM}$ is 25-50 MeV. Both indicate a lower distance limit from where on pure asymptotic behavior could be expected. However, it is well possible that the behavior is already almost asymptotic at much smaller distances, but in terms of correlation functions rather the converse seems to be the case \cite{Fischer:2007pf,Cucchieri:2008fc,Cucchieri:2007rg,Bogolubsky:2009dc,Sternbeck:2007ug}.\\

\no{\bf Scaling behavior}\label{sir:scaling}\\

One hallmark of a scaling-type behavior is that diagrams can be classified as leading or non-leading when all momenta are asymptotically small \cite{Alkofer:2004it}. It is then actually possible to provide an analytic solution to the whole tower of equations on the level of the scaling dimensions \cite{Alkofer:2004it,Huber:2007kc,Huber:2009wh}, which under certain weak assumptions is unique \cite{Fischer:2009tn,Fischer:2006vf}. In particular, for the propagator equations it can be shown that only those one-loop diagrams are relevant which contain a ghost loop, and that in general ghost dominance \cite{Zwanziger:2001kw,Zwanziger:2003cf,vonSmekal:1997is,vonSmekal:1997vx,Dudal:2005na,Sobreiro:2004us} holds \cite{Alkofer:2004it}. To illustrate the method of this so-called infrared analysis again the propagator equations will be useful.

The relevant part of both equations then read \cite{Lerche:2002ep,Zwanziger:2001kw,vonSmekal:1997vx}
\bea
D^{ab-1}_G(p)&=&-\widetilde Z_3\delta^{ab}p^2\label{ir:gheq}\\
&&+gf^{abc}\int\frac{d^dq}{(2\pi)^d}ip_\mu D^{ef}_{\mu\nu}(p-q)D^{dg}_G(q)\Gamma^{c\bar cA\indexsep bgf}_\nu(-p,q,p-q)\nn\\
D^{ab-1}_{\mu\nu}(p)&=&-gf^{abc}\int\frac{d^dq}{(2\pi)^d}ip_\mu D_G^{cf}(q) D_G^{de}(p+q) \Gamma_\nu^{c\bar c A\indexsep feb}(-q,p+q,-p)\nn,
\eea
\no and all contributions which turn out later to be self-consistently sub-leading have already been neglected. Comparing to figures \ref{fig:gheq} and \ref{fig:gleq}, these are just all diagrams containing at least one ghost-line on the right-hand side.

The only further input needed is the then the full ghost-gluon vertex $\Gamma_\nu^{c\bar c A\indexsep abc}$, or actually only its infrared-leading part. The consistency of the scaling solution actually demands it to be infrared finite \cite{Fischer:2009tn,Fischer:2006vf}, though variations as a function of the angle between the two independent momenta are still permissible, but will be ignored for the sake of the argument. Such possibilities have been explored and found to be only a quantitative effect \cite{Lerche:2002ep,Huber:2012td}. Furthermore, the results shown in section \ref{szerot:vertices} indicate that any such variations are rather small. Note that an additional color tensor proportional to the symmetric color tensor $d^{abc}$ would get lost due to the contraction with the antisymmetric tree-level vertex. Hence, at least the leading infrared part will not have color off-diagonal elements for either propagator. Thus, the ansatz is to replace the full ghost-gluon vertex by its bare form.

Making then the scaling ansatz \cite{vonSmekal:1997is,vonSmekal:1997vx,Lerche:2002ep,Zwanziger:2001kw,Watson:2001yv}
\bea
D^{ab}_G(p)&=&-\delta^{ab}A_Gp^{-2-2\kappa_{\bar{c}c}}\label{ir:ghpeq}\\
D^{ab}_\mn(p)&=&\delta^{ab}\left(\delta_\mn-\frac{p_\mu p_\nu}{p^2}\right)A_Zp^{-2-2\kappa_{AA}}\label{ir:gpeq},
\eea
\no it is possible to absorb the remaining tree-level term in \pref{ir:gheq} by implementing the boundary condition of an infrared divergent ghost dressing function \cite{vonSmekal:1997vx}, i.\ e., setting $1/B=0$.

The integrals \pref{ir:ghpeq} and \pref{ir:gpeq} can now be evaluated exactly, using the dimensional regularization formula \cite{Peskin:1995ev}
\be
\int\frac{d^dq}{(2\pi)^d}q^{2\alpha} (q-p)^{2\beta}=\frac{1}{(4\pi)^{\frac{d}{2}}}\frac{\Gamma(-\alpha-\beta-\frac{d}{2})\Gamma(\frac{d}{2}+\alpha)\Gamma(\frac{d}{2}+\beta)}{\Gamma(d+\alpha+\beta)\Gamma(-\alpha)\Gamma(-\beta)}y^{2\left(\frac{d}{2}+\alpha+\beta\right)},\label{dimregrule}
\ee
\no which entails a definite minimal subtraction scheme in odd dimensions \cite{Maas:2004se}. The essential point for the validity of this formula is that, because of the absence of other scales, the integral is dominated by the contributions around the external momenta\footnote{Attempts to determine corresponding general formulas for the case in the presence of further scales can be found in \cite{Alkofer:2008jy}, see also \cite{Fischer:2009tn}.} \cite{vonSmekal:1997vx}. Since this in turn depends on the singularity structure of the propagators, and in general of the vertices \cite{Lerche:2002ep}, this yields that the integration turns, up to a constant pre-factor, into a map from the topology of the diagram to the exponent of the external momentum scale \cite{Alkofer:2008jy}.

The integrals can then be performed analytically to yield
\bea
p^{2\kappa_{\bar{c}c}}&=&g^2 C_A A_G^2 A_Z I_G(\kappa_{\bar{c}c},\kappa_{AA},d) p^{-(4-d)-2\kappa_{\bar{c}c}-2\kappa_{AA}}\label{ir:scal1}\\
p^{2\kappa_{AA}}&=&g^2 C_A A_G^2 A_Z I_Z(\kappa_{\bar{c}c},d) p^{-(4-d)-2\kappa_{\bar{c}c}}\label{ir:scal2}.
\eea
\no The expressions $I_G$ and $I_Z$  \cite{Zwanziger:2001kw,Lerche:2002ep}
\bea
I_G(\kappa_{\bar{c}c},\kappa_{AA},d)&=&-\frac{(d-1)\Gamma\left(\frac{d}{2}-\kappa_{\bar{c}c}\right)\Gamma\left(\frac{d}{2}-1-\kappa_{AA}\right)\Gamma\left(2-\frac{d}{2}+\kappa_{\bar{c}c}+\kappa_{AA}\right)}{2^{1+d}\pi^\frac{d}{2}\Gamma\left(1+\kappa_{\bar{c}c}\right)\Gamma\left(d-1-\kappa_{\bar{c}c}-\kappa_{AA}\right)\Gamma\left(2+\kappa_{AA}\right)}\nn\\
I_Z(\kappa_{\bar{c}c},d)&=&\frac{4^{\kappa_{\bar{c}c}-d}\pi^\frac{1-d}{2}\Gamma\left(\frac{d}{2}-\kappa_{\bar{c}c}\right)\Gamma\left(1-\frac{d}{2}+2\kappa_{\bar{c}c}\right)}{\Gamma\left(\frac{1+d-2\kappa_{\bar{c}c}}{2}\right)\Gamma\left(1+\kappa_{\bar{c}c}\right)^2}\nn
\eea
\no are functions depending solely on the exponents $\kappa_{\bar{c}c}$ and $\kappa_{AA}$, and the structure of the underlying space-time manifold, symbolized by the dependence on $d$.

Counting powers of momenta in \prefr{ir:scal1}{ir:scal2} yields the fundamental scaling relation \cite{vonSmekal:1997is,Lerche:2002ep,Zwanziger:2001kw}
\be
0=2\kappa_{\bar{c}c}+\kappa_{AA}+\frac{d-4}{2}\label{ir:conc}.
\ee
\no This basic relation is the single most important property of the scaling-type solution \cite{Fischer:2009tn,Fischer:2006vf}. If it is not fulfilled, the solution will not be of scaling-type, irrespective of the behavior of the individual propagators or vertices, as long as the ghost-gluon vertex is infrared finite.

The remaining consistency condition
\be
I_Z(\kappa_{\bar{c}c},d)=I_G(\kappa_{\bar{c}c},\kappa_{AA}(\kappa_{\bar{c}c},d),d)\label{ir:consistencycond}
\ee
\no implies that the exponents depend only on the space-time manifold. The solution of this consistency condition for a bare ghost-gluon vertex yields $\kappa_{\bar{c}c}\approx 0.595$ in four dimensions \cite{Lerche:2002ep,Zwanziger:2001kw}, $\kappa_{\bar{c}c}=1/2$ or $\kappa_{\bar{c}c}\approx 0.39$ in three dimensions, and $\kappa_{\bar{c}c}=1/5$ or $\kappa_{\bar{c}c}=0$ in two dimensions \cite{Zwanziger:2001kw}. As a consequence of \pref{ir:conc}, the gluon exponent satisfies $\kappa_{AA}<-1$, and the gluon propagator is infrared vanishing. A formal second solution $\kappa_{\bar{c}c}=1$ of \pref{ir:consistencycond} in four dimensions is not admissible, as it cannot be brought into agreement with renormalizability \cite{Fischer:2002hna,RodriguezQuintero:2010wy}. If the ghost-gluon vertex has a non-trivial angular dependence at zero momentum, the values of $\kappa$ can change quantitatively \cite{Huber:2012td}. Furthermore only the product of the pre-factors $A_g^2 A_z$ is fixed by the equations, and their individual values cannot be determined \cite{Maas:2004se,Maas:2005rf}. Also this product will be affected by angular variations of the ghost-gluon vertex \cite{Huber:2012td}.

A direct consequence of the setup of this infrared analysis is that the exponent $\kappa_{\bar{c}c}$ must satisfy the inequality
\be
\kappa_{\bar{c}c}\ge\frac{d-2}{2}\label{ir:condkappa},
\ee
\no otherwise the basic assumptions made, i.\ e., that the ghost loop in the gluon propagator equation is the leading term, is invalid. If the equality applies, indeed the ghost loop is as leading as all other loops \cite{Watson:2001yv}, and the gluon propagator becomes infrared constant, and its therefore appearing screening mass receives contributions from all other loops \cite{Bloch:2003yu}. This furthermore implies that the relation \pref{ir:consistencycond} receives corrections on top of any corrections from the ghost-gluon vertex. This option has not yet been widely explored, though it may be relevant. Note that furthermore arguments exist, which suggest a lower upper bound in four dimensions of less than 1 \cite{Eichhorn:2010zc}.

A much more stable consequence of the scaling solution than the numerical value of $\kappa_{\bar c c}$ is that the renormalization-group-invariant running coupling \cite{vonSmekal:1997is,vonSmekal:1997vx,vonSmekal:2009ae}
\be
\alpha(p^2)=\frac{g^2(p)}{4\pi^2}=p^{2+d}\alpha(\mu^2)\frac{1}{N_A^3(d-1)}D_G^{aa2}(p^2,\mu^2)D_{\mu\mu}^{aa}(p^2,\mu^2)\label{ir:alpha},
\ee
\no where $N_A$ is the size of the adjoint representation, is constant at zero momentum, as this follows directly from \pref{ir:conc}. Of course, in dimensions smaller than four there is no dependency on the renormalization scale $\mu$. It should be noted that, except for an overall scale factor of the adjoint Casimir in the running coupling, the underlying gauge algebra has not entered into this result \cite{vonSmekal:1997is,Maas:2005ym,Maas:2010qw}. Thus, the infrared in the scaling case should be qualitatively independent of the gauge algebra. The only possible exception occurs, if again equality holds in \pref{ir:condkappa}, in which case contributions from the two-loop diagrams could, in principle, enter \cite{Bloch:2003yu}.

Furthermore, the relation \pref{ir:alpha} also holds at finite momenta, and can be deduced from the ghost-gluon vertex \cite{vonSmekal:1997vx}. It is thus sufficient to know both propagators to calculate the full running coupling in Landau gauge. However, to ensure the correct renormalization behavior of the running coupling, in particular the relation
\be
Z_gZ_3^\frac{1}{2}\tilde{Z}_3=\tilde{Z}_1\label{ir:rencond},
\ee
\no which follows from multiplicative renormalization alone, requires that the renormalization conditions for both propagators are locked as \cite{vonSmekal:1997vx}
\be
Z(\mu,\mu)G(\mu,\mu)^2=1\label{ir:rencondlock}.
\ee
\no Herein, $Z_g$ denotes the renormalization constant of the gauge coupling, $Z_3$ the wave-function renormalization of the gluon, and it has entered that in the renormalization scheme employed, see section \ref{sec:renorm}, the renormalization constant of the ghost-gluon vertex $\tilde{Z}_1$ has been set to one. Of course, an alternative is to introduce a scheme with an explicit (finite) renormalization constant $\tilde{Z}_1$ of the ghost-gluon vertex, which will then appear as a factor on the right-hand side of \pref{ir:rencond} \cite{Boucaud:2008ji,Boucaud:2008ky}.

In dimensions smaller than four, the simplest possibility to ensure \pref{ir:rencondlock} is to formally set $\mu$ to infinity, in which case $Z(\infty)=G(\infty)=1$, just because of asymptotic freedom, as has been discussed in section \ref{ssuv} \cite{Maas:2004se,Maas:2009se}.

Based on the insight gained with the propagator equations, it is possible to turn the results into a general rule \cite{Fischer:2006vf,Fischer:2009tn,Alkofer:2008jy}, akin to the power-counting rules used to determine the superficial degree of divergence in the renormalization process \cite{Collins:1984xc}. In particular, it is possible to calculate the superficial exponent $\delta$ of a diagram, which determines the dependency on the single momentum scale at the symmetric point, i.\ e.\ the momentum configuration $p_1^2=...=p_n^2=p^2$ \cite{Alkofer:2004it,Fischer:2006vf,Huber:2009wh}.

To obtain this result, the first step is to calculate the scaling behavior of a diagram in the DSEs (or FRGs) at this symmetric point in the infrared. Similar to corresponding formulas in the renormalization program, a formula can be constructed which maps such a diagram onto this exponent. For a diagram having $n_{\phi_i}$ external legs of field type\footnote{For the case of mixed propagators, see \cite{Huber:2009tx}.} $\phi_i$ it is given by \cite{Huber:2009wh}
\bea
\delta&=&c+\frac{ld}{2}+\sum_i n_{\phi_i}(\kappa_{\phi_i\phi_i}-1)+\sum_{\mathrm{bare\;vertices}\;r}n_{\phi_{j_1}...\phi_{j_r}}c_{\phi_{j_1}...\phi_{j_r}}\nn\\
&&+\sum_{\mathrm{dressed\;vertices}\;r}n_{\phi_{j_1}...\phi_{j_r}}(\kappa_{\phi_{j_1}...\phi_{j_r}}+c_{\phi_{j_1}...\phi_{j_r}})\label{scal:genexp},
\eea
\no where $n_{\phi_i}$ is the number of external propagators with their corresponding exponents $\kappa_{\phi_i\phi_i}$ signifying their power-law behavior, and the $n_{\phi_{j_1}...\phi_{j_r}}$ count the number of bare and full vertices, respectively, appearing in the diagrams having infrared exponents $\kappa_{\phi_{j_1}...\phi_{j_r}}$ and canonical exponents $c_{\phi_{j_1}...\phi_{j_r}}$, respectively. Finally, $c$ is the overall canonical dimension of the diagram, and $l$ is the number of loops. Similar to the case of the superficial degree of divergence \cite{Collins:1984xc} it is also possible to recast this expression using graph theory into one involving only the critical exponents $\kappa$ and the type of internal lines, but no longer the canonical dimensions \cite{Huber:2009wh}. For the purpose here, it is sufficient that such a mapping exists.

With this mapping at hand, and the assumption that the superficial infrared exponent is indeed the correct one, it is possible to determine a recurrence relation between the infrared exponents \cite{Alkofer:2004it,Fischer:2006vf,Fischer:2009tn,Huber:2009wh}. It follows from the fact that the exponent of the left-hand-side of a DSE must be the same as the dominating, i.\ e.\ most singular, one on the right-hand-side. The simplest example for this is the gluon equation keeping only the tree-level term and the ghost-loop. The corresponding statement is then for a bare ghost-gluon vertex
\be
\kappa_{AA}=\min\left(0,-\frac{(4-d)}{2}-2\kappa_{\bar{c}c}\right)\label{scal:maxrel}.
\ee
\no Since the ghost equation implies that $\kappa_{\bar{c}c}\ge 0$ \cite{Watson:2001yv}, the second option is the dominating one, and the tree-level term is subleading. In analogy, such a hierarchy of relations can be obtained for all equations. To actually resolve them requires a further piece. This comes from comparing relations of type \pref{scal:maxrel} for the same quantities in DSEs and FRGs \cite{Fischer:2006vf}. Since in FRGs all vertices are always dressed, while in the DSEs always one vertex is undressed, both hierarchies are not identical. This is sufficient to find an explicit resolution of the infinite hierarchy of inequalities \cite{Fischer:2009tn}.

This result yields the infrared critical behavior of all correlation functions at the symmetric point for Yang-Mills theory in the scaling case as \cite{Alkofer:2004it,Huber:2007kc}
\be
\Gamma^{(n,m)}(p^2) \sim (p^2)^{(m-n)\kappa_{\bar{c}c} + (n-1)\left(\frac{d}{2}-2\right)}\label{ir:IRsolution},
\ee
\no where $n$ is the number of external ghost legs (identical to the number of external anti-ghost legs) and $m$ is the number of external gluon legs. An important consequence of this result is that any coupling constant defined from the higher vertices is also finite \cite{Alkofer:2004it,Huber:2007kc}, by a construction similar to \pref{ir:alpha}.

Note that the construction presented here is given for Yang-Mills theory only. Various investigations have been performed when matter fields are included \cite{Alkofer:2008tt,Schwenzer:2008vt,Fister:2010yw,Macher:2010ad}, but it is possible that some assumptions necessary for the construction then break down in the matter sector \cite{Fischer:2009tn}. This has not yet been finally clarified. There are also investigations how to extend this analysis to non-symmetric momentum configurations \cite{Alkofer:2008jy}, though these encounter in general ambiguities \cite{Fischer:2009tn}.\\

{\bf Finite-ghost behavior}\\

The finite-ghost case was first discussed in \cite{Cornwall:1981zr,Aguilar:2004sw} and has afterwards been investigated in, e.\ g., \cite{Binosi:2009qm,Boucaud:2008ji,Boucaud:2008ky,Dudal:2008sp,Dudal:2008rm,Alkofer:2008jy,RodriguezQuintero:2010wy,Aguilar:2010zx,Dudal:2007cw,Fischer:2008uz,Aguilar:2009ke,Aguilar:2008xm,Aguilar:2009nf,Aguilar:2006gr,Boucaud:2010gr,Boucaud:2007va,Boucaud:2006if,Tissier:2010ts,Kondo:2009wk,Gracey:2010cg,Sauli:2009se,Oliveira:2010xc,Boucaud:2011ug,Binosi:2012sj}. The finite-ghost case is characterized by an infrared finite gluon propagator and an infrared massless ghost,
\bea
D_G^{ab}&=&-\delta^{ab}A_Gp^{-2}\nn\\
D_\mn^{ab}&=&\delta^{ab}\left(\delta_\mn-\frac{p_\mu p_\nu}{p^2}\right)A_Z\nn.
\eea
\no In this case, it is more complicated to make a general infrared analysis as in the scaling case. Taking the ghost equation, and choosing a finite-ghost behavior by setting $B$ to a finite value, the tree-level term is no longer canceled in \pref{ir:gheq}. Inserting then the finite-ghost ans\"atze in a truncation at the same level as in the scaling case automatically yields that the ghost is infrared massless and the gluon is infrared screened \cite{Fischer:2008uz,Binosi:2009qm,RodriguezQuintero:2010wy,Boucaud:2011ug}. In particular, in the infrared limit the appearing mass scale acts as a bare mass, and only finite momentum corrections make it sufficiently fast vanishing in the ultraviolet to ensure the correct ultraviolet behavior \cite{Fischer:2008uz,Binosi:2009qm,Aguilar:2009ke}.

The reason for this is rather simple. Investigating the appearing integrals in detail show that the diagrams in general harbor a mass-like term \cite{Maas:2004se,Maas:2005rf}, even if not introduced by assumption \cite{Binosi:2009qm}. In perturbation theory, these terms cancel exactly \cite{Peskin:1995ev,Bjorken:1979dk}. Beyond perturbation theory, this is in general no longer the case. However, in the scaling case, they are infrared sub-leading \cite{Maas:2004se,Maas:2005rf,Fischer:2003zc}, since they behave like $m^2/p^2$, where $m$ is some scale, generically proportional to $\Lambda_{\mathrm{YM}}$. The contribution due to the ghost-loop for a scaling ghost and $\kappa_{\bar{c}c}>(d-2)/4$, e.\ g.\ in the gluon propagator DSE \pref{ir:gpeq}, provides a stronger infrared divergence than just $p^{-2}$, therefore dominating the screening behavior and leading to the scaling behavior. This option is, by construction, switched off in the finite-ghost case, and thus the screening behavior becomes leading. This does not occur for the ghost due to the factorization of the external ghost momentum of the ghost-gluon vertex. This makes terms automatically more infrared divergent, promoting the screening mass to a correction of the wave-function renormalization \cite{Fischer:2007pf}. Note that therefore at least all one-loop contributions in the gluon equation contribute equally \cite{Fischer:2008uz,Aguilar:2011aa}.

Thus, a finite-ghost behavior emerges naturally, as soon as any other infrared-dominating behavior is switched off by choosing $B$ finite\footnote{This can also be achieved, e.\ g., by an explicit infrared regulator like a finite volume, see \cite{Fischer:2007pf}.}. Similar effects are expected to also occur in the vertex equations, closing the system self-consistently \cite{Cornwall:2009ud}. However, this has not yet been investigated in detail on the level of a self-consistent solution of the vertex equations, but the investigations are ongoing \cite{Binosi:2011wi}.

On the other hand, this mechanism implies the possible existence of a scaling window in the correlation functions' behavior even for the finite-ghost case \cite{Fischer:2007pf,Fischer:2008uz}: Define the momentum $p_B$ to be the one at which the normalized ghost dressing function $p_B^2 D_G(p_B^2)/(\mu^2 D_G(\mu^2))$ becomes similar of size as $1/B$. Then screening becomes manifest at momenta at or below $p_B$. If $B$ is chosen such that $p_B$ is much smaller than the scale $\Lambda_{YM}$, then there exists a window where the correlation functions show a scaling behavior, up to sub-leading corrections, in the momentum range
\be
p_B\ll p\ll\Lambda_\mathrm{YM}\label{dse:conformalwindow}.
\ee
\no Since the power-laws exhibited in this energy regime are the same as expected for the scaling case, this will be called the scaling window. The existence of this window follows from the fact that for momenta within the range \pref{dse:conformalwindow} no sensitivity to the deep infrared yet exists.

The one technical challenge associated with the finite-ghost solution is that any such screening mass inevitably mixes with the spurious divergences discussed in the perturbative case when a non-gauge-invariant cut-off regularization is used. Therefore, disentangling the different contributions, in principle, requires in this case a resolution of the corresponding broken Slavnov-Taylor identities \cite{Fischer:2008uz,Ellwanger:1994iz,Pawlowski:2005xe}, to obtain the screening mass, e.\ g., of the gluon propagator. In general, this is complicated, and instead usually the necessary counterterms are modeled such that corresponding lattice and perturbative results are reproduced \cite{Fischer:2008uz,Cucchieri:2007ta,Maas:2004se,Fischer:2002hna,Fischer:2005en,Aguilar:2008xm}, or an appropriate truncation is chosen to eliminate these artifacts \cite{Fischer:2008uz,Binosi:2009qm,Aguilar:2009ke,Huber:2012td}. Finally, as is clear from the fact that all free parameters of Yang-Mills theory are fixed once $\Lambda_\mathrm{YM}$ is fixed, the gluon screening mass is not an independent parameter of the theory. Furthermore, the perturbative cancellations enforce that the screening terms have to drop faster than $1/p^2$ at large momenta, thus recovering the masslessness of the gluon in the high-energy limit.

Note that the effective wave-function renormalization of the ghost is less arbitrary. Since its value is uniquely fixed to the $B$ parameter, no problem would emerge even if spurious divergences were encountered in the ghost equations.

\subsubsection{Intermediate momenta}\label{sec:trunc}

At the current time, in order to obtain solutions at all momenta, it is necessary to truncate the system of equations\footnote{Very few systems, like the 1+1-dimensional Schwinger model, are known where this is not necessary \cite{Alkofer:2000wg}.}. In general, the main ingredient is to drop all equations for higher order correlation functions from a certain, usually low, order $n$. This requires to make assumptions for the various higher $n$-point correlation functions still appearing inside the equations, and which are no longer determined self-consistently. Various schemes have been employed for this purpose. Such schemes include modeling the remaining vertices along some guiding principles \cite{vonSmekal:1997vx,Fischer:2008uz,Fischer:2002hna,Maas:2004se,Pawlowski:2003hq,Alkofer:2008tt,Binosi:2011wi}, truncations based on the desired properties of the solutions inferred from other sources \cite{Binosi:2009qm,Boucaud:2008ji,RodriguezQuintero:2010wy,Aguilar:2011ux}, methods based on a background-field formulation \cite{Binosi:2009qm,Aguilar:2009ke,Fischer:2004uk,Gies:2002af,Zwanziger:2002ia}, effective actions \cite{Huber:2009tx}, analyticity concepts \cite{Alkofer:2000wg,Cornwall:2009ud}, lattice input \cite{Huber:2012td}, and others \cite{Alkofer:2000wg,Pawlowski:2005xe,Fischer:2006ub}. The possibilities are essentially only limited by the necessities imposed to answer a particular question.

As an example, the truncation, with slight variations, to be used below in the chapters \ref{szerot} and \ref{sfinitet} will be discussed. Alongside the guiding principles will be presented which motivate this truncation.

The aim in the present case will be results for the propagators. This requires their respective equations, which are then truncated at one-loop level. Thereby models for the three-point functions are required.

The first central requirement is to establish contact with resummed perturbation theory \cite{vonSmekal:1997vx,Fischer:2002hna,Maas:2004se,Cucchieri:2007ta,Fischer:2008uz}. The main motivation to make this the central guide-line for the truncation is that this is the fastest way to make contact to experimentally accessible quantities, instead of using non-perturbative quantities like bound-states as a guiding principle. The second advantage is that this makes the non-perturbative approach here a direct extension of the perturbative case, guaranteeing a smooth transition from one to the other.

The goal is to have agreement to perturbation theory at one-loop order. A consistent truncation scheme therefore requires to keep the equations for the propagators and include all one-loop diagrams, but to drop all two-loop diagrams and all higher-order equations. This level of truncation is the currently most commonly used\footnote{See for first progress in going beyond this level \cite{Alkofer:2008tt,Fischer:2005en}.}. Thus, it is necessary to specify the ghost-gluon vertex and the three-gluon vertex.

One possibility to achieve this is to leave the ghost-gluon vertex bare, which even turns out to be a rather good approximation of its full behavior, see section \ref{szerot:vertices}. This is not the case for the three-gluon vertex. As discussed in section \ref{ssuv}, reproducing leading-order perturbation theory already requires a modification such as to provide the correct self-consistent behavior for the gluon propagator in four dimensions to order $g^2$. A possible asymptotic dressing realizing such a behavior for the three-gluon vertex is, e.\ g., \cite{Cucchieri:2007ta}
\bea
\Gamma^{A^3 abc}_{\mu\nu\rho}(p,q,k)&=&f^{abc}\Gamma^{(\mathrm{tl})}_{\mu\nu\rho}(p,q,k)\Gamma^{\mathrm{uv}}(p^2,q^2,k^2)\label{fm:g3v}\\
  \Gamma^{\mathrm{uv}}(p^2,q^2,k^2)& = & a G\left(\frac{1}{2}\left(q^2+p^2+k^2\right)\right)^{a_G}
        Z\left(\frac{1}{2}\left(q^2+p^2+k^2\right)\right)^{a_Z}\nn \\
a_G&=&\frac{2\kappa_{AA}(1+3\delta)}{\kappa_{AA}\delta+\kappa_{\bar{c}c}+2\delta\kappa_{\bar{c}c}}\nn\\
a_Z&=&\frac{-2-6\delta+a_G\delta}{1+2\delta}\nn\\
a&=&\frac{-9-19\delta}{25\delta Z_1}\nn
\eea
\no with $k^2=(p+q)^2$, $\Gamma^{(\mathrm{tl})}_{\mu\nu\rho}$ the tree-level vertex, and $Z_1$ the renormalization constant of the three-gluon vertex, which is introduced to ensure the correct renormalization of the vertex. This ansatz explicitly preserves the Bose symmetry of the vertex. This is the construction which will be used in section \ref{sfinitet} for finite temperatures. The choice of the constant $a_Z$ is uniquely fixed  by requiring the correct renormalization properties and resummed perturbation theory \cite{Cucchieri:2007ta}. The parameter $a_G$ is then free and chosen such as to make the vertex freeze in the infrared. This establishes a truncation scheme which is self-consistent in the perturbative domain. Note that in three dimensions the choice $\Gamma^{\mathrm{uv}}=1$ is sufficient to ensure the correct perturbative behavior. The only remaining problem is the presence of spurious quadratic divergences due to the cutoff regularization. For this truncation, these can be dealt with by explicit counter-terms \cite{Maas:2005hs,Cucchieri:2007ta}, or by explicit and implicit subtraction \cite{Maas:2004se,Maas:2005hs}.

A bit more elegant is the possibility to provide an additional dressing of both the ghost-gluon and the three-gluon vertex \cite{Fischer:2008uz}. In case of the three-gluon vertex, it is given by\footnote{Note that $f_{UV}$ is not the same as the one in \cite{Fischer:2008uz}, and it is more generally applicable. I am grateful to Markus Huber for pointing out a potential problem with the original version.}
\bea
\Gamma^{A^3 abc}_{\mu\nu\rho}(p,q,k)&=&f^{abc}\Gamma^{(\mathrm{tl})}_{\mu\nu\rho}(p,q,k)\Gamma^{\mathrm{uv}}(q,p)\times\label{zerot:ag3v}\\
&&\times\left(1-\left[\frac{140}{51} - \frac{52}{17} \frac{p^2}{k^2} +    \frac{89}{51} \frac{p^2}{q^2} + \frac{52}{17} \frac{q^2}{k^2} -    \frac{26}{17} \frac{k^2}{q^2} +\frac{104}{17} \frac{(q.k)^2}{q^2      k^2}\right]f_{UV}(p,q;k)\right)\nn\\
  f_{UV}(p,q;k) &=&\left(\frac{q^2}{q^2+\Lambda_{UV}^2}\right)^3 \nn,
\eea
\no where $p$ is the external momentum and $q$ is the loop momentum. This form completely eliminates the spurious quadratic divergences. The formfactors $f_i$ with their corresponding parameters ensure that at large momenta only the ultraviolet completion $\Gamma^{\mathrm{uv}}$ contributes to ensure the correct perturbative behavior. This completion can be chosen, e.\ g., as \pref{fm:g3v} or as \cite{Fischer:2008uz,Fischer:2002hna}
\be
\Gamma^{\mathrm{uv}}(q,p)  = \frac{1}{Z_1}\frac{[G(p^2+\Lambda_{IR}^2) \,G((p+q)^2+\Lambda_{IR}^2)]^{(1-a/\delta-2a)}}{[Z(p^2+\Lambda_{IR}^2)Z((p+q)^2+\Lambda_{IR}^2)]^{(1+a)}}\label{zerot:aguv},
\ee 
\no with $a$ free, but conventionally chosen as $3\delta$, and $Z_1$ the three-gluon vertex renormalization constant. This choice breaks Bose symmetry, and also does not reproduce the results for the three-gluon vertex obtained using lattice gauge theory, shown in section \ref{szerot:vertices} below. However, the difference at the level of the propagators is very small, in particular in the scaling case \cite{Fischer:2008uz}.

This choice also requires a modified ghost-gluon vertex to fully cancel the spurious quadratic divergences. Furthermore, in the scaling case the only contribution in the far infrared are those containing ghosts, and thus the transversality of the gluon propagator must be guaranteed by the ghost loop alone. This is not necessary in the finite-ghost case. This is ensured by the construction \cite{Fischer:2008uz}
\bea
\Gamma^{\bar{c}cA}_\mu (p,q,p+q) &=& i p_\mu A(p,q,k)-i(p+q)_\mu B(p,q,k)\label{zerot:aghglvtx}\\
A(p,q,k)&=&\left( 1 -  \frac{q^2}{p^2} \,f_{UV}(p,q;k)\right)\nn\\
B(p,q,k)&=&\left(\frac{kq}{k^2} \,f_{IR}(p,q;k) \right)\nn\\
  f_{IR}(p,q;k) &=&\frac{\Lambda_{IR}^6}{(p^2+\Lambda_{IR}^2)(q^2+\Lambda_{IR}^2)(k^2+\Lambda_{IR}^2)}\nn
\eea
\no with the same form factor $f_{UV}$ as for the three-gluon vertex. Note that, in principle, the longitudinal part should be determined by the STIs, but generically depends on the four-point ghost-gluon scattering kernel, unknown at this level of truncation \cite{vonSmekal:1997vx}. Making an ansatz to obtain transversality in the scaling case is therefore the only possibility here. However, as emphasized, this longitudinal contribution is completely irrelevant when calculating the non-amputated correlation functions, and in particular physical observables. It should be noted that the transversality of the gluon propagator is guaranteed by this ansatz only at asymptotically large and small momenta. However, the residual longitudinal component at intermediate momenta, being a truncation artifact, is negligible \cite{Fischer:2008uz}. Truncations which ensure transversality at all momenta have been proposed in \cite{Binosi:2009qm,Binosi:2011wi}, though at the expense of disagreement with perturbation theory at large momenta. A truncation having both virtues is in principle possible to construct, though great care is necessary to avoid the introduction of infrared-ultraviolet mixing \cite{Maas:2005rf,Fischer:2003zc}. Such a truncation would make use of the not yet included longitudinal tensor structures of the three-gluon vertex, which could be chosen to ensure transversality at all momenta. Since for the present case the effects appear negligible, this will not be done here. How well transversality can be achieved once the three-point vertices are included self-consistently has not yet been explored. However, for a final construction the four-point vertices and the two-loop terms in the gluon equation could play a role.

\begin{figure}
\includegraphics[width=\textwidth]{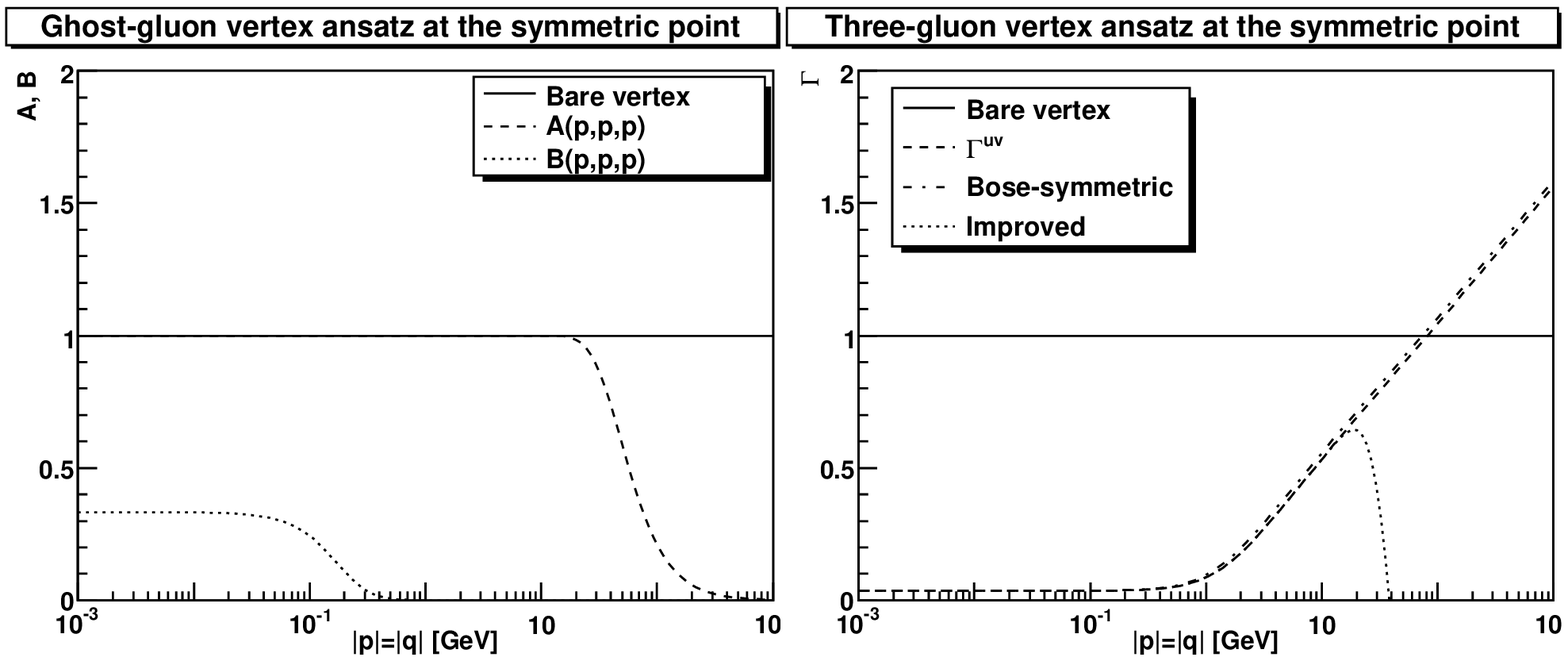}
\caption{\label{fig:tvtx}The two truncation ans\"atze for the ghost-gluon vertex (left panel) and the three-gluon vertex (right panel) in four dimensions. The functions $A$ and $B$ for the ghost-gluon vertex are given in \pref{zerot:aghglvtx}, the function $\Gamma^{\mathrm{uv}}$ in \pref{zerot:aguv}, the improved vertex in \pref{zerot:ag3v}, and the Bose-symmetric one in \pref{fm:g3v}. The parameter $\Lambda_{IR}$ is 300 MeV, and $\Lambda_{UV}$ 20 GeV, the dressing functions are for the scaling case, using the fits in \cite{Fischer:2002hna}.}
\end{figure}

These various ans\"atze for the vertices at the symmetric point are shown in figure \ref{fig:tvtx}, to provide a comparison to the results from lattice gauge theory presented in section \ref{szerot:vertices}. It is visible that the vertices which have the ultraviolet cancellations of the quadratic divergences built in drop off sharply at large momenta, since there the divergent contribution from the integral kernels are needed to generate the correct ultraviolet running. The Bose-symmetric ansatz, which requires explicit counter-terms, shows directly the logarithmic running, which is also encoded in the ultraviolet dressing part of the improved three-gluon vertex \pref{zerot:aguv}. Note that this logarithmic rise may be altered if the ansatz were enlarged to include the other three possible transverse (and ten more longitudinal) tensor structures of the three-gluon vertex \cite{Ball:1980ax}, for which first studies have been performed in \cite{Alkofer:2008dt,Alkofer:2008tt,Binosi:2011wi}.

On the other hand, compared to the results shown in section \ref{szerot:vertices} the bare ghost-gluon vertex, depicted also in figure \ref{fig:tvtx}, shows at the symmetric point a rather decent resemblance of the results obtained below in lattice calculations, and is in agreement with self-consistency studies \cite{Lerche:2002ep,Schleifenbaum:2004id,Fischer:2009tn,Fischer:2006vf,Alkofer:2008jy}, but the three-gluon vertex fails this test. Nonetheless, at the level of propagators this turns out to be of minor importance, in both the scaling and finite-ghost case \cite{Fischer:2008uz}. Therefore, this departure from the presumably better results of lattice calculations seems to be of minor importance at the level of propagators. For other quantities this may not be the case, and for a systematic improvement of the truncation three-point vertices would have to be the first to be treated, see e.\ g.\ \cite{Alkofer:2008tt} for such an investigation.

These are only two examples of employed vertex constructions. Various alternatives and modifications have been explored \cite{vonSmekal:1997vx,Lerche:2002ep,Binosi:2009qm,Maas:2005hs,Maas:2004se,Pennington:2011xs,Huber:2012td}. The results turn out to depend only quantitatively on the employed version of the vertices, after fixing $B$. The only major exception is finite temperature, and there only for one polarization of the gluon \cite{Cucchieri:2007ta}. This will be discussed in more detail in chapter \ref{sfinitet}. Reiterating, the guiding principle here has been to obtain the correct resummed perturbation theory, renormalization \cite{Fischer:2008uz}, and for the scaling case infrared self-consistency. Conceptually different guiding principles have also been employed, in particular those guided by reproducing a particular infrared behavior \cite{Binosi:2009qm,Boucaud:2008ji,RodriguezQuintero:2010wy}. Going beyond such modeling of the vertices is an ongoing endeavor \cite{Alkofer:2008tt,Fister:2011ym}. It should be noted that using FRGs instead of DSEs, self-optimizing flows and regulators can be used to at least partly capture the effects of four-point functions \cite{Fischer:2008uz,Pawlowski:unpublished,Fister:2011ym}. This yields the currently best quantitative agreement between functional and lattice methods on the level of propagators \cite{Fischer:2008uz}.

\subsubsection{Numerical methods}

This fully fixes the equations to be solved here. However, this leaves a coupled set of non-linear integral equations. Solving them is still a non-trivial numerical problem. Various methods exist to solve these equations, mainly based on iterative methods \cite{Aguilar:2004sw,Fischer:2002eq}, as well as local \cite{Fischer:2003zc,Bloch:1995dd} and global \cite{Maas:2005xh} Newton algorithms.

In solving the DSEs, there are two main challenges. One is the numerical integration of the loop integrals. The second is finding a solution for the self-consistency equations.

Concerning the loop integrals, it turns out that these are somewhat non-trivial numerically, especially for the scaling solution. The main problem is that the relevant integration range features three distinct problems. One is that a very large range of momenta is contributing to the integrals, when it comes to a quantitative precise determination of the pre-factors of the power-laws in the ans\"atze \prefr{ir:scal1}{ir:scal2}. Thus, an exponentially large momentum range has to be covered. The second is that the integral kernels feature singularities in the external momenta. In fact, this is the original reason why the infrared analysis discussed before is working. The third is the necessary cancellation of divergences, which are of two kinds. The first is the subtraction of the spurious and physical logarithmic divergences. The second kind are phantom divergences, i.\ e., divergences which appear when the integral in the absolute value of the loop momenta are made, but are exactly canceled when the integral over the relative angle between external and internal momenta are performed.

To deal with all of these problems, the following procedures have been found useful \cite{Bloch:1995dd}. Take the case where divergences are subtracted by a counter-term. The integral kernels can be separated into a part $f$ depending only on the loop momenta's size $q$, one part $h$ depending only the internal momenta and the external momenta $p$, and one part $g$ depending also on the relative angle $\theta$. A suitable rewriting of the integral is then given by
\bea
\Pi(p)-\Pi(s)&=&\int dq\int d\theta f(p,q)h(q)g(p,q,\theta)-\int dq\int d\theta f(s,q)h(q)g(s,q,\theta)\nn\\
&=&\int dq h(q)\left(f(p,q)\int d\theta g(p,q,\theta)-f(s,q)\int d\theta g(s,q,\theta)\right)\nn.
\eea
\no In the case where $f$ should be one, which actually sometimes occurs, the subtraction of the two $g$ functions should furthermore be done before the integration. Note that this combination inside the integral is formally part of the regularization process \cite{Collins:1984xc}, and can be thought of as related to a BPHZ scheme \cite{Maas:2005hs}.

Furthermore, the integral in $q$ can never be extended to real infinity in a numerical treatment, and it is often also not possible to evaluate it at exactly zero momentum. Thus, in practice a numerical upper cutoff  $\Lambda$ and lower cutoff $\epsilon$ are included. To actual perform the numerical integration, the $q$ integral is then split into
\be
\int_\epsilon^\Lambda\to\int_\epsilon^p+\int_p^\Lambda\nn
\ee
\no to deal with the singularities of the integral kernel\footnote{For vertices \cite{Schleifenbaum:2004id} or in lower dimensions \cite{Huber:2012td} additional singularities may occur also in the angular integral, requiring a similar splitting there.}. It is then only required to choose the integrator. A useful choice \cite{Bloch:1995dd} is a Gauss-Legendre integrator, which samples the boundaries of the integration domain well, and thus the singularities appearing at the external momenta. To deal with the exponentially distributed strength of the integral, it is furthermore useful to spread the points instead of on a linear scale on a logarithmic scale. This type of integrator, with probably adapted domains of integration \cite{Nickel:2006vf}, has so far been sufficient to deal with the numerical integrals encountered in such studies.

It remains to solve the self-consistency problem for the sought for function $D$, e.\ g.\ symbolically for a single equation given by
\be
\frac{1}{D(p)}=1+\Pi(p,D)\nn,
\ee
\no neglecting counter terms. To deal with the fact that $D$ are functions, two approaches have been found useful. One is approximating the logarithm of the function by an expansion up to a fixed order in an Chebychev basis \cite{Fischer:2003zc,Bloch:1995dd,Maas:2005xh}. The second is to to calculate the function only at a fixed number of points and interpolate between them, e.\ g.\ with cubic splines \cite{Aguilar:2004sw,Fischer:2002eq,Huber:2012td}. It has been found useful \cite{Maas:2005xh} to split the sought for function into a function which embodies some known properties, like the perturbative behavior or the qualitative infrared behavior, and another function multiplying this fit function. By this, only a function of order one has to be calculated, which stabilizes the numerics.

To find the self-consistent solution with such an approximation an iterative procedure is used. In the case of a function basis, a Newton method can be used \cite{Bloch:1995dd,Fischer:2003zc} to find the self-consistent solutions. In the second case a fixed-point iteration \cite{Aguilar:2004sw,Fischer:2002eq} is appropriate. In both cases, adaptive global versions \cite{Maas:2005xh,Huber:2012td} are useful to stabilize the iterations if no good starting guess is available, and the radius of contraction of the iteration procedure is not yet known.

However, there are two drawbacks with this approach. One is that such an approximation is never exact, and in particular the results are usually only continuous, but not continuous differentiable over the whole momentum range. Thus, the final accuracy which can be obtained in solving the equations is necessarily limited, though given sufficient computing power can still be rather good \cite{Maas:2005xh}. Secondly, in both cases the solutions are only known in a certain interval, and extrapolations outside this interval are necessary when performing the integrals, which due to the sum of internal and external momenta will necessarily also probe outside this domain. At momenta where perturbation theory is applicable, this is not a problem as the unfitted part can then be substituted by perturbation theory. However, at other momenta this is a problem, as extrapolation is usually not very stable. Here it has been found useful \cite{Fischer:2002eq,Bloch:1995dd} to employ analytical knowledge, like the scaling behavior \prefr{ir:scal1}{ir:scal2}, with free coefficients to extrapolate. The free coefficients are then chosen such as to guarantee continuity at the point where the numerical solution ends, and are then adapted during the iteration process, possibly including damping \cite{Maas:2005xh}.

If a system of equations should be solved, it is either possible to iterate all equations simultaneously or one after the other. However, the numerical stability may differ, depending on the details of the system.

With these methods, it was so far always possible to solve this type of DSEs. However, one should be wary as these integral equations are rather forgiving concerning erroneous coding or inconsistent truncations. Secondly, it has rather often be encountered that a truncation lead to inconsistencies which surfaced by a non-convergence of the iteration process. A particular example \cite{Maas:2004se} is e.\ g.\ that if the gluon loop in the equation for the gluon propagator is not sufficiently suppressed at mid-momenta due to a not appropriately chosen three-gluon vertex, it dominates and tries to lead to a sign change in the gluon propagator. When using an expansion of the logarithm of the propagators, this is impossible, leading to a failure of convergence.

Last, but not least, most of these steps are rather similar, and can thus be automatized \cite{Huber:2012cd}, though still great care is required when dealing with both the physical and numerical particulars of a given system, as discussed here.

\subsubsection{Dyson-Schwinger equations on a lattice}\label{sfvdse}

One of the main advantages of DSEs and FRGs is that the can treat all momentum scales equally well. Thus, a natural possibility is to use them to extrapolate lattice results beyond the limited range of momenta accessible in lattice calculations \pref{lat:limrange}. It is in particular interesting to determine the results of the functional equations at larger and larger volume, as well as finer discretizations, to ensure that for the range accessible to lattice calculations the correct behavior is reproduced.

In practical calculation, the finite volume is usually introduced as a sphere enclosing a discretized hyper-cube, to simplify calculations \cite{Fischer:2002eq,Fischer:2005ui,Fischer:2007pf}. Furthermore, the discretization is performed in momentum space, rather than in position space, to take the formulation of functional equations into account. Thus, the volume-dependence and discretization-dependence cannot be expected to be quantitatively the same as in lattice calculations. Nonetheless, good agreement at finite volume and discretization, as well as an understanding of the qualitative behavior, make the functional equations then a useful tool to perform these extrapolations. This type of extrapolations becomes more important when investigating matter, in particular chiral fermions \cite{Fischer:2005nf}.

\begin{figure}
\includegraphics[width=0.5\textwidth]{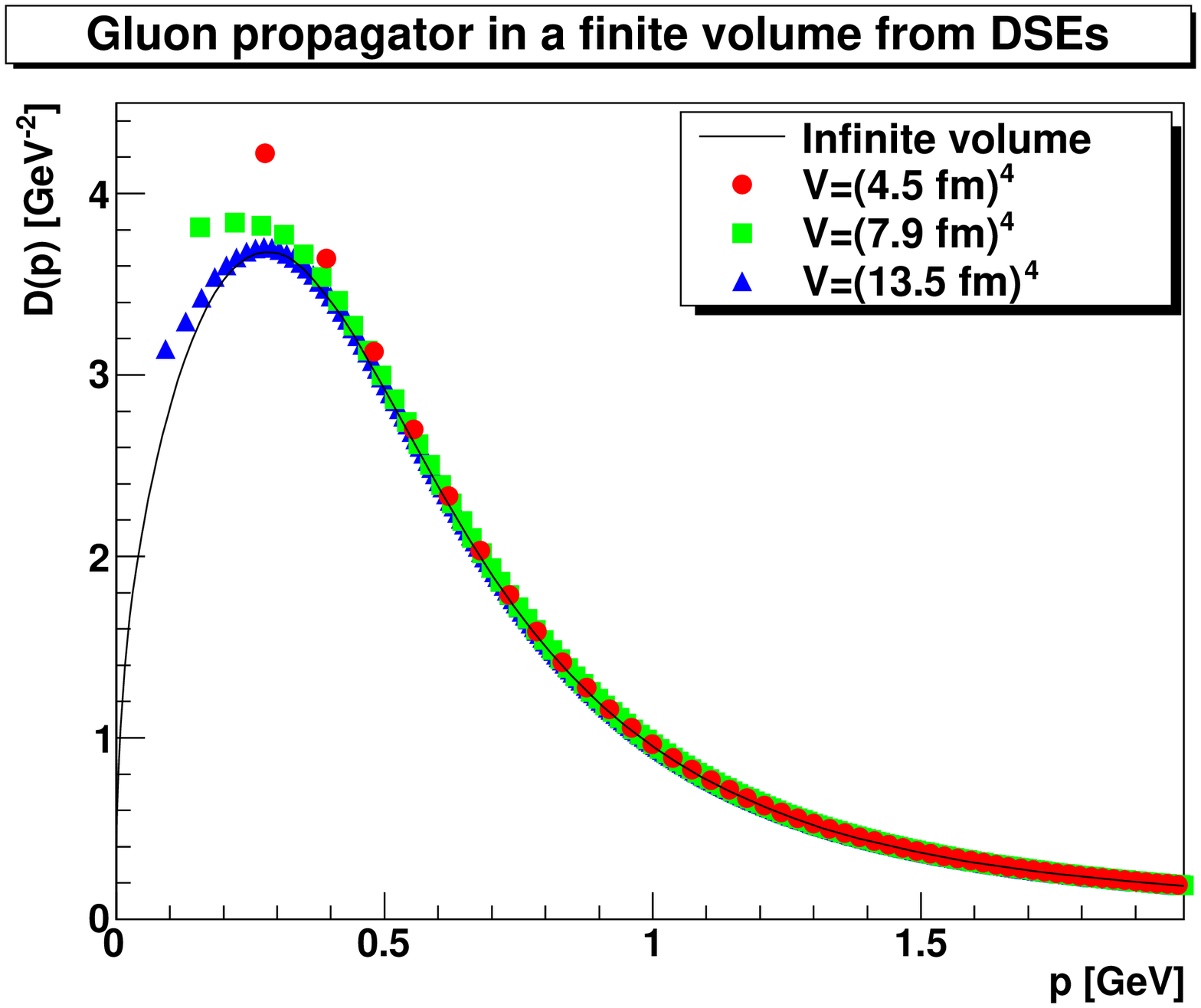}\includegraphics[width=0.5\textwidth]{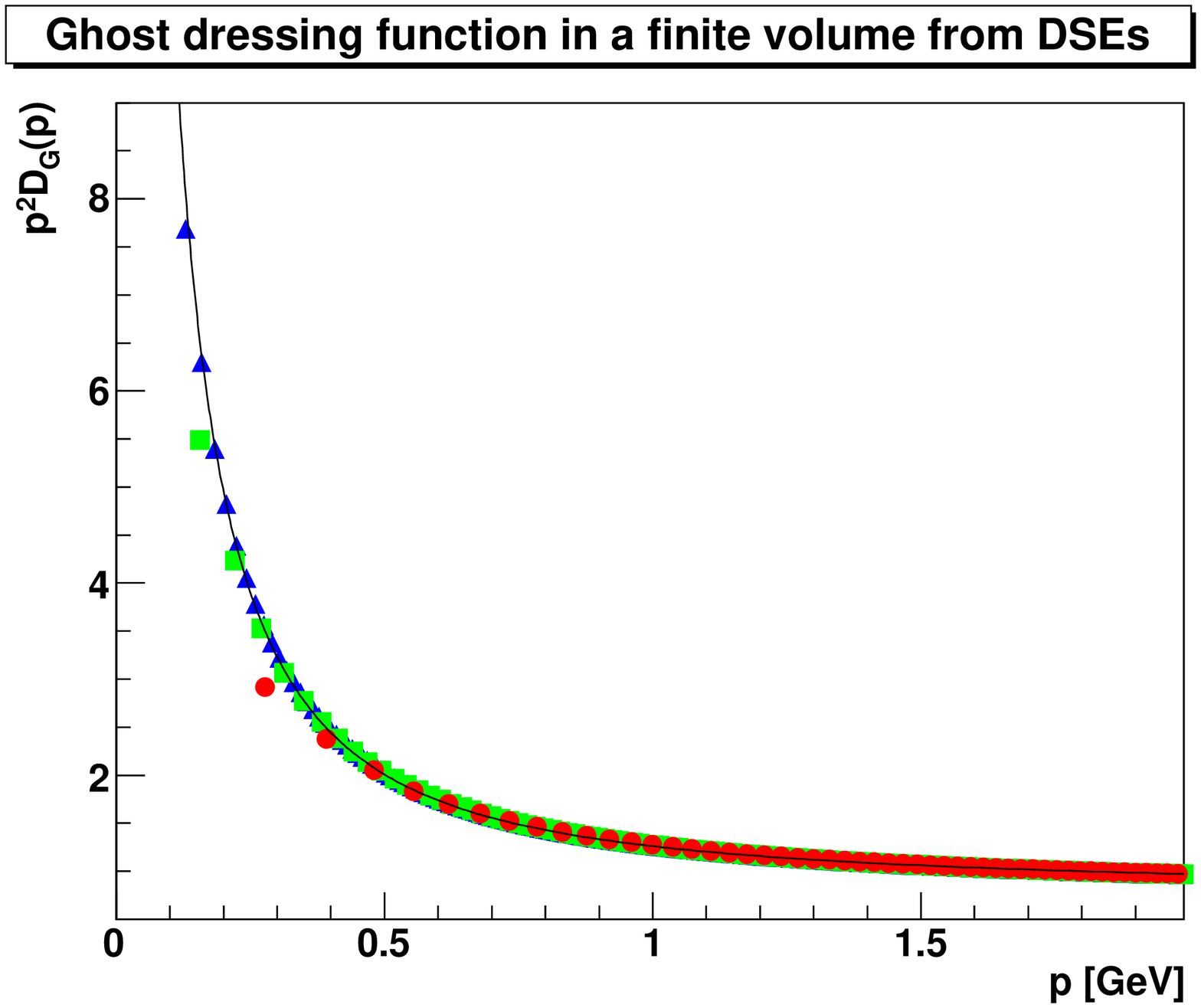}\\
\includegraphics[width=0.5\textwidth]{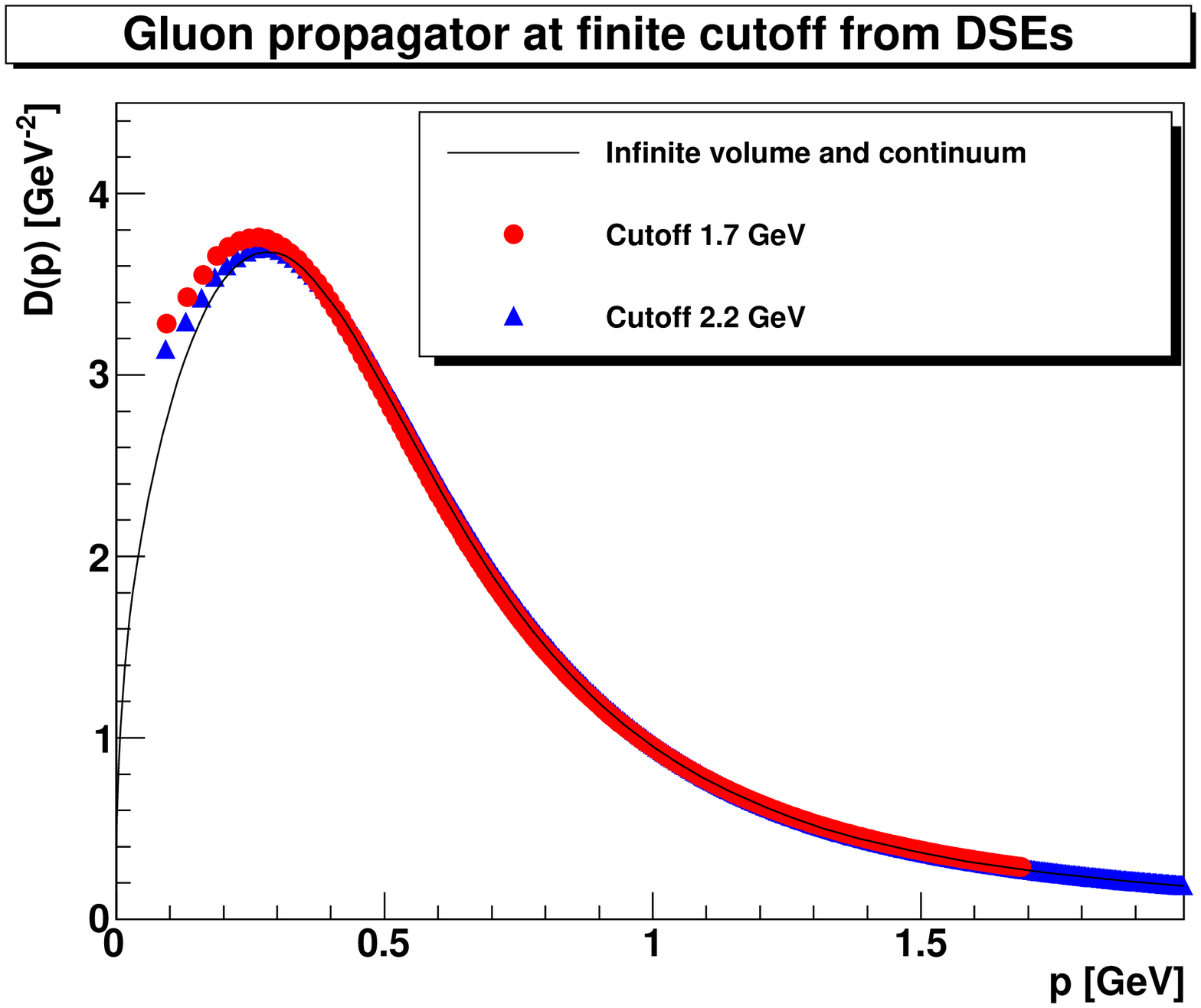}\includegraphics[width=0.5\textwidth]{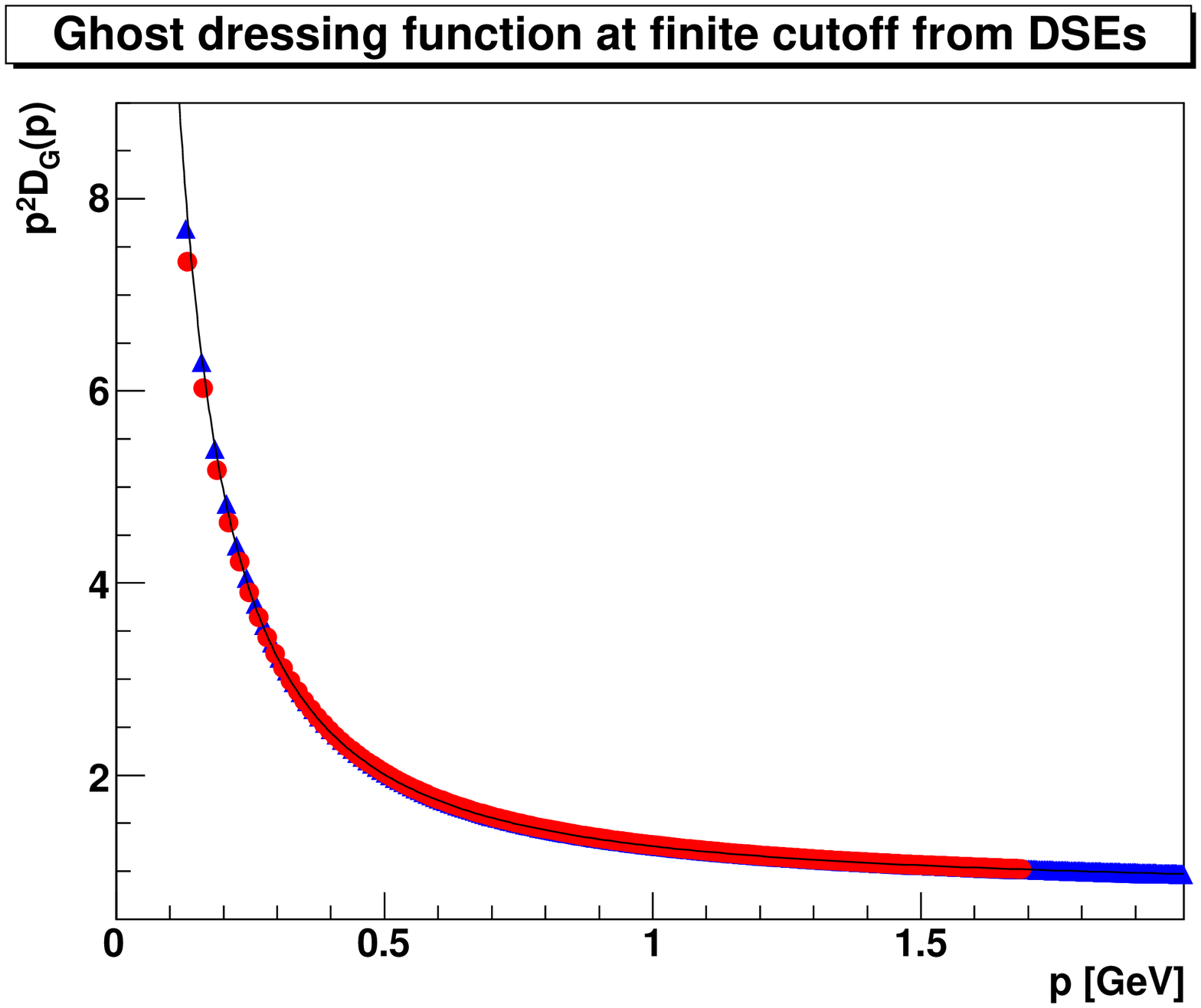}
\caption{\label{fig:fvdse}The results for the gluon propagator (left panels) and ghost dressing function (right panels) as a function of volume (top panels) and of discretization at fixed volume of (13.5 fm)$^4$ (bottom panels), from \cite{Fischer:2007pf}. This is only to illustrate the possibilities of DSEs. A detailed discussion of the solution properties is given below in section \ref{szerot}.}
\end{figure}

So far, this has only been performed to investigate the volume-dependence of the scaling case to predict at which volumes a scaling behavior can be observed instead of the finite-ghost case which will appear necessarily on any finite volume \cite{Fischer:2007pf}. Such results are illustrated in figure \ref{fig:fvdse}. This gave also rise to the notion of effective infrared exponents, which are determined under the assumption of a scaling window \pref{dse:conformalwindow}, and taken as function of volumes \cite{Fischer:2007pf,Maas:2007uv}. Investigations in two dimensions indicate that especially for the gluon propagator the volume dependence is rather well reproduced \cite{Maas:2007uv}.

\subsection{Slavnov-Taylor identities}\label{zerot:sti}

Though fixing the gauge is a difficult problem non-perturbatively, it also has an advantage. In particular, it restricts the structure of the vertex functions. A particular example of such a restriction is that the longitudinal part of the gluon propagator has to vanish in Landau gauge. This follows immediately from \cite{Cucchieri:2008zx}
\be
p_\mu p_\nu D^{AA}_\mn=p_\mu p_\nu <A_\mu A_\nu>=\mathrm{FT}<(\pd_\mu A_\mu)(\pd_\nu A_\nu)>=0,\label{zerot:dlongpart}\\
\ee
\no where $\mathrm{FT}$ denotes Fourier transform, and the Landau gauge condition in Fourier space, $p^\mu A_\mu^a=0$, has been used. Another example is that the part of the non-amputated ghost-gluon vertex longitudinal in the gluon momentum has to vanish upon contraction with the gluon momentum
\be
0=p_\mu <A_\mu(p)\bar{c}(q)c(p-q)>=<p_\mu A_\mu(p)\bar{c}(q)c(p-q)>\label{zerot:stiggv}.
\ee
\no Similarly, it can be shown that in the limit of $q\to 0$, the vertex self-energy is not divergent, yielding a finite\footnote{Finite in the sense that the renormalization constant is finite.} ghost-gluon vertex for all momenta \cite{Taylor:1971ff,Fischer:2005qe}.

These results follow from the constraints imposed by the gauge fixing on the gauge freedom. They can be generalized, yielding the Slavnov-Taylor identities (STIs) \cite{Bohm:2001yx}. In fact, these can be formulated just based on the presence of the gauge-symmetry of the original Lagrangian. After fixing a gauge, they yield a hierarchy of algebraic conditions on the vertex functions, which can be obtained from the Zinn-Justin equations \cite{ZinnJustin:2002ru,Pawlowski:2005xe,Ellwanger:1994iz}
\bea
\int_x\,\left( \frac{\delta \Gamma}{\delta K^a_\mu} \frac{\delta \Gamma}{\delta A^a_\mu}+ \frac{\delta\Gamma}{\delta L^a} \frac{\delta \Gamma}{\delta c^a}+\frac{\partial_\mu A_\mu^a}{\xi}\frac{\delta \Gamma}{\delta\bar c^a}\right) &=& \mbox{cut-off\ terms}\nn\\
\int_x \left(\frac{\delta \Gamma}{\delta\bar c^a}-\partial_\mu\frac{\delta \Gamma}{\delta K^a_\mu}\right)&=&0.\nn
\eea
\no Herein $K$ and $L$ are sources for the gluon and ghost fields in the effective action $\Gamma$, respectively. The cut-off terms on the right-hand side represent contributions which are non-generic, and depend on the regulator used. In case of gauge-symmetry preserving regulators, they will be zero, but non-zero for a cut-off regulator. However, similar to the functional equations, these are a coupled infinite hierarchy of conditional equations. Only in perturbation theory the hierarchy, as for the functional equations, decouples \cite{Bohm:2001yx,Rivers:1987hi}. Therefore, it is generally not possible to decouple these additional conditions. This can be illustrated \cite{Fischer:2008uz} by the STI for the three-gluon vertex, which is given by \cite{Ball:1980ax}
\bea
0&=&-<A_\nu^b\bar{c}^c\pd_\mu c^a>-<A_\mu^a\bar{c}^c\pd_\nu c^b>+\frac{1}{\xi}<A_\mu^a A_\nu^b \pd_\rho A_\rho^c>\nn\\
&&-g\left(f^{ade}<A_\mu^e A_\nu^b c^d \bar{c}^c>+f^{bde}<A_\mu^a A_\nu^e c^d\bar{c}^c>\right).\label{eq:stig3v}
\eea
\no Here, terms from the cut-off regulator have been dropped. The last term is evidently always of higher order in the coupling constant than the other terms, and thus does not contribute in perturbation theory. However, in a non-perturbative setting it cannot be ignored in general. So far, however, all non-perturbative investigations of this STI have kept this term either at best approximately \cite{BarGadda:1979cz} or not at all \cite{vonSmekal:1997vx,Boucaud:2008ji,Boucaud:2008ky,Ball:1980ax}, since as a scattering kernel its structure is enormously complicated. Furthermore, it appears to be consistent to truncate the STIs at the same level as the DSEs \cite{vonSmekal:1997vx,Fischer:2008uz}. Thus, for the here employed truncation scheme, it would be self-consistent to drop it.

The question naturally arises whether these conditions further restrict the solution manifold for the functional equations. In general, the answer is affirmative \cite{Bohm:2001yx}. However, Landau gauge is special in this respect\footnote{It is possible that very severe infrared divergences, proportional to $\pdm A_\mu$, could appear in the vertices when taking the Landau-gauge limit, which spoil this argument. That this is not the case is a rather weak regularity assumption, and there are no indication that this occurs \cite{Fischer:2008uz,Pawlowski:unpublished}.} \cite{Fischer:2008uz}.

As already indicated by \pref{zerot:dlongpart}, \pref{zerot:stiggv}, and \pref{eq:stig3v}, the hierarchy of STIs involve the contributions longitudinal with respect to the gluon momenta of vertex functions. On the other hand, the non-amputated correlation functions have all gluon legs contracted with a transverse projector from the gluon propagators. Hence, no longitudinal tensor structures contribute to these. Since the DSE hierarchy of equations for the non-amputated correlation functions is closed, the longitudinal parts of the correlation functions cannot contribute\footnote{On the lattice, this is manifest, since only full non-amputated correlation functions are accessible.}. The STIs, on the other hand, involve both the transverse and the longitudinal contributions. Since the transverse contributions are completely determined, the STIs are then constraint equations for the longitudinal contributions as a function of the transverse contributions only \cite{Fischer:2008uz,Pawlowski:unpublished}. In particular, the STIs can be solved for the longitudinal tensor structures as functionals of the transverse tensor structures.

This is to be expected: The STIs are relations manifesting the constraints due to the gauge symmetry. As such, they are conditions which are automatically fulfilled by the correct solutions, and thus do not contain any independent information compared to the hierarchy of DSEs. In particular, when determining physical observables, all correlation function are fully contracted with transverse projectors\footnote{Which in a perturbative construction is known as the LSZ formalism, if the asymptotic state space is available \cite{Peskin:1995ev}.}, and hence the longitudinal tensor structures do not carry any physical information. Thus, the STIs cannot constrain physical observations. However, when attempting to solve the functional equations for the amputated correlation functions this will only be possible in general when, at least at the same level of truncation, the longitudinal tensor structures will fulfill the corresponding STIs. Otherwise additional truncation artifacts may be introduced.

Two remarks must be made here. First, the hierarchy of equations for the amputated correlation functions requires the determination of both transverse and longitudinal tensor structures, and for the latter the STIs provide additional information. Second, any kind of truncation potentially violates the hierarchy of DSEs at one point, and therefore also the STIs in general are not fulfilled anymore, even at the same order. From these, the only available information which can be gained is on properties of truncations, but not on the correct result. An example is perturbation theory, in which case the violations of the STIs are well present, though always of one order higher in the coupling constant than the employed order of perturbation theory. Thus, it is possible to show that the violations occur because the expansion is truncated, but it is not possible to infer, e.\ g., the size of higher order contributions just from these violations.

\subsection{Renormalization}\label{sec:renorm}

An important technical issue to be returned to is that besides the spurious divergences discussed earlier there are the divergences which are physical, i.\ e., they result from the fact that Yang-Mills theory (at least in four dimensions) can only be a low-energy effective theory, but not a consistent theory of its own \cite{Weinberg:1995mt,Weinberg:1996kr}. This can be directly seen in perturbation theory \cite{Bohm:2001yx}, but lattice calculations also confirm this beyond perturbation theory \cite{Montvay:1994cy,Rothe:2005nw}.

This fact can be hidden in the renormalization process, which permits to encode this unknown physics in the renormalization constants \cite{Bohm:2001yx}. In the perturbative expansion of a renormalizable theory, like Yang-Mills theory, it can be shown that this is possible with a finite number of independent renormalization constants. For covariant gauges, like the Landau gauge, this is equivalent to multiplying correlation functions by appropriate chosen renormalization factors\footnote{This not true for all gauges, in particular not for non-covariant gauges \cite{Burnel:2008zz}.}, at least as long as no matter fields are involved.  If this is also possible beyond perturbation theory has not yet been proven, though no evidence to the contrary exists. Since due to asymptotic freedom the process of renormalization, once multiplicative renormalizability is assumed, can be performed essentially in the same way as in perturbation theory, it appears likely that this is correct.

As a consequence, for the purposes here the renormalization can essentially be performed as in perturbation theory \cite{vonSmekal:1997vx,Alkofer:2000wg,Fischer:2006ub,Fischer:2008uz}, and will not be discussed in detail other than to state the renormalization conditions. In particular, the renormalization can be performed using counter-terms \cite{vonSmekal:1997vx,Fischer:2003zc,Huber:2012td} or even more conveniently using a BPHZ \cite{Collins:1984xc} regularization prescription \cite{Maas:2005rf}. The only important constraint is that the truncated system of DSEs shows the correct qualitative transformation behavior under the renormalization group. Enforcing the coincidence with leading-order perturbation theory in the ultraviolet in the present truncations guarantees this actually even quantitatively \cite{Fischer:2008uz}. Results from lattice calculations automatically show the correct renormalization behavior, up to lattice artifacts \cite{Montvay:1994cy,Rothe:2005nw}.

It should be noted that Landau gauge is one of the gauges having optimal renormalization properties in the sense that the least number of independent divergent renormalization constants occur, and all of them are at most logarithmically divergent with the cut-off in four dimensions \cite{Bohm:2001yx,ZinnJustin:2002ru,Collins:1984xc}. Indeed, there are at most two independent renormalization factors \cite{Bohm:2001yx,vonSmekal:1997vx}, which can be chosen, e.\ g., to be the one of the running coupling and the wave-function renormalization of the ghost\footnote{Since possibly the resolution of the Gribov-Singer ambiguity may be related to the specification of at least one of the two correlators to some extent, it may be that one of the two renormalization constants can be related to the non-perturbative gauge-fixing procedure. This would be satisfying, as it would leave only one independent condition for the one independent parameter of the theory. However, at the present time, this is pure speculation.}. Furthermore, in dimensions lower than four all renormalization constants, and thus correlation functions, are finite, though still some loop-graphs are divergent and require regularization \cite{Collins:1984xc}. In perturbative diction the theory becomes super-renormalizable, though from the point of view of correlation functions it is actually finite \cite{Maas:2004se}.

A last consideration concerns the scheme dependence \cite{Bohm:2001yx} of the results. The standard scheme used in the calculations here is the miniMOM scheme\footnote{Which is in spirit very close to the conformal scheme proposed in \cite{Brodsky:2011ig}.}, given in \cite{vonSmekal:1997vx,vonSmekal:2009ae}. It essentially implements the condition \pref{ir:rencondlock} to lock the gluon and ghost wave-function renormalizations, and a condition on either propagator at an arbitrary, non-zero momentum can be used to yield the remaining wave-function renormalization. Furthermore, the finite ghost-gluon vertex renormalization constant is set to one. The finite three-gluon vertex renormalization constant, by virtue of the ans\"atze \pref{fm:g3v} and \pref{zerot:aguv}, cancels out, and the finite four-gluon vertex renormalization condition is dropped anyway. This leaves only the renormalization condition for the running coupling, which is implicitly defined by the relation \pref{ir:alpha}.

There are two additional remarks to this scheme. First, if the different non-perturbative gauges indeed exist as described in section \ref{quant:resgrisin}, the propagators will differ for different choices, though the difference at very large momenta will be power-like suppressed \cite{Maas:2008ri}. Still, this implies that the renormalization conditions will differ between different non-perturbative versions of the Landau gauge, introducing a gauge-dependence. The second is that the definition \pref{ir:alpha} for the running coupling is not unique. In particular, it is possible to define a running coupling which will have an infrared fixed-point irrespective of whether the propagators show a finite-ghost or a scaling behavior \cite{Fischer:2008uz,Aguilar:2009nf}. This is in contrast to the definition \pref{ir:alpha}, for which the running coupling is only infrared finite for the scaling case, but not for the finite-ghost case. Since this implies a difference between the corresponding $\beta$-functions, this yields that both correspond to different renormalization schemes. In fact, since the former always produces an infrared fixed-point while the latter can imply either a fixed-point or a zero crossing of the $\beta$ function, as demonstrated in the next chapter, it follows that the scheme transformation between both schemes is singular. Such singular scheme transformations are possible, and can affect as here the presence of fixed points \cite{Creutz:2011hy}.

\section{Gluons at zero temperature}\label{szerot}

With the chapters \ref{squant} and \ref{smethods}, the stage is set to actually determine correlation functions. In the following the results of such calculations will be presented. Also, the interpretation of these results in terms of physical properties will be briefly discussed.

\subsection{Correlation functions}

\subsubsection{Propagators}\label{zerot:prop}

The simplest correlation functions which are non-zero in Landau gauge are the propagators. Before investigating their momentum dependency, a first important question is the justification of whether it is permitted to factorize the color structure in the form of a unit matrix, as it is usually done. The so-called primitiveness assumption of group theory \cite{Cvitanovic:2008} suggests this, as does perturbation theory \cite{Bohm:2001yx}. Furthermore, the DSE for the ghost propagator shows that the non-tree-level color structure of the ghost-gluon vertex is not relevant for the ghost propagator. Still, a general proof is lacking.

\begin{figure}
\includegraphics[width=0.45\textwidth]{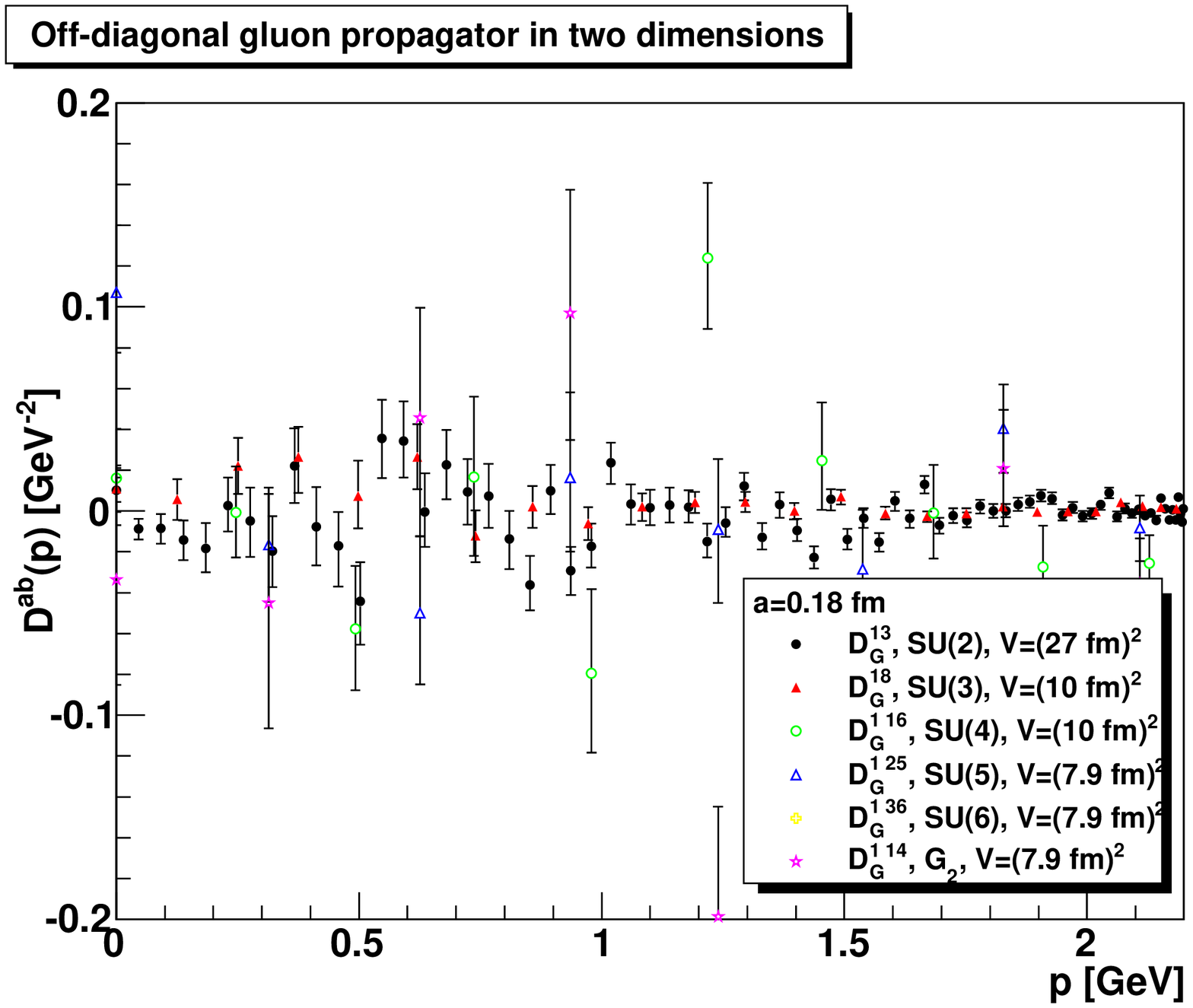}\includegraphics[width=0.45\textwidth]{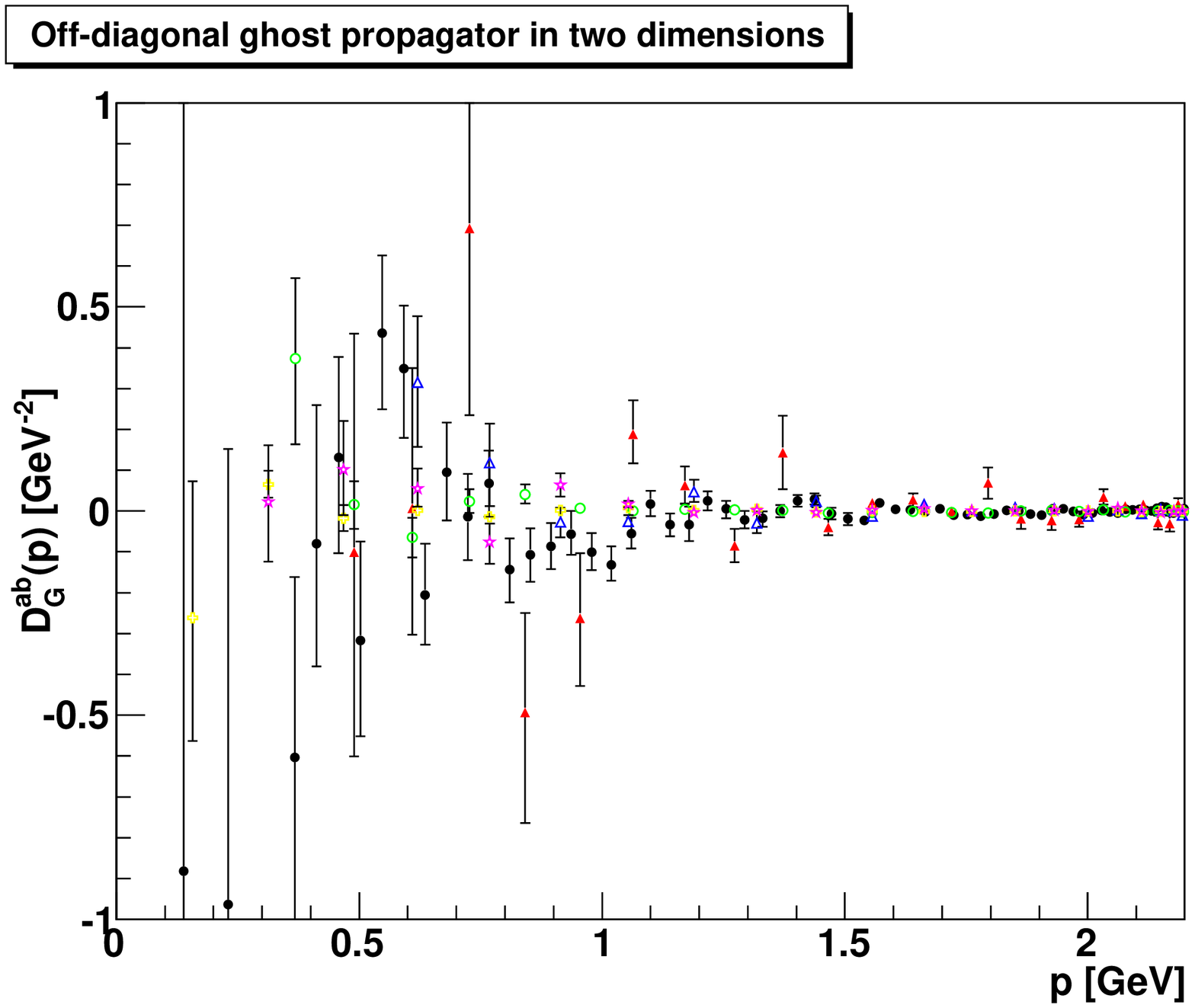}\\
\includegraphics[width=0.45\textwidth]{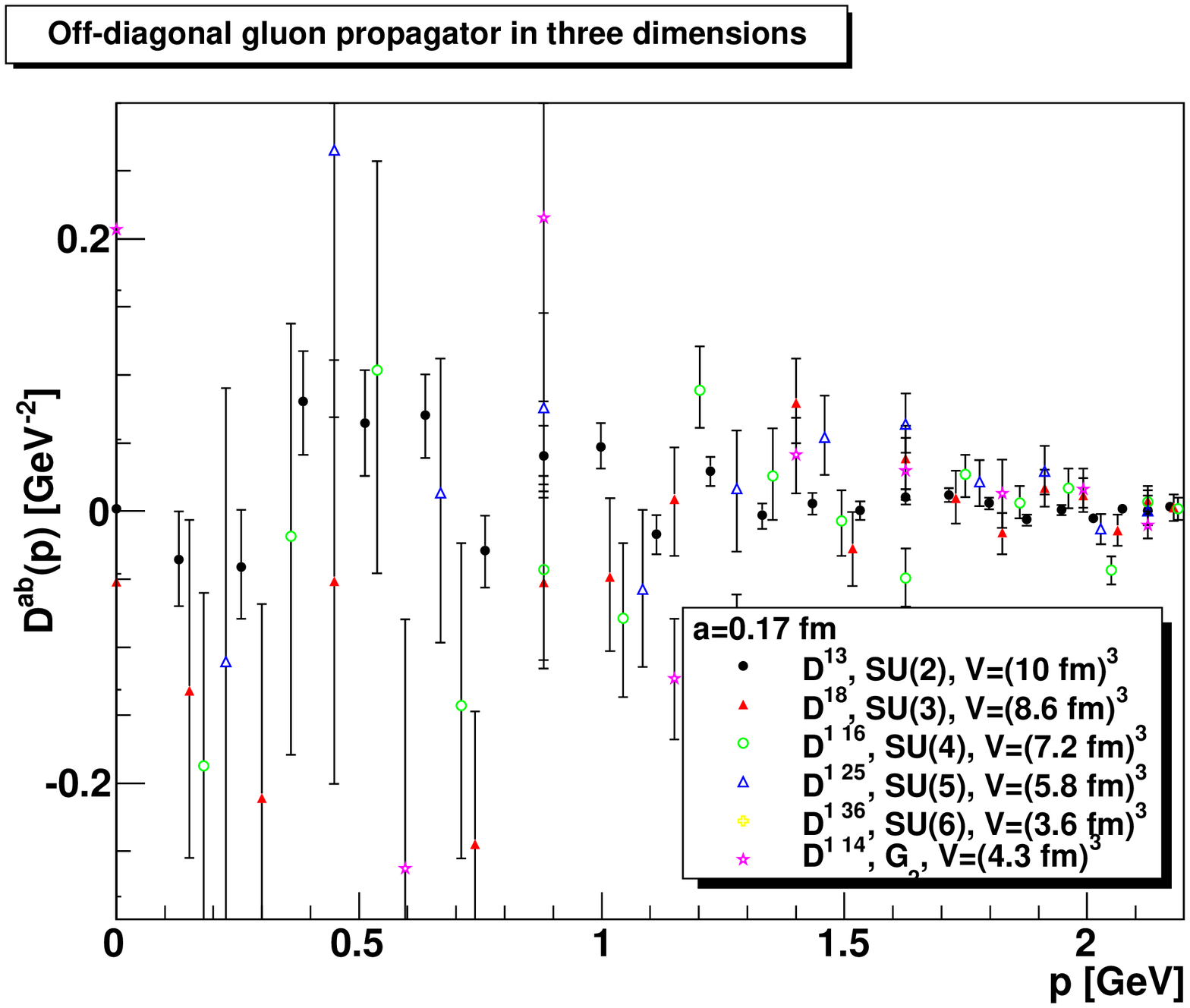}\includegraphics[width=0.45\textwidth]{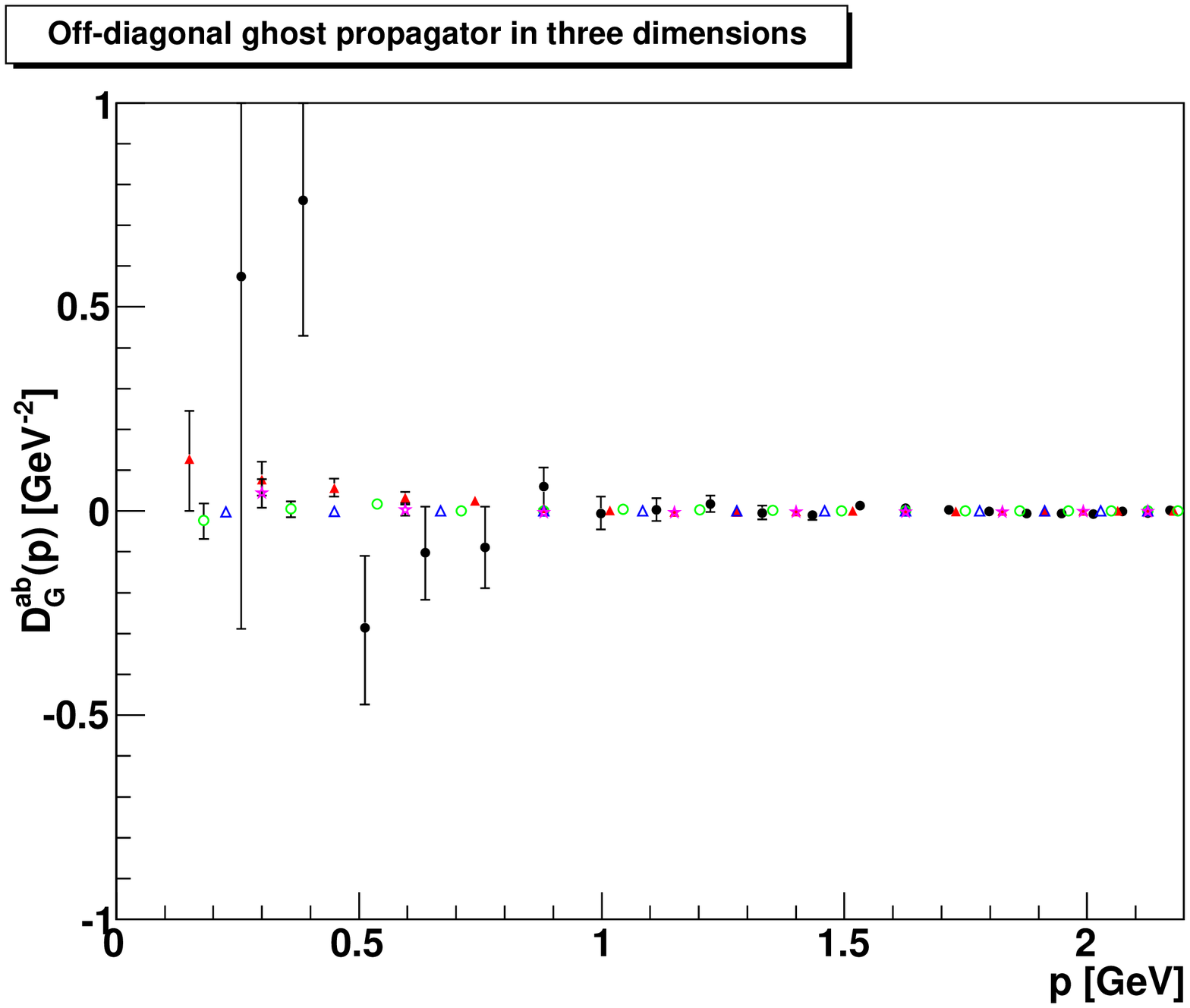}\\
\includegraphics[width=0.45\textwidth]{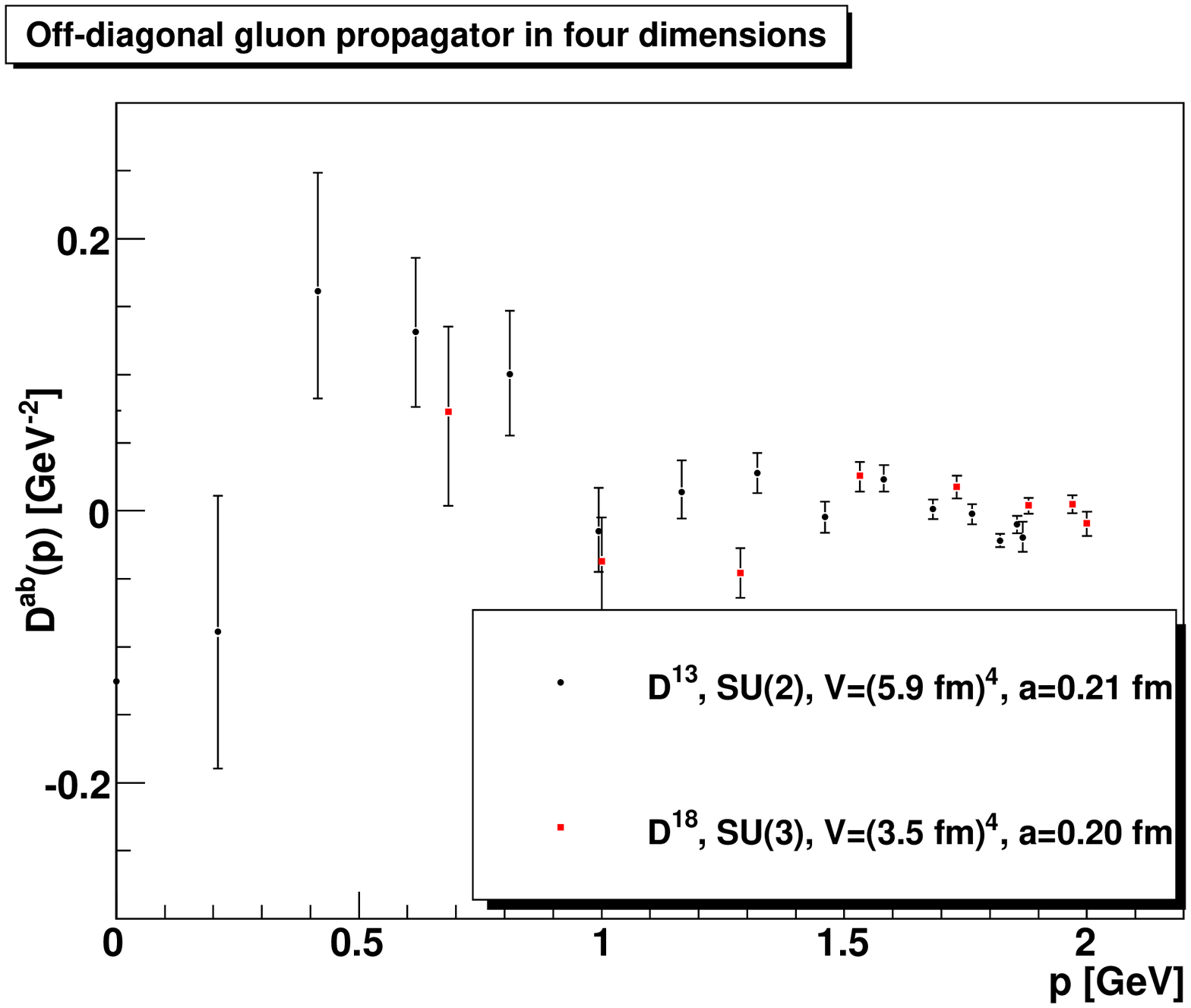}\includegraphics[width=0.45\textwidth]{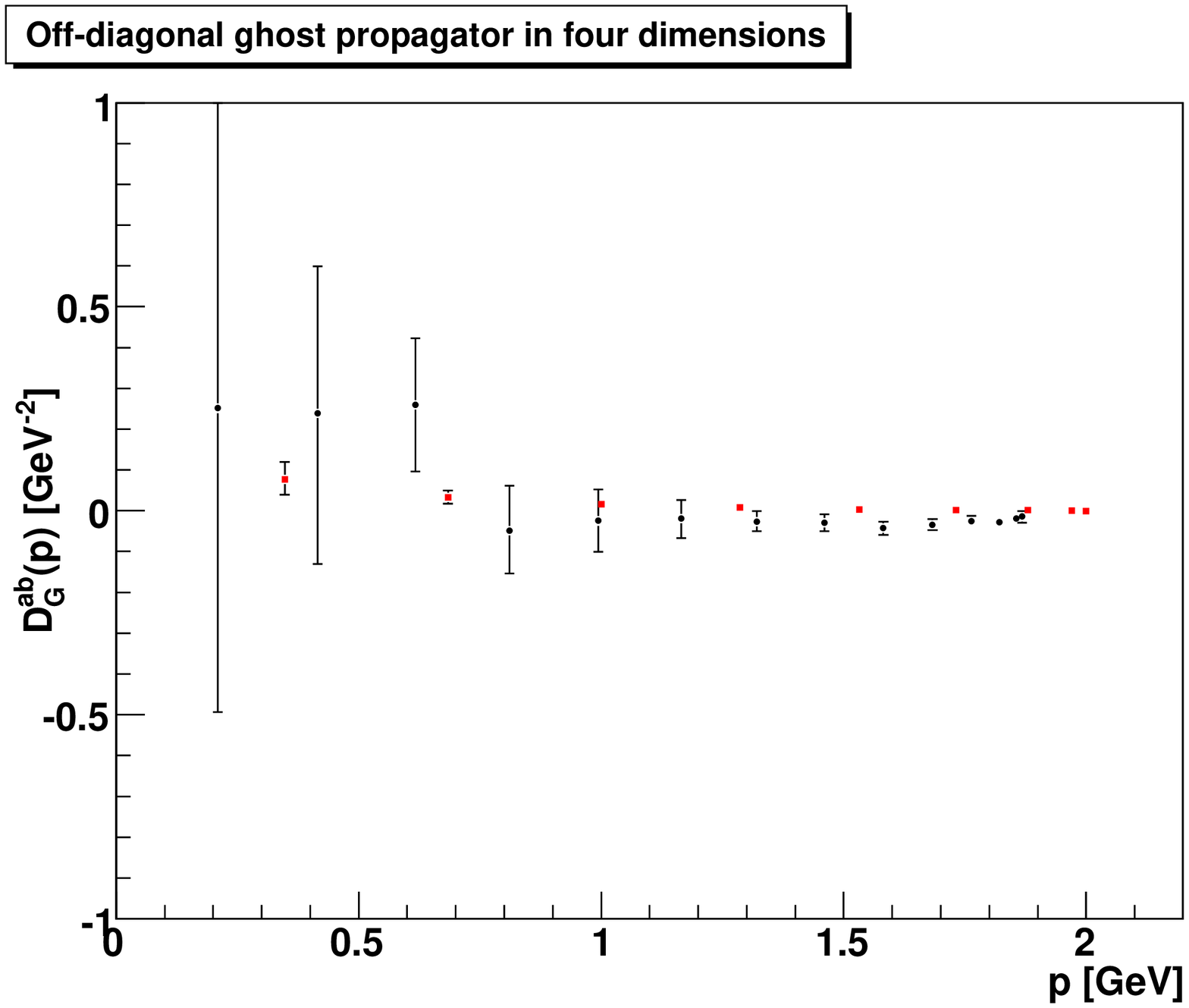}
\caption{\label{fig:off}Some components of the off-diagonal gluon propagator (left panels) and ghost propagator (right panels) in two (top panels) \cite{Maas:2007uv,Maas:2010qw}, three (middle panels) \cite{Cucchieri:2006tf,Maas:2010qw}, and four (bottom panels) \cite{Cucchieri:2008qm,Fischer:2010fx} dimensions, for various gauge groups. The results in four dimensions are renormalized at 2 GeV. Here and later in the legends the groups used for the lattice implementations are given.}
\end{figure}

Explicit lattice calculations have also provided substantial support for color-diagonal propagators \cite{Cucchieri:2006tf,Maas:2007uv,Cucchieri:2008qm,Maas:2010qw}. As an example in figure \ref{fig:off} some of the color-off-diagonal components of the propagators in two, three, and four dimensions for various gauge algebras are shown. In all cases, within statistical errors, the off-diagonal propagators are compatible with zero. Thus, assuming the propagators to be color-diagonal seems to be justified, and therefore will be done henceforth.

\begin{figure}
\includegraphics[width=\textwidth]{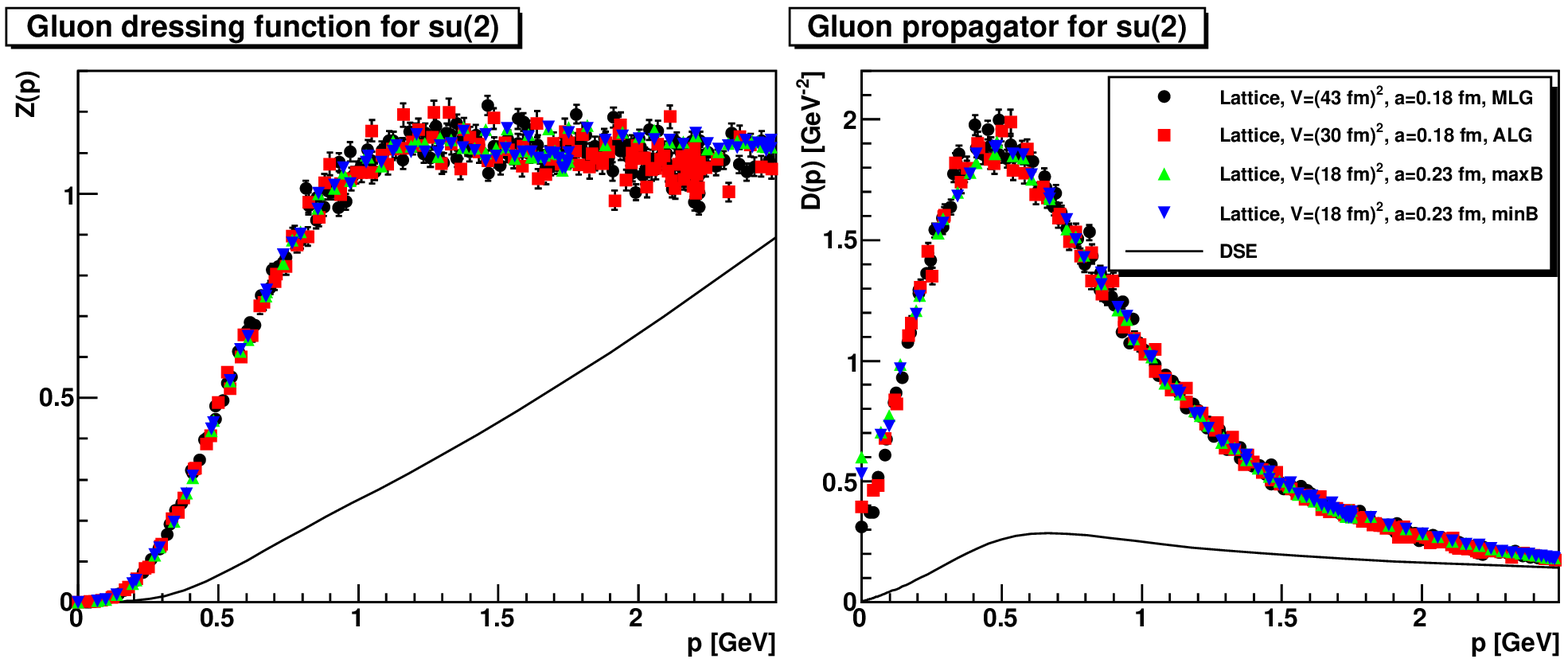}\\
\includegraphics[width=\textwidth]{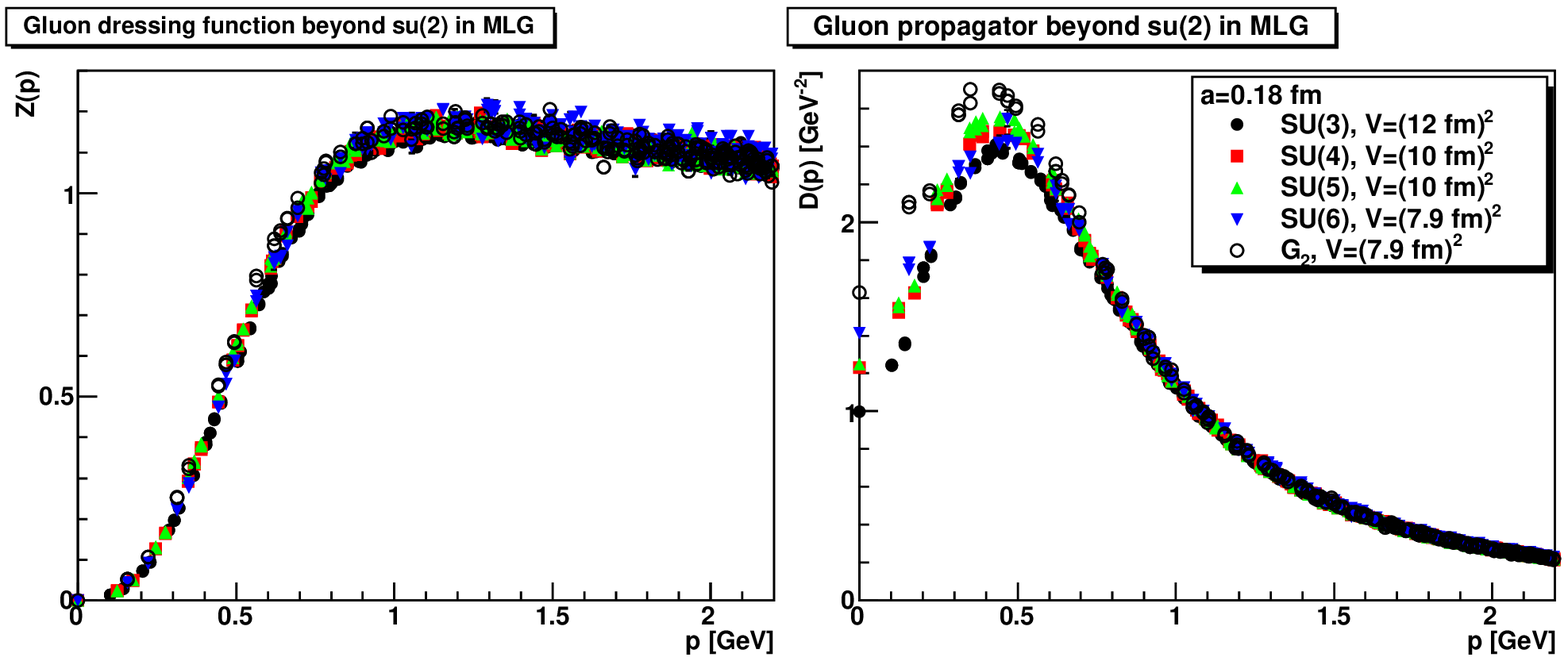}
\caption{\label{fig:gp-2d}The su(2) gluon propagator (top left panel) and its dressing function (top right panel) in two dimensions. Shown are results for minimal Landau gauge (MLG) \cite{Maas:2007uv}, absolute Landau gauge (ALG) \cite{Maas:2008ri}, and the maxB and minB gauges \cite{Maas:2009se,Maas:unpublished}, compared to DSE results in the scaling case \cite{Huber:2012td}. Note that in two dimensions the finite-ghost case seems not to exist \cite{Huber:2012td,Dudal:2012td,Zwanziger:2012se}. The bottom panels show the same, but for gauge algebras su(3-6) and g$_2$ in MLG \cite{Maas:2010qw}. Further results in two dimensions can be found in \cite{Dudal:2008xd,Cucchieri:2007rg,Maas:2009ph,Maas:2007af,Cucchieri:2011ig}.}
\end{figure}

\begin{figure}
\includegraphics[width=\textwidth]{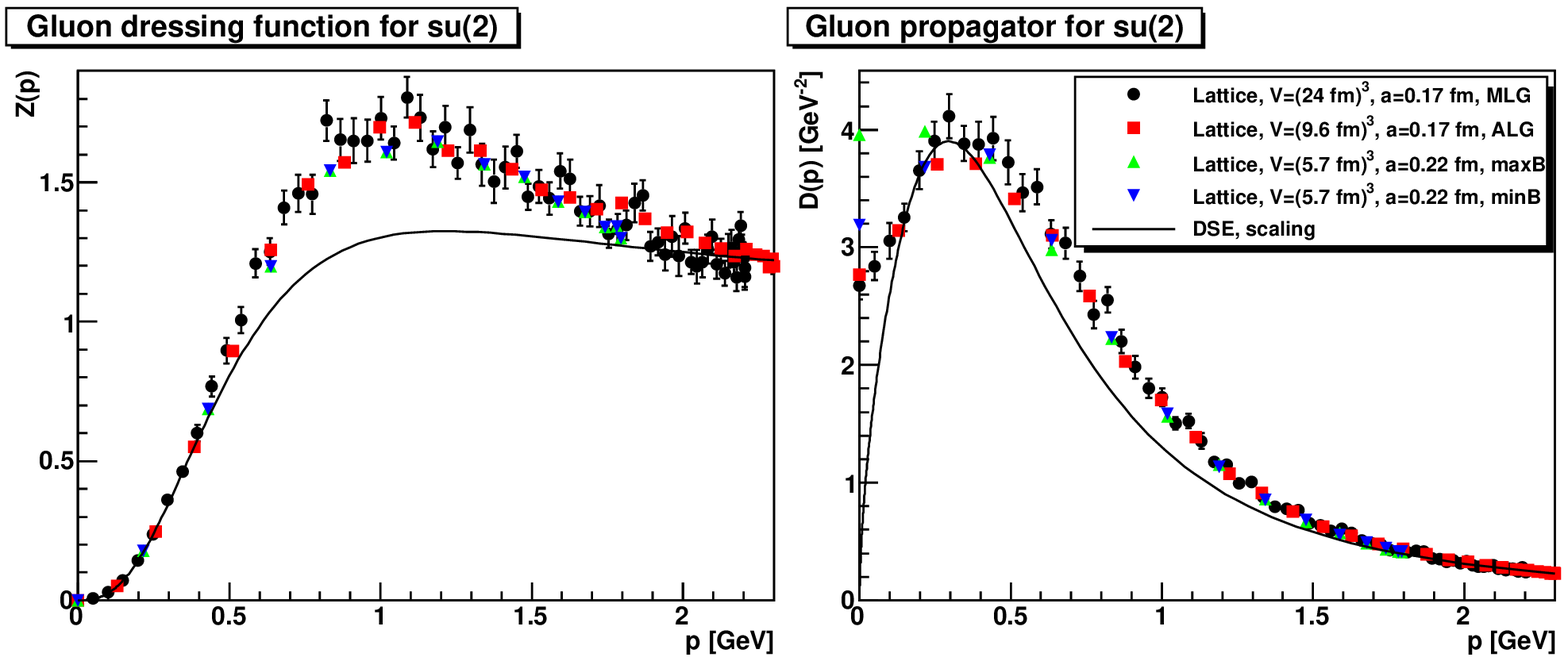}\\
\includegraphics[width=\textwidth]{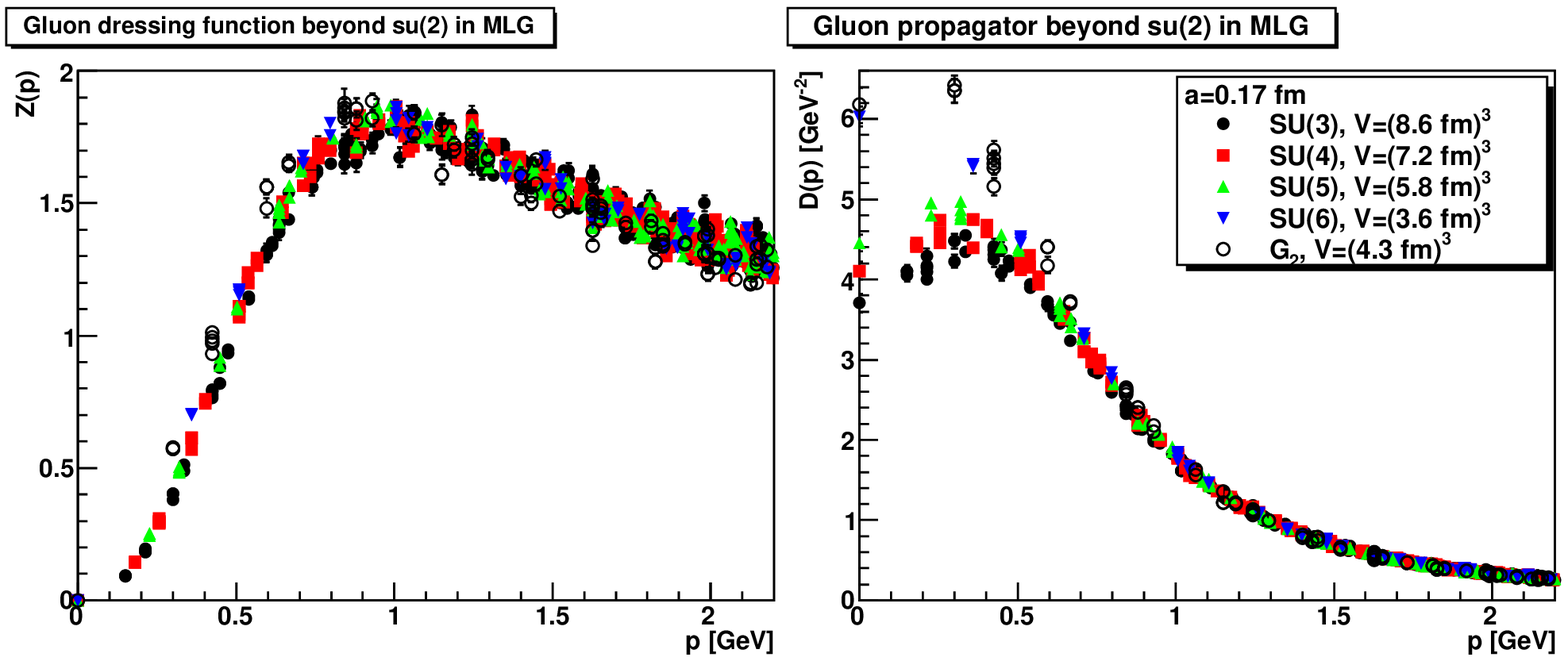}
\caption{\label{fig:gp-3d}The su(2) gluon propagator (top left panel) and its dressing function (top right panel) in three dimensions. Shown are lattice results for minimal Landau gauge (MLG) \cite{Cucchieri:2003di}, absolute Landau gauge (ALG) \cite{Maas:2008ri}, the maxB and minB gauges \cite{Maas:2009se,Maas:unpublished}, and DSE results for the scaling case \cite{Maas:2004se}.  A finite-ghost-type DSE solution can be found in \cite{Aguilar:2010zx}. The bottom panels show the same, but for gauge algebras su(3-6) and g$_2$ in MLG \cite{Maas:2010qw}. Further lattice results can be found in \cite{Maas:unpublished,Maas:2009ph,Cucchieri:2007rg,Cucchieri:2009zt,Cucchieri:2006tf,Cucchieri:2008qm,Maas:2007af,Cucchieri:1999sz,Cucchieri:2011ig,Bornyakov:2011gp}.}
\end{figure}

\begin{figure}
\includegraphics[width=\textwidth]{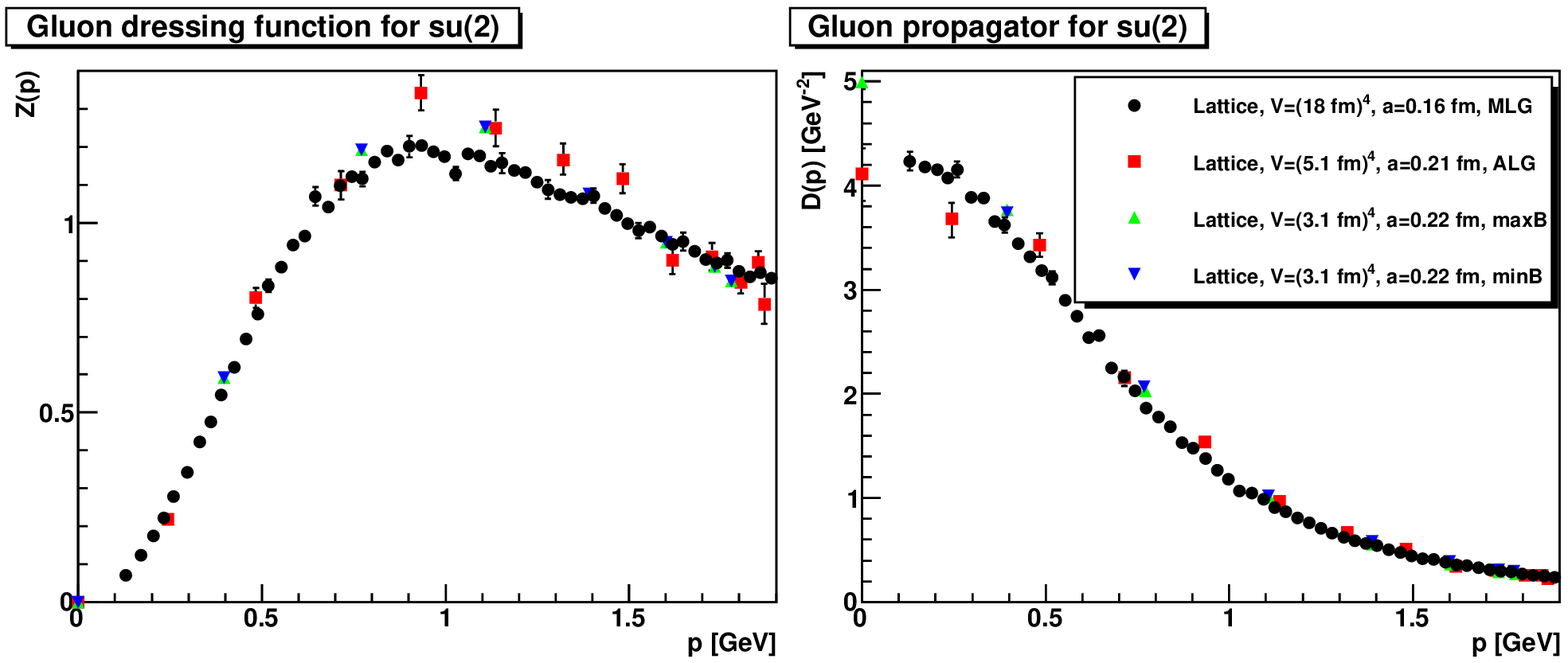}\\
\includegraphics[width=\textwidth]{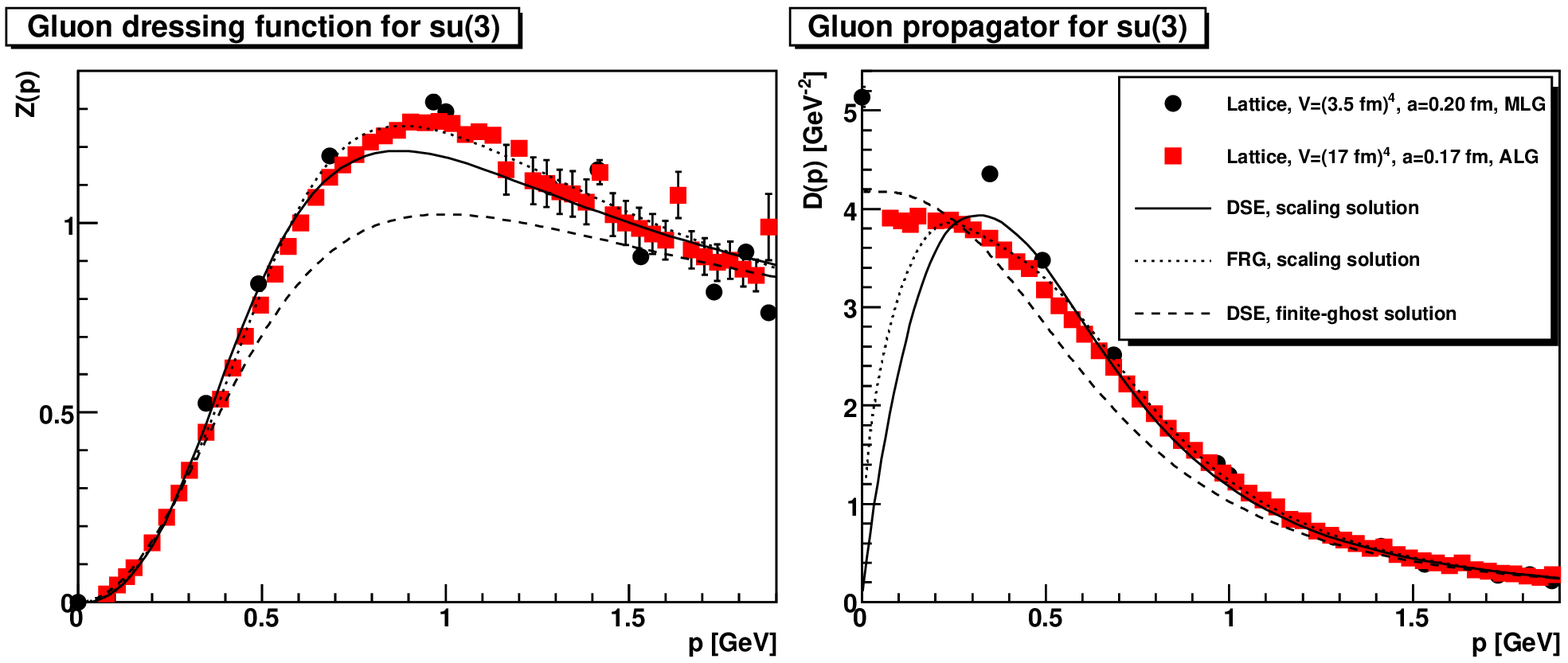}
\caption{\label{fig:gp-4d}The su(2) gluon propagator (top left panel) and its dressing function (top right panel) in four dimensions, renormalized to Z(1.5 GeV$^2$)=1. Shown are lattice results for minimal Landau gauge (MLG) \cite{Sternbeck:2007ug}, absolute Landau gauge (ALG) \cite{Maas:unpublished}, and the maxB and minB gauges \cite{Maas:unpublished}. The bottom panels show the same, but for gauge algebra su(3) in MLG \cite{Fischer:2010fx} and ALG \cite{Bogolubsky:2009dc}, compared to scaling and finite-ghost DSE and scaling FRG results \cite{Fischer:2008uz}. Further results can be found in \cite{Aguilar:2009ke,Alkofer:2000wg,Binosi:2009qm,Bloch:2003sk,Bogolubsky:2005wf,Bornyakov:2008yx,Bornyakov:2009ug,Boucaud:2010gr,Cucchieri:2007rg,Cucchieri:2007zm,Dudal:2008sp,Dudal:2007cw,Fischer:2006ub,Fischer:2004uk,Oliveira:2008uf,Oliveira:2009eh,Pawlowski:2003hq,Pawlowski:2009iv,RodriguezQuintero:2010wy,vonSmekal:2009ae,Sobreiro:2004us,Tissier:2010ts,Gong:2008td,Cucchieri:2009zt,Oliveira:2007dy,Silva:2005hb,Cucchieri:2006za,Furui:2003jr,Cucchieri:2011ig,Oliveira:2012eh}.}
\end{figure}

The first propagator to be studied is the gluon propagator, also as function of the dimensionality, and for the various gauges and cases. The results are shown for both the propagator and the corresponding dressing function in figure \ref{fig:gp-2d} for two dimensions, in figure \ref{fig:gp-3d} for three dimensions, and in figure \ref{fig:gp-4d} for four dimensions, also comparing different gauge algebras. Most of the results are from lattice calculations, though there are also some results provided from the available functional calculations using DSEs and FRGs in two, three and four dimensions. For the lattice calculations four different choices of the non-perturbative gauges are shown, the minimal Landau gauge, the absolute Landau gauge, and the maxB and minB gauge, to explore both a typical choice of $B$, and extreme possibilities. Since the implementation of gauges beyond the minimal Landau gauge are numerically very expensive until now only much smaller volumes are available.

The first result visible is that there is very little dependency on the gauge algebra, at least up to finite volume effects, and quantitative dependencies\footnote{Note, however, \cite{Oliveira:2009eh,Oliveira:2008uf} for su(2) and su(3) in four dimensions.}. The former turn out to be very similar for all cases \cite{Maas:2010qw,Sternbeck:2007ug,Cucchieri:2007zm}. It thus suffices to stay with the su(2) case for now, for which most data is available. In case of DSEs, the gauge algebra dependency is entirely through the combination $g^2C_A$ at this level of truncation, and therefore the dependency can be absorbed in a dependency on the renormalization scale, and perfect 't Hooft scaling is manifest \cite{vonSmekal:1997vx,Maas:2005ym}, except for perturbative corrections in four dimensions. Of course, by construction, leading-order (resummed) perturbation theory is reproduced at sufficiently large momenta for all cases.

\begin{figure}
\includegraphics[width=0.5\textwidth]{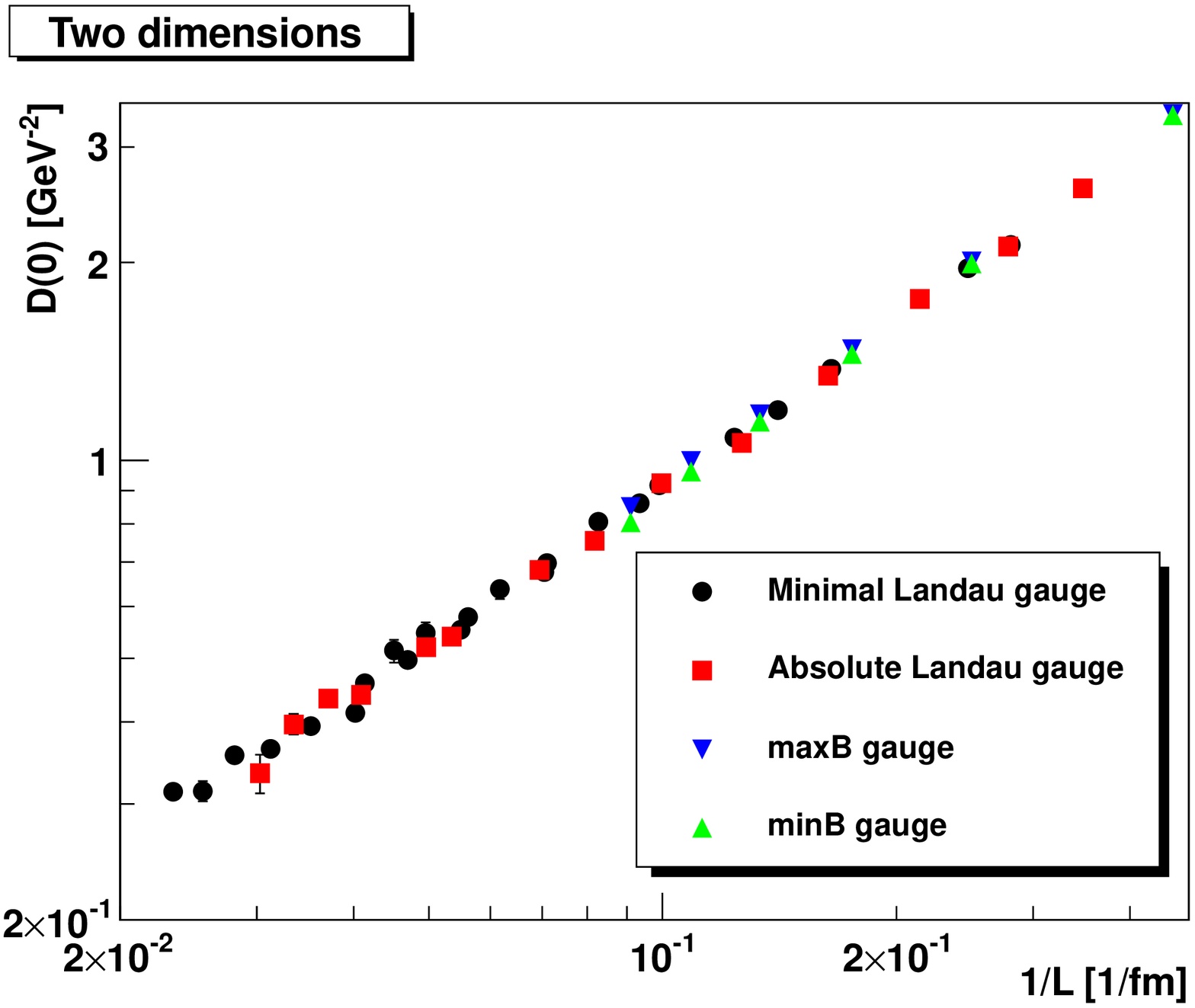}\includegraphics[width=0.5\textwidth]{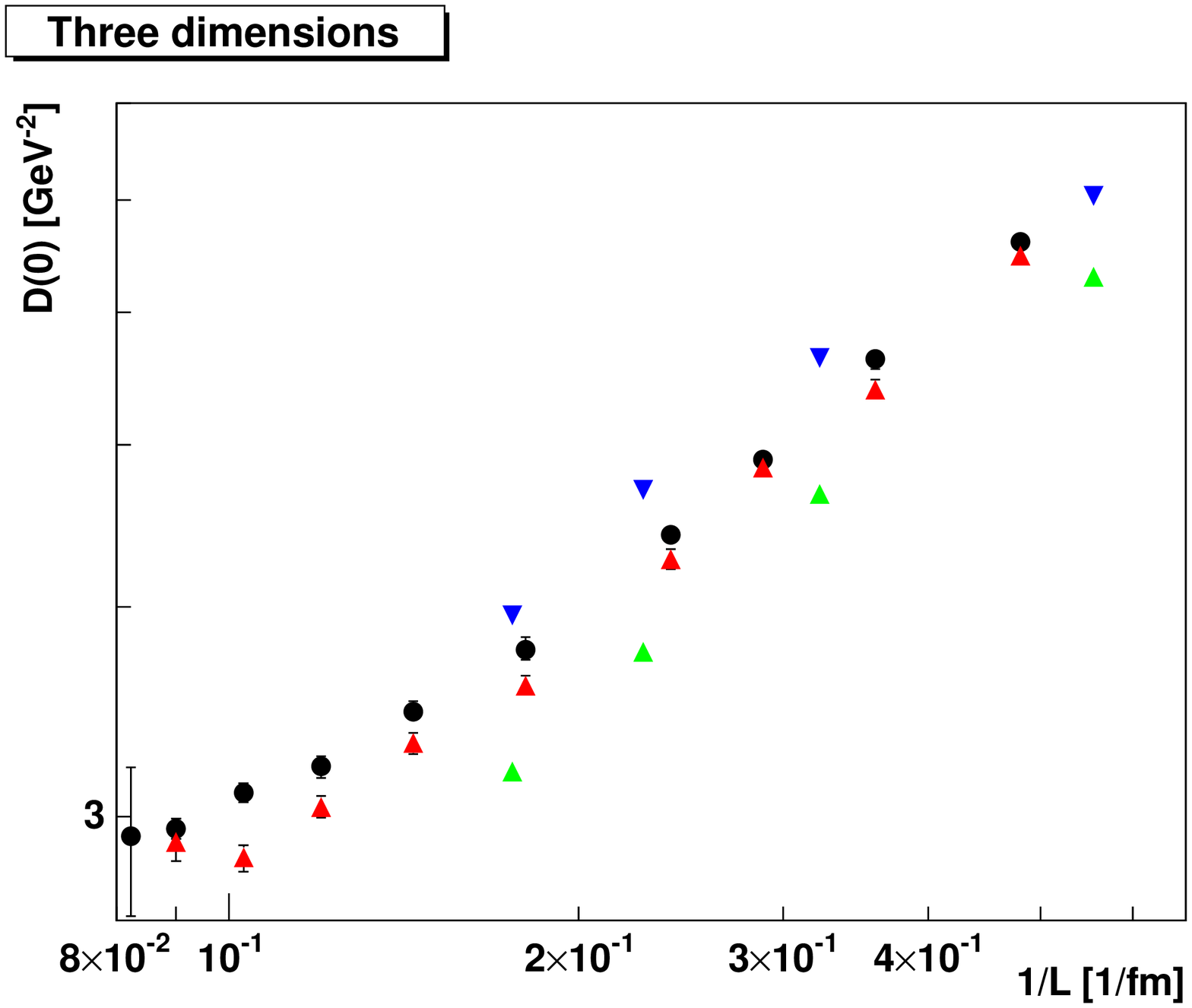}\\
\begin{minipage}{0.5\linewidth}
\begin{center}\includegraphics[width=\textwidth]{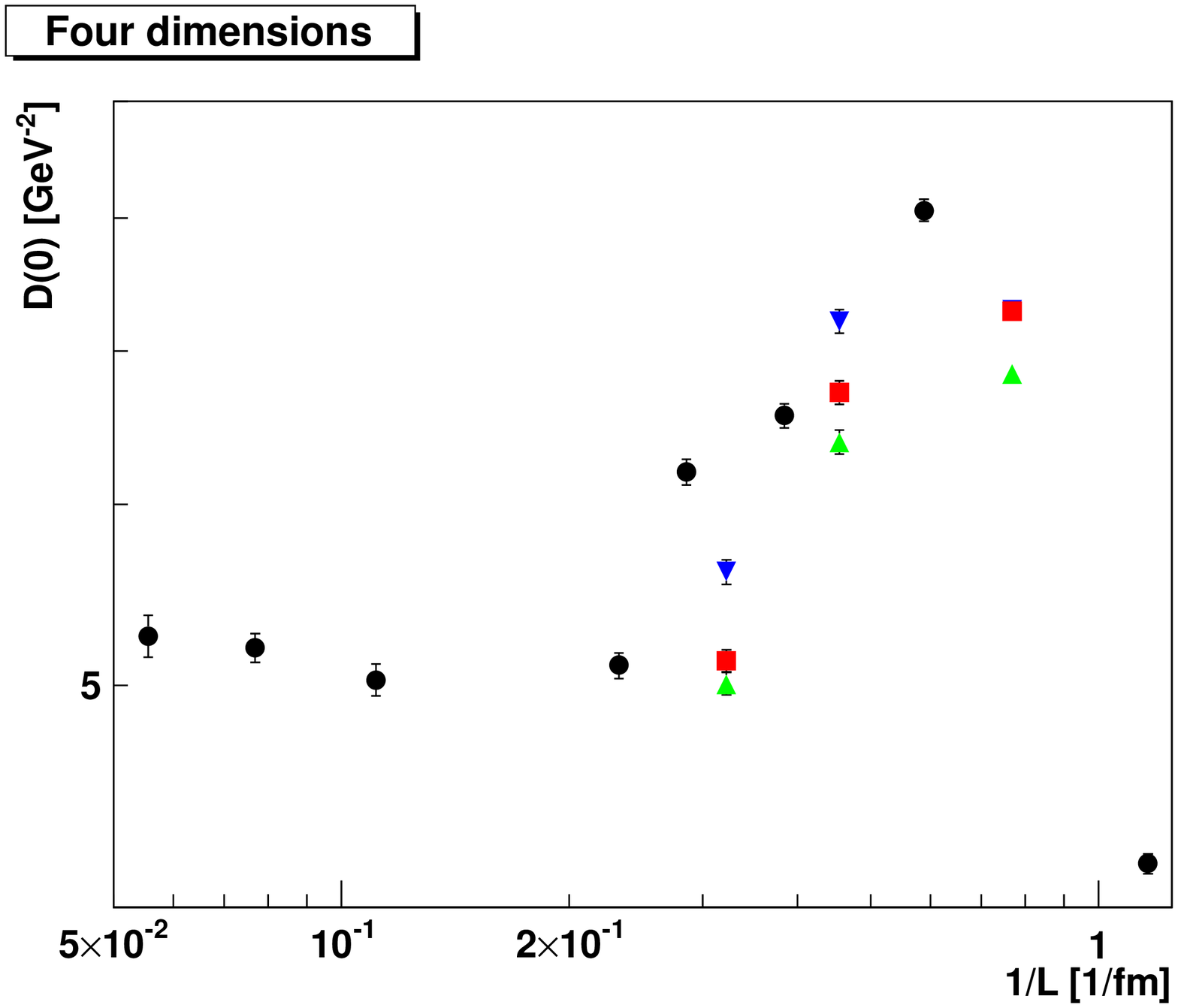}\end{center}
\end{minipage}
\begin{minipage}{0.5\linewidth}
\caption{\label{fig:d0}The gluon propagator at zero momentum \cite{Maas:unpublished,Maas:2008ri,Maas:2009se,Maas:2007uv,Cucchieri:2003di,Sternbeck:2007ug,Bogolubsky:2009dc} for su(2) for maximal Landau gauge, absolute Landau gauge, minB gauge, and maxB gauge. Two dimensions are shown in the top-left panel, three dimensions in the top-right panel, and four dimensions (renormalized at 2 GeV) in the bottom panel. Note the remarks on the reliability of the largest volume in the absolute Landau gauge in \cite{Maas:2008ri}, and that the results could potentially be affected by discretization effects \cite{vonSmekal:2009ae,Bloch:2003sk,Lepage:1992xa,Maas:unpublished}.}
\end{minipage}
\end{figure}

At first glance, there is very little difference between the propagator for the different gauges. This is in line with the expectations from figure \ref{fig:raw-d}, which already indicates that the total integral of the gluon propagator is rather constrained. However, since the measure in \pref{quant:abslg} strongly suppresses the infrared, this is not a necessary result, and large differences in the infrared could be possible \cite{Maas:2008ri}. This is indicated by the DSE results, which show a qualitative difference between the scaling and finite-ghost case, with a vanishing gluon propagator at zero momentum in the former case. Concerning this statement, there are two points to be cautious about. One is that in every finite volume, even in the scaling case, the gluon propagator has to be non-zero at zero momentum \cite{Fischer:2007pf}. As shown in figure \ref{fig:d0}, however, only in the two-dimensional case the gluon propagator at zero momentum appears to extrapolate to zero \cite{Maas:2007uv,Maas:unpublished,Cucchieri:2007rg,Maas:2008ri,Maas:2009se} and in this case for all gauges. On the other hand, the gluon propagator appears to be non-zero in three and four dimensions  in all gauges \cite{Sternbeck:2007ug,Bogolubsky:2009dc,Cucchieri:2007rg,Maas:2008ri,Maas:2009se,Cucchieri:2011ig,Bornyakov:2011gp}. The fact that the result of the minB gauge undershoots the one of the absolute Landau gauge demonstrates again the irrelevance of the zero-momentum behavior for the absolute-Landau-gauge condition. A comparison at finite momenta shows that the integrated weight of the propagator in absolute Landau gauge at the same lattice setting is smaller than for the minB gauge \cite{Maas:unpublished}.

This seems to be ruling out the scaling case in four and three dimensions in the first Gribov region, but, as discussed in section \ref{sir:scaling}, the actual value of the exponent $\kappa_{AA}$ is dependent on the truncation, and its value changes if the ghost-gluon vertex is different from a bare vertex in the infrared. For both ans\"atze employed here, the ghost-gluon vertex becomes constant in the infrared. In this case \cite{Zwanziger:2001kw,Lerche:2002ep} the value of $\kappa_{\bar{c}c}$ can be determined to be approximately 0.595 in four dimensions. Thus, the value of $\kappa_{AA}$ is then about 1.19. In three and two dimensions, two possible solutions are found. In three dimensions the values are 1/2 and about 0.39, yielding $\kappa_{AA}$ as 3/2 and 1.28, respectively. The functional equations yield consistent solutions for both values \cite{Maas:2004se}. In two dimensions, the possibilities are 1/5 and 0. The latter value, yielding $\kappa_{AA}=1$ with an infrared finite gluon propagator, appears to be ruled out by the lattice data, which prefer $\kappa_{AA}\approx 1.4$ \cite{Maas:2007uv,Cucchieri:2011ig}. However, in three and four dimensions it is possible to obtain a scaling solution with an infrared finite gluon propagator with a ghost-gluon vertex, which differs quantitatively from a bare one in the infrared \cite{Huber:2012td}. Hence, at most, the conclusion which can be drawn from the gluon propagator alone is that the truncation in three and four dimensions is possibly not viable, if scaling is realized in the first Gribov region. Since the ghost-gluon vertex indeed shows infrared variations, as will be shown in the next section, this is a possibility to reconcile scaling with the gluon propagator results, but a full understanding requires further investigations for both lattice and continuum methods. Furthermore, the ghost propagator also has to be taken into account, which will be done below. In addition, there are further subtle issues concerning this question, which will be detailed in section \ref{sec:kugo}. Thus, up to this point, no final conclusion can be drawn.

The results in three dimensions in minimal Landau gauge illustrate the concept of the scaling window, \pref{dse:conformalwindow}, rather nicely. After reaching the maximum, the gluon propagator first decreases, over a momentum range, showing an approximate scaling behavior with a local $\kappa_{AA}$ larger than one, before it finally reaches an infrared finite value when the screening sets in. In four dimensions, this window seems to be closed, i.\ e.\ $p_B\approx\Lambda_\mathrm{YM}$. If in two dimensions at very small momenta still screening sets in, for which there is no sign \cite{Cucchieri:2011ig}, then the scaling window in this case would be very large.

\begin{figure}
\includegraphics[width=\textwidth]{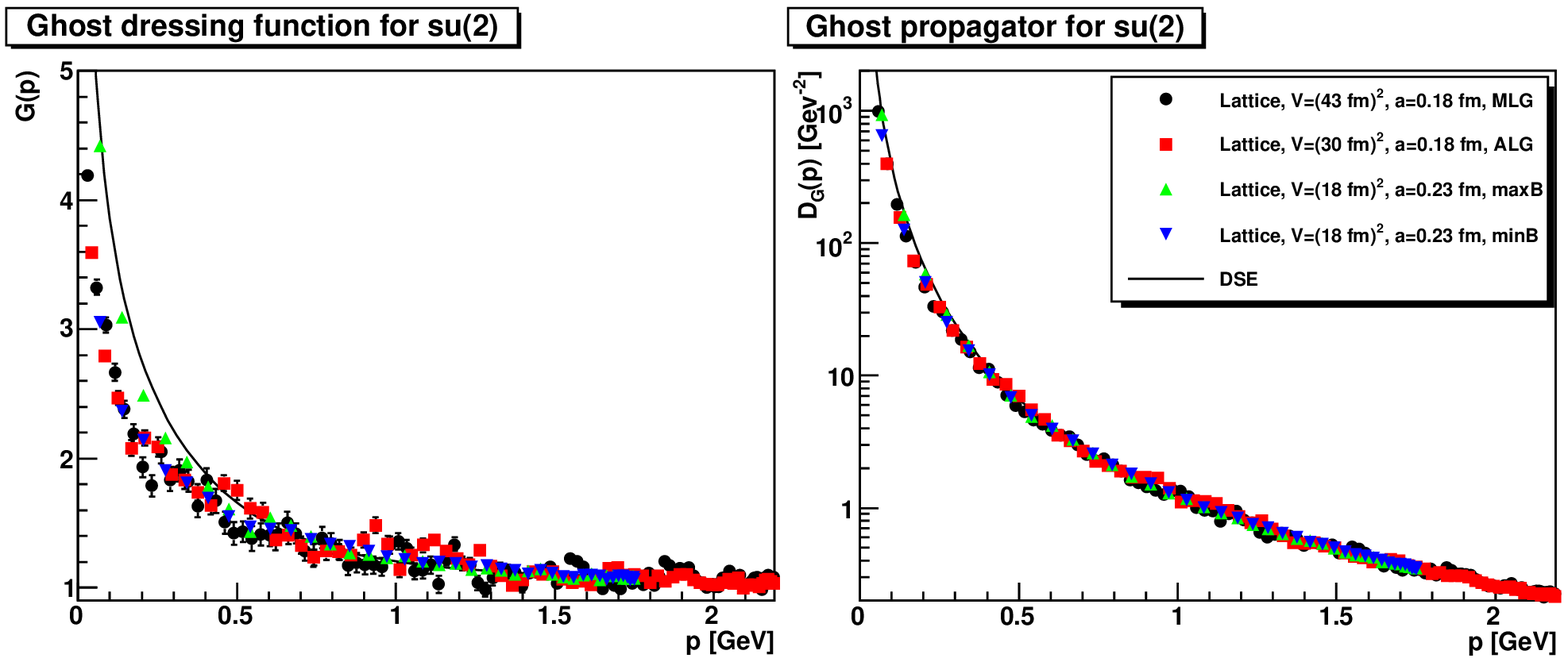}\\
\includegraphics[width=\textwidth]{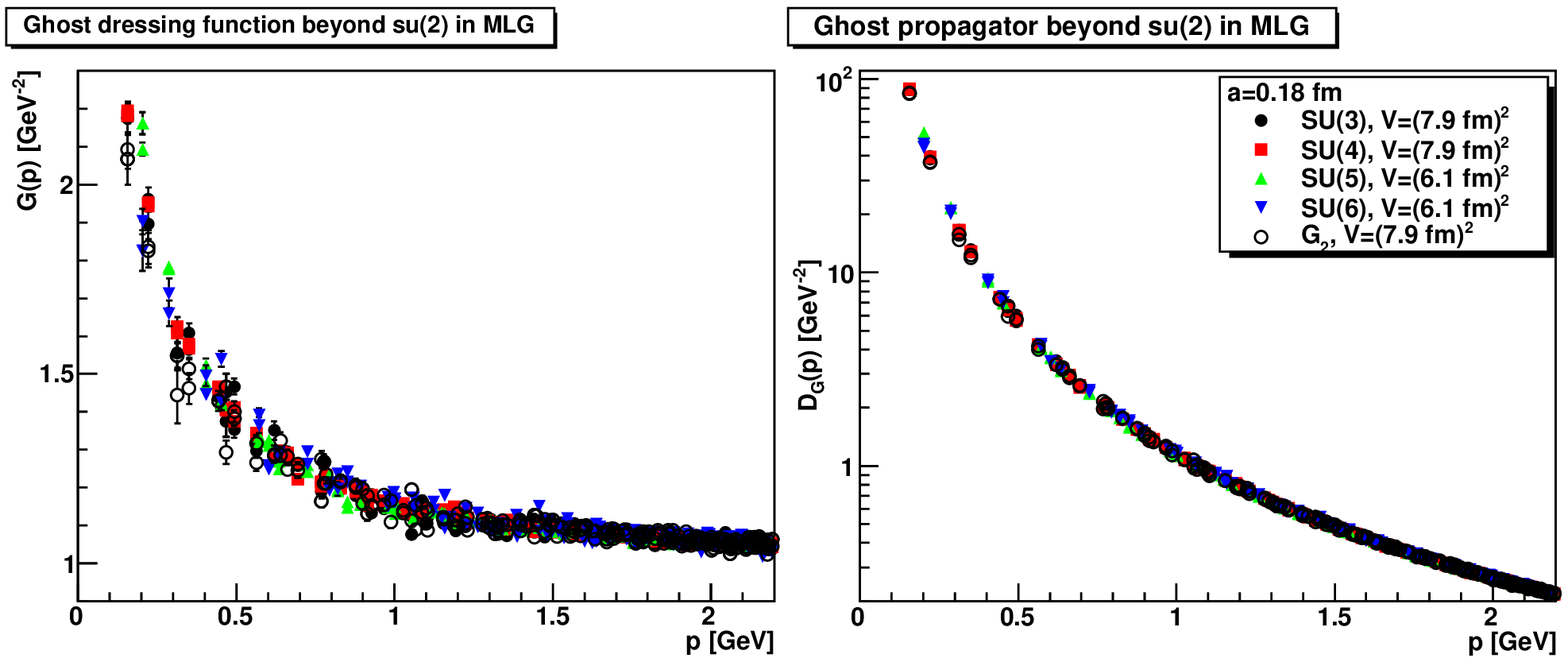}
\caption{\label{fig:ghp-2d}The su(2) ghost propagator (top left panel) and its dressing function (top right panel) in two dimensions. Shown are results for minimal Landau gauge (MLG) \cite{Maas:2007uv}, absolute Landau gauge (ALG) \cite{Maas:2008ri}, and the maxB and minB gauges \cite{Maas:2009se}, compared to DSE results in the scaling case \cite{Huber:2012td}. Note that in two dimensions the finite-ghost case seems not to exist \cite{Huber:2012td,Dudal:2012td,Zwanziger:2012se}. The bottom panels show the same, but for gauge algebras su(3-6) and g$_2$ in MLG \cite{Maas:2010qw}. Further results in two dimensions can be found in \cite{Dudal:2008xd,Cucchieri:2008fc,Maas:2009ph,Maas:2007af}.}
\end{figure}

\begin{figure}
\includegraphics[width=\textwidth]{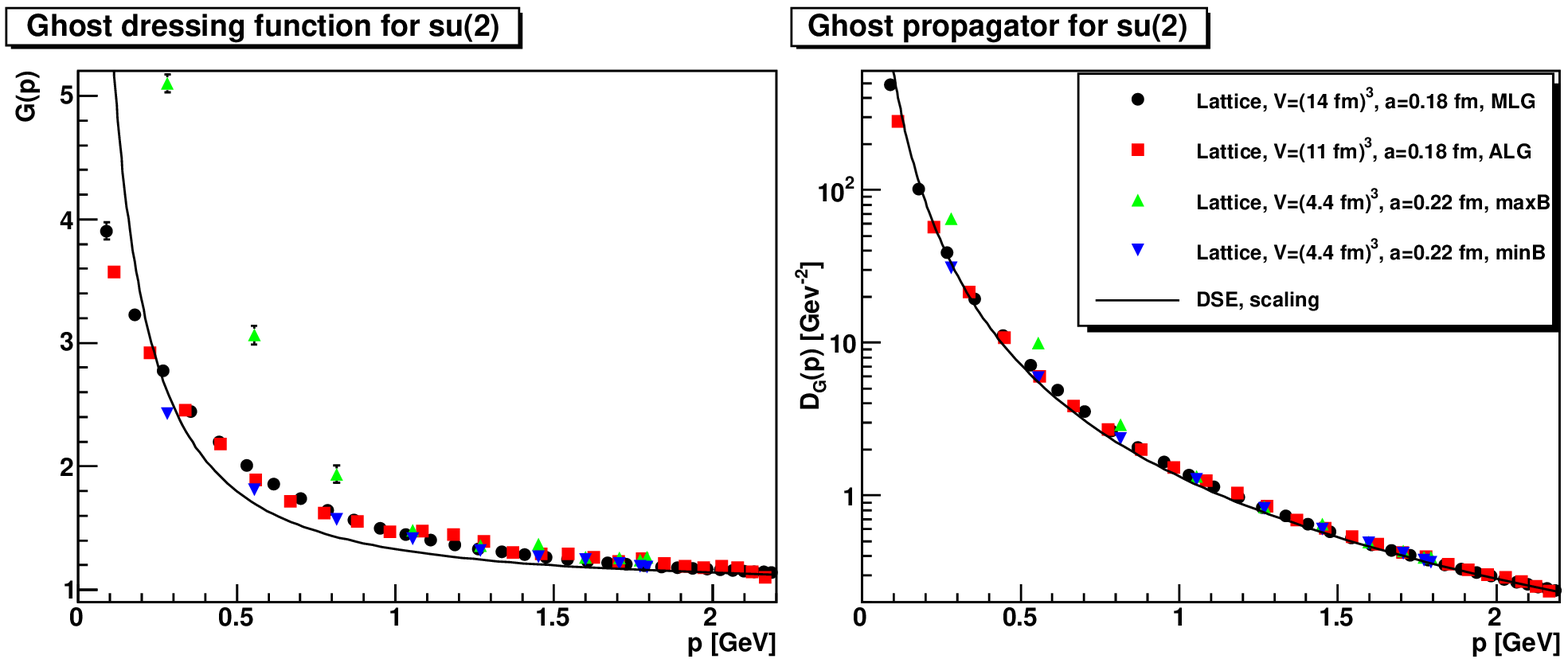}\\
\includegraphics[width=\textwidth]{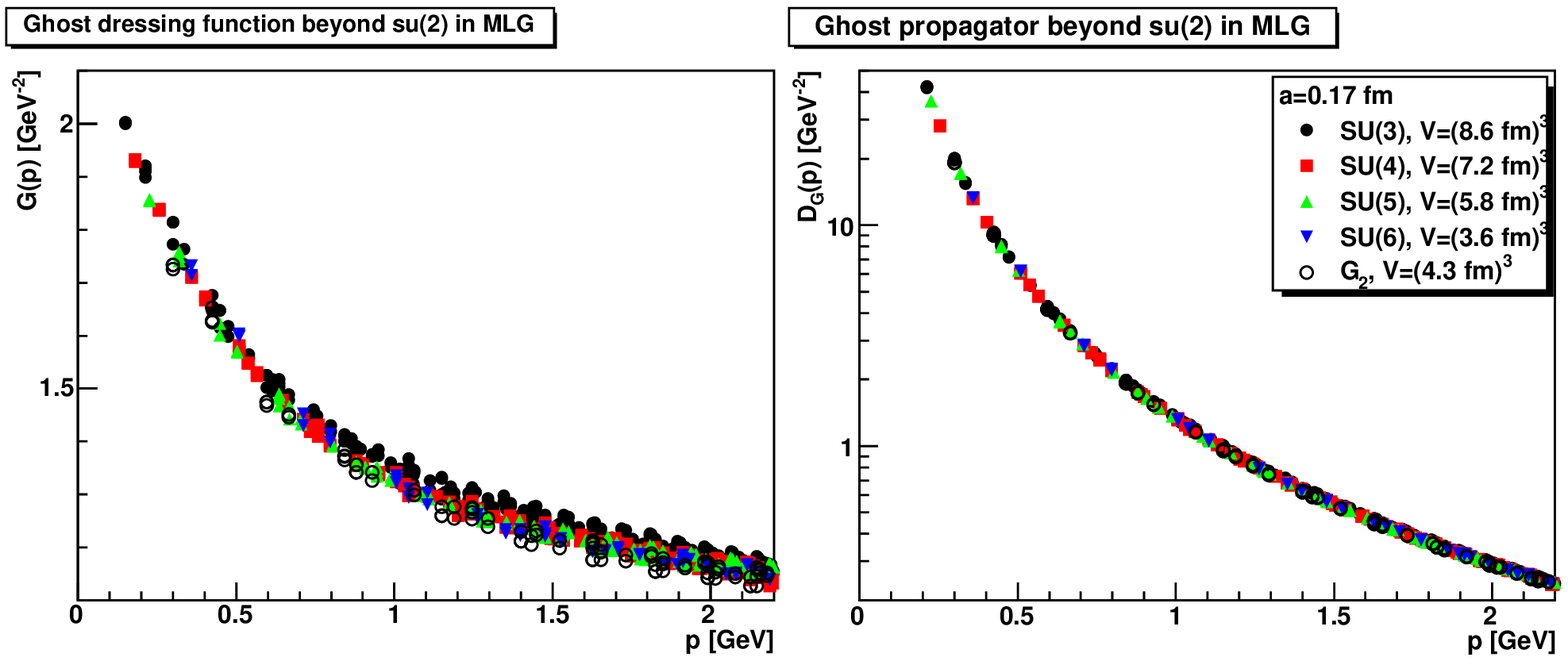}
\caption{\label{fig:ghp-3d}The su(2) ghost propagator (top left panel) and its dressing function (top right panel) in three dimensions. Shown are lattice results for minimal Landau gauge (MLG) \cite{Maas:2007uv}, absolute Landau gauge (ALG) \cite{Maas:2008ri}, the maxB and minB gauges \cite{Maas:2009se}, and DSE results for the scaling case \cite{Maas:2004se}. A finite-ghost-type DSE solution can be found in \cite{Aguilar:2010zx}. The bottom panels show the same, but for gauge algebras su(3-6) and g$_2$ in MLG \cite{Maas:2010qw}. Further lattice results can be found in \cite{Maas:unpublished,Maas:2009ph,Cucchieri:2008fc,Cucchieri:2009zt,Cucchieri:2006tf,Cucchieri:2008qm,Maas:2007af}}
\end{figure}

\begin{figure}
\includegraphics[width=\textwidth]{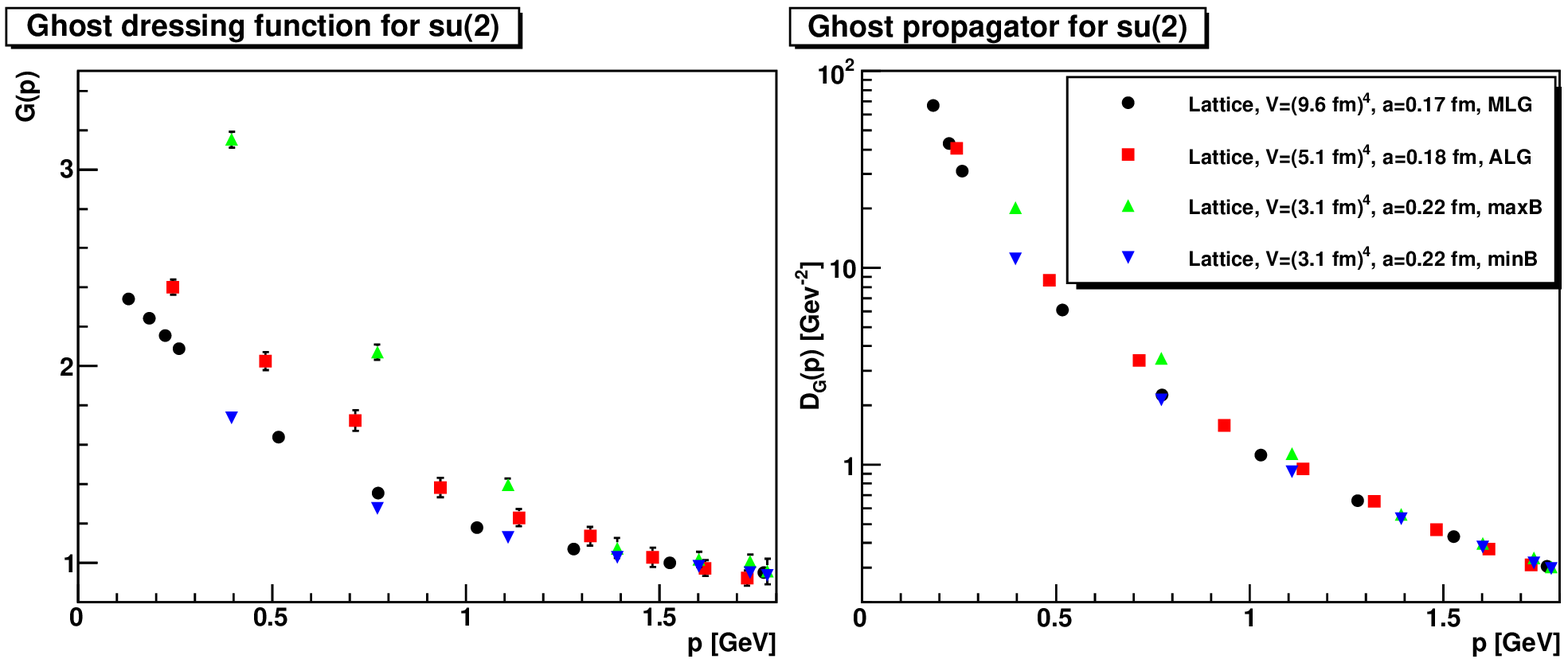}\\
\includegraphics[width=\textwidth]{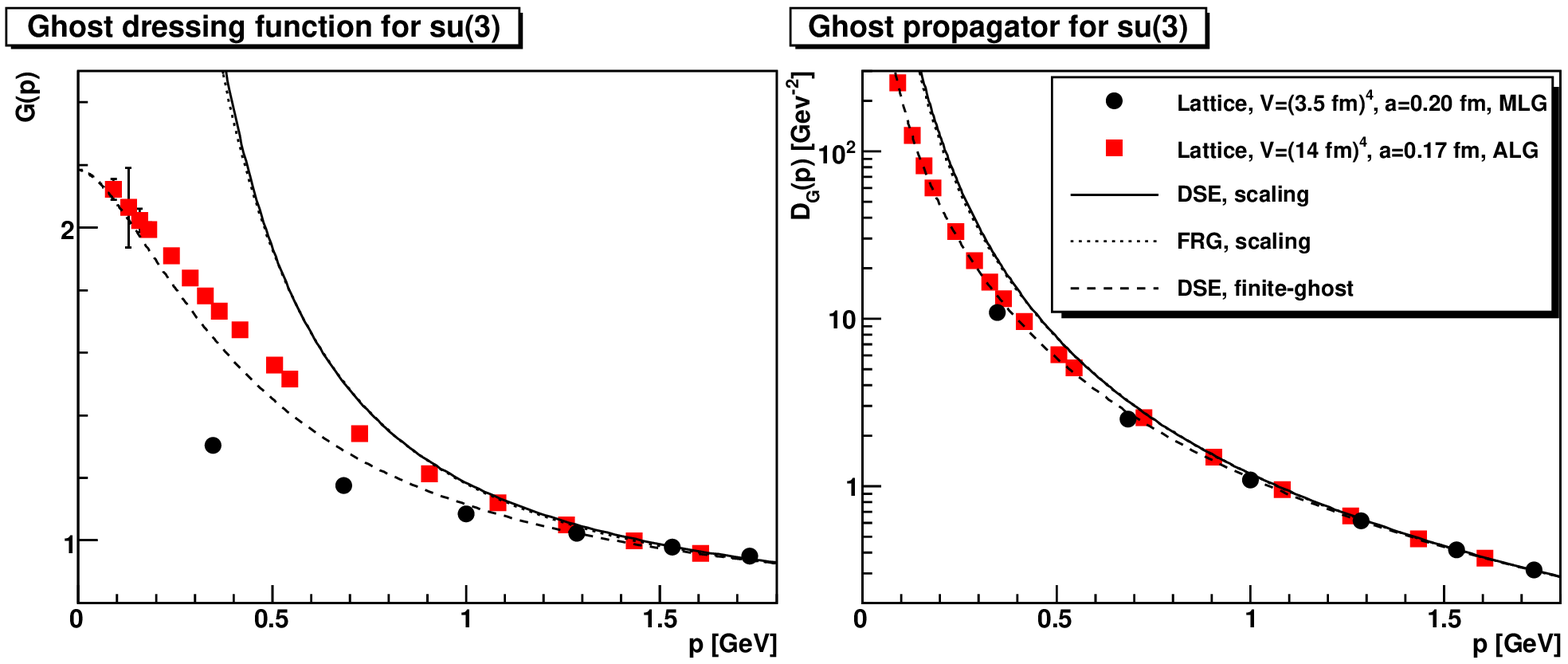}
\caption{\label{fig:ghp-4d}The su(2) ghost propagator (top left panel) and its dressing function (top right panel) in four dimensions, renormalized to G(1.5 GeV$^2$)=1. Shown are lattice results for minimal Landau gauge (MLG) \cite{Sternbeck:2007ug}, absolute Landau gauge (ALG) \cite{Maas:unpublished}, and the maxB and minB gauges \cite{Maas:unpublished}. The bottom panels show the same, but for gauge algebra su(3) in MLG \cite{Fischer:2010fx} and ALG \cite{Bogolubsky:2009dc}, compared to scaling and finite-ghost DSE results and scaling FRG results \cite{Fischer:2008uz}. Further results can be found in \cite{Aguilar:2009ke,Alkofer:2000wg,Binosi:2009qm,Bloch:2003sk,Bogolubsky:2005wf,Bornyakov:2008yx,Bornyakov:2009ug,Boucaud:2010gr,Boucaud:2008ji,Boucaud:2008ky,Cucchieri:2008fc,Cucchieri:2007zm,Dudal:2008sp,Dudal:2007cw,Fischer:2006ub,Fischer:2004uk,Oliveira:2008uf,Oliveira:2009eh,Pawlowski:2003hq,Pawlowski:2009iv,RodriguezQuintero:2010wy,vonSmekal:2009ae,Sobreiro:2004us,Cucchieri:2009zt,Oliveira:2007dy,Cucchieri:2006za,Furui:2003jr}.}
\end{figure}

The second elementary propagator is the ghost propagator. It, and again its dressing function, is shown in figure \ref{fig:ghp-2d} for two dimensions, in figure \ref{fig:ghp-3d} for three dimensions, and in figure \ref{fig:ghp-4d} for four dimensions. Again, there is little dependency on the gauge algebra, and thus the following will be restricted to the su(2) case \cite{Maas:2010qw,Sternbeck:2007ug,Cucchieri:2007zm}. The ghost exhibits a rather strong  dependency on the gauge. The propagator is least divergent in the minB Landau gauge \cite{Maas:2008ri,Maas:2009se,Cucchieri:1997dx,Bogolubsky:2005wf,Bogolubsky:2007bw}, at least for the same lattice settings. Different volumes and discretization artifacts in the renormalization in four dimensions can lead to an apparent different ordering. In the case of minimal Landau gauge the dressing function is infrared finite in three and four dimensions \cite{Cucchieri:2008fc,Sternbeck:2007ug,Bogolubsky:2009dc}, but does not appear to be so in two dimensions \cite{Maas:2007uv,Cucchieri:2008fc}. It is also quite visible that the infrared enhancement strongly depends on the volume \cite{Maas:2007uv,Cucchieri:2006tf,Fischer:2007pf}. The features of the finite-ghost case is rather well reproduced by the finite-ghost solution of the DSEs \cite{Fischer:2008uz}. The behavior is thus that of a massless particle\footnote{Actually, the lattice data cannot exclude the possibility of a logarithmic divergence, which has been proposed for the finite-ghost case \cite{Cucchieri:2008fc,Aguilar:2008xm}. It will be assumed here henceforth that this is not the case, as the results from most calculations are much better in agreement with an infrared finite ghost dressing function in the finite-ghost case.}.

The situation for the maxB gauge is drastically different \cite{Maas:2009se}. In this case, the ghost propagator is much more divergent than even the scaling case, and more divergent than the results in any of the other gauges. Such an over-scaling can actually also be observed in functional calculations in a finite volume at finite cut-off for the scaling case \cite{Fischer:2005ui}. In the latter case the over-scaling originates from a mismatch between continuum and finite-volume regularization, and can thus be regarded as a kind of lattice artifact \cite{Fischer:2007pf}. Given that in the other gauges the ghost propagator also becomes less divergent with increasing volume, it is to be expected that this also applies to the maxB gauge. If then in the infinite-volume case the ghost propagator becomes the scaling one or a finite-ghost one is currently an open question \cite{Maas:2009se}. Since the functional calculations naturally provide such a scaling case, there exists a motivation that this should be the case. However, as will be discussed in more detail in section \ref{sec:kugo}, it may also be that it is not possible to achieve scaling within a single Gribov region. This will require further investigations.

It should be noted that the various gauges only start to differ significantly below about 1 GeV. As to be expected \cite{Maas:2008ri}, in the essentially perturbatively dominated energy range above 1 GeV the non-perturbative gauge dependency does not alter the results. This indicates that the contributions decay faster than the leading-order perturbative contributions. However, when performing an operator-product expansion of the propagators \cite{Boucaud:2003xi,Boucaud:2008gn}, the alterations would contribute.

\begin{figure}
\begin{center}
\includegraphics[width=0.45\textwidth]{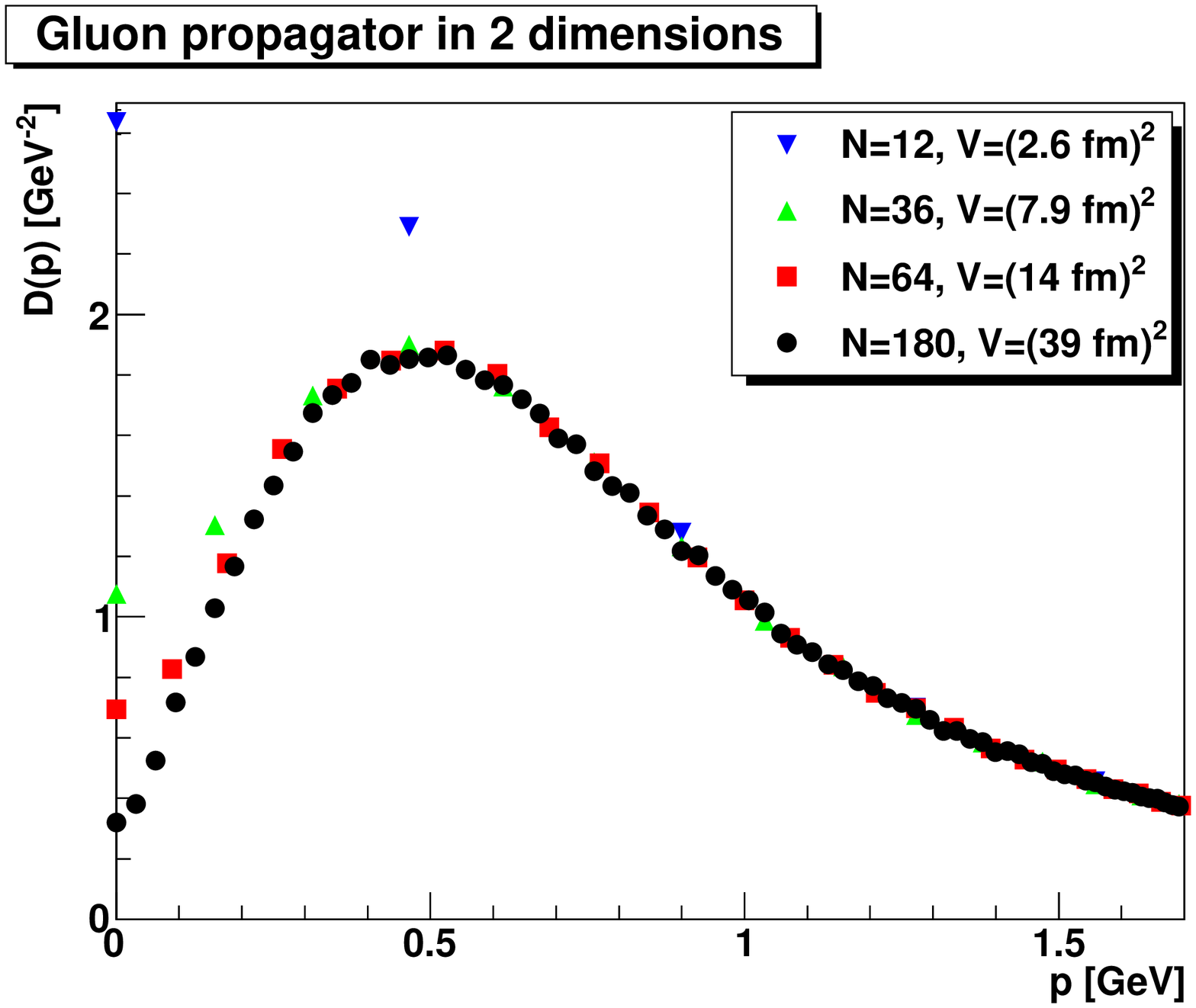}\includegraphics[width=0.45\textwidth]{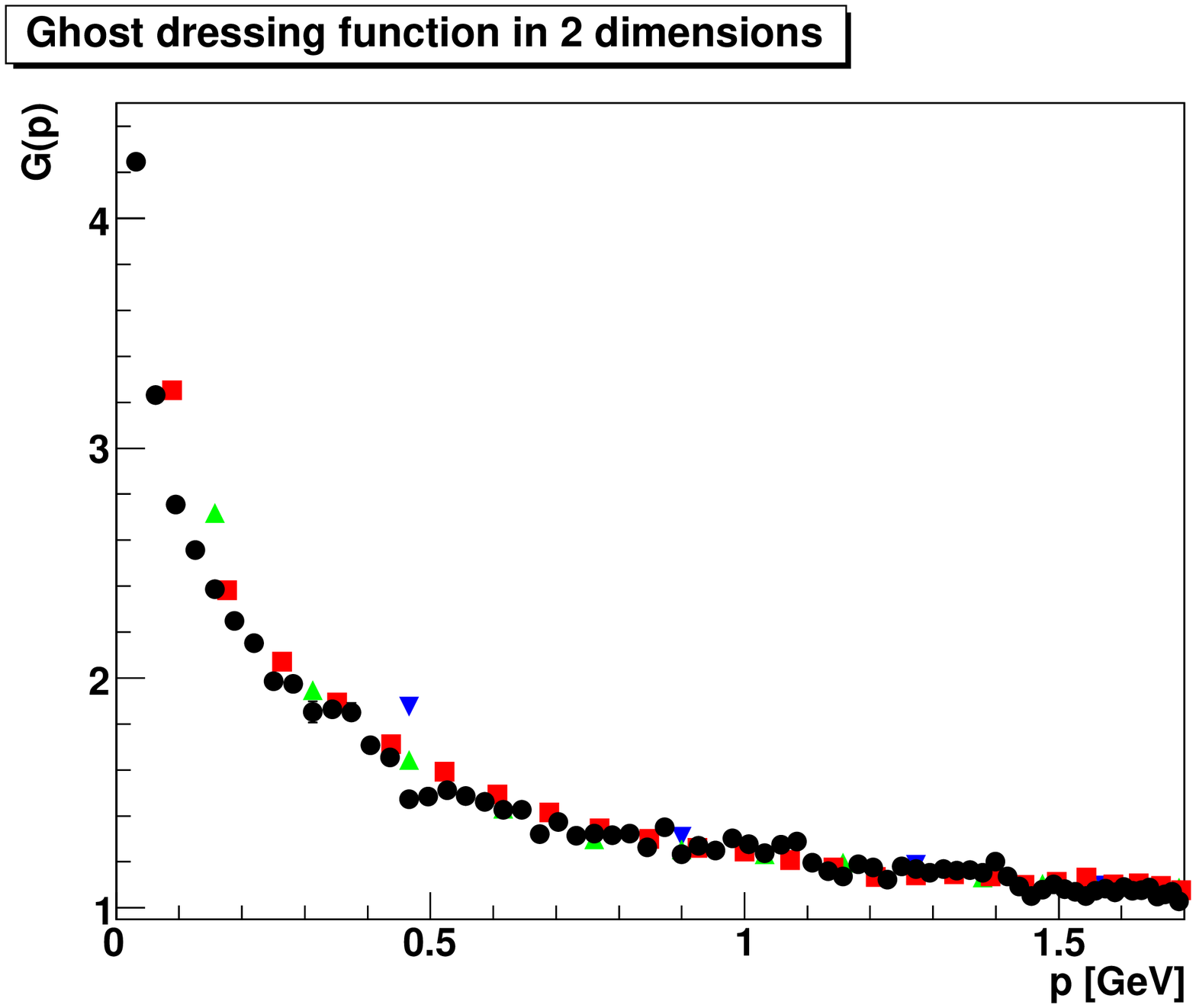}\\
\includegraphics[width=0.45\textwidth]{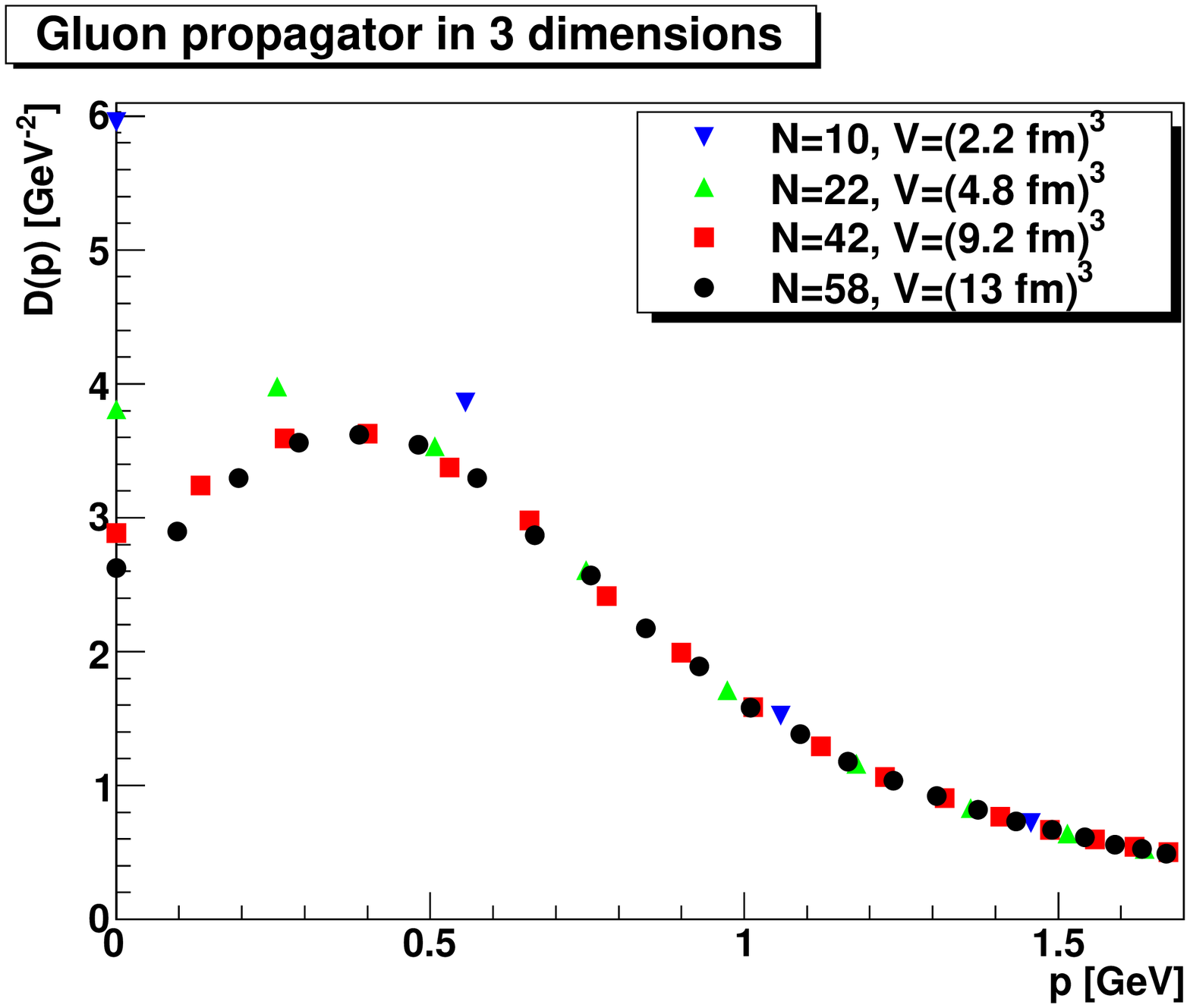}\includegraphics[width=0.45\textwidth]{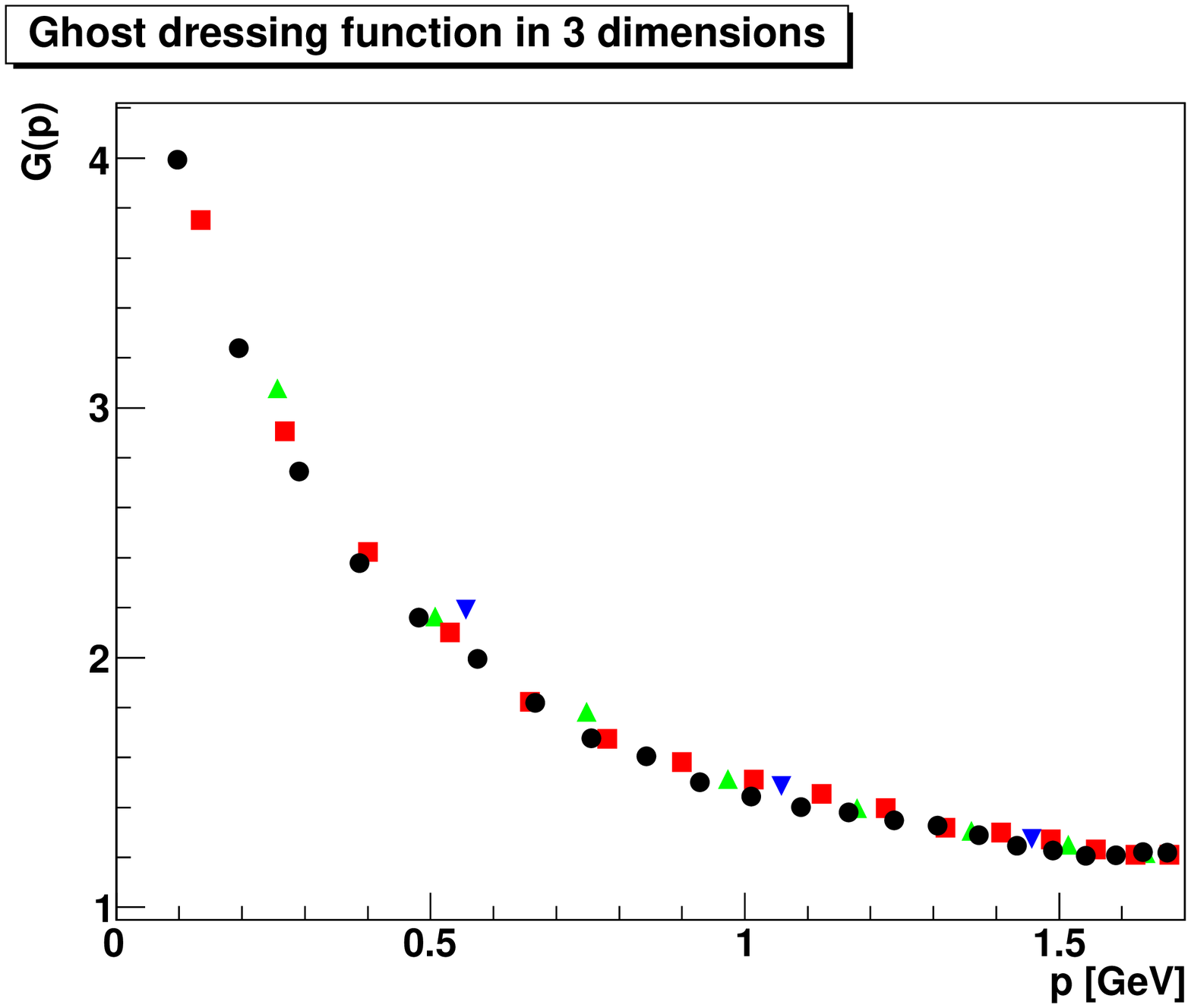}\\
\includegraphics[width=0.45\textwidth]{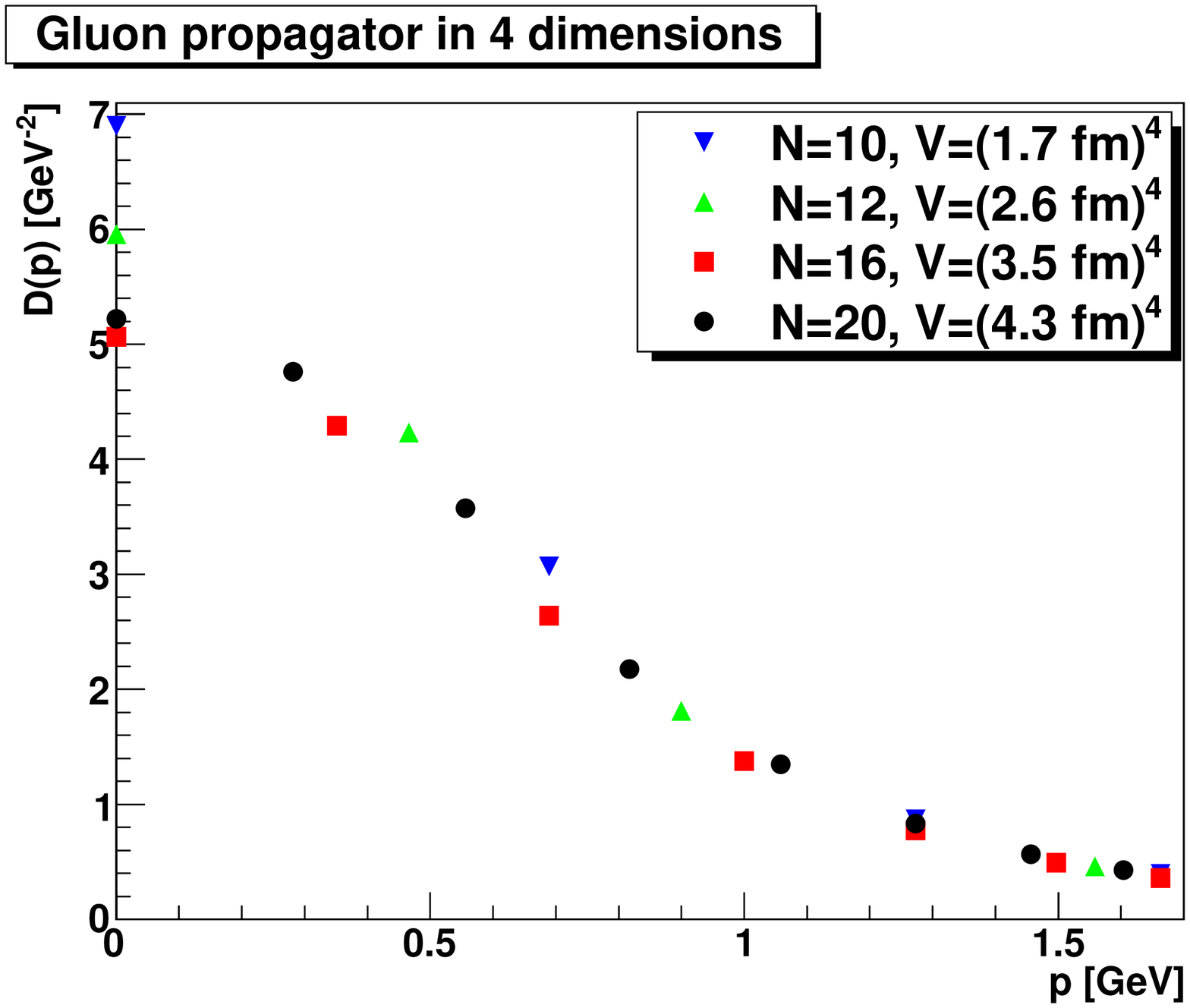}\includegraphics[width=0.45\textwidth]{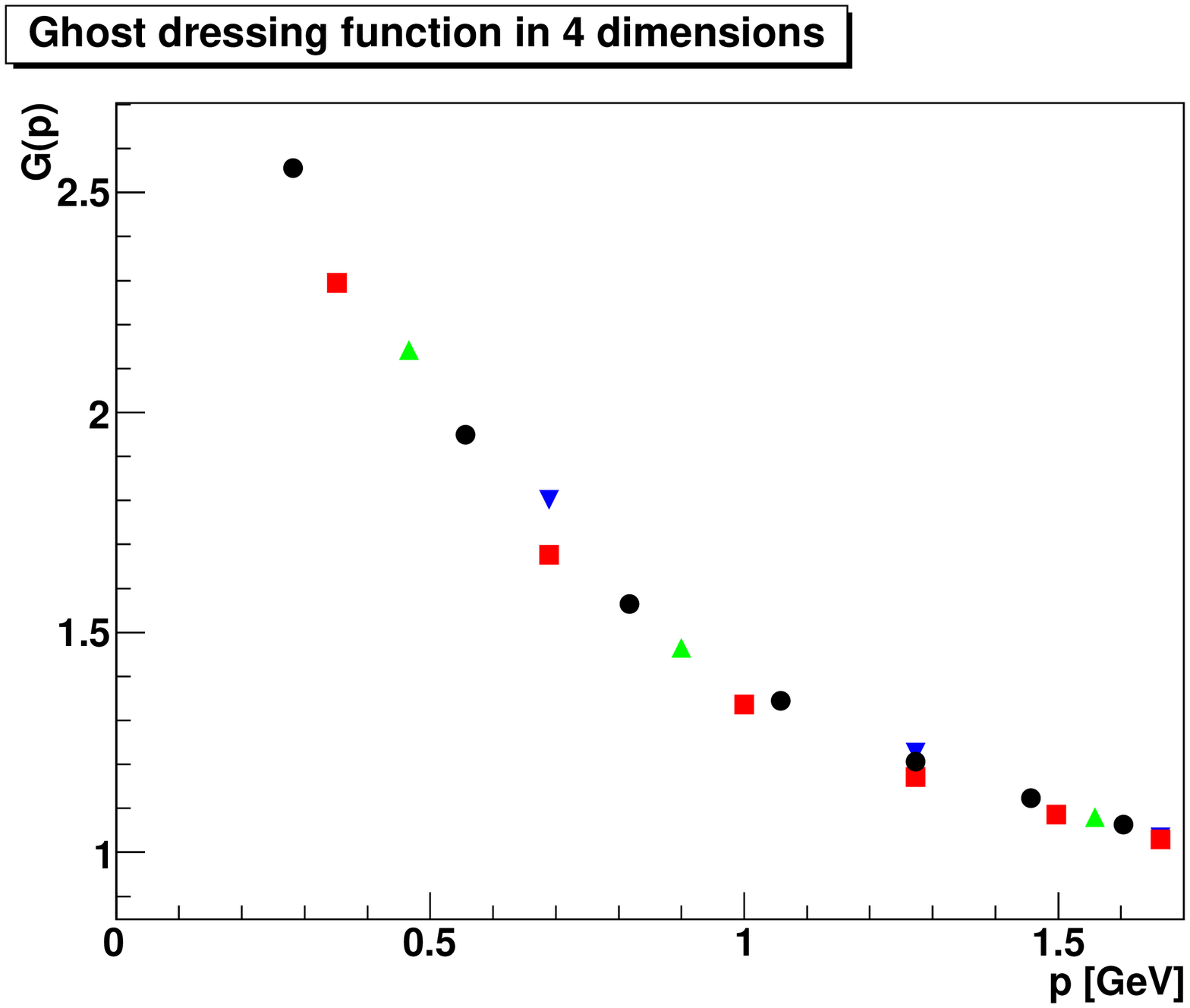}
\end{center}
\caption{\label{fig:volume}Dependency of the gluon propagator (left panels) and the ghost dressing function (right panels) on the volume in two (top panels), three (middle panels), and four dimensions (bottom panels) \cite{Maas:unpublished}. All results are from lattice calculations at the same lattice spacing $a=0.22$ fm and in minimal Landau gauge.}
\end{figure}

As repeatedly stated, these results have a significant dependency on lattice artifacts. Studies of such artifacts due to volume effects and discretization effects can be found, e.\ g., in \cite{Cucchieri:2008fc,Cucchieri:2007rg,Cucchieri:2006tf,Maas:2007uv,Maas:unpublished,vonSmekal:2009ae,Bornyakov:2009ug}. The major contribution of the artifacts is due to finite-volume effects. In the figures \ref{fig:gp-2d} to \ref{fig:gp-4d} and \ref{fig:ghp-2d} to \ref{fig:ghp-4d} always very large volumes have been used. To illustrate the dependency on the volume, results for different volumes are shown in figure \ref{fig:volume}. It is visible that the volume-dependency for the gluon propagator is apparently stronger than for the ghost dressing function, making the infinite-volume extrapolation of the latter a rather subtle problem \cite{Cucchieri:2008fc,Cucchieri:2011ig}. It is also possible to perform studies using functional methods in a finite volume and at finite discretization, which also show sizable finite volume artifacts \cite{Fischer:2007pf}, as discussed in section \ref{sfvdse}.

\begin{figure}
\begin{center}
\includegraphics[width=0.42\textwidth]{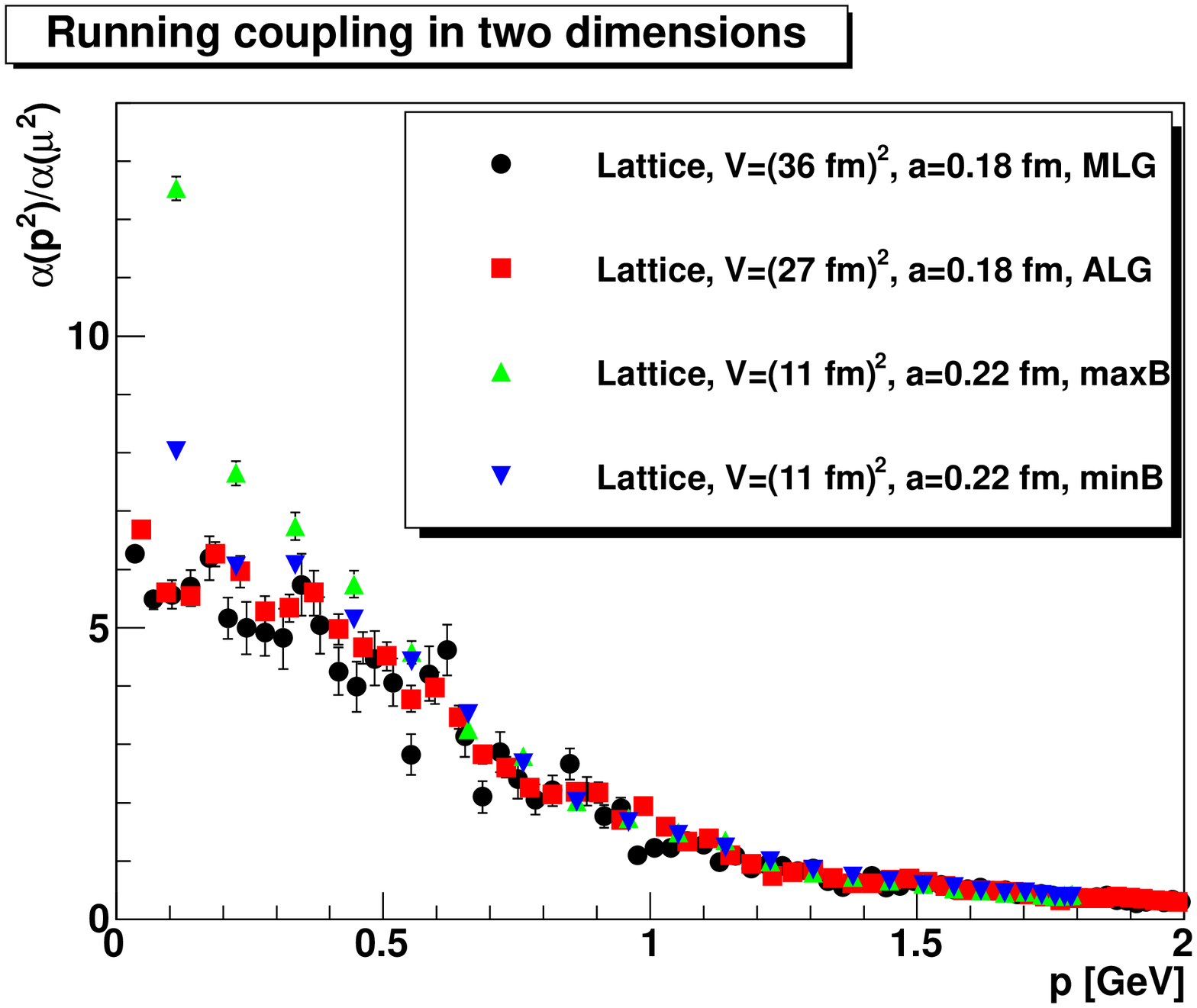}\includegraphics[width=0.42\textwidth]{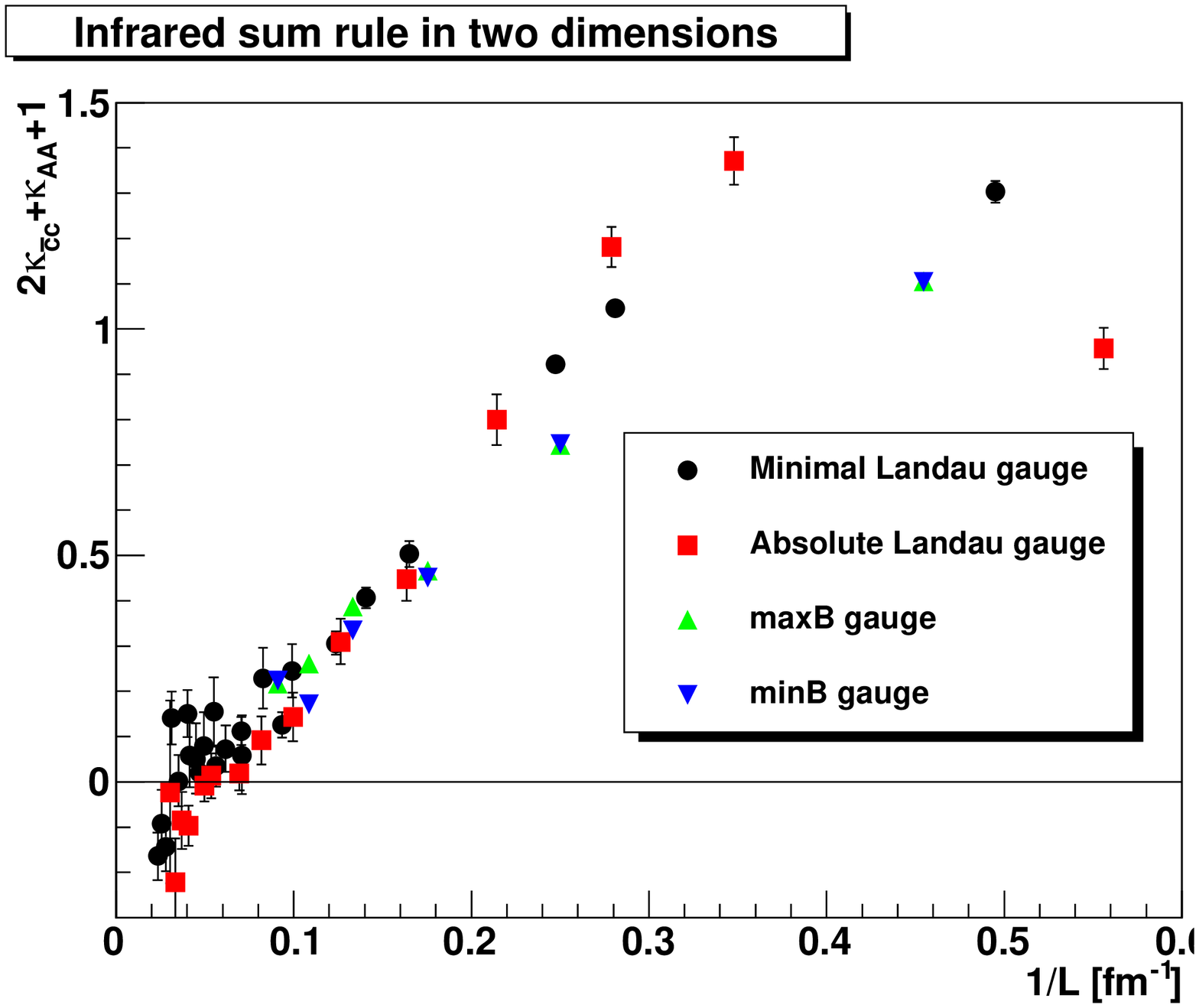}\\
\includegraphics[width=0.42\textwidth]{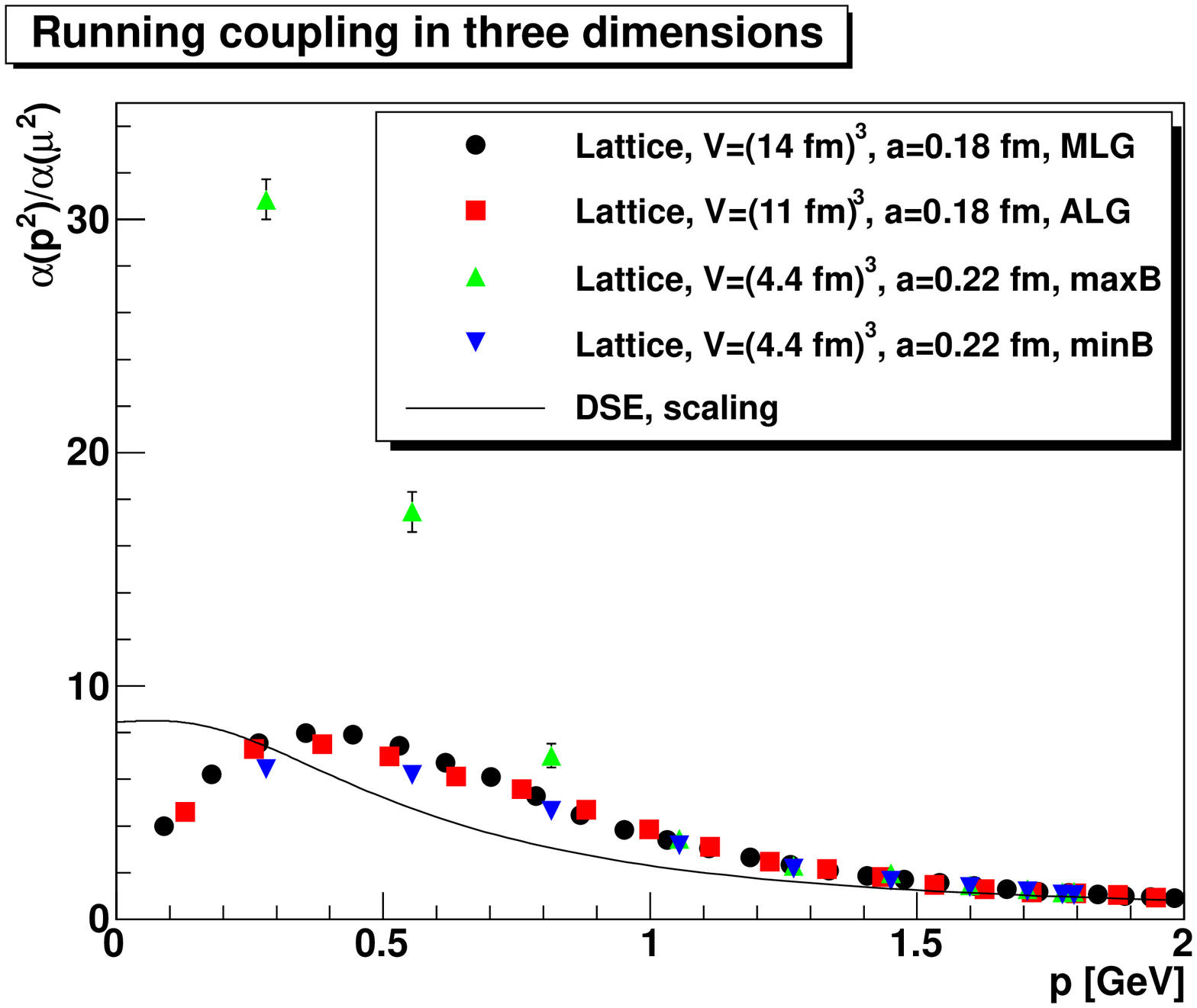}\includegraphics[width=0.42\textwidth]{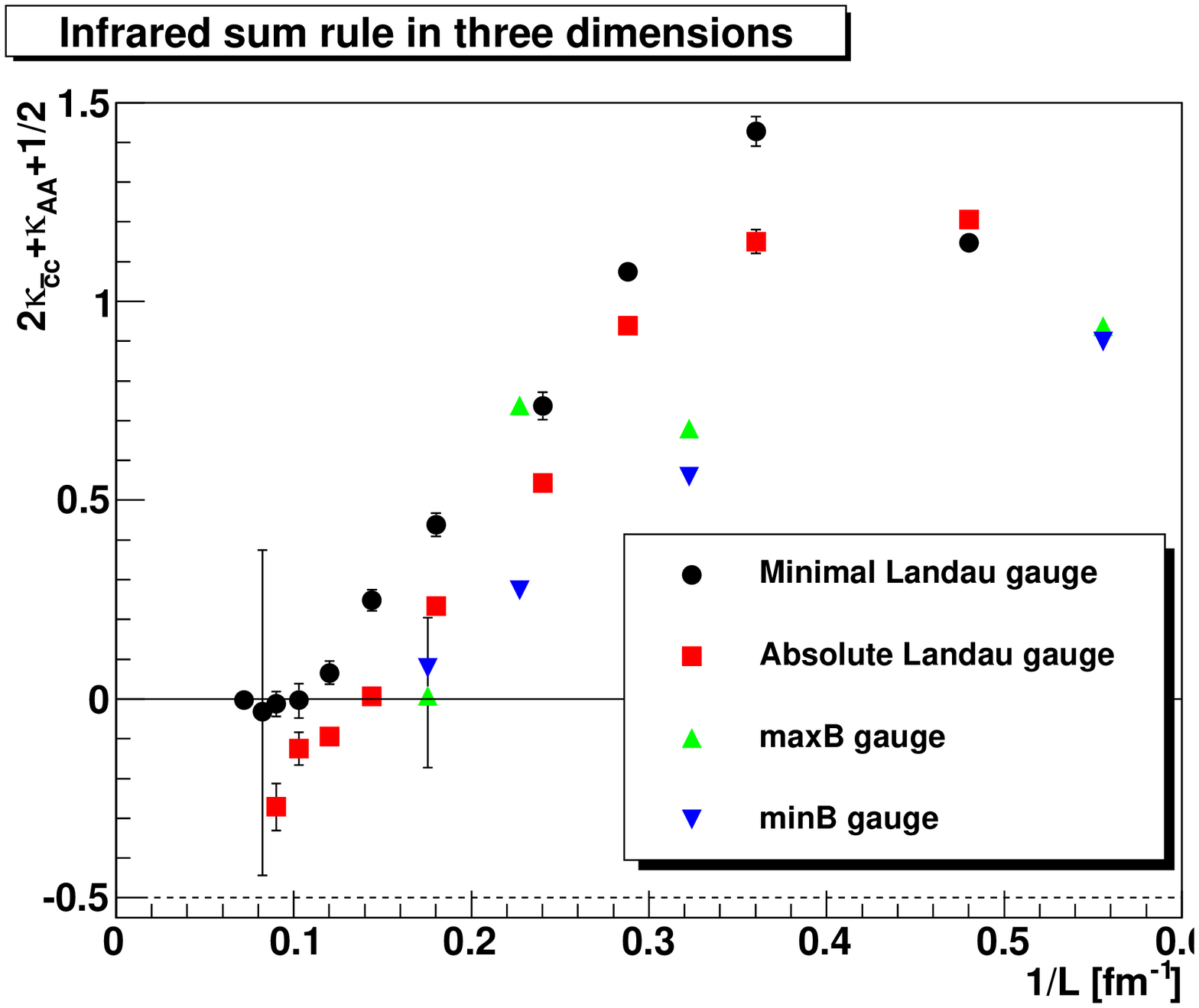}\\
\includegraphics[width=0.42\textwidth]{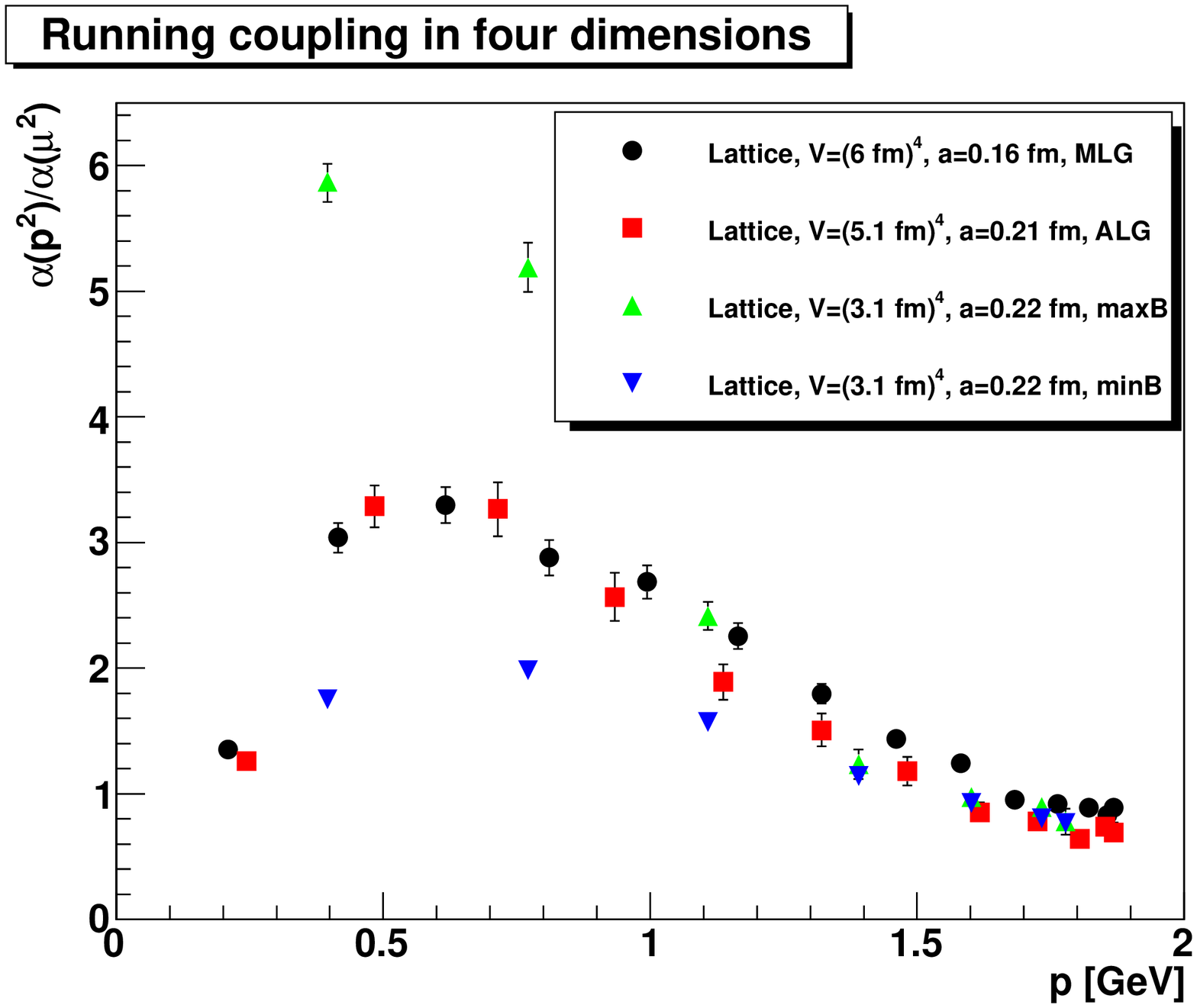}\includegraphics[width=0.42\textwidth]{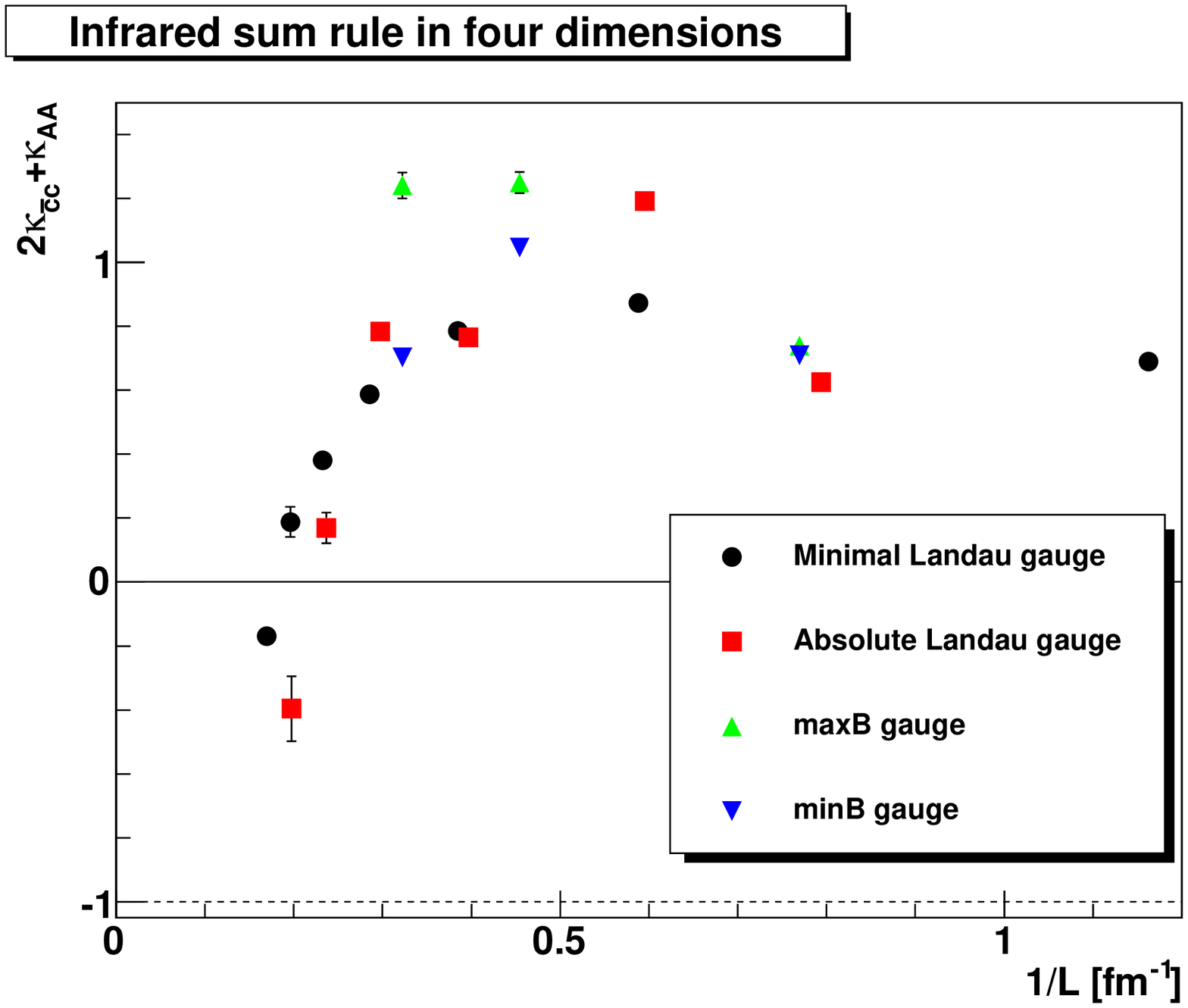}
\end{center}
\caption{\label{fig:alpha}The running coupling \pref{ir:alpha} for the gauge algebra su(2) in two (top-left panel) \cite{Maas:2007uv,Maas:2008ri,Maas:2009se}, three (middle-left panel) \cite{Maas:2008ri,Maas:2009se}, and four (bottom-left panel) \cite{Maas:unpublished} dimensions for the minimal, absolute, and maxB and minB gauges. For comparison, results from DSEs for scaling in three dimensions are also shown \cite{Maas:2004se}. In the right-hand plots the corresponding behavior of the sum-rule \pref{ir:conc} is shown. Note that the lowest momentum points are not used to calculate the exponents \cite{Maas:2007uv}. Solid lines give the value of the sum rule in the scaling case and dashed lines for the finite-ghost case. In two dimensions both lines coincide.}
\end{figure}

Even if the gluon propagator and the ghost propagator would be infrared vanishing and infrared divergent, respectively, this is not sufficient for the existence of the scaling case of the functional results. The characteristic property of scaling is the sum rule \pref{ir:conc}, and thus the infrared constancy of the running coupling \pref{ir:alpha}. This coupling is also of generic interest, as it is the characteristic strength parameter of the theory \cite{Bohm:2001yx}, and the most direct contact to perturbation theory \cite{vonSmekal:2009ae}, and thus experiment \cite{Bohm:2001yx}. For the su(2) case, this coupling is shown for the four gauges and three dimensionalities in figure \ref{fig:alpha}. Shown alongside is the volume-dependent behavior of the left-hand-side of the sum rule \pref{ir:alpha}, using volume-dependent effective infrared exponents \cite{Maas:2007uv,Fischer:2007pf}. The results show a finite-ghost behavior in three and four dimensions for minimal and absolute Landau gauge as well as for minB gauge, at least for the given volumes. This apparently rules out the original idea that the absolute Landau gauge could be connected to the scaling case \cite{Maas:2008ri,Fischer:2008uz,Zwanziger:2003cf,Bornyakov:2011gp}, at the very least for the volumes investigated so far. However, in any finite volume a finite-ghost-like behavior is expected for the running coupling \cite{Fischer:2007pf}. Hence, only a detailed analysis and extrapolation of the sum rule \pref{ir:conc}, shown in the right-hand side of figure \ref{fig:alpha}, can distinguish between the different cases. Indeed, the sum rule appears to confirm the statement that scaling is at best seen only in two dimensions, but neither in three nor in four dimensions, as has also been established indirectly on large lattice volumes including only relatively few points for the ghost propagator at low momenta \cite{Cucchieri:2008fc,Sternbeck:2007ug,Bogolubsky:2009dc,Cucchieri:2007rg}. In two dimensions the situation is rather special, since the sum rule coincides for both the finite-ghost case and scaling. Thus here the propagators, showing in all cases a scaling-like behavior, have to be included in the discussion as well, as detailed above.

The situation looks somewhat different when going to the maxB gauge. In both two and three dimensions the coupling shows over-scaling, i.\ e., it appears to diverge. As noted above, this may be a lattice artifact. Still, it implies that because the distribution of $b$ is continuous over the gauge orbit, see figure \ref{fig:raw-b}, in all volumes studied the value of $B$ can be adjusted such that a scaling-like behavior is obtained \cite{Maas:2009se,Maas:2009ph}. It remains to be seen whether this is also possible on significantly larger volumes. In four dimensions, the coupling seems to saturate, though this requires also a more systematic investigation of the systematic errors: Since in four dimensions the number of Gribov copies at fixed lattice extension is largest, see figure \ref{fig:gc}, it cannot yet be excluded that over-scaling occurs also in four dimensions. Given the experience with including more and more Gribov copies in the maxB gauge, see figure \ref{fig:copydep}, this appears to be at least a viable option to be studied further. Nonetheless, over-scaling in the infinite-volume limit would be an unexpected result.

\begin{figure}
\includegraphics[width=0.5\textwidth]{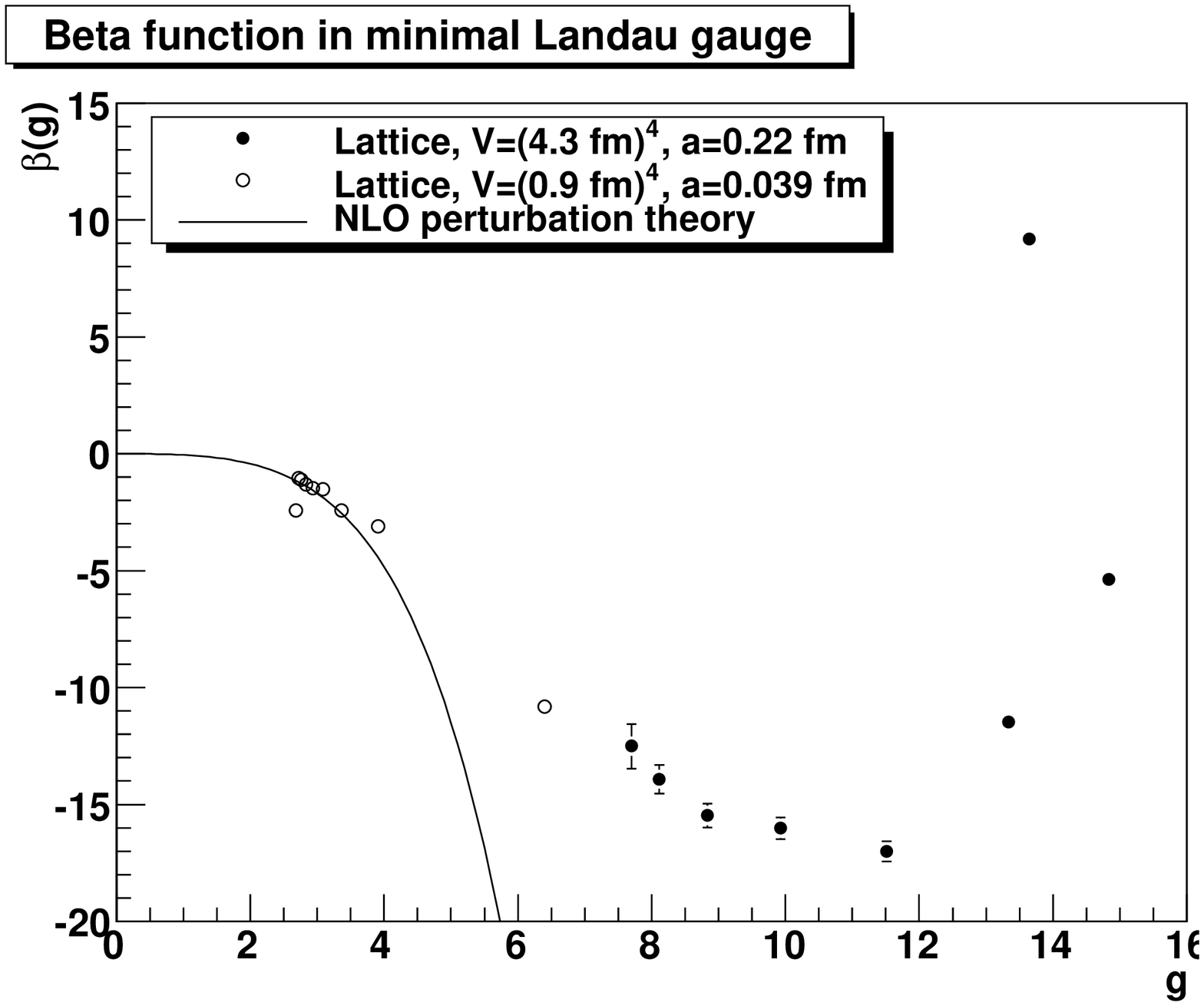}\includegraphics[width=0.5\textwidth]{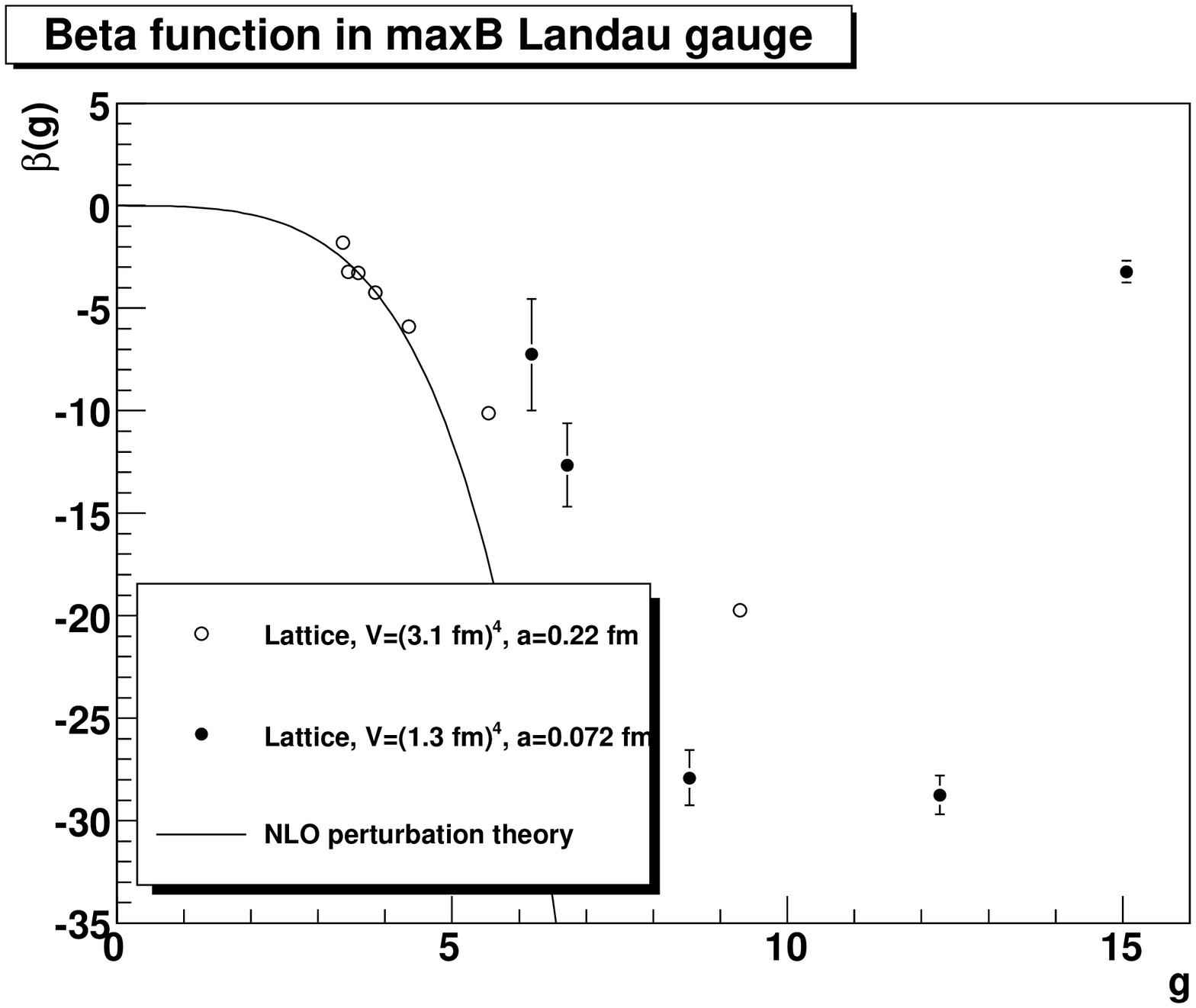}\\
\includegraphics[width=0.5\textwidth]{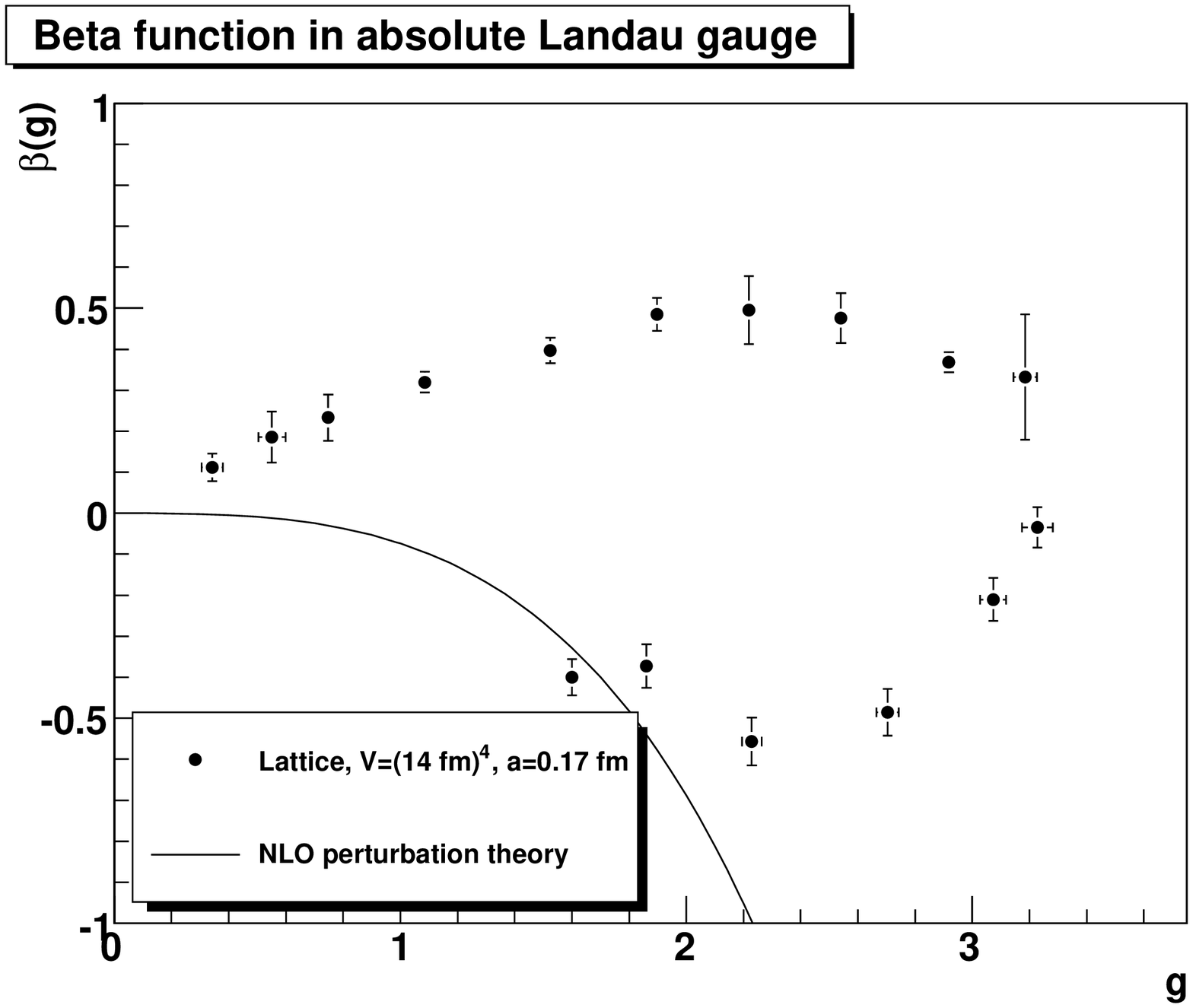}\includegraphics[width=0.5\textwidth]{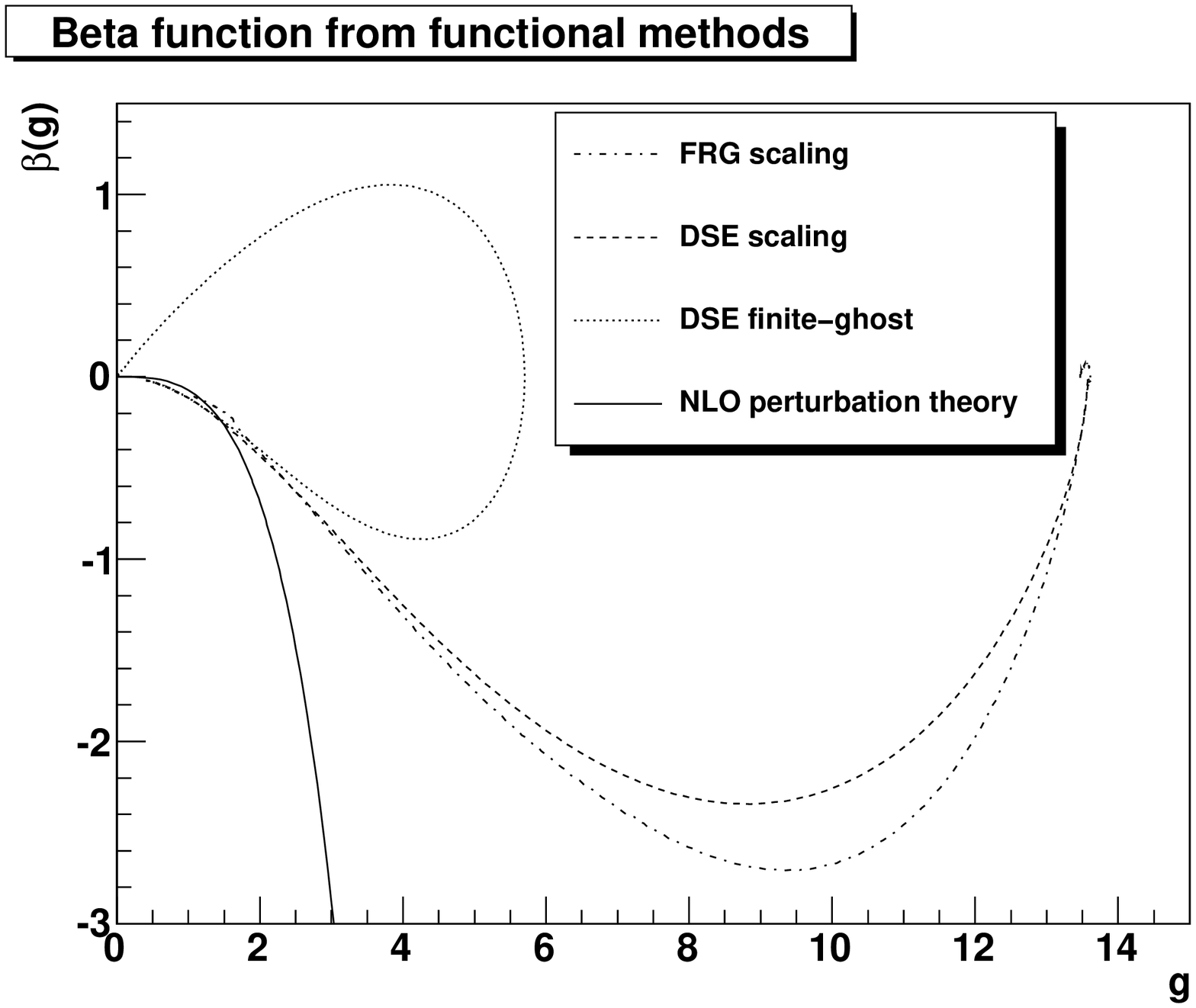}
\caption{\label{fig:beta}The $\beta$-function in four dimensions for su(2) in the minimal Landau gauge (top-left panel) and the maxB gauge (top-right panel) \cite{Maas:unpublished}, and in absolute Landau gauge for su(3) in the bottom-left panel \cite{Bogolubsky:2009dc}, compared to NLO perturbation theory \cite{Bohm:2001yx}, and the functional continuum results for su(3) for the finite-ghost and scaling case in the bottom-right panel \cite{Fischer:2008uz}. Both axis have been rescaled by an arbitrary constant for visibility, and therefore the absolute values should not be taken from the figures.}
\end{figure}

Since the running coupling is both gauge-dependent and scheme-dependent, it is interesting to also investigate the underlying $\beta$-function. The latter has the advantage that it, at least in principle, can be connected to scattering amplitudes \cite{Rosten:2010vm}, though a scheme dependency still remains. The $\beta$-function is defined as
\be
\beta(g(\mu))=\mu\frac{\pd g(\mu)}{\pd\mu}\nn,
\ee
\no where $g$ is connected to the running coupling \pref{ir:alpha} by $g^2=4\pi^2\alpha$. The resulting $\beta$-functions are shown in figure \ref{fig:beta}. It is well visible that next-to-leading order (NLO) perturbation theory describes the running of the coupling rather well up to $g$ around five, correspondingly $\alpha$ around 0.6, for the lattice data. The functional results are only up to leading order (LO) consistent with perturbation theory, and therefore start to deviate quicker from the NLO result. It is also well visible how the $\beta$ function approaches the Gaussian fix-point in the ultraviolet. In the infrared for the minimal and absolute Landau gauge lattice data and the finite-ghost functional a zero of the $\beta$ function emerges at a finite $g$, corresponding to the maximum in the running coupling. At this point the coupling, however, does not stop running, but decreases once more, and the $\beta$-function becomes positive. This indicates that the derivative of the $\beta$ function at this point is non-zero, and becomes dominant. In the very far infrared it appears that the $\beta$-function reaches another, likely trivial Gaussian, fix-point when the running coupling becomes zero. It would be highly interesting whether this occurs in an analytic or non-analytic way, but this has not yet been investigated. In the maxB gauge a zero-crossing is not yet seen, but much larger volumes are required to check whether this remains true. Of course, for the scaling solution, the $\beta$-function has indeed a non-trivial infrared fixed-point, but the running coupling also exhibits a shallow maximum, and thus the $\beta$-function arrives at its infrared fixed point from above. This can essentially not be resolved on the scale of the figure \ref{fig:beta}, and could be a truncation artifact \cite{Fischer:2002hna}.

Thus, from these investigations it should be concluded that in the infinite-volume limit both the finite-ghost case and scaling case can be found using functional methods. In the minimal Landau gauge, within systematic errors, in three and four dimensions a finite-ghost behavior is obtained, and in two dimensions a behavior which is at least very close to scaling. The same appears to be the case for the absolute Landau gauge and the minB gauge, though in these cases the systematic uncertainties are larger due to the higher computational costs. Finally, the results for the maxB case are not yet conclusive, and require further systematic studies. Of course, the lattice statements apply to calculations in the first Gribov region. Whether this marks the existence of the scaling case in three and four dimensions an artifact of the functional equations is a subtle question, and will be discussed in more details in section \ref{sec:kugo}, and requires that the non-perturbative gauge-fixing procedure for functional equations described in section \ref{squant:dse} and \ref{zerot:dse} indeed lead to an equivalent gauge in both functional and lattice calculations.

Before continuing, it should be noted that the functional results show by construction the correct perturbative behavior. By comparing it to the lattice results, which automatically contain up to lattice artifacts all orders of perturbation theory, it shows that already at momenta around 1.5 GeV resummed leading-order perturbation theory becomes an adequate description of the propagators within a few percent. Given that there are processes for which a perturbative description is not valid even for substantially higher energies \cite{Brodsky:2009bg}, this is rather early. On the other hand, the onset of Gribov effects at momenta starting from around 1 GeV shows that significant non-perturbative effects come into play already very close to the perturbative domain, leaving only a rather small window where higher-order perturbation theory can substantially improve the description of the propagators.

In total, the results for the propagators exhibit clearly different behaviors, depending on how the non-perturbative gauge-fixing is treated, being it either in the form of explicit Gribov copies in lattice calculations or of undetermined parameters in the functional equations. The presence of the finite-ghost case is definitely established for the minimal Landau gauge, and rather well reproducible by functional calculations for the same choice of $B$ in both three and four dimensions.

\subsubsection{Vertices}\label{szerot:vertices}

The next-simple correlation functions beyond propagators are the three-point vertices. Two of them exist in Landau-gauge Yang-Mills theory. One is the ghost-gluon vertex, the other one is the triple-gluon vertex.

Of both, the ghost-gluon vertex $\Gamma^{A\bar{c}c abc}_\mu$ is the significantly simpler one. It is related to the corresponding correlation function by\footnote{Connected and full correlation functions coincide in Landau gauge for three-point functions \cite{Cucchieri:2006tf}.} \cite{Cucchieri:2004sq}
\be
<A_\mu^a(p)\bar{c}^b(q) c^c(k)>=D_\mn^{A^2ad}(p)D^{\bar{c}cbe}(q)D^{\bar{c}ccf}\Gamma^{A\bar{c}c def}_\nu(p,q,k)\delta(p+q+k)\label{zerot:uggv}.
\ee
\no Because of Lorentz-symmetry, the ghost-gluon vertex can be expressed in terms of two tensor structures \cite{Taylor:1971ff}
\be
\Gamma^{A\bar{c}c abc}_\mu(p,q,k)=ig\left(q_\mu A^{abc}(p,q,k)+p_\mu B^{abc}(p,q,k)\right)\label{zerot:ggvtensors},
\ee
\no which are chosen here conveniently to be one along the anti-ghost momentum, and one along the gluon momentum. At tree-level, $A^{abc}=f^{abc}$ and $B^{abc}=0$. Due to the explicit (transverse) gluon propagator, of the two possible Lorentz structures only the one not longitudinal in the gluon momentum appears in the unamputated correlation function \pref{zerot:uggv}, and in particular in functional equations projected transversely. So there is only one relevant tensor-structure to be kept. At tree-level, this structure has a color-structure of $f^{abc}$. In principle, others can appear beyond tree-level. This has so far only been investigated to some extent, indicating the appearance of a $d^{abc}$ structure for su($N>2)$ gauge algebras \cite{Macher:2010ad}, i.\ e., for the only ones with non-vanishing $d^{abc}$ \cite{Cvitanovic:2008}. However, since only the component proportional to $f^{abc}$ is relevant for the functional equations at propagator level, here only the component proportional to the tree-level color structure will be discussed. This leaves one single scalar dressing function. In all gauges implementing the Landau gauge condition \pref{quant:lg}, it can be shown that the associated renormalization constant is finite \cite{Taylor:1971ff}. Furthermore, because of a ghost-anti-ghost symmetry in Landau gauge \cite{Alkofer:2000wg}, which is presumably intact \cite{Cucchieri:2005yr}, the ghost and anti-ghost legs can be exchanged without changing the value of the dressing function, i.\ e., $A(p,q,k)=A(p,k,q)$.

\begin{figure}
\includegraphics[width=0.85\textwidth]{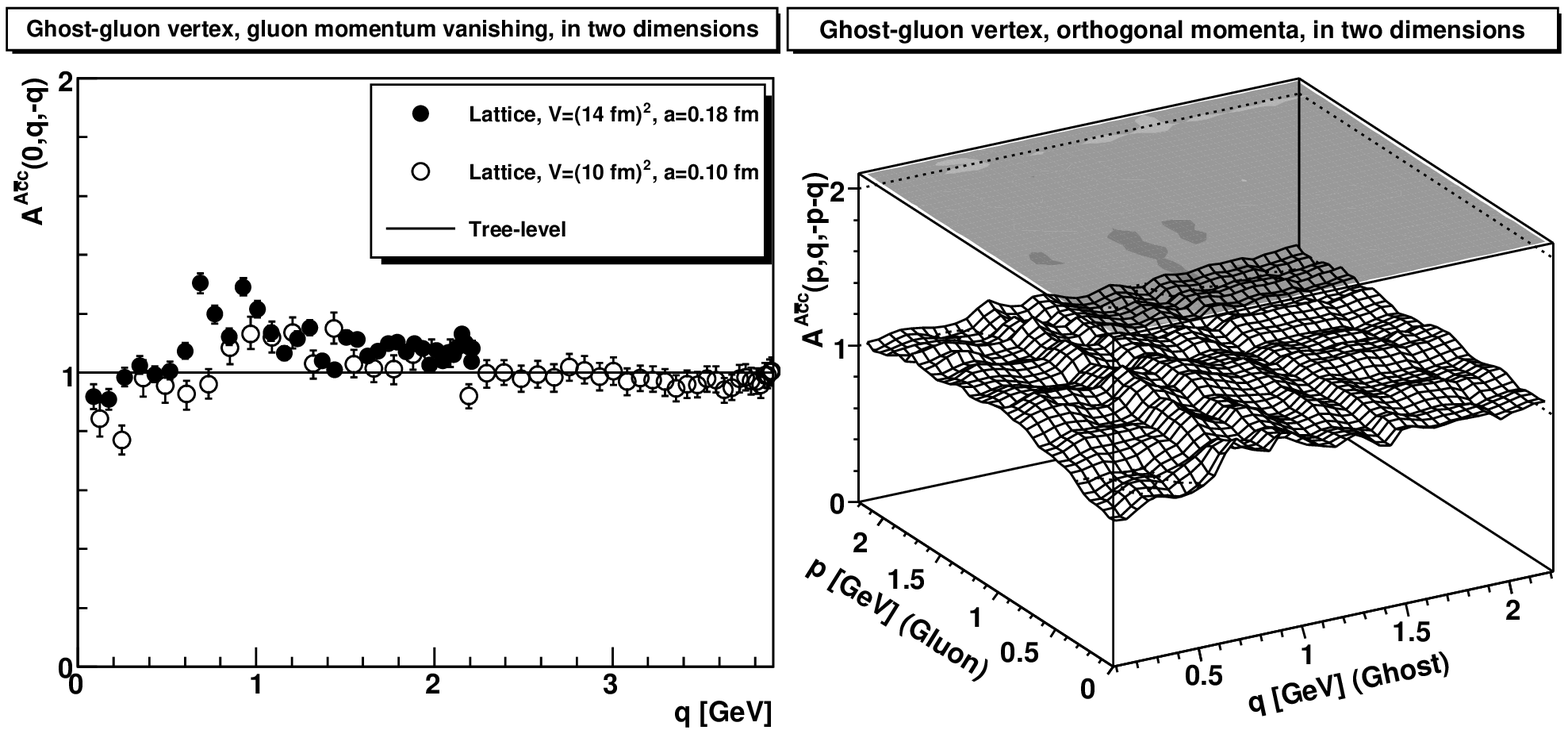}\\
\includegraphics[width=0.85\textwidth]{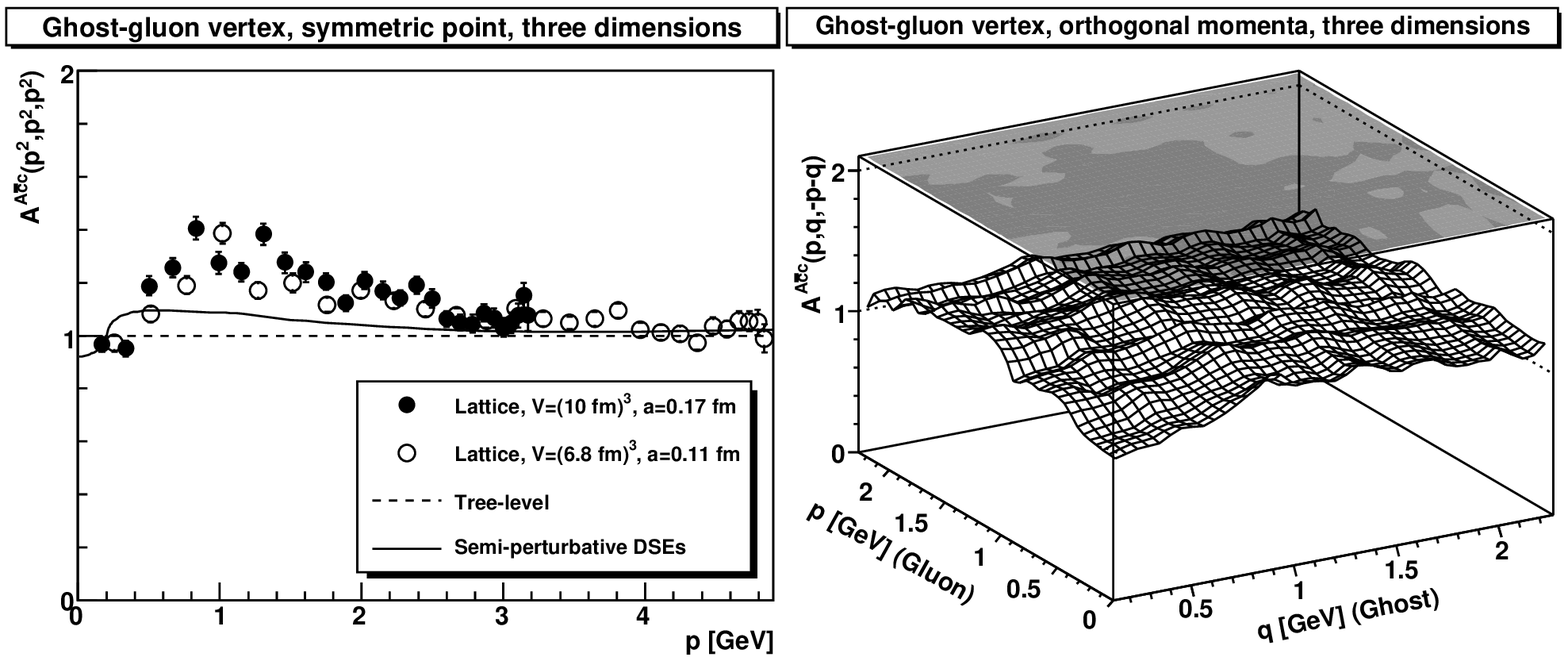}\\
\includegraphics[width=0.85\textwidth]{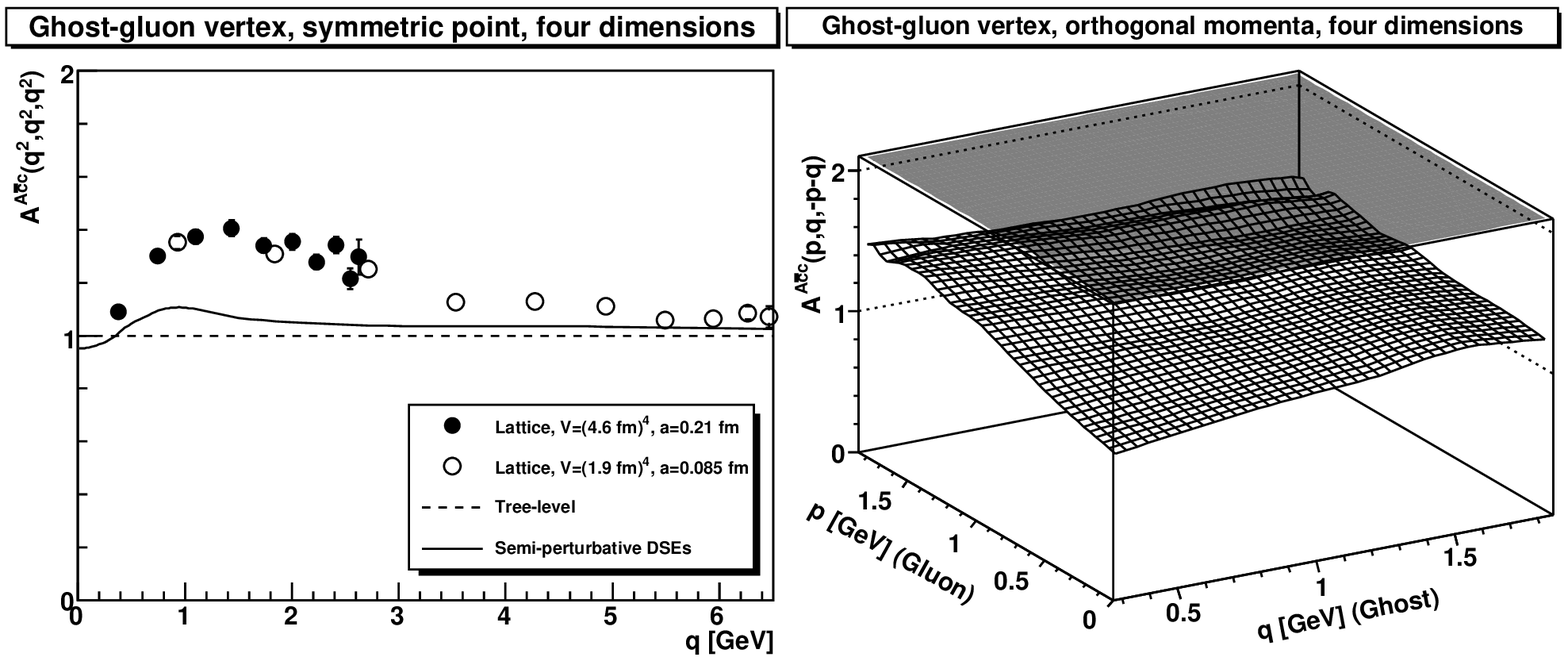}
\caption{\label{fig:ggv}The tensor component $A$ of the ghost-gluon vertex for two (top panels) \cite{Maas:2007uv}, three (middle panels) \cite{Cucchieri:2008qm}, and four (bottom panels) \cite{Cucchieri:2008qm} dimensions, in minimal Landau gauge. In three and four dimensions, results from a semi-perturbative evaluation of the DSEs in the scaling case are also shown \cite{Schleifenbaum:2004id}. See \cite{Huber:2012td} for DSE results in two dimensions. The left panel shows the result in the symmetric momentum configuration $p^2=q^2=k^2$ for three and four dimensions. In two dimensions, this configuration is not possible on a quadratic lattice except at special points \cite{Maas:2007uv}, and the result for zero gluon momentum is shown instead. The right panels show the results for the gluon momentum orthogonal to the ghost momentum for the largest volume shown in the left panel.}
\end{figure}

Lattice results in two, three, and four dimensions for su(2) for this dressing function in various kinematic configurations are shown in figure \ref{fig:ggv} for the minimal Landau gauge\footnote{There are not yet any results in other gauges available.}. In all cases, the vertex dressing only slightly deviates from tree-level at mid-momentum, and becomes comparable to tree-level in both the infrared and ultraviolet, though might be slightly smaller than tree-level in the infrared. The enhancement at mid-momentum increases with dimensionality, but never exceeds about 50\%. Thus, the ghost-gluon vertex is indeed very close to tree-level, as commonly assumed in functional calculations. However, a comparison with figure \ref{fig:ghglv} immediately shows that the ansatz to remove quadratic divergences from the ghost loop is different in the ultraviolet. From this it can immediately be deduced that the removal of these divergences in the gluon equation has, in principle, to be done with the necessary counter-terms for the non-gauge-invariant regularization, instead of some modification of the vertices.

For the scaling case, functional results are available, which are obtained in a semi-perturbative manner, i.\ e., by using a skeleton expansion \cite{Alkofer:2004it} but replacing the propagators with the non-perturbative ones from the previous section \cite{Schleifenbaum:2004id}. Thus, these results are not self-consistent, and will only give a first estimate of the non-perturbative corrections. The result is also included in figure \ref{fig:ggv}. These are found to not deviate significantly from tree-level as well. Even when modifying the input vertices drastically \cite{Lerche:2002ep}, this does not change \cite{Schleifenbaum:2004id,Schleifenbaum:2004di}.

In four dimensions, results for su(3) are also available, from both lattice \cite{Sternbeck:2006rd,Ilgenfritz:2006he} and continuum calculations \cite{Schleifenbaum:2004id,Dudal:2012zx}. They show the same qualitative behavior, and almost the same quantitative behavior.

In summary thus the ghost-gluon vertex is essentially constant. Since this is also the case in two dimensions, it can be expected to be the case in both the scaling and the finite-ghost case. As noted, this is also anticipated in the scaling case from very general consistency arguments between the DSEs and the FRGs \cite{Fischer:2009tn,Fischer:2006vf}. The slight angular variations at low momenta may however be of quantitative importance, as they could alter the value of the exponent $\kappa_{\bar{c}c}$ \cite{Huber:2012td}. In particular, such changes can decrease this exponent, and therefore move it closer to one generating an infrared finite gluon propagator even in the scaling case \cite{Lerche:2002ep}.

The three-gluon vertex is a much more complicated object, due to its much more involved Lorentz structure \cite{Cucchieri:2006tf,Cucchieri:2008qm,Parrinello:1994wd,Alles:1996ka,Boucaud:1998bq,Boucaud:2000ey,Alkofer:2008dt}. It is related to its full correlation function by \cite{Cucchieri:2006tf,Parrinello:1994wd}
\be
<A_\mu^a(p)A_\nu^b(q) A_\rho^c(k)>=D_{\mu\alpha}^{A^2ad}(p)D_{\nu\beta}^{A^2be}(q)D_{\rho\gamma}^{A^2cf}\Gamma^{A^3def}_{\alpha\beta\gamma}(p,q,k)\delta(p+q+k)\nn.
\ee
\no The correlation function harbors four independent Lorentz structures, which are transverse in all gluon momenta \cite{Ball:1980ax}, and on top one or more color structures. The latter have been investigated, and it has been found that at least one other non-trivial structure can exist in principle \cite{Macher:2010ad,Gracey:2009mj}. The various structures have been investigated using functional studies in the far infrared in the scaling case \cite{Alkofer:2008dt}, showing differences not too large for the various Lorentz structures, though this remains a subject of further investigation.

\begin{figure}
\includegraphics[width=0.9\textwidth]{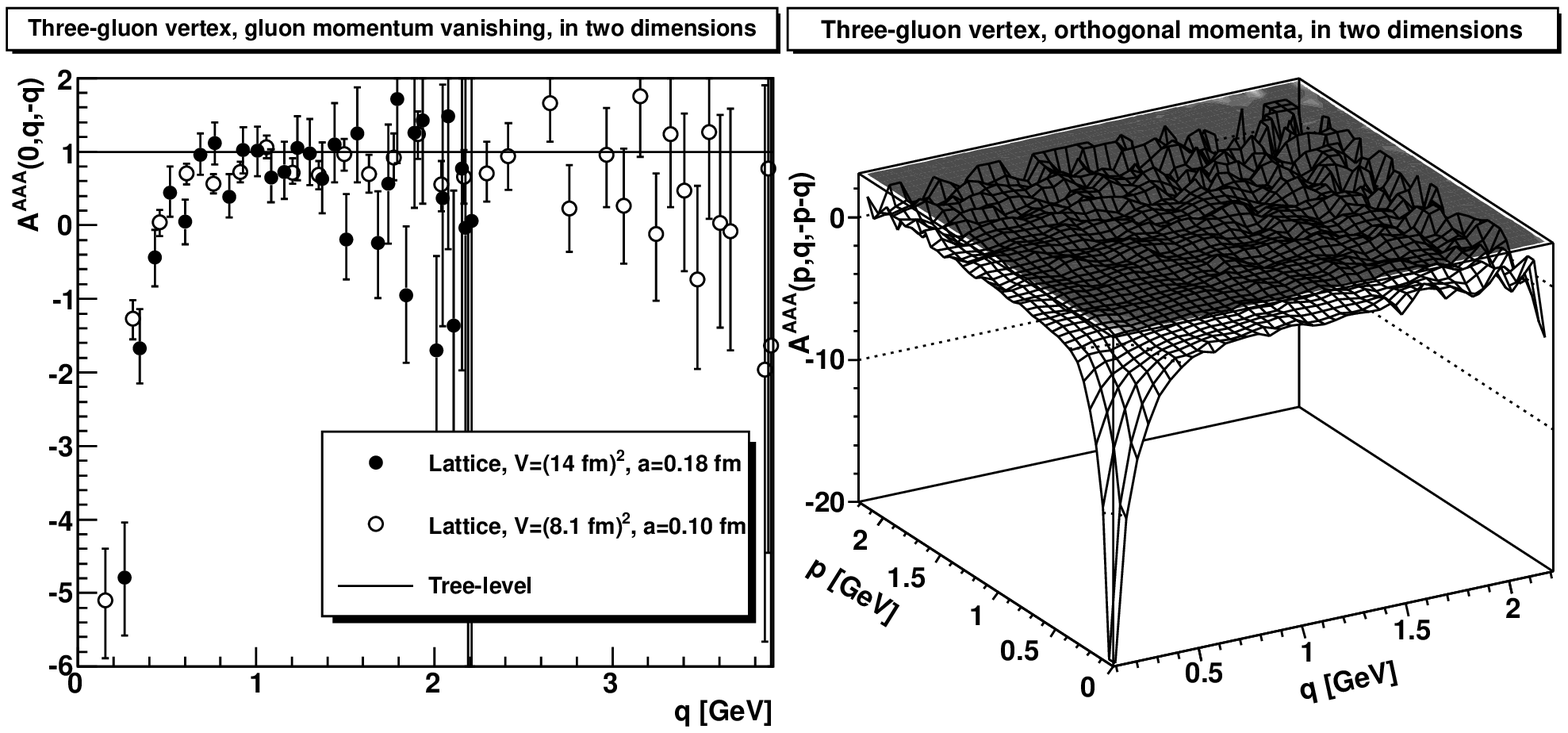}\\
\includegraphics[width=0.9\textwidth]{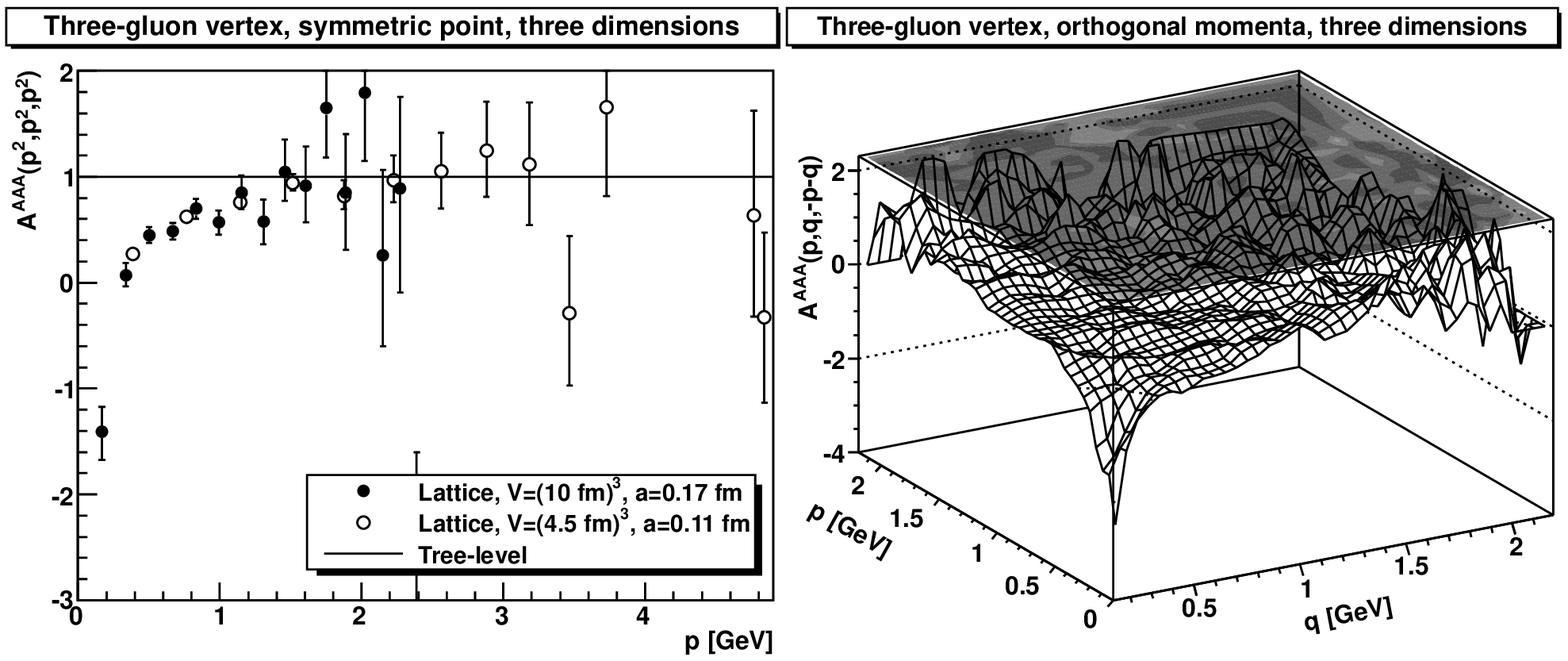}\\
\includegraphics[width=0.9\textwidth]{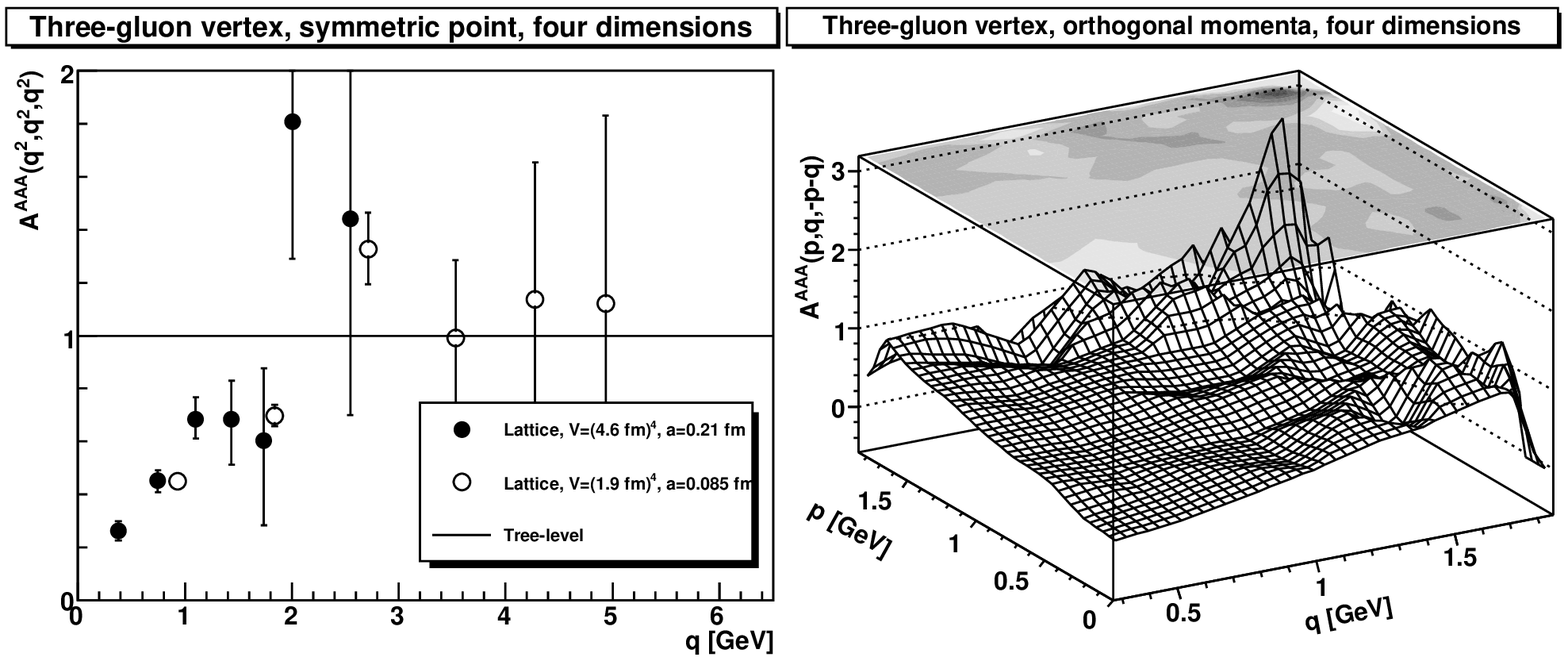}
\caption{\label{fig:g3v}The tensor component \pref{zerot:g3v} of the three-gluon vertex for two (top panels) \cite{Maas:2007uv}, three (middle panels) \cite{Cucchieri:2008qm}, and four (bottom panels) \cite{Cucchieri:2008qm} dimensions, in minimal Landau gauge. Momentum configurations are as in figure \ref{fig:ggv}. Points with an error larger than 1 have been suppressed. Some results in four dimensions for gauge algebra su(3) can be found in \cite{Parrinello:1994wd,Alles:1996ka,Boucaud:1998bq,Boucaud:2000ey}, for one gluon momentum vanishing. See \cite{Huber:2012td} for DSE results in two dimensions.}
\end{figure}

In lattice calculations, only the projection on the tree-level color and Lorentz structure has been investigated so far \cite{Cucchieri:2006tf,Cucchieri:2008qm,Parrinello:1994wd,Alles:1996ka,Boucaud:1998bq,Boucaud:2000ey}, which is given in general by \cite{Cucchieri:2006tf}
\be
A^{AAA}(p,q,k)=\frac{\Gamma_{\mu\nu\rho}^{\tl\indexsep L\indexsep A^3\indexsep abc}(p,q,k)<A_\mu^a(p)A_\nu^b(q) A_\rho^c(k)>}{\Gamma_{\mu\nu\rho}^{\tl\indexsep L\indexsep A^3\indexsep abc}(p,q,k) D_{\mu\lambda}^{ad}(p) D_{\nu\sigma}^{be}(q) D_{\rho\omega}^{cf}(k)\Gamma^{\tl\indexsep L\indexsep A^3\indexsep def}_{\lambda\sigma\omega}(p,q,k)}\label{zerot:g3v}.
\ee
\no The index $L$ at the tree-level vertex denotes the usage of the lattice version of the tree-level vertex \cite{Rothe:2005nw}, to reduce artifacts from violation of rotational symmetry \cite{Cucchieri:2006tf}. By definition, the function \pref{zerot:g3v} would be one if the full and the tree-level vertex coincided. It should be noted that this is the only contribution of the three-gluon vertex which contributes in the gluon loop of the gluon propagator's DSE \cite{Cucchieri:2006tf}. Results in two, three, and four dimensions are shown in figure \ref{fig:g3v} for su(2) in the minimal Landau gauge. The results are very noisy, and thus very high statistics are necessary for an accurate result \cite{Cucchieri:2006tf,Maas:2007uv,Cucchieri:2008qm}, making the determination of this vertex a computational challenge. The results finally show that the qualitative behavior is rather similar for all dimensions. In particular, in all cases a suppression at momenta around the typical scale $\Lambda_\mathrm{YM}$ of a few hundred MeV is found. In two and three dimensions a sign change occurs at about 100-300 MeV. In four dimensions, data at these momenta are not available yet, but the tendency is of the same type.

\begin{figure}
\begin{center}
\includegraphics[width=0.6\textwidth]{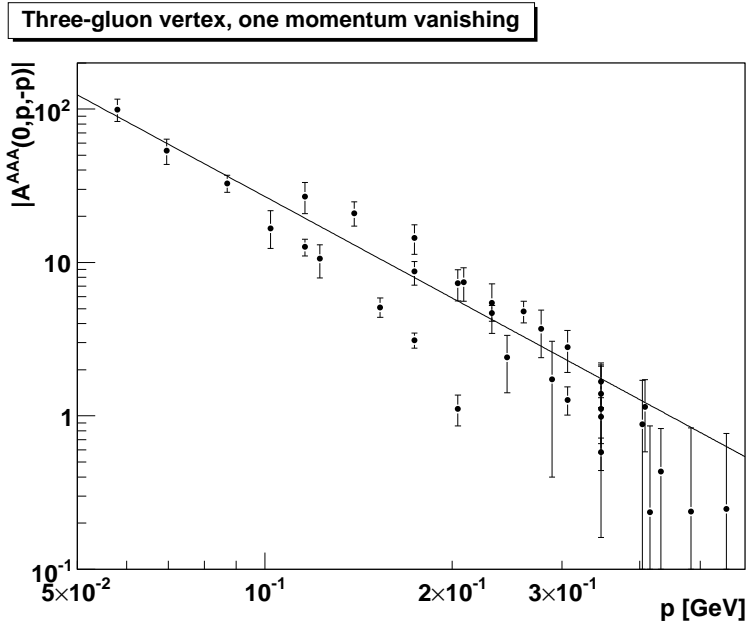}
\end{center}
\caption{\label{fig:g3v-ir}The tensor component \pref{zerot:g3v} of the three-gluon vertex in two dimensions for the gauge algebra su(2) in the minimal Landau gauge at low momenta \cite{Maas:2007uv}. The points come from different lattice volumes and discretizations. The line corresponds to a function $0.17p^{-2.2}$, the exponent coming from \pref{ir:IRsolution}.}
\end{figure}

In the very far infrared, in two and three dimensions a very strong enhancement (with negative sign) is found \cite{Maas:2007uv,Cucchieri:2006tf,Cucchieri:2008qm}. In fact, the behavior is very much reminiscent of the infrared divergence found in the scaling case for the vertex in functional studies \cite{Alkofer:2004it,Huber:2007kc}. Since again in the two-dimensional case minimal Landau gauge is at the very least very close to the scaling case, it is even possible to quantify the infrared exponent \cite{Maas:2007uv}, which turns out to be in agreement in with the formula \pref{ir:IRsolution}, confirming the analysis of all vertex functions in the infrared \cite{Alkofer:2004it,Huber:2007kc}. This is shown explicitly\footnote{Note that the momentum configuration is strictly speaking not the same as used in the corresponding calculations using functional methods, which would be at the symmetric point. However, for the Bose-symmetric three-gluon vertex, this could be of minor importance.} in figure \ref{fig:g3v-ir}. Remarkably, the fitted pre-factor is about an order of magnitude smaller than the one of the propagator power-law \cite{Maas:2007uv}. This has also been found in semi-perturbative functional calculations to be the case \cite{Alkofer:2008dt,Kellermann:2008iw}. Moreover, when determining the four-gluon vertex with the same method, the pre-factor is found to be a further order of magnitude smaller \cite{Kellermann:2008iw}, while at the same time showing the self-consistent exponent of \pref{ir:IRsolution} \cite{Alkofer:2004it}.

Though in higher dimensions it cannot be expected that the enhancement is a genuine divergence due to the screening occurring in the propagators at very low momenta, at the very least, the triple-gluon interaction strength in this channel is strongly enhanced at low momenta. In the scaling case, it has been found to be infrared divergent in all dimensions \cite{Alkofer:2004it,Huber:2007kc}, as would be all other tensor structures, as an infrared analysis of the functional equations \cite{Alkofer:2004it}, semi-perturbative studies \cite{Alkofer:2008dt}, and approximate self-consistent treatments have shown \cite{Alkofer:2008tt}.

Calculations of the four-point functions using lattice gauge theory have not yet been performed due to the strong statistical noise in such higher-order correlation functions \cite{Cucchieri:2006tf,Maas:2007uv}. An even more restricting problem is that disconnected contributions have to be removed explicitly, beginning with the four-point functions. Functional calculations have started to explore higher $n$-point functions using a semi-perturbative ansatz \cite{Kellermann:2008iw}. Beyond this point in both lattice and continuum calculations usually only bound-state properties have been calculated, either directly from the correlation functions in lattice calculations from their exponential decay in Euclidean time \cite{Montvay:1994cy,Gattringer:2010zz}, or using Bethe-Salpeter or Faddeev equations in functional methods \cite{Alkofer:2000wg,Fischer:2006ub,Roberts:2007jh,Eichmann:2009zx}. Rather recently, dynamic hadronization has also received attention for an effective possibility to include higher-order correlation functions in way of a self-consistent effective theory \cite{Pawlowski:2005xe,Braun:2009gm}. For the only physical bound-states in Yang-Mills theory, glueballs \cite{Mathieu:2008me,Crede:2008vw}, functional methods have met quite severe technical problems, which are still a matter of research \cite{Hauck:1998ir,Schaden:1994gb,Dudal:2009zh,Dudal:2010cd}, while lattice studies have been quite successful in determining their spectrum \cite{DeGrand:2006zz,Bali:1993fb}. Further objects of interest so far are scattering processes and bound-state coupling constants, though their direct determination in terms of correlation functions becomes increasingly complicated \cite{Gattringer:2010zz,Montvay:1994cy,Alkofer:2000wg,Bicudo:2001jq,Roberts:2007jh}.

Even with the yet limited range of higher order $n$-point functions results, conclusions can be drawn both for physics and for technical aspects. From the technical point of view, the fact that the ghost-gluon vertex should be enhanced at mid-momentum and this particular tensor structure of the three-gluon vertex should be suppressed, has been inferred from functional studies under the assumption that the two-loop terms are negligible \cite{Maas:2004se,Pennington:2011xs}. The reason is that otherwise the opposite signs of the ghost and the gluon loop destabilize the system, and in particular tends to drive the gluon dressing function into the negative, which is by construction not permitted. This is now confirmed a-posterior to be correct by the results shown in this section.

The sign change of this particular tensor structure of the three-gluon vertex also has quite far-reaching implications. Since it is the only contribution relevant for the gluon loop in the gluon propagator equation at one-loop order, it can provide a sign change of the whole loop in the infrared. Thus, the gluon physics changes from being anti-screening at large momenta to screening at small momenta. In accordance with the observations in section \ref{sir:scaling}, the screening mass from the ghost-loop in the critical limit of $\kappa_{\bar{cc}}=(d-2)/2$ is then enlarged by the gluon-loop, as both now have the same sign. This is also of profound importance for the finite-ghost case, where the gluon loop is one of the leading contributions, and the sign change helps to provide a positive screening mass.

The second consequence is even more profound. The three-gluon vertex behavior shows that at the hadronic energy scale the emission of a gluon from a gluon is strongly suppressed, at least for the tree-level tensor. However, it becomes strongly enhanced at momenta close to zero, i.\ e., close to the the light-cone. At these small momenta the gluons are screened by either a finite or an infinite mass, thus not contributing anymore. This could be an important hint into how the infrared enhancement threatening to be unitarity-violating in parton structure functions could be remedied by non-perturbative effects \cite{Dissertori:2003pj}.

\subsection{Schwinger functions, mass, and asymptotic states}\label{quant:schwinger}

Having the propagators now available, it is an interesting question what kind of insight can be obtained from them on the properties of the elementary particles. In particular, their masses and widths are of great interest, as is their analytic structure. This information is encoded in the propagators, and can be obtained by determining their space-dependence.

The most direct access to properties like masses and widths is granted by the Schwinger function \cite{Alkofer:2003jj}. This function is defined as\footnote{The Schwinger function can also be evaluated directly in position space without the detour over momentum space \cite{Gattringer:2010zz,DeGrand:2006zz}.}
\be
\Delta(t)=\frac{1}{\pi}\int_0^\infty dp_0\cos(tp_0)D(p_0^2)\label{cschwing}\\
\ee
\no for a particle with propagator $D(p^2)$. The notation $p_0$ indicates that the dressing function is evaluated at zero spatial momentum. On the lattice, the corresponding expression is given by \cite{Cucchieri:2004mf}
\be
\Delta(t)=\frac{1}{a\pi}\frac{1}{N_t}\sum_{P_0=0}^{N_t-1}\cos\left(\frac{2\pi tP_0}{N_t}\right)D(P_0^2)\nn,
\ee
\no where both $t$ and $P_0$ are in lattice units, i.\ e., integer, and $N_t$ is the lattice extension in time direction. Note that the sum extends over the whole momentum range, and includes the parts of the propagator reproduced by periodicity.

In case of a stable particle with a simple pole mass $M$, having an Euclidean propagator
\be
D(p)=\frac{1}{p^2+m^2}\label{zerot:simpleprop},
\ee
\no and thus a pole at $p=\pm im$, the Schwinger function is given by
\be
\Delta(t)=\frac{1}{2m}e^{-mt}\nn.
\ee
\no Therefore, a simple exponential decay is expected in Euclidean time for a massive, stable particle. It is worthwhile to note for the investigations at finite temperature in chapter \ref{sfinitet} that such an exponential decay for a bosonic, stable particle with zero width is also expected in the spatial directions for the soft mode, i.\ e., for $\Delta(z)$, since its propagator is $1/(\vec{p}^2+m^2)$. Here, the energy is zero, as it is a soft mode, as are the other spatial momentum components orthogonal to the direction along which the integral to determine the Schwinger function is performed.

Beyond tree-level, the propagator \pref{zerot:simpleprop} is modified by the appearance of a cut, starting at $p=\pm2im$ \cite{Peskin:1995ev}. At lowest order in perturbation theory, a possible analytic form for such a propagator in four dimensions is given by
\bea
D(p)&=&\frac{1}{p^2+m^2+\Pi(p^2,m^2)}\label{zerot:stablecut}\\
\Pi(p^2,m^2)&=&-g^2\left(\frac{\pi}{2\sqrt{3}}+\sqrt{1+\frac{4m^2}{p^2}}\atanh\left(\sqrt\frac{p^2}{4m^2+p^2}\right)\right)\nn,
\eea
\no where $g$ has dimension of mass. This form is motivated by leading-order perturbation theory, and occurs, e.\ g., for a scalar theory with a three-point coupling \cite{Gerhold:2010wy}.

If the particle is not stable, the poles are moved off the first Riemann sheet onto the second Riemann sheet at $m+i\Gamma/2$. The cut then starts at $\pm 2iM$, where $M$ is the mass of the particles in which the original particle can decay, assuming for the moment only this two particles in the theory \cite{Peskin:1995ev}. In this case, the propagator \pref{zerot:stablecut} is modified to \cite{Gerhold:2010wy}
\bea
D(p)^{-1}&=&p^2+\left(m+i\frac{\Gamma}{2}\right)^2-g^2\left(\Pi\left(p^2,m^2,\Lambda^2\right)-\Pi\left(\left(m+i\frac{\Gamma}{2}\right)^2,m^2,\Lambda^2\right)\right)\nn\\
&&-h^2\left(\Pi\left(p^2,M^2,\Lambda^2\right)-\Pi\left(\left(m+i\frac{\Gamma}{2}\right)^2,M^2,\Lambda^2\right)\right)\label{zerot:unstablecut}\\
\Pi(p^2,m^2,\Lambda^2)&=&-\sqrt{1+\frac{4m^2}{p^2}}\atanh\sqrt{\frac{p^2}{4m^2+p^2}}+\frac{\left(2\Lambda^2+4m^2+p^2\right)\atanh\sqrt{\frac{p^2}{4\Lambda^2+4m^2+p^2}}}{\sqrt{p^2}\sqrt{4\left(\Lambda^2+m^2\right)+p^2}}\nn\\
&&+\frac{1}{2}\ln\left(1+\frac{\Lambda^2}{m^2}\right)\nn,
\eea
\no where there are now two couplings of dimension mass, $g$ and $h$, describing\footnote{Note that not all parameters are independent. The reality of the Euclidean propagator fixes one of the parameters $m$, $\Gamma$, $M$, $g$, and $h$ as a function of the others, since the decay width is not an independent quantity.} a self-interaction $g$ and a decay channel with strength $h$, and $\Lambda$ is the cutoff, of which this renormalized propagator is independent.

However, since neither the gluon (nor the ghost) behaves necessarily like a physical particle, after all it is not gauge-invariant and thus not physical, its propagator may not be of either form \pref{zerot:stablecut} or \pref{zerot:unstablecut}. There have been various proposals, which form it may have instead. One, the so-called Gribov-Stingl type \cite{Gribov:1977wm,Habel:1989aq,Habel:1990tw,Stingl:1994nk}, has complex poles on the first Riemann sheet, but may have vanishing residues at the poles \cite{Habel:1989aq,Habel:1990tw,Stingl:1994nk}. This behavior can be described by a meromorphic function \cite{Alkofer:2003jj}
\be
D(p)=\frac{e^2+f p^2}{p^4+2m^2\cos\left(2\phi\right)p^2+m^4}\label{zerot:unstable}.
\ee
\no Its Schwinger function is given by
\be
\Delta(t)=\frac{e^2}{2m^3\sin\left(2\phi\right)}e^{-tm\cos\left(\phi\right)}\left(\sin\left(\phi+tm\sin\left(\phi\right)\right)+\frac{fm^2}{e^2}\sin\left(\phi-tm\sin\left(\phi\right)\right)\right).\label{zerot:unstableft}
\ee
\no The Schwinger function therefore exhibits oscillation with a period determined by the angle $\phi$ and the mass parameter $m$. Its positivity violations are signaling therefore the instability of the particle. The propagator \pref{zerot:unstable} illustrates at the same time the difference between the concept of screening mass and pole mass. The screening mass is defined as the inverse square-root of the propagator at zero momentum, and given by
\be
D(0)^{-\frac{1}{2}}=\frac{m^2}{e}\nn.
\ee
\no Since $e$ depends on the wave-function renormalization, this immediately shows that a screening mass is never renormalization-group-invariant, and can therefore not be a physical observable quantity. The pole mass, i.\ e., the location of the poles of \pref{zerot:unstable} is given by
\be
i m e^{\pm i\phi}\nn,
\ee
\no and is thus a set of complex conjugate poles, but in general with non-zero residuum. This mass is renormalization-group invariant, if there is no additive mass-shift to $m$, and can thus be, at least in principle, a physical mass, though in the present case it may be gauge-dependent. Note that in the limit of real masses, $e$ has to go to $m$ and $f$ to one, or otherwise a double-pole would emerge.

There is another concept which appears when using the Euclidean correlation function. Assume that the propagator can be more generally written as
\be
D(p)=\frac{Z}{p^2+M(p^2)^2},\label{zerot:mdmass}
\ee
\no with some wave-function renormalization constant $Z$. At first sight, the condition $p^2=-M(p^2)^2$ could look like an indicator for the pole mass of the particle described by the propagator. However, this is only possible if the function $M(p^2)$ only depends on the real part of $p^2$. This is certainly the case for \pref{zerot:simpleprop}, but is not necessarily the case for \pref{zerot:mdmass}. Thus, the point $-M(p^2)^2$ only gives a would-be pole mass, and the correct pole mass is only obtained when taking into account the full dependence of $M(p^2)$ on complex momenta \cite{Alkofer:2003jj,Fischer:2008uz}.

None of these forms can actually fit a scaling type behavior. The results from functional studies suggest a form which has a cut on the imaginary momenta axis starting at zero momentum. Functional forms  which provide such a structure are given by\footnote{Note that similar functional forms are already encountered in QED \cite{Maris:1996zg,Alkofer:2000wg}, but also appear in (near-)conformal theories, as are considered for unparticles \cite{Sannino:2008ha}, and to some extent technicolor models in beyond-the-standard-model scenarios \cite{Sannino:2009za}.} \cite{Alkofer:2003jj,Maas:2004se}
\bea
Z_1(p)&=&\frac{A_z p^{2\kappa_{AA}}}{1+f+A_z p^{2\kappa_{AA}}}f_\mathrm{UV}(p^2)\nn\\
Z_2(p)&=&\frac{A_z p^{2\kappa_{AA}}}{(1+f+A_z p)^{2\kappa_{AA}}}f_\mathrm{UV}(p^2)\label{zerot:4dscaling},
\eea
\no where $f_\mathrm{UV}$ encodes the perturbative tail. While the first version works better in three dimensions \cite{Maas:2004se}, the second is more adequate for four dimensions \cite{Alkofer:2003jj}. Both have no simple complex or real poles on the first Riemann sheet, but have additional poles on further Riemann sheets. If one assumes that the finite-ghost case is a modification of the scaling case, a fit form like
\be
Z_3(p)=\frac{p^2(m^2+p^2)^{\kappa_{AA}}}{\sigma(1+fp^2+gp^{2+2\kappa_{AA}}}f_\mathrm{UV}(p^2),\label{imscaling}
\ee
\no is suggested. This form can retain the intermediate scaling behavior in a scaling window \pref{dse:conformalwindow} to some extent, and has the same analytic structure as the scaling case. I.\ e., the only singularities in the complex plane is a cut along the real axis \cite{Alkofer:2003jj}. This structure is indeed retained in the finite-ghost case when solving the DSEs in the full complex momentum plane \cite{Strauss:2012as}. Thus, currently this appears to be the most likely analytic structure for the gluon propagator.

\begin{figure}
 \includegraphics[width=\textwidth]{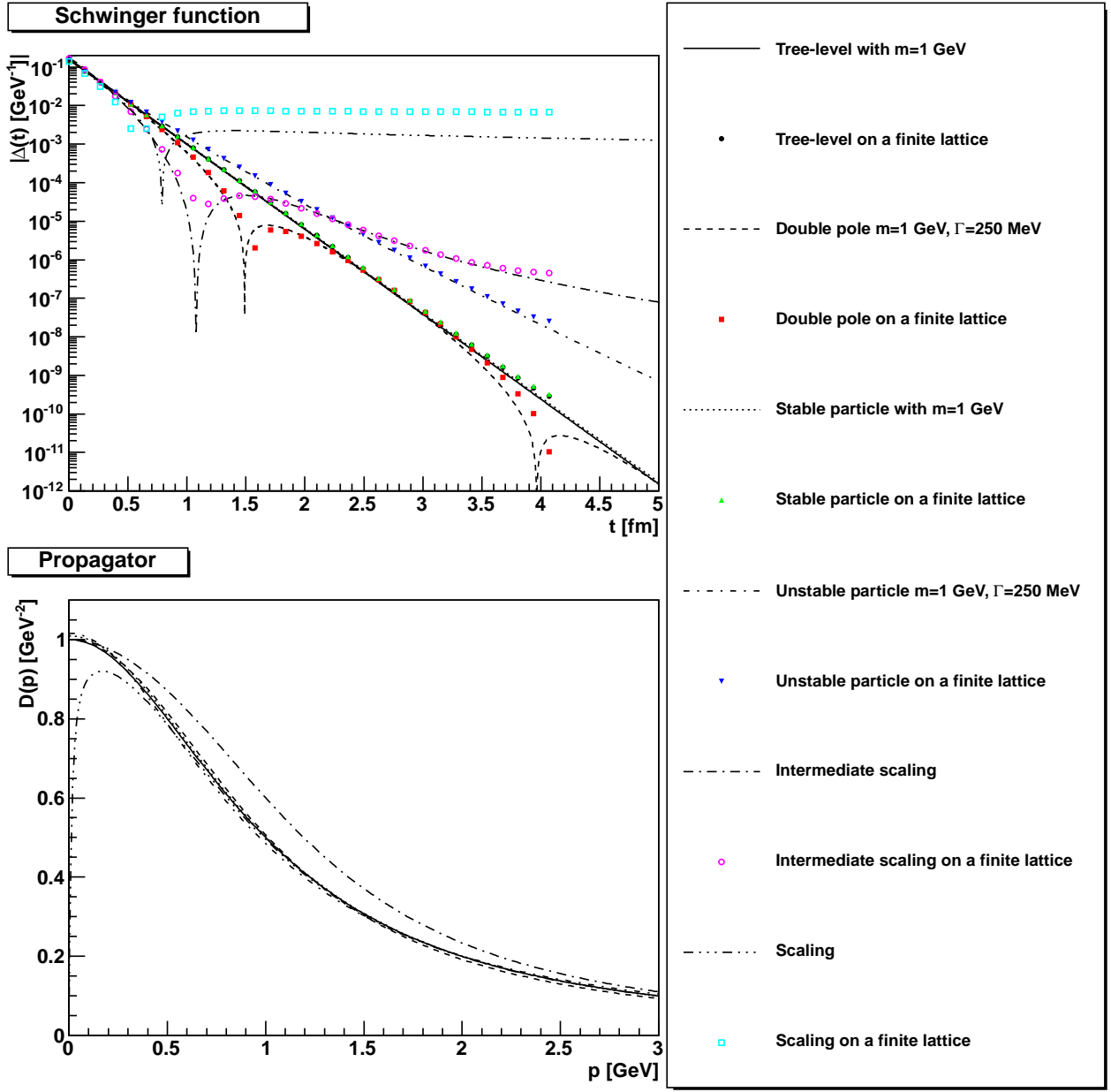}
 \caption{\label{fig:schwinger}Various forms of propagators in the Euclidean domain and of the corresponding Schwinger functions. To illustrate lattice artifacts, the Schwinger functions are also shown for the case of a $64^4$ lattice with $a=0.13$ fm. The parameters for the propagators are: tree-level propagator \pref{zerot:simpleprop} $m=1$ GeV (full line); double pole \pref{zerot:unstable} $m=1.03$ GeV, $e=(1.03$ GeV$)^2$, $f=0.9$, $\phi=0.245$ (dashed line); stable particle with cut \pref{zerot:stablecut} $m=1$ GeV, $g=70$ MeV (short-dashed line); unstable particle with cut \pref{zerot:unstablecut} $m=1$ GeV, $\Gamma=250$ MeV, $M=300$ MeV, $g^2=(1.44$ GeV$)^2$, $h^2=(0.528$ GeV$)^2$ (short-dashed-dotted line; intermediate scaling form \pref{imscaling} $m=1$ GeV, $\kappa_{AA}=1.19$, $\sigma=(1$ GeV$)^{2+2\kappa_{AA}}$, $f=(1.3$ GeV$)^2$, $g=1$ GeV$^{-2-2\kappa_{AA}}$, $f_{UV}=1$ (dashed-dotted line); scaling \pref{zerot:4dscaling} $\kappa_{AA}=1.19$, $f=-0.996$, $A_z=(1$ GeV)$^{-2\kappa_{AA}}$, $f_{UV}=1$ (dashed-triple-dotted line).}
\end{figure}

Most importantly, all of the forms permit Wick rotation \cite{Alkofer:2003jj}, essential to transfer the results back to Minkowski space. To illustrate the differences, both the propagators and the corresponding Schwinger functions for all of these forms are shown in figure \ref{fig:schwinger}. It is nicely visible that the form of the Euclidean propagator is very similar in all cases, and only the Schwinger function shows significant differences. The stable form, with or without cut, shows an essentially exponential decay. The unstable particle shows first a decay with the heavier mass of the parent state and at later time a slower with the mass of the daughter state. Still, the long-time decay is essentially single exponential. The double-pole structure shows an oscillatory behavior, superimposed with an exponential decay. The cases with full or intermediate scaling both also show a zero-crossing, but no oscillations, and a decay which is not exponential in time. Thus, it is the long-time behavior which ultimately is the most powerful possibility to distinguish between the analytic structure of the possible propagator forms.

\begin{figure}
\includegraphics[width=0.5\textwidth]{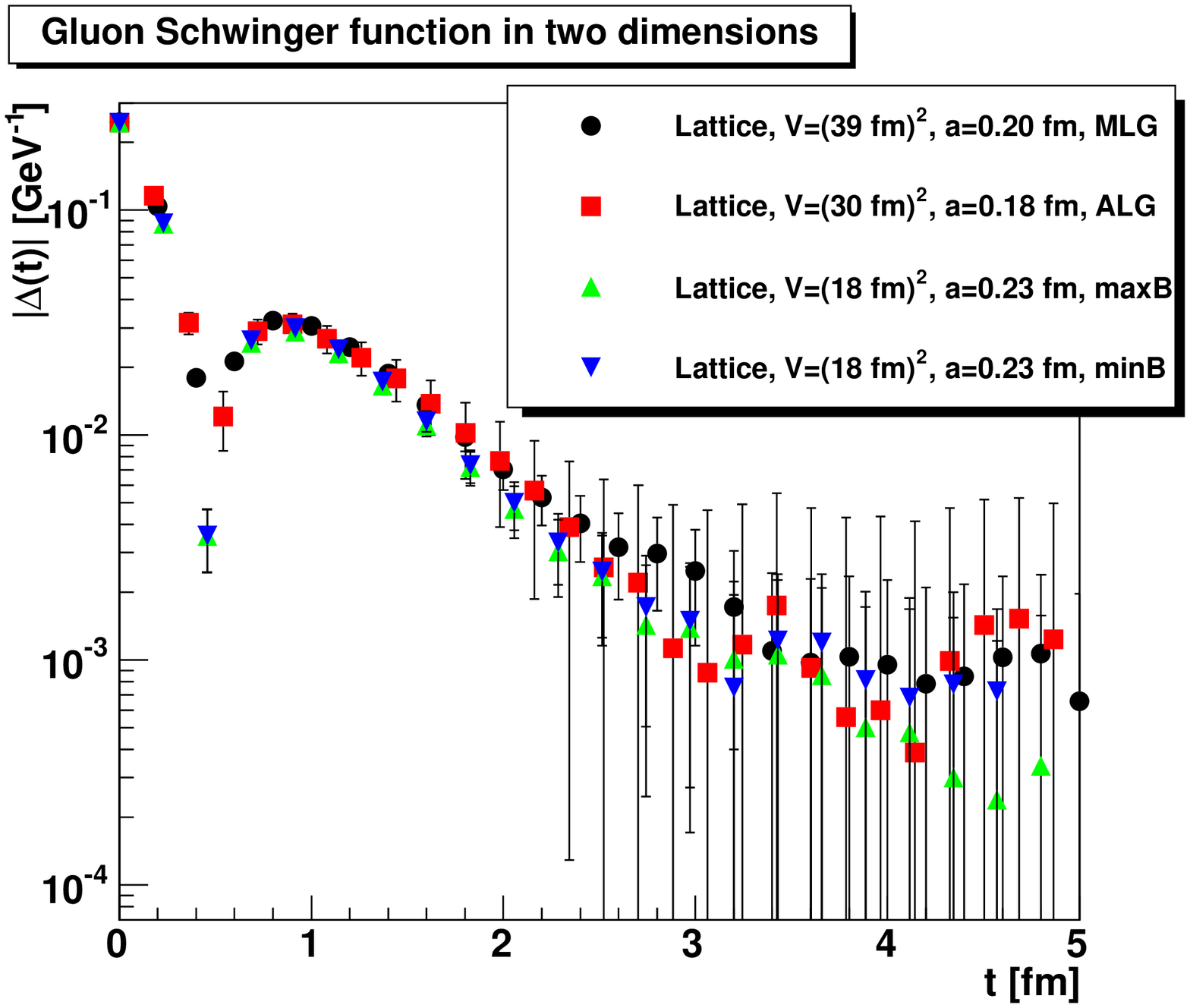}\includegraphics[width=0.5\textwidth]{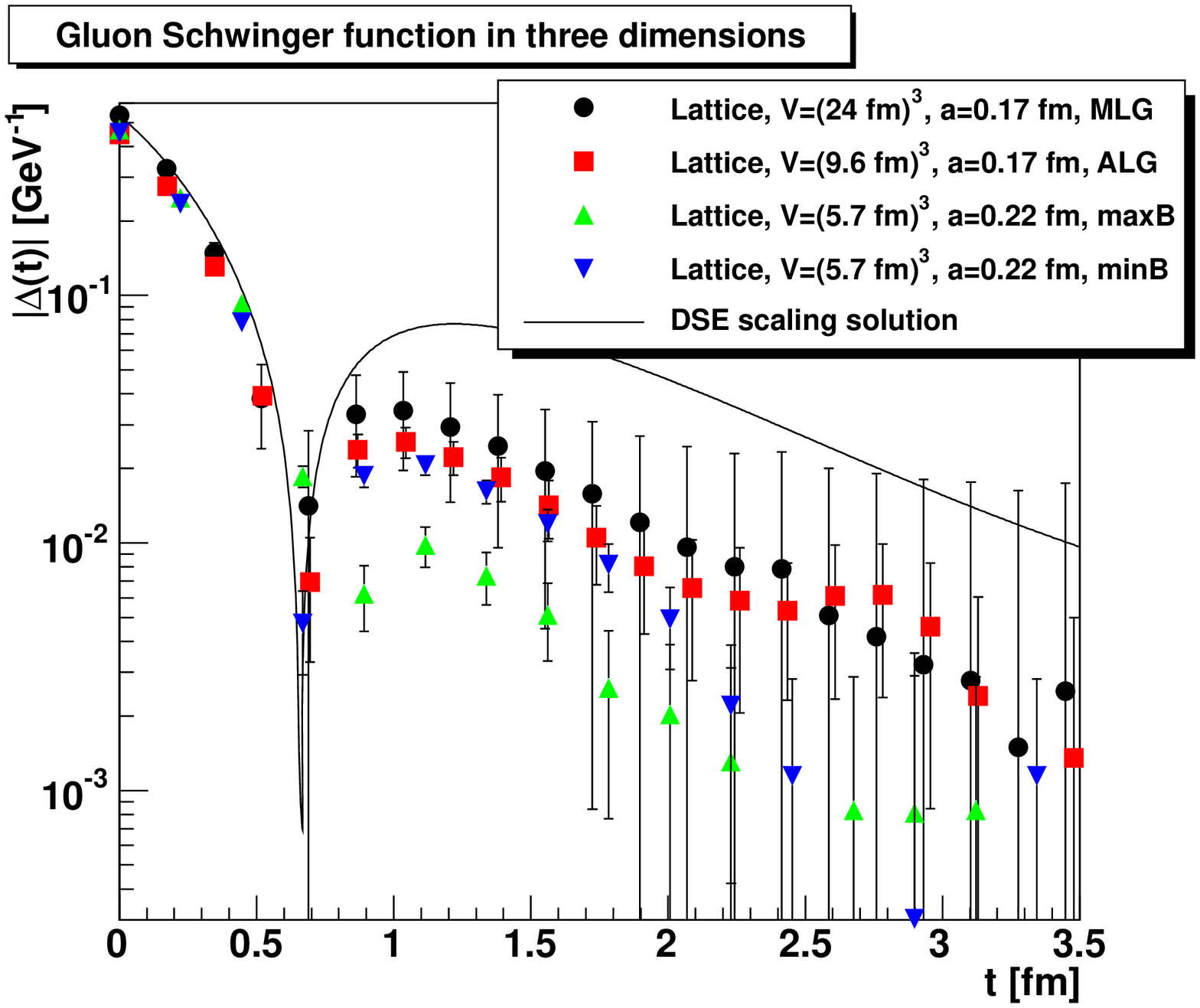}\\
\begin{minipage}[c]{0.5\linewidth}
\includegraphics[width=\textwidth]{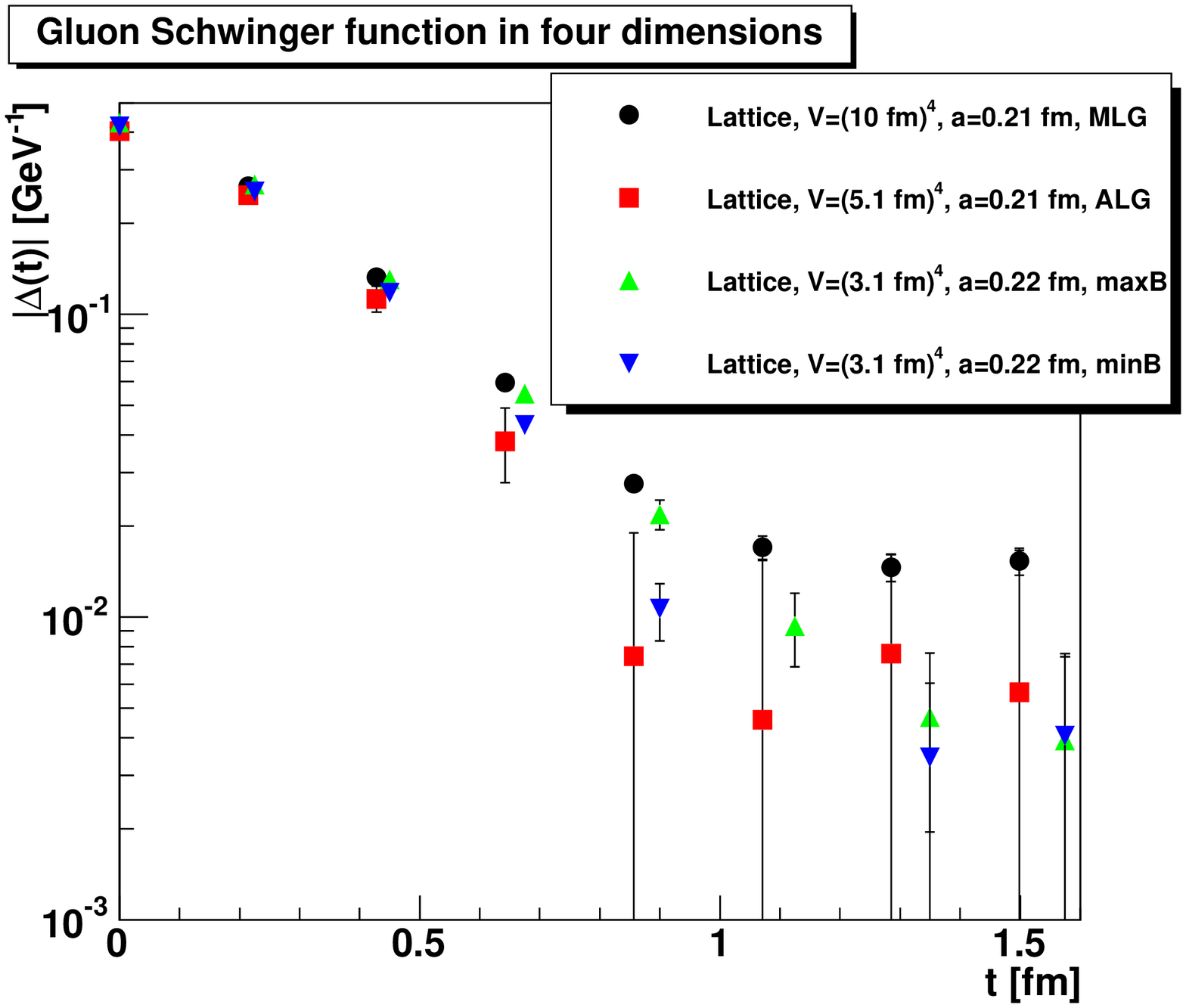}
\end{minipage}
\begin{minipage}[c]{0.5\linewidth}
\caption{\label{fig:gp-schwinger}The Schwinger function for the gluon in two (top-left panel) \cite{Maas:2007uv,Maas:2009se,Maas:2008ri}, three (top-right panel) \cite{Cucchieri:2003di,Cucchieri:2004mf,Maas:2008ri,Maas:2009se}, and four (bottom panel) dimensions \cite{Maas:unpublished,Cucchieri:2008qm}, for the minimal Landau gauge, the absolute Landau gauge, the maxB gauge, and the minB gauge for su(2). Also results from DSEs are shown for the scaling case in three dimensions \cite{Maas:2004se}. Functional results for four dimensions can be found for both the finite-ghost case and scaling in \cite{Fischer:2008uz} and further lattice results also in \cite{Bowman:2007du,Langfeld:2001cz}. These are all qualitatively and quantitatively very similar to those shown.}
\end{minipage}
\end{figure}

After this rather general introduction, the Schwinger function can be obtained for the different propagators in Yang-Mills theory. The first object to apply this formalism to is the gluon propagator. As figure \ref{fig:gp-schwinger} shows, the Schwinger functions for the various cases are qualitatively identical. Furthermore, the DSE results for both the finite-ghost and scaling cases are also qualitatively identical \cite{Fischer:2008uz}. However, the results are strongly affected by finite-volume corrections, and it requires a certain minimal volume to observe the most characteristic feature, the zero crossing of the Schwinger function at a range of a half to a few Fermi, depending on the number of dimensions, the gauge algebra, and whether the behavior is of finite-ghost type or scaling\footnote{Similar results have also been obtained using an effective action approach \cite{Dudal:2008sp,Dudal:2008rm}.} \cite{Cucchieri:2004mf,Maas:unpublished}. Besides this first zero crossing, no further crossings have been observed in functional calculations, and thus there is no oscillatory behavior, as would be expected for a double pole structure. Furthermore, the decay at large times is not found to be of exponential time for both cases, thus making an analytic structure like \pref{imscaling} the best description. For the lattice calculations, statistical uncertainty makes it yet impossible to decide which of the possible forms with a zero crossing, \pref{zerot:unstablecut}, \pref{zerot:unstable}, or \pref{imscaling}, is most appropriate. In any case, the gluon Schwinger function is negative above a certain distance.

This implies that positivity is manifestly violated for the gluon\footnote{Note that a gluon propagator which vanishes at zero momentum is implying (maximal) positivity violation \cite{Alkofer:2003jj}.}. This implies the absence of the gluon from the asymptotically physical Hilbert space \cite{Haag:1992hx,Alkofer:2003jj}. However, positivity violation is not equivalent to confinement. Whether confinement is implying positivity violation is not known. However, a confined particle is necessarily absent from the asymptotic physical Hilbert space. Thus, the positivity violation implies already this necessary condition.

Note that this does not prevent the existence of a stable asymptotic gluon state in the unphysical part of the Hilbert space of the theory; it is just excluded from the physical part of the Hilbert space\footnote{This observation can already be indirectly inferred from the violation of the Oehme-Zimmermann superconvergence relation in perturbation theory \cite{Alkofer:2000wg,Oehme:1979ai}.} \cite{Kugo:1979gm}. Furthermore, the screening mass found in the finite-ghost case is not implying the existence of a pole mass of the gluon\footnote{If one exists, it would be be gauge-parameter-independent, as described by the Nielsen identities \cite{Nielsen:1975fs}, though in general not gauge-independent \cite{Nielsen:1975fs}.}, as illustrated by the form \pref{imscaling}, compatible with the results from functional calculations. The only implication of the positivity violation is thus just that the gluon will not appear as an asymptotic state in the physical part of the Hilbert space.

\begin{figure}
\includegraphics[width=0.5\textwidth]{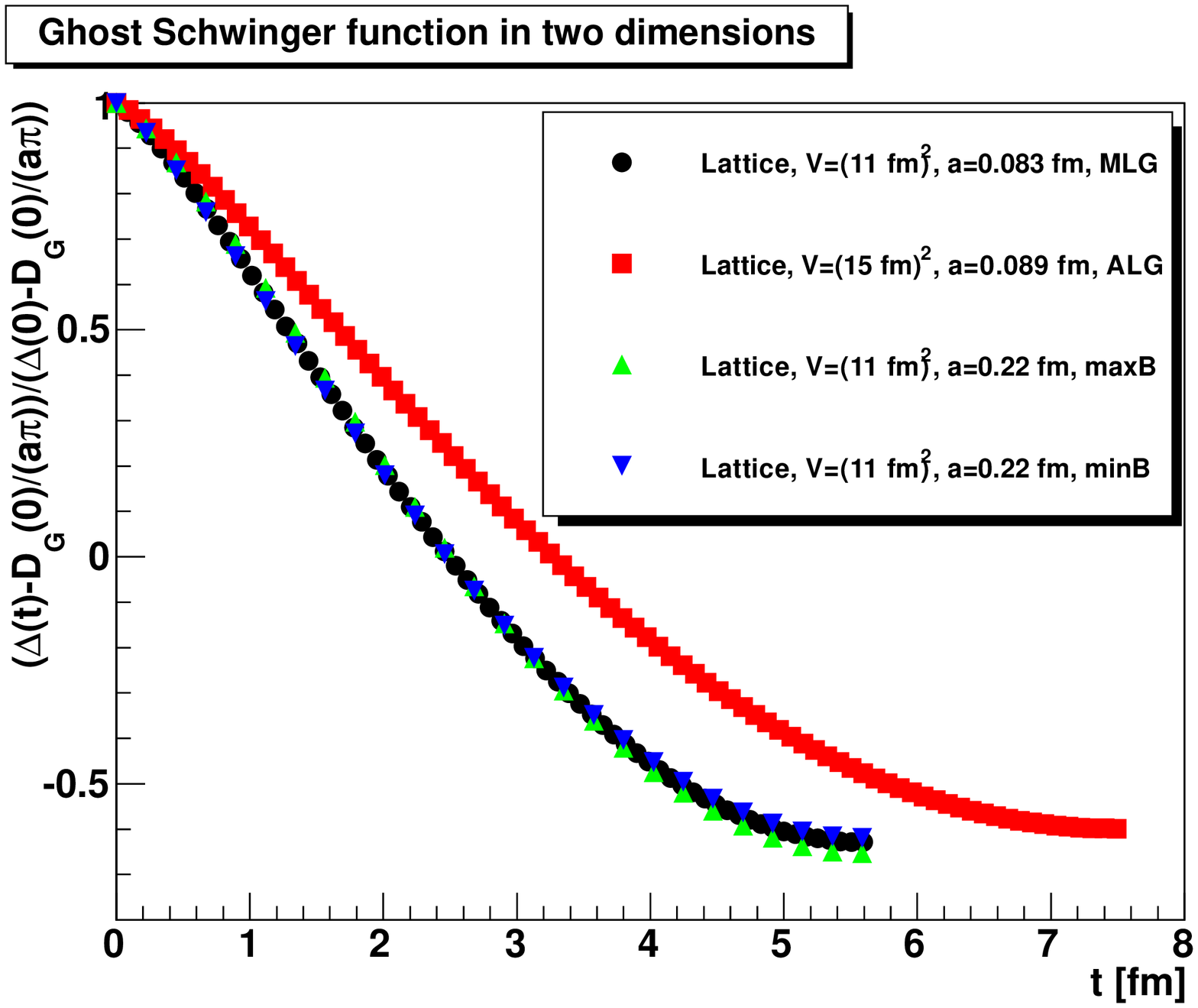}\includegraphics[width=0.5\textwidth]{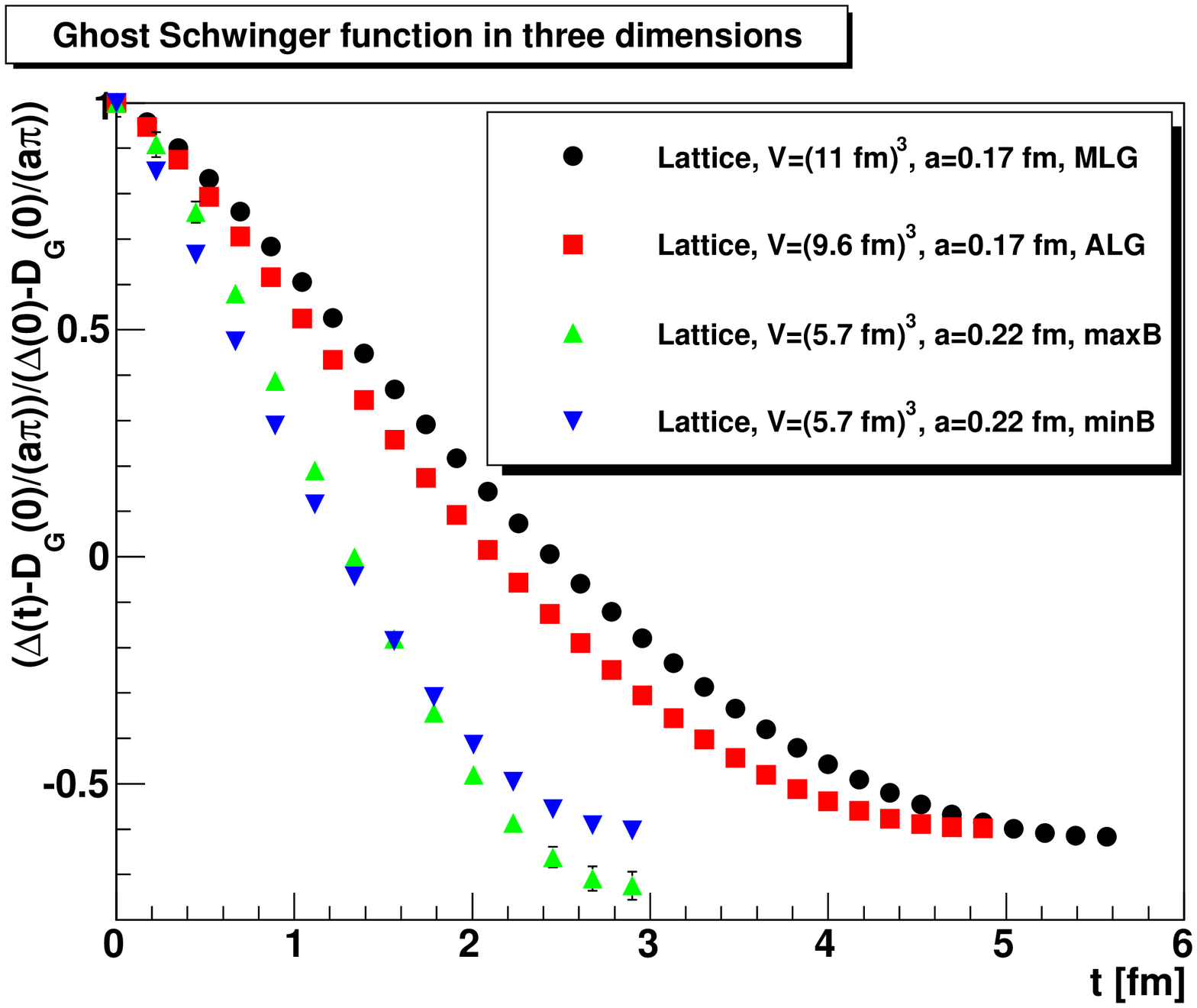}\\
\begin{minipage}[c]{0.5\linewidth}
\begin{center}\includegraphics[width=\textwidth]{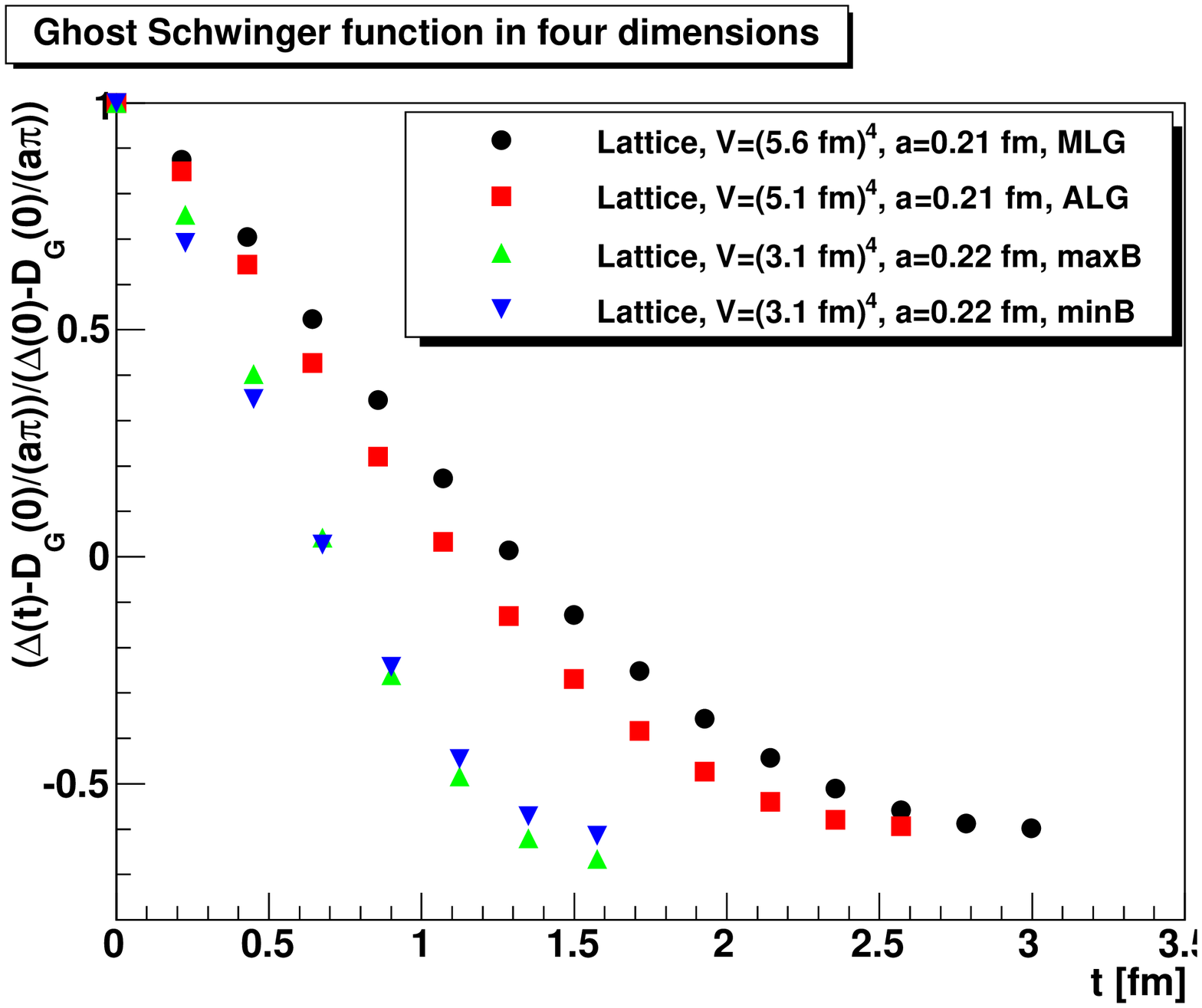}\end{center}
\end{minipage}
\begin{minipage}[c]{0.5\linewidth}
\caption{\label{fig:ghp-schwinger}The renormalized Schwinger function for the ghost \cite{Maas:unpublished} in two (top-left panel), three (top-right panel), and four (bottom panel) dimensions, for the gauge algebra su(2), and for the minimal Landau gauge, the absolute Landau gauge, the maxB gauge, and the minB gauge.}
\end{minipage}
\end{figure}

Though not linked to a particle, it is also interesting to investigate the Schwinger function of the ghost. A complication in its determination is that the propagator diverges at least as fast as the one of a massless particle. Thus, its Schwinger function has to be renormalized. It is shown in figure \ref{fig:ghp-schwinger}, and it is visible that is suffers from quite strong lattice artifacts and/or has a significant gauge dependence. This has to be investigated further \cite{Maas:unpublished}. Nonetheless, after taking into account the divergence at zero momentum by appropriately normalizing it, it is found that the Schwinger function becomes constant at long distances, instead of decaying to zero as for ordinary particles or the gluon. Thus the ghost implies correlations over arbitrarily large distances. This automatically implies the absence of a mass gap in the spectrum of the full Hilbert space \cite{Alkofer:2000wg,Haag:1992hx}, despite the presence of one in the physical subspace. In particular, this implies the failure of cluster decomposition for Yang-Mills theory on the level of gauge-dependent elementary particles and the absence of a 'behind-the-moon'-problem \cite{Kugo:1979gm}. The presence of these infinite-range correlations are then a consequence of the necessity to maintain gauge invariance over long, and also space-like, distances. The Schwinger function itself is then also negative at large distances. Thus, the ghost trivially violates positivity, and also does not belong to the physical state space.

\begin{figure}
\includegraphics[width=0.5\textwidth]{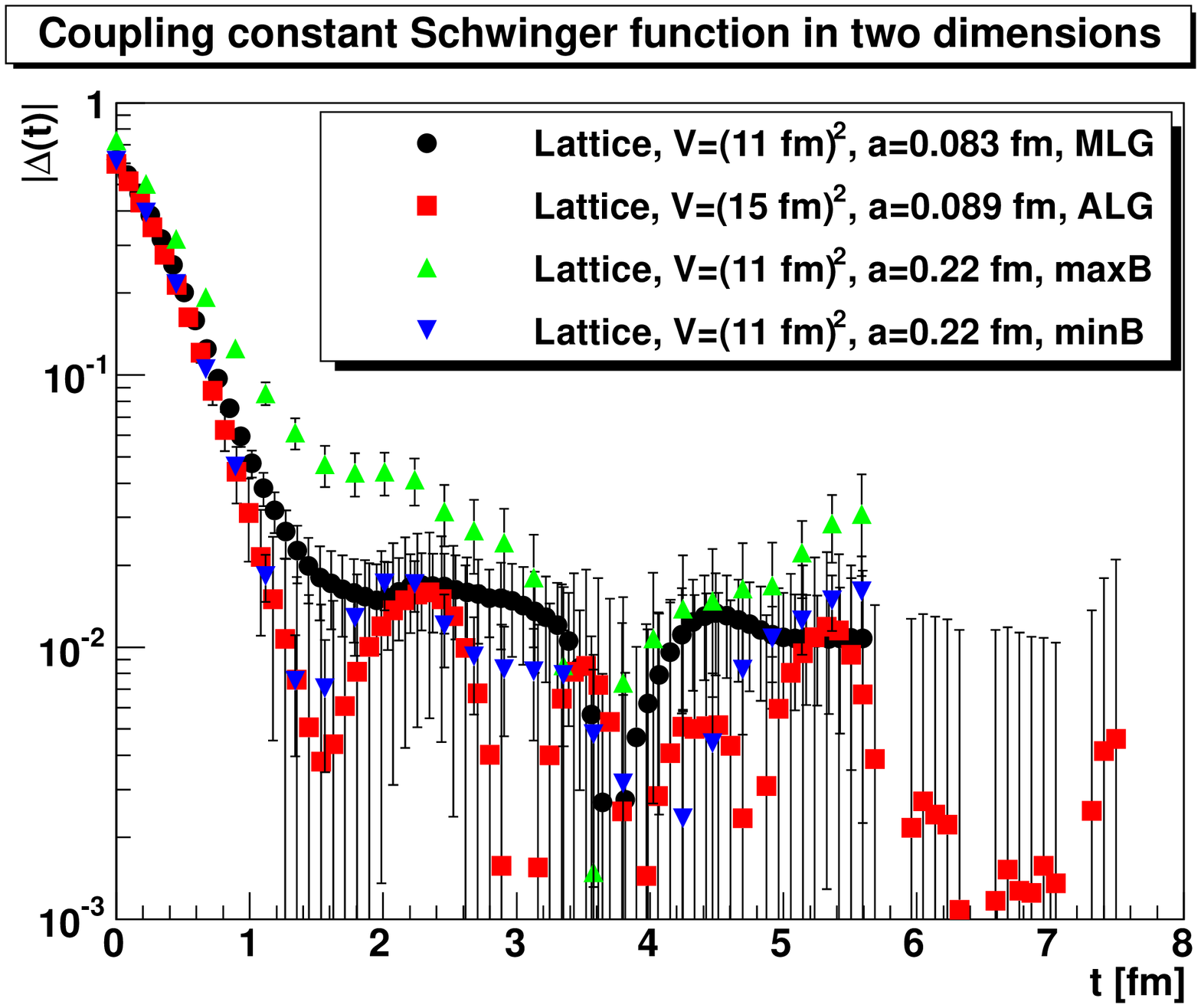}\includegraphics[width=0.5\textwidth]{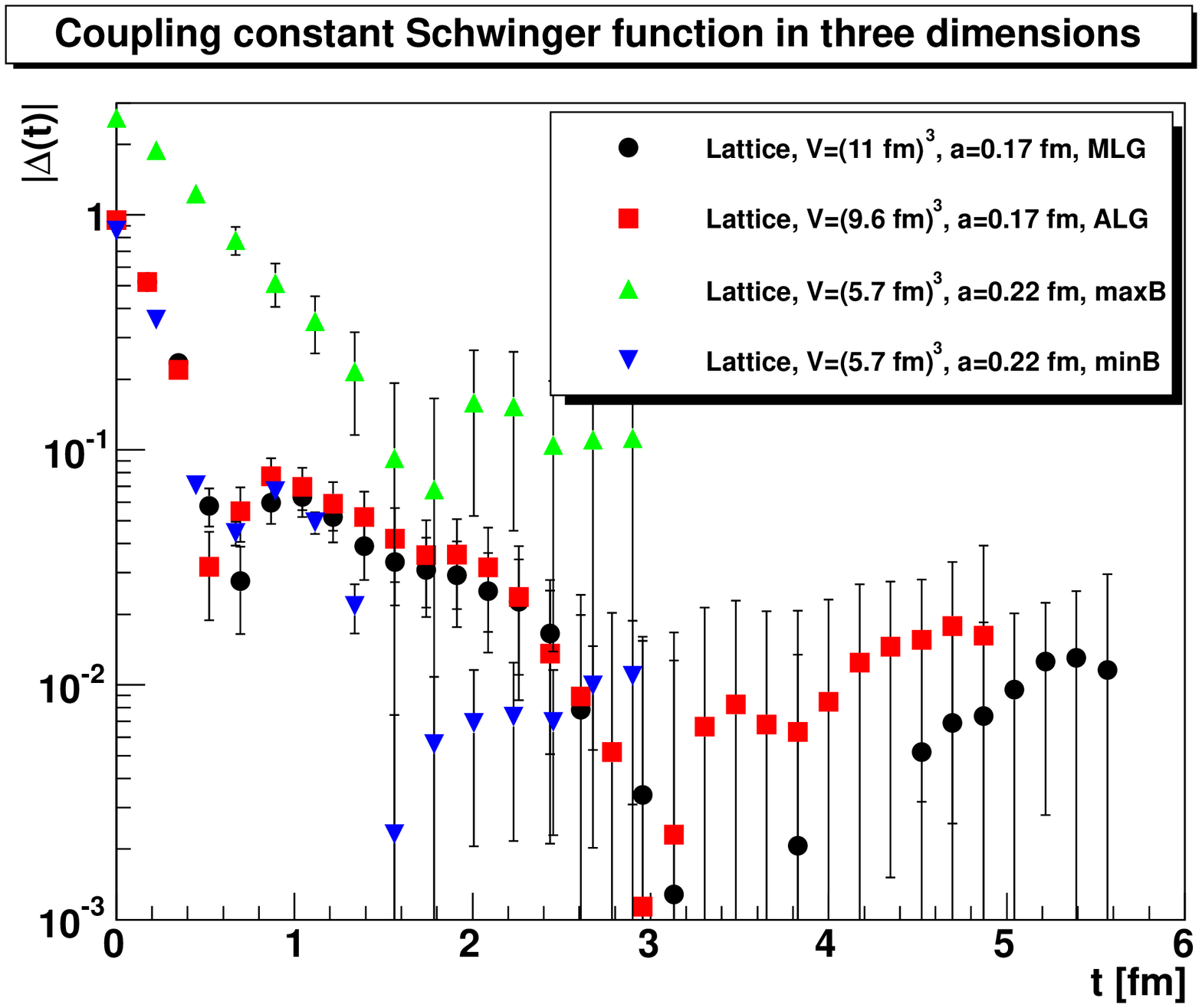}\\
\begin{minipage}[c]{0.5\linewidth}
\begin{center}\includegraphics[width=\textwidth]{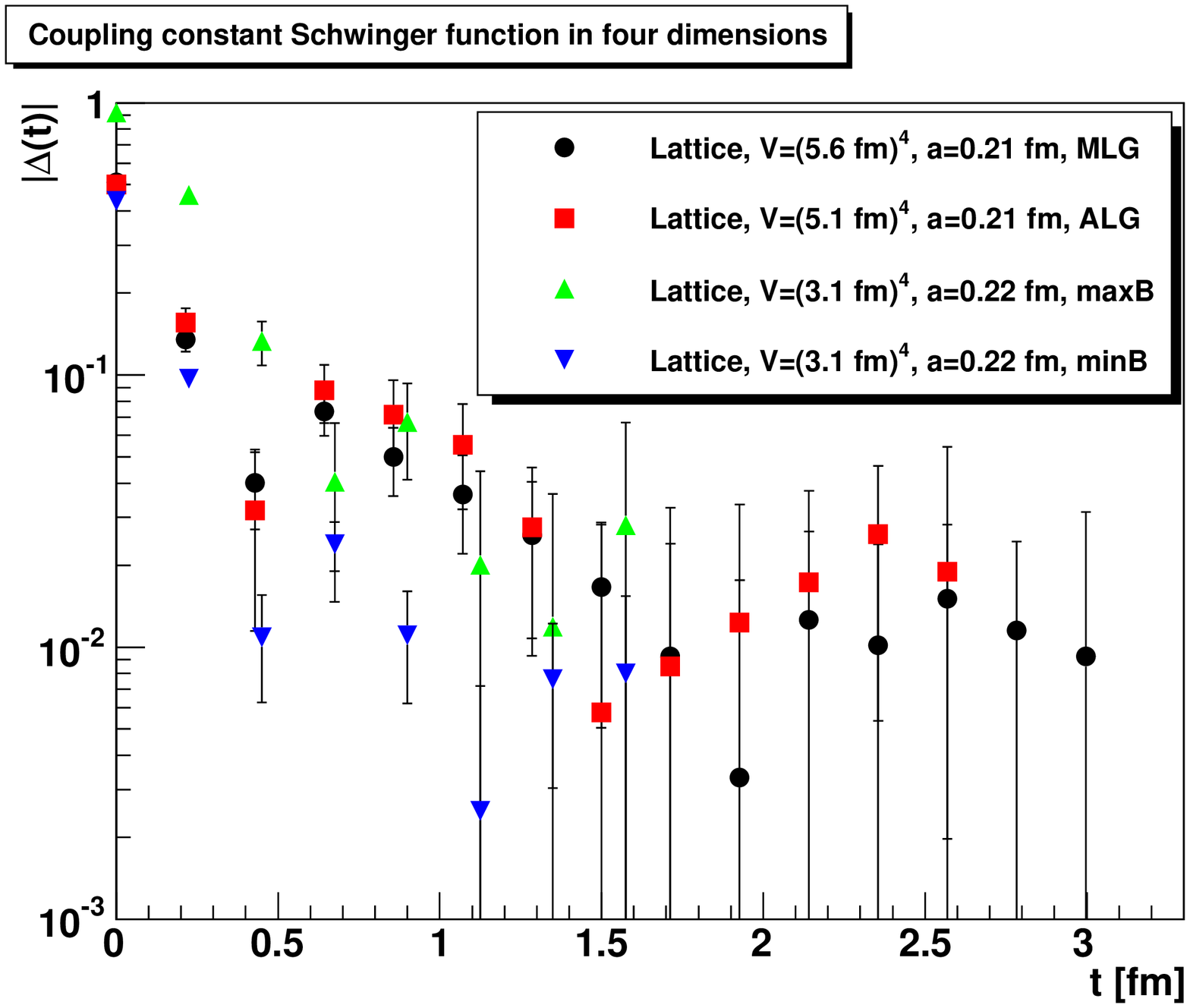}\end{center}
\end{minipage}
\begin{minipage}[c]{0.5\linewidth}
\caption{\label{fig:alpha-schwinger}The Schwinger function for the coupling constant \cite{Maas:unpublished} in two (top-left panel), three (top-right panel), and four (bottom panel) dimensions, for the gauge algebra su(2), and the minimal Landau gauge, the absolute Landau gauge, the maxB gauge and the minB gauge.}
\end{minipage}
\end{figure}

It is another interesting question to investigate the Schwinger function of the coupling constant. Though it does not describe a particle, this is essentially a measure of the interaction strength as a function of distance\footnote{In QED, it corresponds to the potential, but this is no longer the case in a non-Abelian theory \cite{Peskin:1995ev}.}. The result, removing trivial momentum factors before forming the Schwinger function, is shown in figure \ref{fig:alpha-schwinger}. At short distances, the Schwinger function is relatively large, while at long distances, even in the two-dimensional case where the coupling freezes out, the corresponding Fourier transform vanishes. However, also here lattice artifacts may play a role. The Schwinger function is always positive, despite that the propagators from which it is formed violate individually positivity. Furthermore, the Schwinger function also exhibits distinct minima, even when taking the relatively large errors into account. Whether these have any meaning is at the current time rather unclear. Finally, the Schwinger function for the maxB gauge differs from the others, which is likely linked to the fact that it is still infrared divergent for all lattice volumes investigated here.

It should be noted that positivity violations of the Schwinger functions is necessarily implying positivity violations of the spectral function $\rho$, defined implicitly by
\be
D(p)=\int_0^\infty dM^2 \frac{\rho(M^2)}{p^2+M^2}\label{spectralfunction},
\ee
\no where any possible one-particle pole is included in the spectral function. The implication is obtained by inserting into \pref{cschwing} the representation \pref{spectralfunction}
\be
\Delta(t)=\frac{1}{\pi}\int_0^\infty dp_0\cos(tp_0)\int_0^\infty dM^2 \frac{\rho(M^2)}{p_0^2+M^2}=2\int dM e^{-M t}\rho(M^2)\nn,
\ee
\no where it is assumed that both integrations can be exchanged.

Any positivity violations of the spectral function implies the absence of a K\"allen Lehmann representation. The spectral function of a unstable but otherwise physical particle, like described by \pref{zerot:unstablecut}, remains positive \cite{Peskin:1995ev}. Thus, so must be its Schwinger function. This is not necessarily the case for unphysical particles. This illustrates how sensitive the Schwinger function is to details of the propagator structure.

From a practical point of view, the Schwinger function yields a more direct access to the analytic properties than the spectral function. Also, the reconstruction of the spectral function is non-trivial, and in case of the propagator being only available on a finite number of (lattice) momenta necessarily not unique, leading to significant systematic uncertainties \cite{Nickel:2006mm}. 

A positivity violation of the spectral function immediately marks a state as unphysical. Such a violation follows also immediately if the propagator either vanishes at zero momentum or its derivatives w.\ r.\ t.\ $p^2$ are not of constant sign, since \pref{spectralfunction} implies
\be
\frac{\pd^n D(p)}{\pd (p^2)^n}=(-1)^n\int_0^\infty dM^2 \frac{\rho(M^2)}{(p^2+M^2)^n}\label{specfuncderiv}
\ee
\no This provides a sufficient, but not necessary, condition to identify a particle as unphysical. The positivity violation in the Schwinger function removes a particle immediately from the physical asymptotic state space, which is a weaker statement.

The condition \pref{spectralfunction} immediately implies that the ghost is unphysical. The condition \pref{spectralfunction} implies that the gluon in two dimensions is unphysical, since when a propagator vanishes at zero momentum, the spectral function cannot be positive. The condition from \pref{specfuncderiv} at $n=1$ implies that the gluon is not a physical particle in three dimensions, since the propagator is non-monotonous and therefore its first derivative changes sign \cite{Zwanziger:2010iz}. However, such a positivity violation may not be simple to detect. In case of the double-pole structure \pref{zerot:unstable} with the parameters of figure \ref{fig:schwinger}, the change of sign only occurs starting from its fifth derivative, which makes an actual computation rather complicated.

\subsection{Confinement scenarios}\label{quant:confinement}

Explicit violation of positivity, manifest in the Schwinger functions, shows that gluons are not part of the physical asymptotic spectrum. This makes any kind of asymptotic gluon states part of the unphysical Hilbert space. Assuming that the $S$ matrix only transforms physical states into physical states, this implies the absence of gluons from the asymptotic observable world. Given that this is a necessary assumption to prevent a gauge anomaly to occur \cite{Henneaux:1992ig}, this assumption appears reasonable.

The emergence of gluons as rather well-defined, quasi-massless states in high-energy scattering experiments is not at odds with this result. Indeed, the gluon propagator behavior at large energies is dominated by the perturbative part, or by the tree-level part in lower dimensions. Thus the corrections to the perturbative picture of a well-defined gluon state are sub-leading at sufficiently high energies, and only manifest themselves at energies where the effective interaction permits to resolve the propagation of the gluons over sizable distances. This duality of a highly non-perturbative, screened state at low energies and an almost perturbative state at high energies is again a consequence of asymptotic freedom. However, even at the shortest distances the gluon has always a non-perturbative dressing, which, though not quantitatively relevant, ensures its confinement as a necessary consequence of having only gauge-invariant observable states: There is no almost local definition of a gluon in contrast to the case of a photon \cite{Haag:1992hx}.

However, this is not showing that gluons are confined. After all, any unstable particle will also never reach a detector sufficiently far away, though its spectral function is positive. Thus, it is worthwhile to compare the results for the gluons to scenarios describing a confinement mechanism, which yield predictions for the correlation functions. Three such scenarios have been of particular relevance recently, the ones of Kugo and Ojima \cite{Kugo:1979gm,Kugo:1995km}, of Gribov and Zwanziger \cite{Gribov:1977wm,Zwanziger:1993dh,Zwanziger:2003cf}, and of Gribov and Stingl \cite{Gribov:1977wm,Habel:1989aq,Habel:1990tw,Stingl:1994nk}. There exist also other scenarios which give predictions on the analytic structure of the correlation functions \cite{Alkofer:2000wg,Alkofer:2006fu}, though those seem not fully compatible with the results obtained, and thus will not be detailed further here.

There are also many other proposed mechanisms, but most operate on different entities, like collective (topological) excitations \cite{Greensite:2003bk}. Relations to them will be detailed more in section \ref{sreltop}.

In all of the following it should be permanently kept in mind that there is no full systematic control of any of the methods employed to determine the correlation functions, and thus the following is an evidence-driven discussion.

\subsubsection{Perturbative BRST}\label{zerot:brst}

Before starting with these scenarios, it is worthwhile to reconsider how in perturbation theory the unphysical degrees of freedom, like ghosts, are removed from the spectrum, and therefore loosely speaking confined \cite{Bohm:2001yx,Kugo:1979gm}. This mechanism at the perturbative level is not sufficient to confine transversely polarized gluons, leaving the illusion that these are physical. Therefore, it cannot be sufficient beyond perturbation theory. The only exception is two dimensions, where the perturbative mechanism works, but this is due to the lack of transverse degrees of freedom of the gluon in two dimensions, rather than a qualitative change \cite{Maas:2007uv}.

In Landau gauge, the perturbative removal of ghosts and longitudinally and time-like polarized gluons is achieved by the quartet mechanism \cite{Kugo:1979gm}. This mechanism utilizes the BRST symmetry of the perturbatively quantized theory to structure the complete space in a physical and an unphysical subspace, with the ghosts and longitudinal and time-like gluons residing in the unphysical subspace \cite{Bohm:2001yx,Henneaux:1992ig}.

The corresponding BRST symmetry transformations are \cite{Bohm:2001yx}
\bea
\delta_\mathrm{BRST}A_\mu^a(x)&=&\delta\lambda D_\mu^{ab}c_b(x)\label{brst:trans1}\\
\delta_\mathrm{BRST}c^a(x)&=&-\delta\lambda\frac{1}{2}g_df^{abc}c_b(x)c_c(x)\\
\delta_\mathrm{BRST}\bar c^a(x)&=&\delta\lambda\frac{1}{\xi}\pdm A_\mu^a(x)\label{brst:trans3},
\eea
\noindent where for the Landau gauge the appropriate limit $\xi\to 0$ has to be taken and $\delta\lambda$ is an infinitesimal constant Grassmann parameter. These transformations leave the gauge-fixed Lagrangian \pref{quant:gfpi} invariant. This defines the BRST-operator $s$ as
\be
\delta_\mathrm{BRST}F=\delta\lambda sF,\nonumber
\ee
\noindent by its action on any field $F$.

It is possible to linearize the transformation by introducing an auxiliary field, the Nakanishi-Lautrup field $B^a$ \cite{Henneaux:1992ig}. This manifestly establishes the nil-potency of the BRST transformation
\be
\delta_\mathrm{BRST}^2=0.\label{nilpot}
\ee
\no Without this field, the nil-potency is manifest only on-shell.

To find the corresponding algebra, it is necessary to include another symmetry: Rescaling the ghost fields by a scale transformation $\exp(s)$ and its anti-field by $\exp(-s)$, leaving all other fields unchanged, is a symmetry of the Lagrangian \cite{Henneaux:1992ig,Kugo:1979gm}. This gives rise to the conserved ghost number $Q_{G}$, in analogy to the fermion number. As the ghosts are the only fields carrying them, it is necessary that all observable final states must have ghost number 0. Then, counting ghosts shows that the BRST charge $Q_\mathrm{BRST}$ has ghost number 1. Together, this establishes a closed algebra
\bea
\left\{Q_\mathrm{BRST},Q_\mathrm{BRST}\right\}&=&0\nonumber\\
\left[iQ_G,Q_\mathrm{BRST}\right]&=&Q_\mathrm{BRST}\nonumber\\
\left[iQ_G,iQ_G\right]&=&0\nonumber
\eea
\no as a global residual of the gauge symmetry in Landau gauge, besides the global color symmetry \cite{Kugo:1979gm} and the ghost-anti-ghost symmetry \cite{Alkofer:2000wg}. The important consequence of \pref{nilpot} is that the BRST charge is also nilpotent.

A well-defined nilpotent charge splits the state space directly into three disjoint parts \cite{Henneaux:1992ig}. The states which are not annihilated by the BRST-transformation form a subspace $Q_1$, carrying BRST-charge. By acting on these states, daughter states in a subspace $Q_2$ are generated which are annihilated by the BRST-charge. The last possibility are states which are also annihilated by the BRST-charge but are not generated from parent states. These form a subspace $Q_0$. Physical states must be gauge invariant and are therefore annihilated by $Q_\mathrm{BRST}$, which is essentially a gauge transformation with the ghost field acting as the transformation parameter. In addition, any states in $Q_2$ do not contribute to matrix elements, since they have zero norm. Therefore the physical subspace is
\be
H_\mathrm{phys}=\overline{\mathrm{Ker} Q_\mathrm{BRST}/\mathrm{Im} {Q_\mathrm{BRST}}}=\overline{Q_0}.\nonumber
\ee
\no It is this subspace in which the perturbatively physical transverse gauge bosons exist, while forward polarized gluons and anti-ghosts belong to $Q_1$ and backward polarized gluons and ghosts belong to $Q_2$. This can be seen directly using the Nakanishi-Lautrup formulation of the gauge-fixed Lagrangian \cite{Bohm:2001yx,Kugo:1979gm}. Due to the relation of $Q_2$ and $Q_1$, the unphysical degrees of freedom are connected by BRST transformations and are thus metric partners. They form a quartet under this charge, and are thus said to be confined by the quartet mechanism. Hence in perturbation theory the physical subspace $Q_0$ contains only transverse gluons, and perturbatively unphysical degrees of freedom are confined\footnote{In principle it is possible to have states in $Q_0$ with non-vanishing ghost number, which would render the theory ill-defined \cite{Kugo:1979gm}. This does not seem to be the case for Yang-Mills theories.}.

\subsubsection{Non-perturbative BRST and the Kugo-Ojima confinement scenario}\label{sec:kugo}

The basic idea of the Kugo-Ojima confinement mechanism is to extend the BRST construction such that not only transverse gluons, but also in general any colored state is moved to the unphysical subspace \cite{Kugo:1979gm,Kugo:1995km}. E.\ g., pairing transverse gluons with gluon-ghost bound states \cite{Kugo:1979gm,Alkofer:2011pe} in a quartet would make them unobservable.

The basic ingredients needed for this construction are threefold.

One is the existence of a mass gap in the physical spectrum. Since the known lowest-lying glueballs all have quite a significant non-zero mass, this appears to be fulfilled \cite{Bali:1993fb,Wellegehausen:2010ai,Teper:1998kw}. For QCD this would actually be known from experiments\footnote{Note that for QCD in the chiral limit with a massless pion this would have to be reconsidered.} \cite{pdg}. At the same time no mass gap is permitted in the complete state space, including the unphysical states. As has been shown in section \ref{quant:schwinger}, the absence of a mass gap can be read off the ghost Schwinger function for all cases investigated. This is therefore also fulfilled.

The second ingredient is that the global color charge after gauge-fixing is well defined and unbroken. If all other preconditions of the construction are fulfilled, this is immediately implied if the ghost propagator has a stronger than massless pole with negative residue, i.\ e., its dressing function must diverge to positive infinity \cite{Kugo:1979gm,Fischer:2008uz}. Evidently, this is satisfied in the scaling case. But to the finite-ghost case the Kugo-Ojima scenario in its original form cannot be applied. There have been several investigations whether it is possible to construct a modified Kugo-Ojima mechanism also in this case \cite{Dudal:2009xh,Kondo:2009ug,Dudal:2010fq,Kondo:2009gc,Boucaud:2009sd,Sorella:2010it}, but this has not been finally settled. Thus, the remainder of this section only applies to the scaling case.

The third, and probably most complicated, requirement is that there exists, in the sense of differential geometry, a globally well defined and unbroken BRST charge \cite{Fischer:2008uz}. Moreover, the field transformations induced by this charge must take the same form as the perturbative ones\footnote{Actually, when the original work of Kugo and Ojima was done \cite{Kugo:1979gm} the existence of Gribov copies was unknown, and the perturbative definition of the BRST charge was used. Since only the algebraic properties of the charge are relevant to the construction, this supplemental condition is sufficient to perform the construction even in presence of Gribov copies \cite{Fischer:2008uz}. If this condition is not fulfilled, the original Kugo-Ojima scenario fails.} \prefr{brst:trans1}{brst:trans3} \cite{Kugo:1979gm}.

The perturbative definition of the BRST charge is trivially not sufficient in the non-perturbative setting. The reason is that perturbatively a BRST transformation mediates between different gauge copies separated infinitesimally in a covariant gauge. It shrinks to the identity transformation in Landau gauge, since perturbatively only one representative of the gauge orbit fulfills the Landau gauge condition. Since there are Gribov copies, connected by large gauge transformations \cite{Gribov:1977wm,Sobreiro:2005ec}, a well-defined BRST transformation cannot simply be just the identity. This is also clear from the fact that massless transverse gluons are put into the physical state space by the perturbative construction, thus being in conflict with experiment. A more formal argument can be found in \cite{Dudal:2008sp,Capri:2010hb}.

The question is then whether there exists a non-perturbative extension of the BRST charge such that its algebra coincides with the perturbative one. In fact, there is one \cite{Neuberger:1986xz,vonSmekal:2007ns,vonSmekal:2008es,vonSmekal:2008ws}, which is based on first performing a lattice regularization of the theory, and then taking the continuum limit. The result is that one obtains a non-perturbative BRST charge in the fully quantized and renormalized theory, i.\ e., with all renormalization constants fixed. This Neuberger-von Smekal construction even leads to the same algebraic form of the non-perturbatively well defined BRST charge. Indeed, together with the assumption of a mass gap in the physical spectrum this is then sufficient to imply the Kugo-Ojima construction if an infrared divergent ghost dressing function, and thus a globally well defined color charge, exists, which appears to be the case in functional calculations of the scaling case \cite{Fischer:2008uz}.

The remaining problem is how to relate the argument to the present situation, in particular concerning lattice calculations. The reason is that the Neuberger-von Smekal construction is based on an average over all Gribov copies satisfying the Landau gauge condition \pref{quant:lg}. This implies an average over all Gribov regions, weighted by the signed Faddeev-Popov determinant \cite{Neuberger:1986xz,vonSmekal:2007ns,vonSmekal:2008es,vonSmekal:2008ws}, and is thus equivalent to a full non-perturbative evaluation of the path integral \pref{quant:gfpi} in the sense of Hirschfeld \cite{Hirschfeld:1978yq}. Since the DSEs and FRGEs are identical whether they are applied to a single Gribov region or to the whole of all Gribov regions \cite{Zwanziger:2003cf,Maas:2011ba}, the existence of the scaling solution implies that it is associated with the sum over Gribov regions by virtue of the Kugo-Ojima construction as outlined above \cite{Fischer:2008uz}.

The question is then how to relate this scaling case to lattice calculations presented here. If this were possible at all, there must exist some prescription to select Gribov copies inside the first Gribov region such that all correlation functions take the same value as in the Neuberger-von Smekal construction an thus be equivalent to the average over all Gribov copies. This may appear hopeless at first sight. However, there are a number of arguments which motivate that such a relation may indeed exist.

The first is the simple fact that the scaling solution has a positive semi-definite ghost dressing function, both in momentum and real space. In general, outside the first Gribov region this does not need to be the case. An explicit counter example is given by 1+1-dimensional Coulomb gauge, which can be solved exactly \cite{Reinhardt:2008ij}. In this case it was also shown that for a finite number of Gribov regions the scaling case is only realized if the restriction is made to the first Gribov region, when only a finite number of Gribov regions is considered.

The second argument is that scaling appears to be realized inside the first Gribov region in two dimensions, for reasons not yet fully understood \cite{Maas:2007uv,Cucchieri:2008fc,Cucchieri:2007rg,Dudal:2008xd,Zwanziger:2001kw,Lerche:2002ep,Dudal:2012td,Zwanziger:2012se}. Since the construction of the global BRST proceeds in two dimensions in the same way as in higher dimensions \cite{Neuberger:1986xz,vonSmekal:2007ns,vonSmekal:2008es,vonSmekal:2008ws}, in this case even the minimal Landau gauge seems to be equivalent to a sum over all Gribov copies.

The third argument is that the finite-volume version of scaling can be realized at least for small volumes inside the first Gribov region using the Landau-$B$ gauges \cite{Maas:2009se,Fischer:2007pf,Maas:2009ph}.

All of these arguments are not sufficient to ensure the possibility of scaling in the first Gribov region, nor can they explain how the appropriate cancellation of Gribov copies should occur such that this is possible. However, it is sufficient motivation to indeed investigate whether such a construction exists. In particular, it requires a realization of the idea of a simultaneous formulation on both the lattice and in the continuum, e.\ g.\ like \pref{quant:bgauge} or \pref{quant:bgauge2}. The Landau-$B$ gauges provide at least a possibility in principle how this could be realized, and an explicit construction description to verify this possibility: Trace the maximal attainable value of $B$ as a function of volume inside the first Gribov region. If it diverges in the infinite-volume (and continuum) limit, scaling could be realized. Check then whether the correlation functions exhibit a scaling behavior. Unfortunately, current lattice calculations can at best give evidence that such a behavior could be correct, and, even worse, could never falsify it. Furthermore, this does not exclude that a different selection procedure for Gribov copies could still yield scaling inside the first Gribov region, even if it were not realized in some Landau-$B$ gauges. An alternative would, e.\ g., be to select Gribov copies such that the infrared behavior of the running coupling is as constant as possible. Finally, arguments from stochastic quantization \cite{LlanesEstrada:2012my} also indicate a connection between Gribov copies and the type of solution. 

It thus remains at the current time a speculation, based on the arguments presented, that such a connection could exist. This working hypothesis is the central element to make contact between the results obtained for the correlation functions and the Kugo-Ojima construction, as well as the possible scaling Landau-$B$ gauge on the lattice and the DSE/FRG scaling case. This is a matter of ongoing research \cite{Maas:unpublished,Maas:2011ba}. However, the realization of the Kugo-Ojima scenario is not necessary for gluon confinement, as noted above, but it would be very desirable, as it would immediately extend the construction to all colored quantities, and would immediately permit an interpretation of Yang-Mills theory as a local quantum field theory beyond perturbation theory \cite{Lerche:2002ep}.

As pointed out above, a vital ingredient is the divergence of the ghost dressing function. Thus, in the finite-ghost cases the Kugo-Ojima construction is not possible, and in fact even a local off-shell version of the standard BRST symmetry is not implementable \cite{Dudal:2007cw,Dudal:2008sp,Tissier:2008nw,Lavrov:2011wb}. Still, proposals exist to provide even in the finite-ghost case a non-local symmetry which could possibly replace BRST symmetry and pave the way to a construction analogous to the Kugo-Ojima one in the scaling case \cite{Sorella:2010it,Sorella:2010fs,Baulieu:2009ha,Sorella:2009vt,Baulieu:2008fy,Kondo:2009qz,Capri:2010hb,Dudal:2010hj,Becchi:1998vv}.

\subsubsection{Gribov-Zwanziger confinement scenario}

A scenario \cite{Vandersickel:2012tg} based on a completely different point of view, but leading to rather similar predictions for the correlation functions, is the one of Gribov and Zwanziger \cite{Gribov:1977wm,Zwanziger:1992qr}.

The original basic idea started by restricting to the first Gribov region \cite{Gribov:1977wm}. In a semi-perturbative way it was then shown that this restriction yields an infrared divergent ghost dressing function with a negative residue. When it was later realized that there are also Gribov copies inside the first Gribov region it became clear that this required further refinement.

The next step was the assumption that it is permissible to implement a selection of Gribov copies inside the first Gribov region by replacing the $\theta$ function in \pref{quant:fgr} by a $\delta$-function \cite{Zwanziger:1992qr}. This reduces the region of field configuration space in the functional integral to the first Gribov horizon. This was motivated by the idea that most Gribov copies are located there, since most of the volume of the hypersurface of Landau gauge in the first Gribov region of the total infinite-dimensional field configuration space is located there \cite{Zwanziger:1993dh,Zwanziger:2003cf}. From the current point of view this corresponds just to a particular choice among Gribov copies, though without a formal justification up to now.

Assuming this procedure to be valid, it was possible to cast this condition with the help of auxiliary fields into a renormalizable and local, albeit complicated, Lagrangian \cite{Zwanziger:1992qr,Capri:2010hb}. Already at tree-level this Lagrangian exhibits an infrared divergent ghost dressing function and infrared vanishing gluon propagator, in qualitative agreement with the results in the scaling case \cite{Zwanziger:1992qr}. This is not altered by perturbative corrections \cite{Dudal:2005na,Gracey:2006dr,Gracey:2009mj,Vandersickel:2011zc} and remains true beyond perturbation theory \cite{Zwanziger:2010iz}.

This qualitative agreement even becomes quantitative beyond tree-level in the non-perturbative domain \cite{Huber:2009tx}. Implementing the same restriction of perfect tree-level cancellation like in the conventional scaling case in certain DSEs in the Gribov-Zwanziger framework, it was found that the correlation functions exhibit the same qualitative behavior. In particular, the ghost dressing function still diverges in the infrared. Depending on the precise behavior of the ghost-gluon vertex, even the same infrared critical exponents are found for the Faddeev-Popov ghost and gluon propagators.

Thus, so far this appears just to reproduce the scaling case. Given the results of lattice calculations in the minimal Landau gauge in three and four dimensions, it has been investigated whether it is possible to implement a finite-ghost behavior also in the Gribov-Zwanziger Lagrangian \cite{Dudal:2008xd,Dudal:2007cw,Dudal:2008sp,Dudal:2008rm,Gracey:2010cg,Vandersickel:2011zc,Chifarelli:2010dh}. This is indeed possible, by imposing condensation of certain combinations of the auxiliary fields needed for localization. An unambiguous determination of the values of the condensates is an intricate problem \cite{Dudal:2008sp,Dudal:2008rm,Dudal:2010tf,Dudal:2011gd}. In fact, a possibility would be that the value of these gauge-dependent condensates could be determined by the treatment of Gribov copies, e.\ g.\ corresponding to the $B$ parameter of Landau-$B$ finite-ghost gauges. This possibility has not yet been explored. Note however that dimension-two condensates, though being employed also beyond the Gribov-Zwanziger scenario \cite{Boucaud:2008gn}, may be problematic in an operator product expansion setting \cite{Greensite:1985vq}.

Irrespective of the actual values of the condensates, the generic result of perturbative calculations in these cases coincide qualitatively with the finite-ghost case \cite{Dudal:2008sp,Dudal:2008rm,Gracey:2010cg}, i.\ e., an infrared photon-like ghost and a screened gluon. In addition, in these cases also a Schwinger function showing the same qualitative behavior for the gluon propagator is found \cite{Dudal:2008sp,Dudal:2008rm}. Thus, the results are in good agreement with lattice results in three and four dimensions. Interestingly, at least the perturbative treatment in two dimensions breaks down \cite{Dudal:2008xd}, i.\ e.\  exactly for the case in which it is not yet clear whether a finite-ghost behavior can be achieved by a choice of Gribov copies in lattice calculations.

If indeed the imposed $\delta$-function on the field configuration space is equivalent to the $\theta$-function, it is also possible to provide a connection to the results in absolute Landau gauge \cite{Zwanziger:2003cf}. The basic assumption is that all correlation functions of a finite number of field variables take the value which is most probable, based on a maximization of the corresponding entropy. This is exactly analogous to what indeed happens in the minimal Landau gauge, thus indicating a possible formal connection. In this case, the results of the Gribov-Zwanziger Lagrangian for these correlation functions should coincide with the ones of minimal Landau gauge. If furthermore all copies become equivalent for such finite-order polynomials of field operators, like discussed in section \ref{sec:abslg}, the result would also agree with the absolute Landau gauge on the lattice.

\subsubsection{Gribov-Stingl scenario}

A more pragmatic scenario is given by the one of Gribov and Stingl \cite{Gribov:1977wm,Habel:1989aq,Habel:1990tw,Stingl:1994nk}. In this scenario, the existence of a double pole structure similar to \pref{zerot:unstable} is postulated. However, the generic propagator must have the form
\be
D_\mathrm{GS}=\frac{Z(p)}{p^4+2m^2\cos(2\phi)p^2+m^4}\nn
\ee
\no where $Z(p)$ satisfies the condition
\be
Z\left(ime^{\pm i\phi}\right)=0\nn,
\ee
\no i.\ e., the residue vanishes at the poles. Under these conditions the appearance of physical gluons can be avoided, and such solutions have been argued for using DSEs \cite{Habel:1989aq,Habel:1990tw,Stingl:1994nk}. However, in case of the full solutions of the functional equations, as presented in section \ref{quant:schwinger}, the corresponding analytic behavior is not compatible with the Gribov-Stingl type. Whether it is compatible with lattice calculations is yet unclear, since though it fits the gluon propagator rather well \cite{Cucchieri:2011ig}, it is not yet possible to discern whether the Schwinger function on the lattice shows the necessary oscillatory behavior.

\subsection{Relations to topological structures}\label{sreltop}

Many popular and powerful confinement scenarios are not based on correlation functions, but rather on topological, i.\ e.\ collective, gluon excitations \cite{Greensite:2003bk,Ripka:2003vv,Negele:2004hs,DiGiacomo:2008nt}. These scenarios in turn are usually based on one particular type of excitations, like vortices \cite{Greensite:2003bk}, monopoles \cite{Ripka:2003vv,DiGiacomo:2008nt}, merons \cite{Negele:2004hs,Zimmermann:2012zi}, and others \cite{Alexandru:2005bn,Horvath:2003yj}. From this non-exhaustive list it is evident that not even the underlying dimensionality of the relevant structure is yet decided. Still, there has been quite a number of investigations to find relations between the various excitations, yielding e.\ g.\ deep connections between vortices and monopoles \cite{Greensite:2003bk,deForcrand:2000pg,Boyko:2006ic,Reinhardt:2001kf,Caudy:2007sf}, even at finite temperature \cite{Chernodub:2011pr}.

Given that these scenarios are quite successful in describing properties of non-local observables like the string tension \cite{Greensite:2003bk,Ripka:2003vv} it is of importance to understand how correlation functions contain the same information\footnote{Of course, there is no guarantee that any kind of truncation in functional equations will preserve any of this. Given that the modern truncation schemes presented here maintain not only the shape of the correlation functions but also effects which are usually attributed to topological excitations, like chiral symmetry breaking \cite{Fischer:2006ub,Fischer:2003rp,Fischer:2003zc}, possibly the formation of a string \cite{Alkofer:2008tt,Fister:2010yw}, or the $\eta'$ mass \cite{Alkofer:2008et}, it is likely that at least a considerable amount of the topological information is captured.}. That this information is contained follows directly from the fact that the exact correlation functions are sufficient to reconstruct the complete generating functional using the equality \pref{quant:recon} \cite{Bohm:2001yx}, and in turn the correlation functions are fully described by the functional equations by virtue of the relation \pref{quant:genfcf} \cite{Rivers:1987hi}. In particular cases, this has even been demonstrated explicitly for solvable models \cite{Alkofer:2000wg,Radozycki:1998xs}.

Besides such formal arguments, there is an explicit possibility to investigate this relation. This is done by obtaining the correlation functions in a topological background field and comparing the results to the ones obtained in the full Yang-Mills vacuum. This has been done using analytical and modeling methods, which indicate that the characteristic features of correlation functions are influenced or even dominated by topological excitations. In particular, the infrared behavior of the ghost propagator has been linked analytically via the Faddeev-Popov operator to topological excitations \cite{Maas:2005qt,Maas:2006ss}, as have been the properties of the gluon propagator \cite{Boucaud:2003xi}.

Complementary to analytical methods, lattice simulations provide a possibility to access the topological content of a field configuration by smearing and projecting, at least in an approximate way \cite{Bruckmann:2006wf,DeGrand:1997ss,Greensite:2003bk}. The so treated field configurations become self-dual, which classifies them as topological \cite{Bohm:2001yx,Felsager:1981iy}. Evaluating the correlation functions in such smeared or projected configurations provides then a possibility to assess which generic features are dominated by topological contributions.

One possibility for such a smearing prescription is the so-called APE smearing \cite{DeGrand:1997ss}. APE smearing on the lattice is implemented by iteratively replacing links by a weighted average over neighboring links
\bea
U_\mu(x)&\to& \alpha U_\mu(x)+\frac{1-\alpha}{2(d-1)}\times\sum_{\nu\neq\mu}\left(U_\nu(x+e_\mu)U^+_\mu(x+e_\nu)U^+_\nu(x)\right.\nn\\
&&\left.\left.+U_\nu^+(x+e_\mu-e_\nu)U_\mu^+(x-e_\nu)U_\nu(x-e_\nu)\right)\right|_\mathrm{projected\;to\;the\;group}\nn,
\eea
\no where ``projected to the group'' implies that the non-group element $U_\mu'$ found after addition is replaced by the group element $U_\mu''$ closest to the result, where the distance is given by $\tr U_\mu'' U_\mu'$, with no summation implied. A typical value for $\alpha$, which is also employed here, is 0.55. Choosing $\alpha$ to be zero is referred to as cooling.

\begin{figure}
\includegraphics[width=0.5\textwidth]{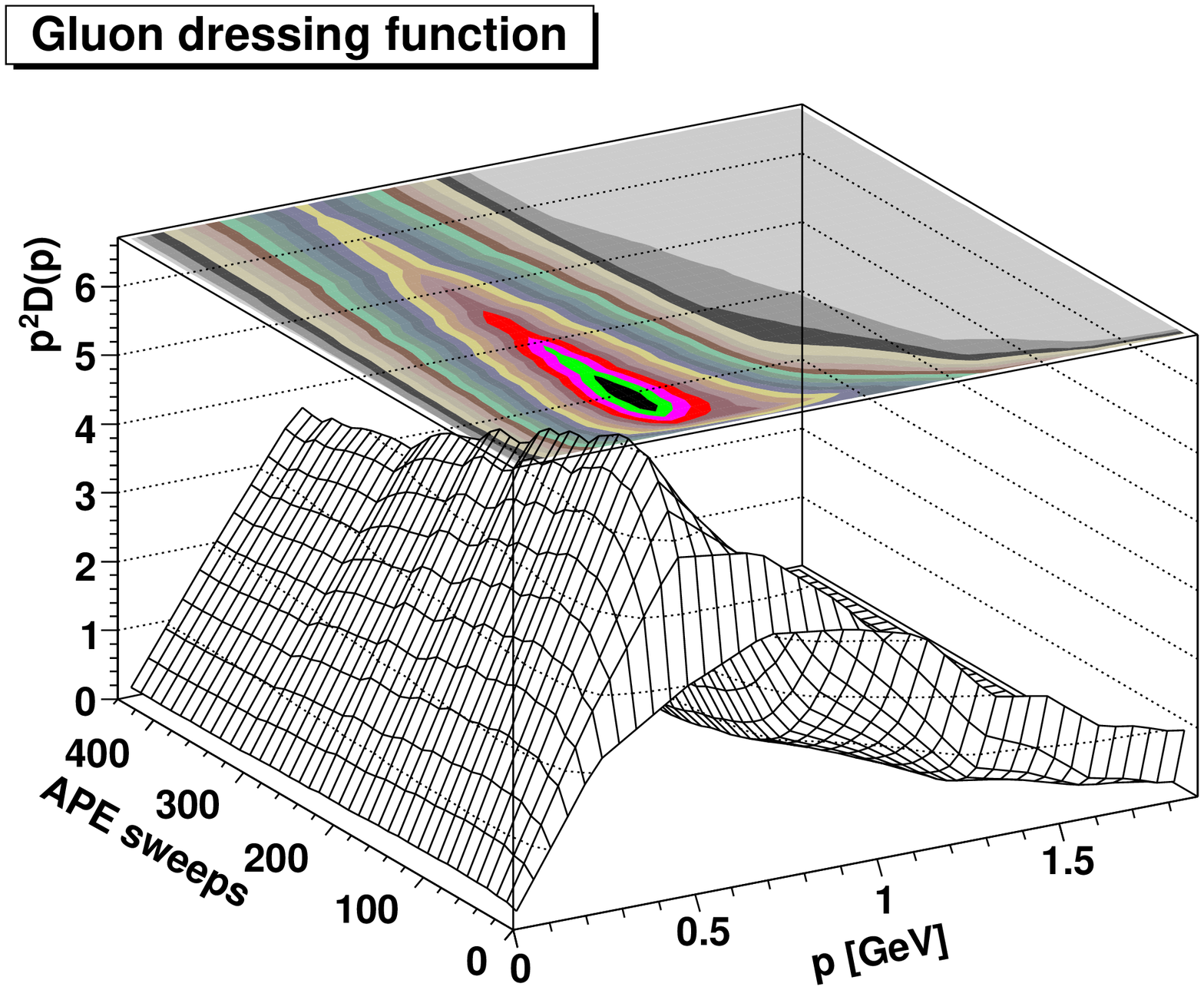}\includegraphics[width=0.5\textwidth]{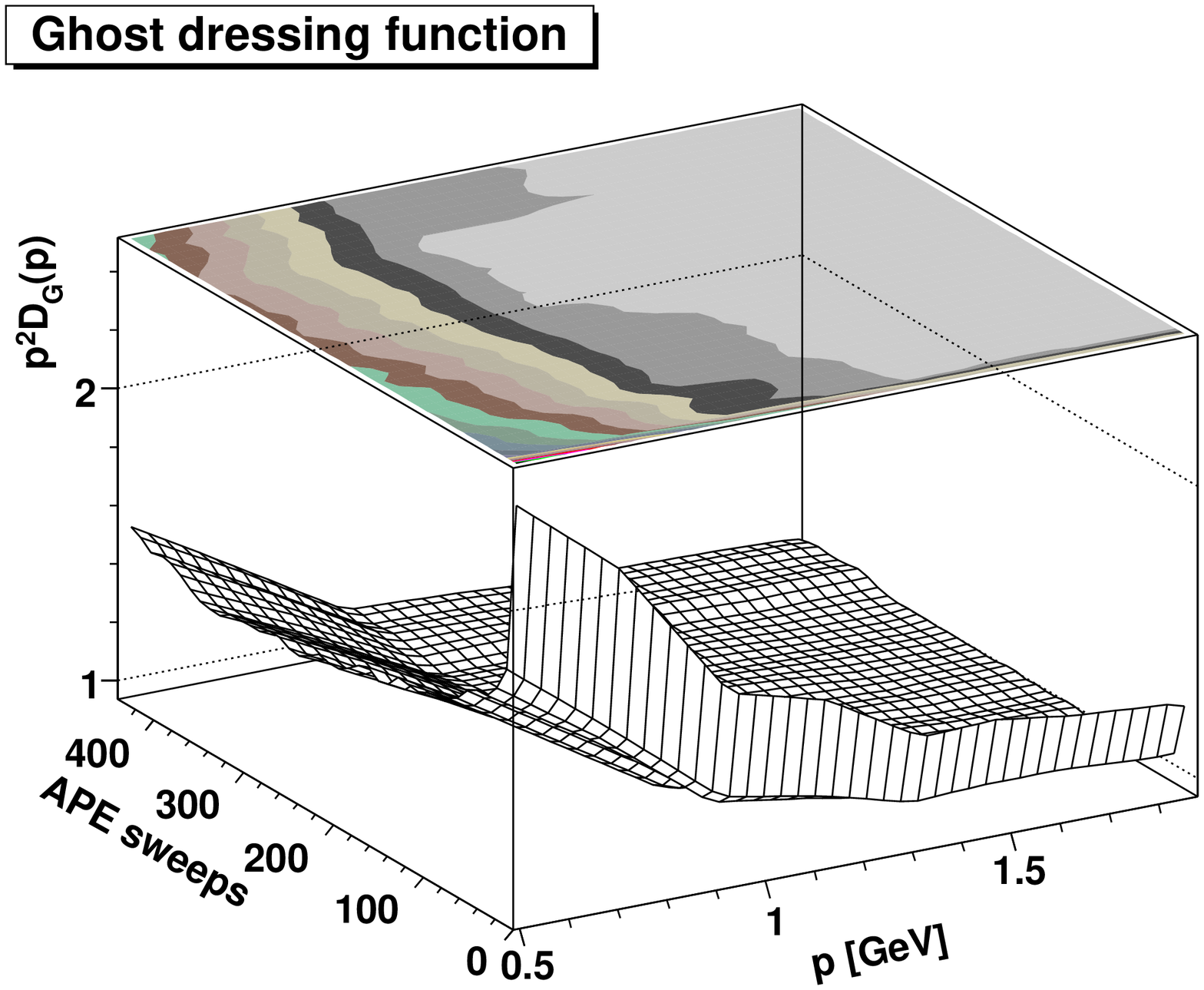}
\caption{\label{fig:top}The gluon dressing function (left panel) and ghost dressing function (right panel) evolving under APE smearing \cite{Maas:2008uz,Maas:unpublished}. Already at about 100 smearing steps, the configurations are essential self-dual \cite{Bruckmann:2006wf}, and thus approximately pure topological configurations. Results are for a $V=(2.5$ fm$)^4$ lattice at $a=0.21$ fm and su(2) in the minimal Landau gauge.}
\end{figure}

Results for such a calculation are shown in figure \ref{fig:top} in minimal Landau gauge \cite{Maas:2008uz}. The absence of the ultraviolet tail of the correlation functions after a few smearing steps results from the removal of the ultraviolet fluctuations by the smearing process. Furthermore, the behavior at or below $\Lambda_\mathrm{YM}$ is qualitatively almost unchanged, showing that indeed the low-momentum behavior of correlation functions does not only contain topological information, but is rather dominated by it\footnote{This also supports that the truncations presented here captures a significant part of the topological information.}. Similar results have been obtained using projection methods, in particular for center vortices \cite{Gattnar:2004bf,Langfeld:2001cz,Quandt:2010yq}, and also in Coulomb gauge \cite{Greensite:2004ur,Quandt:2010yq}.

This indicates that the collective excitations are indeed reflected in the properties of the correlation functions at low momenta. Even though more exact analytical mappings have yet to be developed fully \cite{Maas:2005qt,Maas:2006ss,Boucaud:2003xi,Maas:unpublished}, this makes clear that topology-based and correlation-function-based approaches to the non-perturbative properties of Yang-Mills theory are just two (equivalent) facets of the same physics \cite{Alkofer:2006fu}. Therefore, also the question whether a description in terms of elementary particles or of collective excitations is more appropriate is rather a question of practicality for a given question than a conceptual difference.

\begin{figure}
\includegraphics[width=0.5\textwidth]{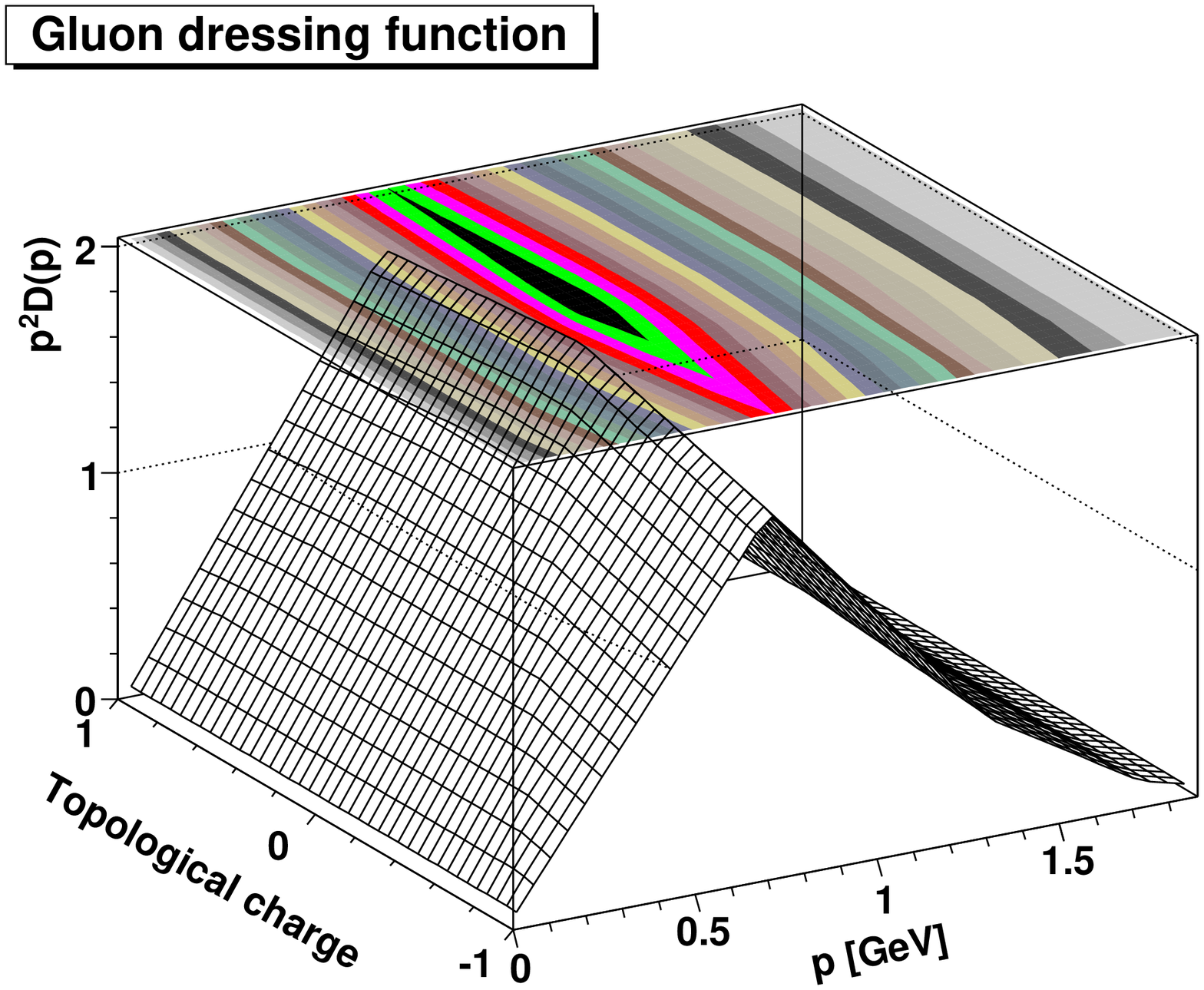}\includegraphics[width=0.5\textwidth]{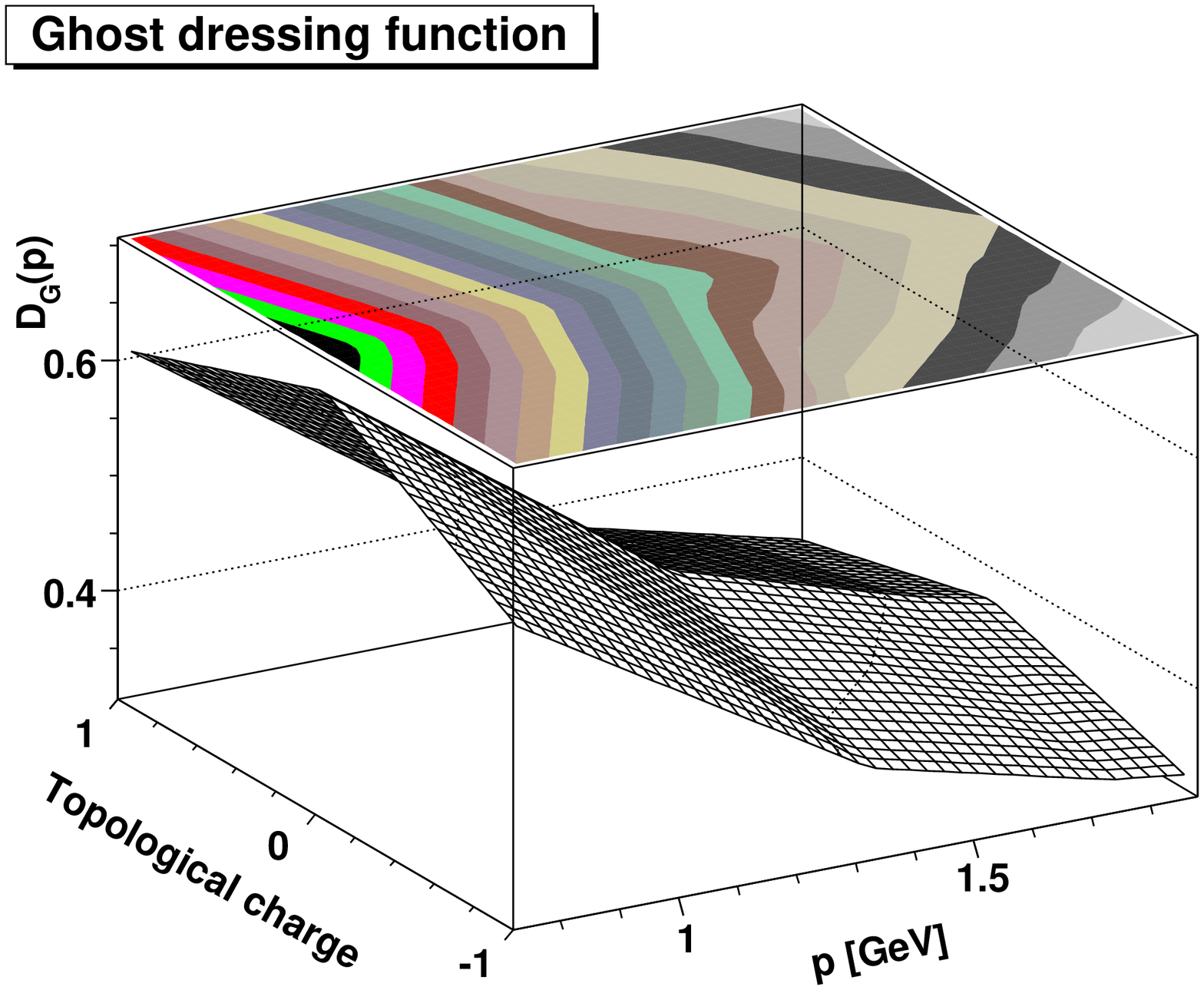}
\caption{\label{fig:top-q}The gluon dressing function (left panel) and ghost dressing function (right panel) as a function of topological charge after ten APE sweeps \cite{Maas:unpublished,Maas:2008uz}. The topological charge has been measured as the volume integral of $\tilde{F}_\mn^a F_\mn^a$ \cite{DeGrand:2006zz}. Results are for a $V=(1.7$ fm$)^4$ lattice at $a=0.21$ fm and su(2) in minimal Landau gauge.}
\end{figure}

As a side-remark, it is also interesting to investigate whether the correlation functions depend on the net topological charge. This is exemplified in figure \ref{fig:top-q}, where it is seen that the dependence is probably only quantitative. This is of considerable practical importance, given that topological properties have a much longer relaxation time in lattice simulations than most other quantities \cite{DelDebbio:2002xa}.

\subsection{Beyond Yang-Mills theory}\label{sbym}

Of course, the approach discussed here is not limited to gauge bosons in Yang-Mills theory. Two rather important extensions are the inclusion of matter fields and the determination of the properties of composite objects, in particular bound states. However, these topics are beyond the scope of this review, and will only be briefly introduced here.

\subsubsection{Matter fields}

Adding matter fields to Yang-Mills theory is rather straight-forward \cite{Bohm:2001yx}. Since Landau gauge, in contrast to e.\ g.\ 't Hooft gauge outside the Landau gauge limit \cite{Bohm:2001yx,Maas:2010nc}, does not involve the matter fields in any active way, all that has been said so far on gauge-fixing remains valid. This does not imply that the process remains completely unchanged. E.\ g., the corridor of the permitted $B$ values may change in the presence of matter fields. An indirect example of this has already been found in the Higgs phase\footnote{Strictly speaking, the Higgs phase is continuously connected to the confinement phase when a (lattice) cut-off is imposed \cite{Caudy:2007sf,Fradkin:1978dv}, but the notation will be kept for the sake of simplicity.} of a fundamental-Higgs-Yang-Mills system in both Coulomb \cite{Caudy:2007sf,Greensite:2004ur} and Landau \cite{Maas:2010nc} gauge. In this case the average properties of the Faddeev-Popov operator change qualitatively compared to the confinement phase. Though this has not yet been investigated in detail, this makes it likely that also the corresponding $B$-corridor is different in both phases \cite{Maas:2010nc}.

The same goes for the viability of the truncations in functional calculations \cite{Fischer:2009tn,Macher:2010ad}, not to mention technical problems in lattice calculations when including fermions \cite{Gattringer:2010zz}. However, these are technical rather than conceptual problems.

Thus, there have been plenty investigations including matter fields. The emphasis is of course on fermions in the fundamental representation, as these are necessary for an investigation of QCD \cite{Fischer:2006ub,Alkofer:2000wg,Alkofer:2008tt,Fischer:2003rp,Blank:2010pa,Fischer:2008wy,Fischer:2005en,Zayakin:2009dy,Bowman:2008qd,Kamleh:2007ud,Parappilly:2005ei,Costa:2010pp,Schrock:2011qp}\footnote{This only includes references which to some extent take into account the Yang-Mills correlation functions and are in Landau gauge and only focusing on the properties of the quarks. There is a wealth of literature on calculations using various ans\"atze for these correlation functions, see \cite{Alkofer:2000wg,Roberts:2007jh,Roberts:2000aa,Bashir:2012cp} for introductory reviews.}. Less often the properties of adjoint and fundamental scalars have been investigated \cite{Maas:2010nc,Fister:2010yw,Fischer:2009tn,Macher:2010ad,Cucchieri:2001tw,Maas:2004se,Karsch:1996aw}, though they are much more amendable to lattice calculations \cite{Montvay:1994cy}. Investigations of fermions in different representations than the fundamental, however, have received significant attention lately \cite{Sannino:2009za}, as these are candidates for theories exhibiting an approximate conformal behavior in the infrared, and thus are possibly relevant for, e.\ g., walking technicolor extensions of the standard model. However, determinations of the corresponding correlation functions are just beginning \cite{Maas:2011jf,Aguilar:2010ad}.

\begin{figure}
\includegraphics[width=0.5\textwidth]{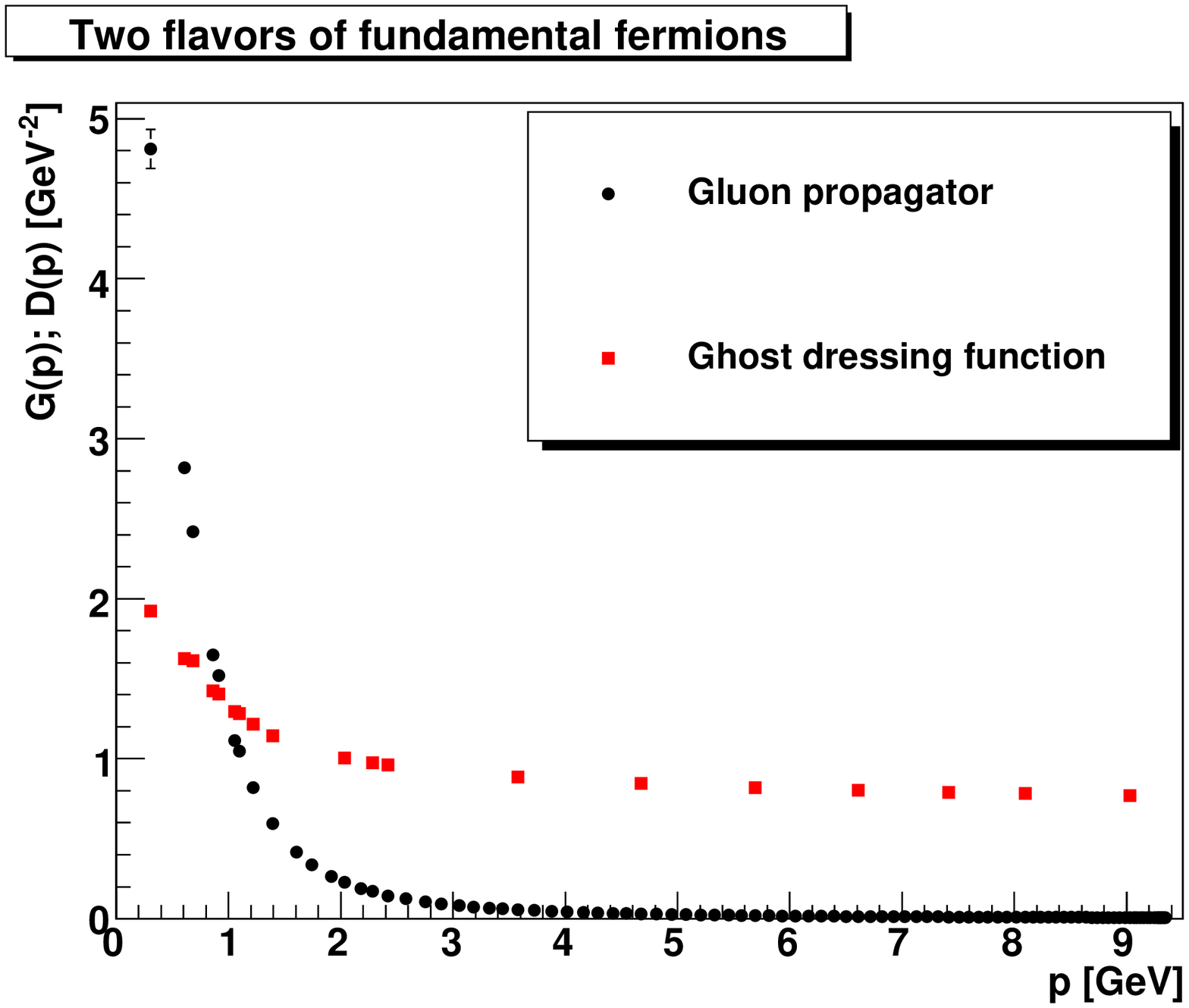}\includegraphics[width=0.5\textwidth]{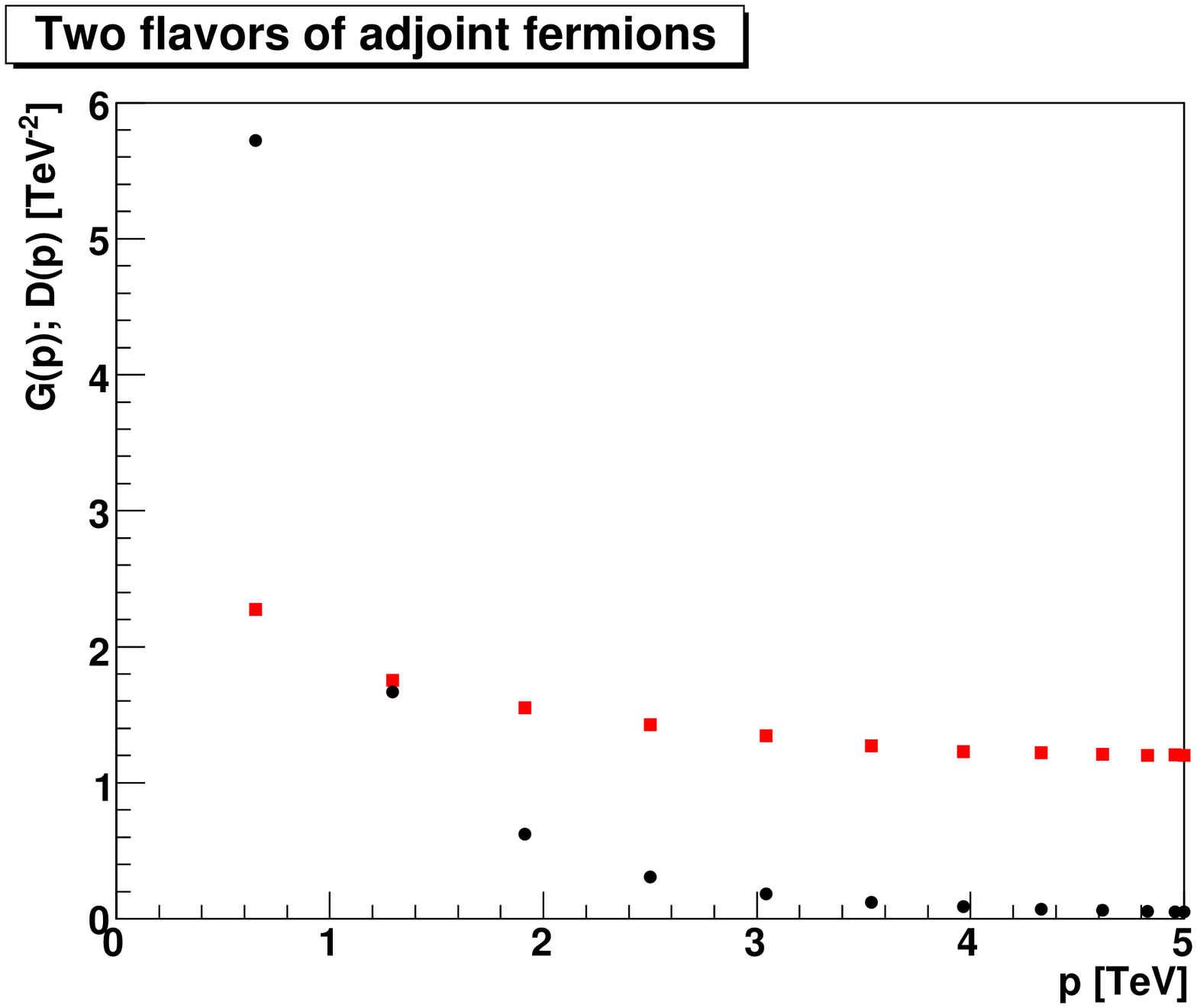}\\
\includegraphics[width=0.5\textwidth]{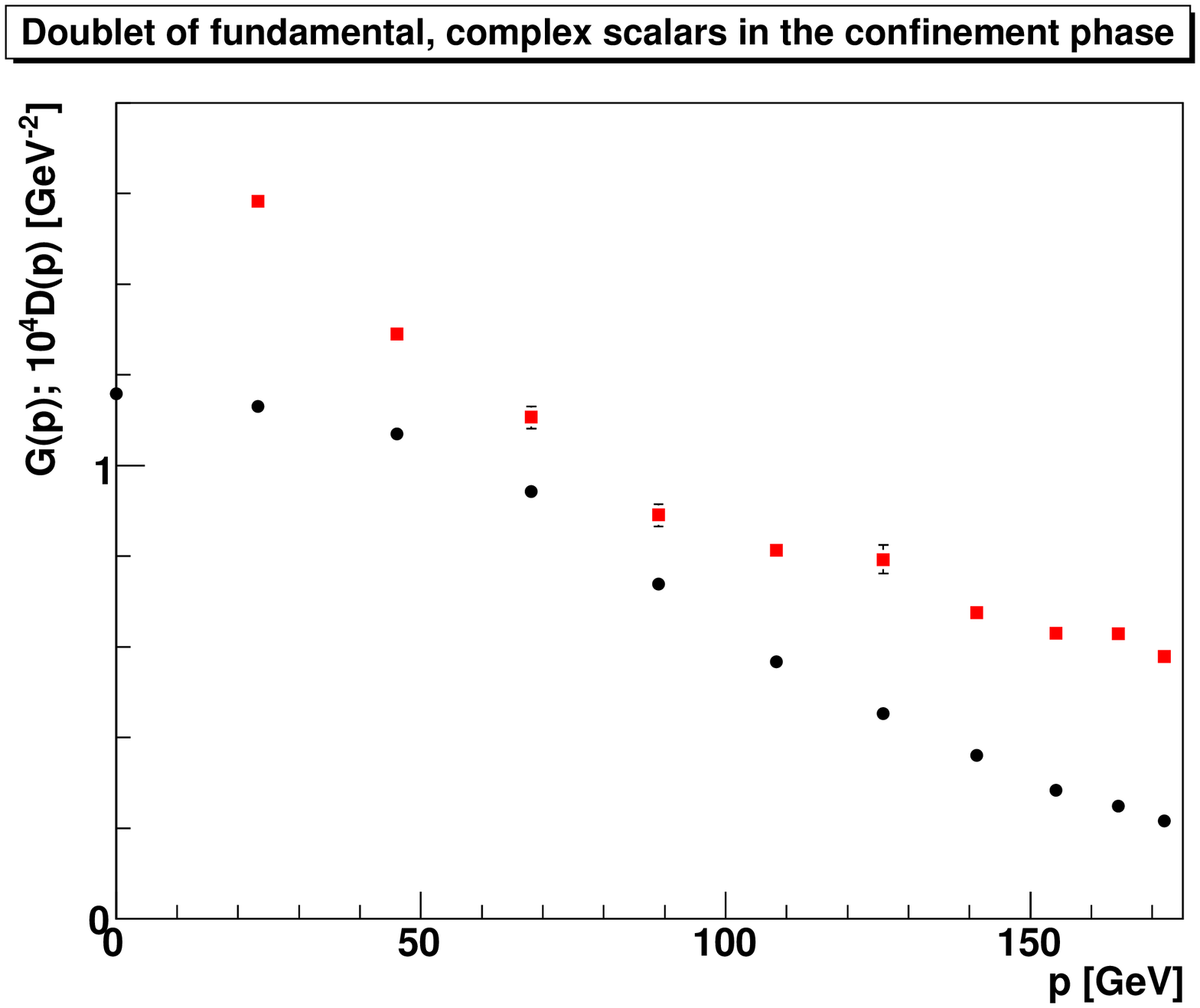}\includegraphics[width=0.5\textwidth]{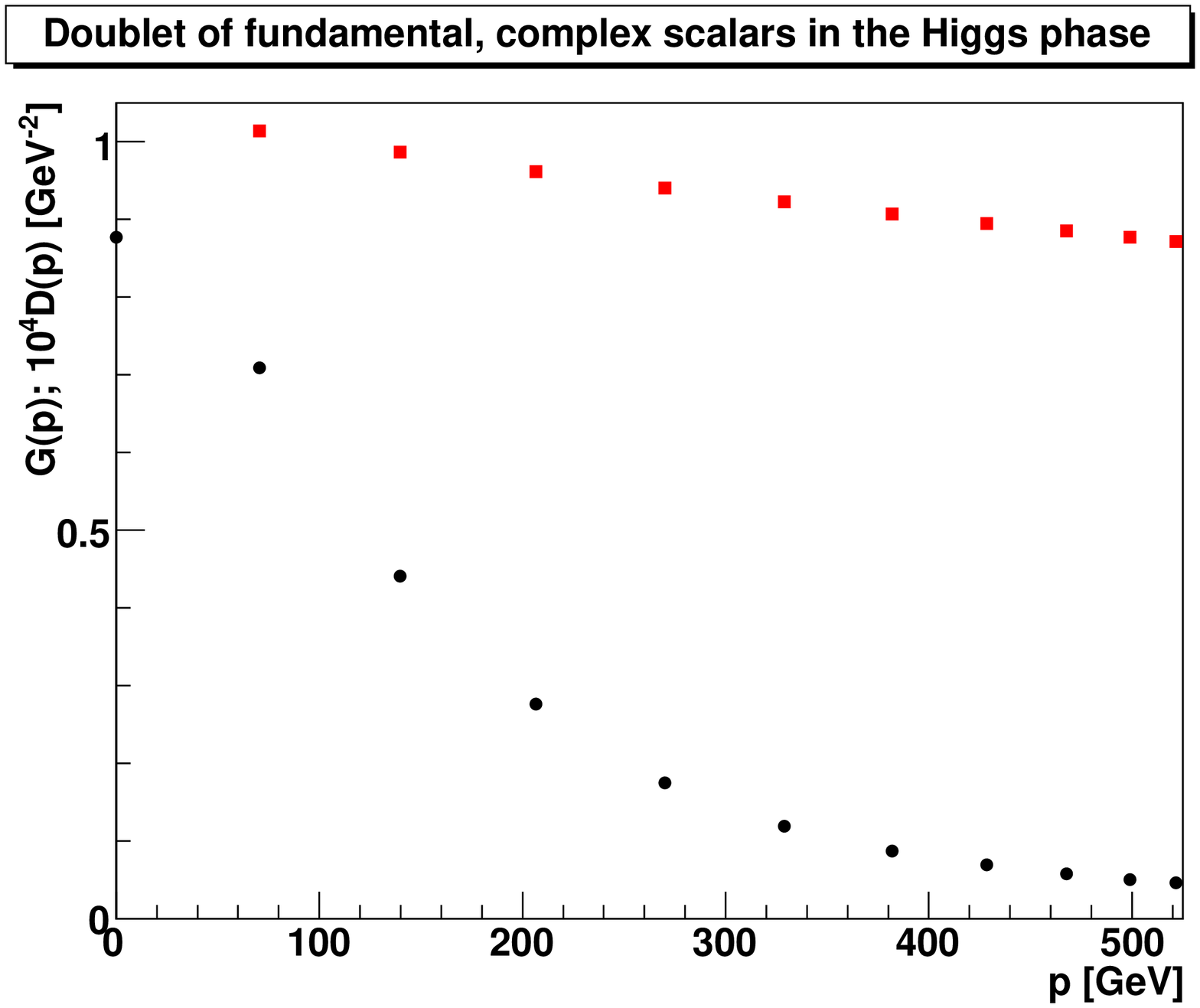}
\caption{\label{fig:other}Examples of the gluon propagator and ghost dressing function for theories with different matter content. Top-left panel: Two flavors of fundamental fermions and gauge group SU(3) \cite{Sternbeck:2006cg,Sternbeck:2006rd,Ilgenfritz:2006he} on a $24^3\times 48$ lattice and Wilson clover fermions with $\kappa=0.13575$. The scale is fixed by the intermediate-distance string tension. Top-right panel: Two flavors of adjoint fermions and gauge group SU(2) \cite{Maas:2011jf,DelDebbio:2008zf} on an $12^3\times 24$ lattice with Wilson fermions and $-am_0=0.95$. Scale is fixed by a 2 TeV techniglueball \cite{Maas:2011jf}. Bottom-left panel: The confinement phase for a doublet of complex fundamental scalars for SU(2) in the confinement phase \cite{Maas:2010nc} on a $24^4$ lattice with $\kappa=0.25$ and $\lambda=0.5$. Bottom-right panel: The Higgs phase for a doublet of complex fundamental scalars for SU(2) in the Higgs phase \cite{Maas:2010nc} on a $24^4$ lattice with $\kappa=0.32$ and $\lambda=1$. In both cases the scale is fixed by a 250 GeV Higgsonium \cite{Maas:2010nc}. All results are in minimal Landau gauge. Results using functional methods can be found, e.\ g., in \cite{Alkofer:2008tt,Fischer:2003rp,Fister:2010yw,Fischer:2009tn}.}
\end{figure}

A sample of results for propagators for various cases are assembled in figure \ref{fig:other}. Going into details or even listing a complete set of references just for Landau gauge is far beyond the scope of this review. It suffices to say that one can investigate from them a phletora of physical phenomena like, e.\ g., the dynamical generation of mass \cite{Fischer:2006ub,Braun:2009gm,Fischer:2007ze,Zayakin:2009dy,Bowman:2008qd,Kamleh:2007ud,Parappilly:2005ei,Bowman:2005vx,Sternbeck:2006cg,Sternbeck:2006rd,Ilgenfritz:2006he,Aguilar:2010ad}, chiral symmetry breaking and restoration at finite temperature \cite{Braun:2008pi,Braun:2009gm,Fischer:2009wc}, the deconfinement transition \cite{Fischer:2010fx,Mueller:2010ah,Fischer:2009wc,Karsch:2007wc,Karsch:2009tp}, center observables like the Polyakov loop \cite{Braun:2007bx,Fischer:2010fx}, conceptual access to the string tension \cite{Alkofer:2008tt,Fister:2010yw,Schwenzer:2008vt,Zayakin:2009jz}, confinement of matter fields \cite{Braun:2007bx,Alkofer:2008tt,Fister:2010yw,Schwenzer:2008vt,Fischer:2008sp}, color-superconducting phases at large densities \cite{Nickel:2008ef,Marhauser:2006hy,Nickel:2006kc,Nickel:2006vf,Hands:2006ve}, the Higgs effect \cite{Maas:2010nc}, and infrared conformality \cite{Maas:2011jf,Braun:2009ns,Sannino:2009za,Braun:2010qs}, only to mention some. However, already from the brief presentation in figure \ref{fig:other}, it can be deduced that the presence of matter fields can be both, almost irrelevant and of qualitative impact for the properties of the Yang-Mills sector.

This rapidly growing field shows the versatility of correlation-functions-based methods. The wealth of possible physics results has yet only been tapped to a small extent, and will certainly be the focus of many investigations in the time to come.

\subsubsection{Composite objects}

The natural quantities to be determined from the point of view of experiments, beside the ultraviolet properties of elementary particles, are the properties of bound states. In particular, quantities like masses and decay constants, but also widths, have therefore received a lot of attention \cite{Alkofer:2000wg,Roberts:2007jh,Roberts:2000aa,Fischer:2006ub}.

In principle, these quantities can be obtained from the corresponding correlation functions, e.\ g.\ a glueball mass from the 4-point correlation function \cite{Hauck:1998ir,Dudal:2009zh,DeGrand:2006zz}. Since the masses of colorless bound states are gauge-invariant and renormalization-group invariant, they can even be obtained without explicit gauge fixing \cite{DeGrand:2006zz,Gattringer:2010zz,Montvay:1994cy,Seiler:1982pw}. In particular for lattice calculations use has been made of this possibility. Despite the numerical costs to simulate quarks with physical masses it was even possible to obtain results essential at the physical parameters of QCD \cite{Aoki:2009ix,Durr:2008zz,Hagler:2009ni}. Using this approach, information on the gauge-dependent internal structure of composite objects is lost, and at best it is possible to expand the states in terms of gauge-invariant operators \cite{Glozman:2009cp,Glozman:2009rn}. In particular, relative momenta between the compounds are not accessible without fixing a gauge.

Similarly, it is also possible to use functional methods to determine the gauge-invariant masses of colorless bound states. Instead of using the DSEs or FRGs for the corresponding correlation functions directly, it is possible to employ the Bethe-Salpeter \cite{Fischer:2005en,Fischer:2008wy,Krassnigg:2009zh,Blank:2010pa} and Faddeev \cite{Eichmann:2009zx,Eichmann:2009qa} equations to obtain the mass spectrum of mesons and baryons, respectively. This has been widely discussed in the literature \cite{Alkofer:2000wg,Roberts:2007jh,Roberts:2000aa,Fischer:2006ub}.

The lattice results suffer primarily from statistical and systematic uncertainties, the latter mostly due to the extrapolation to the continuum and infinite volume, as well as to physical quark masses. The functional calculations suffer in contrast from the errors introduced by the truncations. Thus, comparing both approaches, the uncertainties have completely different origins. Thus, once more, the comparison of both methods can give mutual support \cite{Luecker:2009bs,Fischer:2005nf}. In fact, the results of both methods are in acceptable agreement with each other and with the experimental values, thus giving even more confidence to have access to a theory like QCD even at low momenta and beyond perturbation theory than a single method alone could.

It should be noted that the results using a hierarchy of correlation functions, as with Bethe-Salpeter equations, are rather insensitive to the details of the infrared behavior of the gauge-dependent correlation functions \cite{Blank:2010pa,Luecker:2009bs,Fischer:2009gk,Fischer:2009jm,Costa:2010pp}. This is as it should be, given that this infrared behavior is essentially dominated by the gauge-fixing procedure. Explicit calculations show that, when taking the gauge-dependence of all involved correlation functions into account, the results for the finite-ghost case and the scaling case, and thus possibly for different non-perturbative gauges, agree within the errors, and with experiment \cite{Blank:2010pa}. This also provides the possibility to understand how gauge-invariance is recovered for physical states, as has been done previously for QED \cite{Kizilersu:2009kg}.

This concludes the construction of the framework at zero temperature, which gives an impression how to start with the elementary degrees of freedom and obtain the correlation functions. A discussion more detailed than the short presentation here how to finally arrive with this approach at gauge-invariant, observable quantities can be found in various reviews and textbooks, e.\ g.\ \cite{Alkofer:2000wg,Roberts:2007jh,Roberts:2000aa,Fischer:2006ub,DeGrand:2006zz,Gattringer:2010zz,Montvay:1994cy}.

\section{Gluons at finite temperature}\label{sfinitet}

In the previous chapters the properties of gluons have been determined using correlation functions. It has long been discussed whether these properties fundamentally change when the system is heated up \cite{Kapusta:2006pm}. In the case of Yang-Mills theory it is rather well established that a phase transition occurs at a non-zero temperature $T_c$ of the order of $\Lambda_\mathrm{YM}$, for any gauge group investigated so far \cite{Kapusta:2006pm,Karsch:2001cy}, including SU($N<9$) \cite{Lucini:2005vg}, Sp(2) and Sp(3) \cite{Holland:2003kg,Braun:2010cy}, E$_7$ \cite{Braun:2010cy}, and G$_2$ \cite{Holland:2003jy,Pepe:2006er,Cossu:2007dk}. Only the order and other details of the phase transition turn out to be dependent on the particular gauge group.

One of the central ideas about this transition was deconfinement \cite{Kapusta:2006pm,Cabibbo:1975ig}. Based on the argument that the running coupling evaluated at the temperature becomes small for increasing temperature it was assumed that confinement as a strong coupling phenomenon should cease. At first sight, this is supported by the fact that thermodynamical quantities, like the pressure, logarithmically approach a Stefan-Boltzmann behavior with increasing temperature \cite{Karsch:2001cy,Kajantie:2002wa}.

It is of course clear that this can only be superficially correct. On the one hand, Haag's theorem forbids the absence of non-perturbative interactions \cite{Haag:1992hx}. On the other hand, this would require a gauge-invariant formulation of asymptotic, almost local colored states at finite temperature. Though the existence of such a formulation has not yet been excluded, the impossibility of such a bleaching at zero temperature for non-Abelian gauge theories  \cite{Haag:1992hx} makes this rather unlikely. Furthermore, the impossibility of stable asymptotic states of elementary particles at finite temperature due to the Narnhofer-Thirring theorem \cite{Narnhofer:1983hp} alleviates this question anyway \cite{Henning:1995sm}.

After introducing in the following the formalism to describe gluons using correlation functions at finite temperature, it will be shown that indeed gluons are not deconfined. In fact, they do not develop an asymptotic physical pole, as will be read off from their Schwinger functions once more. However, it will be necessary to distinguish the possibilities of transversely and longitudinally polarized gluons with respect to the heat bath four-momentum. It will furthermore be argued that persisting gluon confinement is not in contradiction to a Stefan-Boltzmann-like behavior of thermodynamical bulk quantities. These will be found to be dominated by hard processes, i.\ e., at energies of the order of the temperature. Since this becomes a hard scale with increasing temperature, indeed their leading behavior will be generated by leading-order perturbative contributions. However, the Linde problem \cite{Linde:1980ts,Kapusta:2006pm} of such a perturbative treatment will be cured by the non-perturbative effects. In addition, the formal infinite-temperature limit will be discussed, showing the validity of the arguments at arbitrarily high temperatures.

Finally, this will be concluded by showing that already from the two-point functions gauge-invariant quantities like the critical temperature and order parameters can possibly be extracted.

It should be noted that the approach has also been applied to the case of finite density, after coupling to fundamental fermions. Since this goes beyond Yang-Mills theory, this will not be described here. Discussions of this topic can be found, e.\ g., in \cite{Nickel:2008ef,Marhauser:2006hy,Nickel:2006kc,Nickel:2006vf,Hands:2006ve,Roberts:2000aa,Fischer:2011mz}.

\subsection{Finite temperature}

In the following only a thermostatic setting, i.\ e.\ equilibrium, will be discussed \cite{Kapusta:2006pm}. The extension of the framework presented here to non-equilibrium states \cite{Calzetta:2008ne} is not yet completely constructed, especially beyond perturbation theory, but progress in this direction is made \cite{Berges:2004yj}.

\subsubsection{Formulation}

An adequate setting for equilibrium calculations is the Matsubara formalism \cite{Kapusta:2006pm}. Without going into detail, it essentially requires to change to Euclidean space\footnote{In a sense, the finite temperature results are therefore much more closer to the real world than the results presented at zero temperature where analytic continuation or the use of the reconstruction theorem are necessary to make contact with Minkowski space.} and then compactify the time direction with size $\beta_T=1/T$. Therefore, equilibrium is manifest as no genuine time dependency remains.

A further consequence of this distinction of the time direction is that energy $p_0^2$ and three momentum $\vec p^2$ become independent variables. Though this may at first seem like a breaking of Lorentz invariance, this is not the case \cite{Weldon:1982aq}. In this standard setting, merely a choice for the four velocity of the heat bath is made such that the latter is at rest. By explicitly including this four velocity in the calculations, Lorentz symmetry can be made manifest once more. Despite the gain in symmetry, the practical calculations become rather cumbersome, and it is therefore simpler to have only hidden Lorentz symmetry. This will be done here as well. However, it should be kept in mind that this is just a choice of frame, without importance for the physics. A further consequence is that at four-momenta with $p^2\gg T^2$ temperature becomes negligible, and Lorentz symmetry, or in Euclidean space-time O(4) symmetry, becomes restored: If all momenta are large compared to the temperature the (perturbative) four-dimensional behavior of correlation functions will be recovered. That is essentially a consequence of the Appelquist-Carrazone theorem \cite{Appelquist:1974tg,Collins:1984xc}, and applies within the same domain of validity.

Furthermore, the reference direction of the heat bath makes it possible to distinguish between a polarization in the direction of the heat bath or transverse with respect to it. As a consequence it is necessary to decompose the gluon propagator as \cite{Kapusta:2006pm}
\be
D_\mn^{ab}(p_0,\vec p)= P_\mn^T(p_0,\vec p)D_{T}^{ab}(p_0,\vec p^2)+ P_\mn^L(p_0,\vec p) D_L^{ab}(p_0,\vec p^2)\nn,
\ee 
\no i.\ e., there are two independent dressing functions. The tensor structures are given by
\bea
P_\mn^T(p_0,\vec p) &=& (1-\delta_{\mu0})(1-\delta_{\nu0})\left(\delta_\mn-\frac{p_\mu p_\nu}{\vec p^2}\right)\nn\\
P_\mn^L(p_0,\vec p) &=& P_\mn(p) -P_\mn^T(p)\nn.
\eea
\no A third possible tensor structure, having a mixture of longitudinal and transverse indices, is due to the requirement of Landau gauge that $D_\mn$ is four-dimensional transverse not independent, and thus does not appear explicitly. In terms of correlation functions, these propagators are given by \cite{Cucchieri:2007ta}
\bea
D_T(p) &=\frac{Z(p)}{p^2}=& \frac{1}{(d-2)N_g} \left\langle\sum_{\mu=1}^3 \, A_\mu^a(p) A_\mu^a(-p)-\frac{p_0^2}{{\vec p\,}^2} \, A_0^a(p)A_0^a(-p)\right\rangle\label{ft:corrt}  \\
D_L(p) &=\frac{H(p)}{p^2}=& \frac{1}{N_g} \left(1+\frac{p_0^2}{{\vec p\,}^2}\right) \langle A_0^a(p)A_0^a(-p)\rangle \label{ft:corrl},
\eea
\no where explicit use has been made of the Landau gauge condition. Of course, the ghost propagator as being a scalar is still represented by a single scalar function, though now also depending separately on energy and three-momentum. Furthermore, due to CP invariance, the correlation functions do not change under the replacement $p_0$ into $-p_0$. Thus, only positive (and zero) energy solutions will be discussed below.

As a consequence of a compactified time direction the energies of the system become quantized. Because of the Kugo-Martin-Schwinger condition gluons as bosons have to have periodic boundary conditions, while fermions would have anti-periodic boundary conditions \cite{Das:1997gg}. Thus, the energies gluons can have are
\be
p_0=2\pi T n\label{ft:matfreq},
\ee
\no with $n$ an integer. To maintain gauge invariance, it is then necessary that ghosts, despite their Grassmannian nature, also have periodic boundary conditions, and thus the same energy levels \cite{Bernard:1974bq}. Since for $n=0$ the energy vanishes there is a natural separation of states in soft states $n=0$ and hard states $n\neq 0$. The latter can never have $p^2=0$, and thus cannot be on the tree-level light cone. In a perturbative setting only the soft modes can go on-shell, though non-perturbatively this statement loses its meaning \cite{Henning:1995sm}. Furthermore, from the perturbative point of view, there are then three different energy scales. One is given by the hard scale $T$, while there is a separation of soft states $g(T)T$ and ultrasoft states $g(T)^2T$. Here, this further distinction will not be made. A more important distinction is to declare a momentum to be infrared if and only if it is not only much smaller than $\Lambda_\mathrm{YM}$, but in addition also much smaller than $T$, as already given in \pref{meth:irscale} \cite{Cucchieri:2007ta}.

With this distinction comes another insight. When the limit of infinite temperature is taken\footnote{This limit has to be taken after renormalization \cite{Maas:2005hs}.}, which will be discussed in detail in section \ref{sfinitet:inft}, only the soft modes have finite energy and are therefore dynamical \cite{Appelquist:1981vg,Maas:2005ym}. Thus, it is formally admissible to remove all hard modes from the Lagrangian. The result is the Lagrangian of a three-dimensional Yang-Mills theory coupled to an adjoint scalar. Formally, the ghost remains unaltered in the process, and the spatial components of the gluon field become the three-dimensional gauge field, while the zero component of the gluon field becomes the adjoint scalar. From the relations \pref{ft:corrt} and \pref{ft:corrl} this is immediately clear, as the soft mode of the transverse propagator is just the corresponding spatial gluon propagator and the adjoint scalar propagator is given by the soft longitudinal propagator. However, in contrast to an arbitrary such theory, the mass and self-coupling of the adjoint scalar are not free parameters, but are fixed by their descent from the four-dimensional theory \cite{Kajantie:1995dw}. In particular, since the parent theory is only logarithmically divergent the mass of the adjoint scalar is protected from large radiative corrections of the order of the cutoff with the four-dimensional gauge symmetry acting as a custodial symmetry. The infinite-temperature limit can therefore be treated in such a reduced three-dimensional setting, which will be made use of below. Since the longitudinal propagator is then entirely given by chromoelectric components of the gluon field and the transverse part entirely by chromomagnetic ones, these notations are also used at finite temperature. Since the ghost in the limit of infinite temperature becomes the one of the three-dimensional Yang-Mills theory, it is usually also associated with the chromomagnetic gluon, and both together are referred to as the magnetic sector of the theory.

\subsubsection{Modifications of the methods}

After these general consequences of introducing finite temperature, the necessary alterations to implement the Matsubara formalism in the methods employed should be noted.

In lattice calculations it is straightforward to implement finite temperature by explicitly compactifying the time direction \cite{Rothe:2005nw,Gattringer:2010zz}. For that purpose, the time extension $aN_t=1/T$ must be made much smaller than the spatial extensions $aN_s=L$. The reason is that otherwise the time direction cannot be considered compactified. Such a difference can be achieved by two possibilities. One is that the discretization in time direction is made different than in the spatial direction, $a_s\gg a_t$. The other is that the number of lattice points in time direction is made much smaller than in space direction, $N_t\ll N_s$ (or a combination of both). While the prior possibility provides less violations of rotational symmetry, the latter is much cheaper. Given the necessity to reach large volumes, it is the method of choice here. Note that in the infinite-volume and continuum limit also $a_t$ must go to zero and $N_t$ to infinity, but while the product $a_sN_s$ has to diverge, in this limit the product $a_t N_t$ must be kept fixed at $1/T$. Other than this, no further modifications in the lattice methods are necessary.

The implementation inside functional methods is also comparatively simple. Besides including the general structure of propagators and arguments explicitly, it amounts to replacing all integrals over energies with discrete sums over Matsubara frequencies \cite{Roberts:2000aa}
\be
\int \frac{dp_0}{2\pi}\to T\sum_n\nn,
\ee
\no where $n$ is the integer in \pref{ft:matfreq}. This is sufficient. The evaluation can be simplified by making use of the fact that all correlation functions remain unchanged under the replacement $p_0\to -p_0$, as long as CP is unbroken.

Due to the discretization, the correlation functions can be regarded as an infinite tower of correlation functions enumerated by $n$. Thus, in principle, an infinite number of independent dressing functions have to be determined. In practice, as shown below, it turns out that the approximation
\be
\Gamma(p_0,\vec p^2)\to\Gamma(0,p_0^2+\vec p^2)\label{ft:restore}
\ee
\no is even for $n=1$ rather well fulfilled \cite{Maas:2005hs,Cucchieri:2007ta,Fischer:2010fx}. This is a consequence of the effective restoration of Lorentz symmetry at large momenta. Therefore, only a small number of Matsubara frequencies have to be treated independently, permitting the development of efficient algorithms \cite{Maas:2005xh}.

It should be noted that all of this and the following can also be transferred qualitatively unaltered to dimensions different than four. However, due to the immediate practical application to heavy-ion experiments and the early universe \cite{Kapusta:2006pm,Das:1997gg}, the presentation here is restricted to four dimensions. Some investigations concerning the three-dimensional case in the present framework are available in \cite{Maas:unpublished2}, and show no qualitative difference to the four-dimensional case.

\subsection{Propagators}\label{sfinitet:props}

As discussed previously, due to \pref{ft:restore}, in the ultraviolet at momenta large compared to both the temperature and $\Lambda_\mathrm{YM}$, just the zero temperature perturbative behavior is obtained. In case of functional methods the truncation has to respect this. An explicit calculation of how this restoration takes place can be found in \cite{Cucchieri:2007ta,Maas:2005ym}.

In case the temperature is comparable or larger than $2\pi \Lambda_\mathrm{YM}$, it is possible to describe most of the temperature effects quite successfully with finite-temperature perturbation theory \cite{Kajantie:2002wa}. This is the domain of hard-thermal loop calculations \cite{Blaizot:2001nr}, which are included in both lattice and functional methods by construction. Both latter approaches, by introduction of non-perturbative effects, also solve the Linde problem. This problem essentially states that from order $g^6$ onwards all orders of perturbation theory contribute equally, leading to a breakdown of perturbation theory \cite{Kapusta:2006pm} even if the factorial growth leading to its ultimate breakdown \cite{Rivers:1987hi} would not yet be a problem. The reason is that in perturbation theory the gluons acquire a screening mass, which is of order $gT$ in case of the longitudinal propagator. By resummation, this produces an inverse dependence on the coupling constant, yielding by power-counting the breakdown of perturbation theory since powers of $g$ in numerator and denominator cancel to one. A further obstacle is that the cancellation of infrared divergences of zero temperature perturbation theory breaks down due to the screening of longitudinal gluons, which then cannot cancel infrared divergences generated by perturbatively unscreened transverse gluons.

To overcome these problems constructively, improved perturbative methods have been developed \cite{Kajantie:2002wa}, relying essentially on one non-perturbative input parameter, which can, e.\ g., be determined on the lattice, at least in principle \cite{Hietanen:2008tv}. This essentially amounts to introducing an effective transverse gluon screening mass \cite{Blaizot:2001nr}, yielding the missing contribution of non-perturbative and higher order contributions. Instrumental in this process is the possibility to map the theory at large temperatures on a three-dimensional theory, as discussed above \cite{Kajantie:1995dw}.

Here, the infrared suppression of the transverse gluon propagator will be generated without an external input. Thus the dangerous infrared divergences no longer arise. This genuine non-perturbative effect is obtained in the non-perturbative methods, which therefore do not suffer from the breakdown of the perturbative series at order $g^6$. Thus, the Linde problem is absent from the calculations presented here \cite{Maas:2005ym}.

\subsubsection{Infrared}\label{sfinitet:ir}

The infrared behavior is most transparent once more in the scaling case. The situation in the finite-ghost case will be discussed afterwards. It should be noted that due to \pref{meth:irscale} the momenta are infrared only when they are much smaller than the temperature. On the other hand, at momenta much larger than the temperature the zero-temperature behavior will remain. In particular, if a momentum window exists with momenta satisfying
\be
T,p_B\ll p\ll\Lambda_\mathrm{YM}\nn,
\ee
\no i.\ e., at very small temperatures, the temperature acts as an infrared cut-off, and in the same way as at large finite volume \cite{Fischer:2007pf} within this momentum domain the four-dimensional infrared behavior, and in particular a power-law dependence of the correlation functions, will be obtained \cite{Cucchieri:2007ta}.

The derivation of the finite-temperature DSEs can be found elsewhere \cite{Maas:2004se,Maas:2005hs}. The important point is that by an appropriate projection, two coupled equations for the dressing functions of the transverse and longitudinal part are obtained \cite{Maas:2004se}. Keeping for now only the contributions from the ghost loop, the equations for the soft modes are given by
\bea
\frac{1}{G(0,\vec k)} &=& \tilde{Z}_3+\frac{g^2TC_A}{(2\pi)^3}\sum_{q_0}\int dq d\theta\label{geq}\\
&&  \left( A_T(0,q_0,\vec k,\vec q) G(q_0,\vec q) Z(q_0,\vec q-\vec k)+A_L(0,q_0,\vec k,\vec q)G(q_0,\vec q)H(q_0,\vec q-\vec k) \right)\nn\\
\frac{1}{Z(0,\vec k)} &=& Z_{3T}+\frac{g^2TC_A}{(2\pi)^3}\sum_{q_0}\int dq d\theta R(0,q_0,\vec k,\vec q)G(q_0,\vec q) G(q_0,\vec q+\vec k)\label{zeq}\\
\frac{1}{H(0,\vec k)} &=& Z_{3L}+    \frac{g^2TC_A}{(2\pi)^3}\sum_{q_0}\int dq d\theta  P(0,q_0,\vec k,\vec q) G(q_0,\vec q) G(q_0,\vec q+\vec k)\label{heq}.
\eea
\no The case of the hard modes will be discussed below. The wave-function renormalization constants for the longitudinal and transverse propagator can become different at finite temperature, depending on the renormalization scheme, and are thus denoted differently \cite{Maas:2005hs,Das:1997gg}. The integral kernels $A_T$, $A_L$, $R$, $P$ for the case of a undressed ghost-gluon vertex can be found in \cite{Maas:2005hs}. By only keeping the $q_0=0$ contribution in the Matsubara sums these equations become equivalent to the ones of the three-dimensional Yang-Mills-adjoint-Higgs system in the same truncation \cite{Maas:2004se}. In that limit the longitudinal equation decouples, the longitudinal propagator becomes tree-level-like, and the transverse and ghost sector behave as a three-dimensional Yang-Mills theory. Thus, in the scaling case they will exhibit the characteristic three-dimensional exponents \cite{Maas:2004se}. On the other hand, in the limit of zero temperature it is possible to show that these equations yield again the four-dimensional behavior \cite{Cucchieri:2007ta}.

Equations \prefr{geq}{heq} can also be written as
\bea
\frac{1}{G(0,\vec k)} &=& \tilde{Z}_3 +\Pi_G(0,\vec k) +\sum_{q_0\neq 0}\Pi_G(q_0,\vec k)\label{eq:geq1} \\
\frac{1}{Z(0,\vec k)} &=& Z_{3T} + \Pi_Z(0,\vec k) +\sum_{q_0\neq 0}\Pi_Z(q_0,\vec k)\label{eq:geq} \\
\frac{1}{H(0,\vec k)} &=& Z_{3L} + \sum_{q_0\neq 0}\Pi_H(q_0,\vec k)\label{eq:HofZ},
\eea
\no where $\Pi_i$ denotes the various self-energies. As can be gleaned from the infinite-temperature limit in the longitudinal equation, the zero-component $q_0=0$ explicitly vanishes, since $P(0,0,\vec k,\vec q)=0$ \cite{Maas:2004se}. This is a direct consequence of the tensor structure of the ghost-gluon vertex, and it would require a highly non-trivial, temperature-dependent dressing to alter this behavior. As a consequence, the result for the longitudinal mode $H(0,\vec k)$ is entirely determined by the hard modes and depends only implicitly on the soft ones.

This requires to consider the hard modes, which have an effective tree-level mass of $2\pi T n$. Since the external momentum is much smaller than this mass, the hard-mode dressing functions become essentially constant in the infrared. Thus, for $\vec k \to 0$, the dressing functions are, respectively, given by constants $A_z(k_0)$, $A_h(k_0)$ and $A_g(k_0)$, which are assumed to be bounded as a function of $k_0$. Actually, because of asymptotic freedom, they will decrease logarithmically with $k_0$ for $k_0\to\infty$ \cite{Cucchieri:2007ta}.

For the soft modes power-law ans\"atze
\bea
G(0,\vec k) &=& B_g\, k^{2\kappa} \label{iransatz1} \\
Z(0,\vec k) &=& B_z\, k^{2t} \\
H(0,\vec k) &=& B_h\, k^{2l} \label{iransatz3}
\eea
\no are made with independent exponents $\kappa$, $t$ and $l$ and constant coefficients $B_g$, $B_z$ and $B_h$. The scaling condition requires $\kappa<0$. Then, the soft self-energies $\Pi_G(0,\vec k)$ and $\Pi_Z(0,\vec k)$ take exactly the same form as in three dimensions \cite{Maas:2004se}, i.\ e.\
\bea
\Pi_G(0,\vec k) &=& -B_g\,B_z\, g^2TC_A \frac{2^{1-4\kappa}\Gamma(2+2\kappa)}{   \kappa(3+4(-2+\kappa)\kappa)\Gamma\left(2\kappa+\frac{3}{2}\right)}y^{t+\kappa-\frac{1}{2}}\nn \\
\Pi_Z(0,\vec k) &=& -B_g^2\, g^2TC_A \frac{2^{-4(1+\kappa)}\Gamma(2+2\kappa)\sec(2\pi\kappa)\sin(\pi\kappa)^2}{\kappa^2(1+\kappa)\Gamma\left(2\kappa+\frac{3}{2}\right)}y^{2\kappa-\frac{1}{2}}\label{eq:PiZ},
\eea
\no where $y = k^2 = \vec k^2$. These are exactly the results obtained in the three-dimensional case \cite{Zwanziger:2001kw}, and thus will yield a three-dimensional behavior as the leading infrared behavior for the transverse gluon and ghost if they are not exactly canceled or dominated by the remaining Matsubara sums. Setting the hard-mode dressing functions equal to the constants $A_g(k_0)$, $A_z(k_0)$ and $A_h(k_0)$, the self-energies in the ghost equation \pref{eq:geq1} can be rewritten as \cite{Cucchieri:2007ta}
\bea
\Pi_G(q_0\neq 0,\vec k)&=&g^2TC_A\label{eq:PiGfinal}\\
&\times&\left(-\left(\frac{A_z(q_0)}{12\pi}+\frac{A_h(q_0)}{96}\right)\frac{1}{|q_0|}+\left(\frac{A_z(q_0)}{80\pi}+\frac{A_h(q_0)}{1920}\right)\frac{\vec k^2}{|q_0|^3}+\vec k^3\pi(q_0,\vec k)\right).\nn
\eea
\no The first term is logarithmically divergent and can be absorbed in the wave-function renormalization constant. The second term is sub-dominant when compared to the 3d-term and is finite after summation over $q_0$. Finally, the function $\pi(\vec k)$ vanishes identically as $\vec k\to 0$ \cite{Cucchieri:2007ta}. Thus, the leading part of the ghost equation is the same as in the 3d-case, i.e.\ it is given by $\Pi_G(0,\vec k)$.

For the transverse equation it is necessary to first get rid of the spurious divergences in both gluon equations. After subtraction of these divergences, the contributions of the hard-modes in the self-energies vanish as $\vec k\to 0$, separately for each Matsubara term \cite{Maas:2005hs}. Thus, the subtracted part cannot contribute to the IR behavior in the transverse equation. On the other hand, the infrared contributions from the unsubtracted self-energies cannot be neglected in general \cite{vonSmekal:1997vx,Fischer:2003zc,Maas:2005rf}. This requires a regularization of the spurious  divergences\footnote{Note that in the transverse case, divergences appear for each hard mode, while in the longitudinal case only the sum over the hard modes is affected by spurious divergences \cite{Cucchieri:2007ta}.}. To achieve this, one can replace the approximately constant dressing functions of the hard modes by $A_g(q_0)(q^2+q_0^2)^{-\tau}$, which are suppressed when $\tau>0$. This is the prescription commonly used for regularizing the divergences at zero temperature with $\tau=-\kappa$ \cite{vonSmekal:1997vx,Zwanziger:2001kw,Lerche:2002ep}.

Then, the integrals can be performed and yield in the limit $\vec k \to \vec 0$ 
\bea
\Pi_Z^D(q_0,\vec k\to 0) &=& -\frac{g^2TC_A}{k^2}\frac{(\zeta-3)\Gamma\left(\tau-\frac{1}{2}\right)}{32\pi^{3/2}\Gamma(2+2\tau)}(2\pi T)^{1-4\tau}\sum_{n\neq 0}A_g(n)^2|n|^{1-4\tau} \label{imt} \\
\Pi_H^D(q_0,\vec k\to 0) &=& \frac{g^2TC_A}{k^2}\frac{\Gamma\left(2\tau+\frac{1}{2}\right)}{8\pi^{3/2}\Gamma(2+2\tau)}(2\pi T)^{1-4\tau}\sum_{n\neq 0}A_g(n)^2|n|^{1-4\tau} \label{iml},
\eea
\no where $n=q_0/2\pi T$. These sums diverge for $\tau\le 1/2$, due to the term $q_0^{1-4\tau}$. For $\tau=1/2$, this exponent becomes equal to $-1$, i.e.\ the sums are logarithmically divergent. Finally, for $\tau>1/2$, the sums are finite and can be resummed analytically. One should also note that the terms in \prefr{imt}{iml} behave like $1/k^2$, i.e.\ like a mass term in the IR limit. In the transverse case - but not in the longitudinal case \cite{Maas:2005hs} - this term may not be renormalized, as this is not allowed by gauge-invariance \cite{Das:1997gg}. If  either a divergence not stronger than logarithmic should be encountered or the same exponents $\tau$ for the transverse and for the longitudinal case should be used, then the only possibility\footnote{Of course, $\tau=1/2$ just corresponds to the limiting scaling case of an infrared finite gluon propagator and an infrared divergent ghost dressing function. This possibility will not be discussed here for simplicity, as this does not add anything new compared to the discussion in section \ref{sir:scaling}.} is to have $\tau>1/2$. For all such values of $\tau$, the contribution \pref{imt} is sub-leading in the transverse equation \pref{eq:geq} \cite{Cucchieri:2007ta,Zahed:1999tg}. At the same time, the screening mass in the longitudinal equation would then be solely due to the regularized contribution \pref{iml}. This result is a consequence of the truncation scheme. In fact, due to the decoupling of the soft modes in the longitudinal equation (see equation \pref{eq:HofZ}), the result is dominated by the hard modes, which are very sensitive to truncation artifacts, since they live on a scale that is effectively mid-momentum. Thus, this truncation scheme is not able to yield a consistent description of the chromoelectric screening mass, and a determination of its value is not possible\footnote{Another, though somewhat simplified, truncation has been investigated in \cite{Chernodub:2007rn}, and yielded a qualitatively similar result.}. 

More sophisticated truncations which eliminate the spurious divergences already at zero temperature, e.\ g.\ the one discussed in section \ref{sec:trunc} or of \cite{Fister:2011ym}, would also eliminate the spurious mixing of physics and truncation artifacts at finite temperature \cite{Fischer:2008uz}, and thus permit to access the electric screening mass. Indeed, the lattice calculations shown below show that the phase transition dynamics is encoded in this screening mass, making its self-consistent determination the most important task at finite temperature. The question to which extent the modeling of the vertices in turn determines the thermodynamic properties then becomes a central question. Thus, taking the vertices self-consistently into account appears to become mandatory at finite temperature. Ultimately, this should resolve whether it is a change of gluon properties or of bound state properties which drives the phase transition \cite{Cucchieri:2007ta}.

Nonetheless, the primary message is that the longitudinal gluon propagator exhibits a non-zero screening mass, giving a qualitatively correct description of the physics involved. It is also this screening mass which prevents the longitudinal gluon propagator from modifying the infrared behavior of either the transverse gluon propagator or of the ghost propagator. Therefore these two propagators should behave at all non-zero temperatures exactly as in the three-dimensional case at momenta much smaller than the temperature $T$ and than $\Lambda_{\mathrm{YM}}$.

A similar discontinuous change from zero to non-zero temperature is also found in background-gauge calculations, using renormalization group equations and a scaling assumption for the running coupling \cite{Braun:2005uj}. It is thus possibly a more generic feature of the scaling case.

The situation for the finite-ghost case is even more involved. In this case, the screening masses obtained at zero temperature now obtain temperature-dependent contributions also for the transverse gluon propagator. Similar modifications apply to the finite dressing function of the ghost. In close correspondence to the longitudinal gluon propagator in the scaling case the transverse propagator's screening mass will now also mix truncation artifacts and temperature effects in a generic truncation. As a consequence, in the finite-ghost case all screening masses will receive a temperature-dependent contribution, but it will require a better understanding of truncation artifacts to separate the spurious and physical contributions. However, this also implies that in the finite-ghost case no qualitative change occurs when a non-zero temperature is introduced.

\subsubsection{Intermediate momenta and temperatures}

The investigation using functional methods provides a qualitative description of the general infrared behavior for the propagators. In particular, they provide direct access to the properties related to the asymptotic state space in section \ref{sfinitet:schwinger}. However, the fact that the longitudinal gluon propagator is sensitive to contributions on the scale of the temperature already indicates that a pure infrared analysis is not sufficient to describe the thermodynamics. That no sign of the well-known phase transition has been found is another serious drawback.

A treatment of the DSEs over the total momentum range has been done to also investigate the intermediate momenta \cite{Maas:2004se,Maas:2005hs,Cucchieri:2007ta}. However, since the truncation employed could remove the quadratic divergences only up to a finite truncation artifact, this provided another source for an electric screening mass: In this case the spurious divergences are formally under control \cite{Maas:2005hs}, since at finite temperature a mass renormalization of the zero mode of the chromoelectric gluon propagator is actually admissible. But this renormalization is in general arbitrary. Therefore, it is not possible to obtain physically useful results from such an approach. It would require at least a truncation scheme, like the one discussed in section \ref{sec:trunc}, which removes the quadratic divergences already at zero temperatures\footnote{There are also alternative truncation schemes proposed for that purpose \cite{Aguilar:2009ke}.}, to give at least the possibility for a correct behavior. However, this is a necessary requirement, but not necessarily a sufficient one. In particular, if, e.\ g., the thermodynamics were driven by glueball dynamics, this would possibly not be captured, except for an ingenious choice of truncation. It remains to solve the DSEs at finite temperature with such an improved truncation to clarify whether more is required. Calculations using FRGs \cite{Fister:2011ym} permitted to have a better control over these artifacts, but are not yet in fully satisfactory agreement at temperatures around the phase transition. Nevertheless, they constitute a major step towards control of the truncation artifacts. An alternative and rather successful approach was to use gauges which directly include the Polyakov-loop dynamics, like Polyakov gauges \cite{Marhauser:2008fz}.

\begin{figure}
\begin{center}
\includegraphics[width=0.45\textwidth]{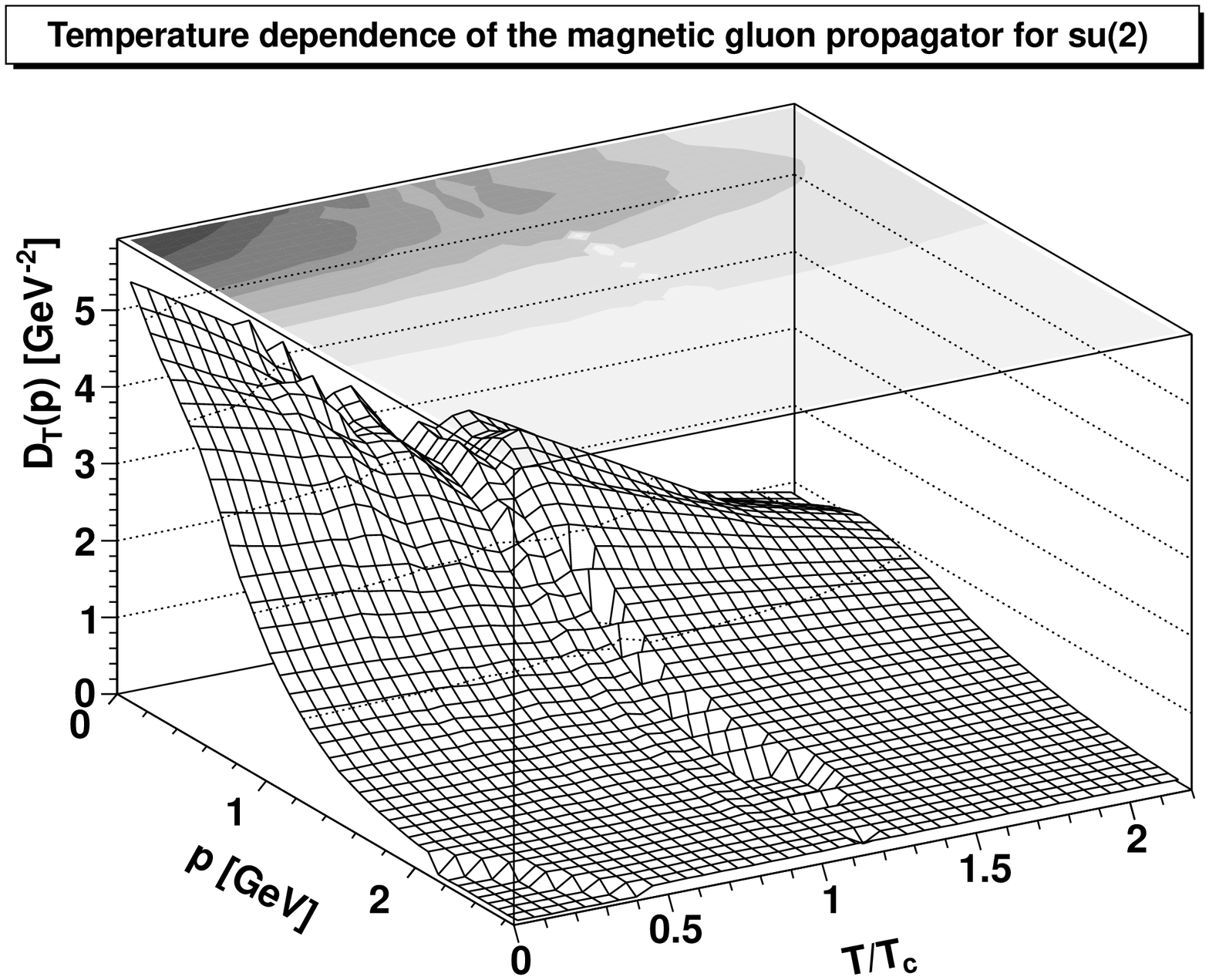}\includegraphics[width=0.45\textwidth]{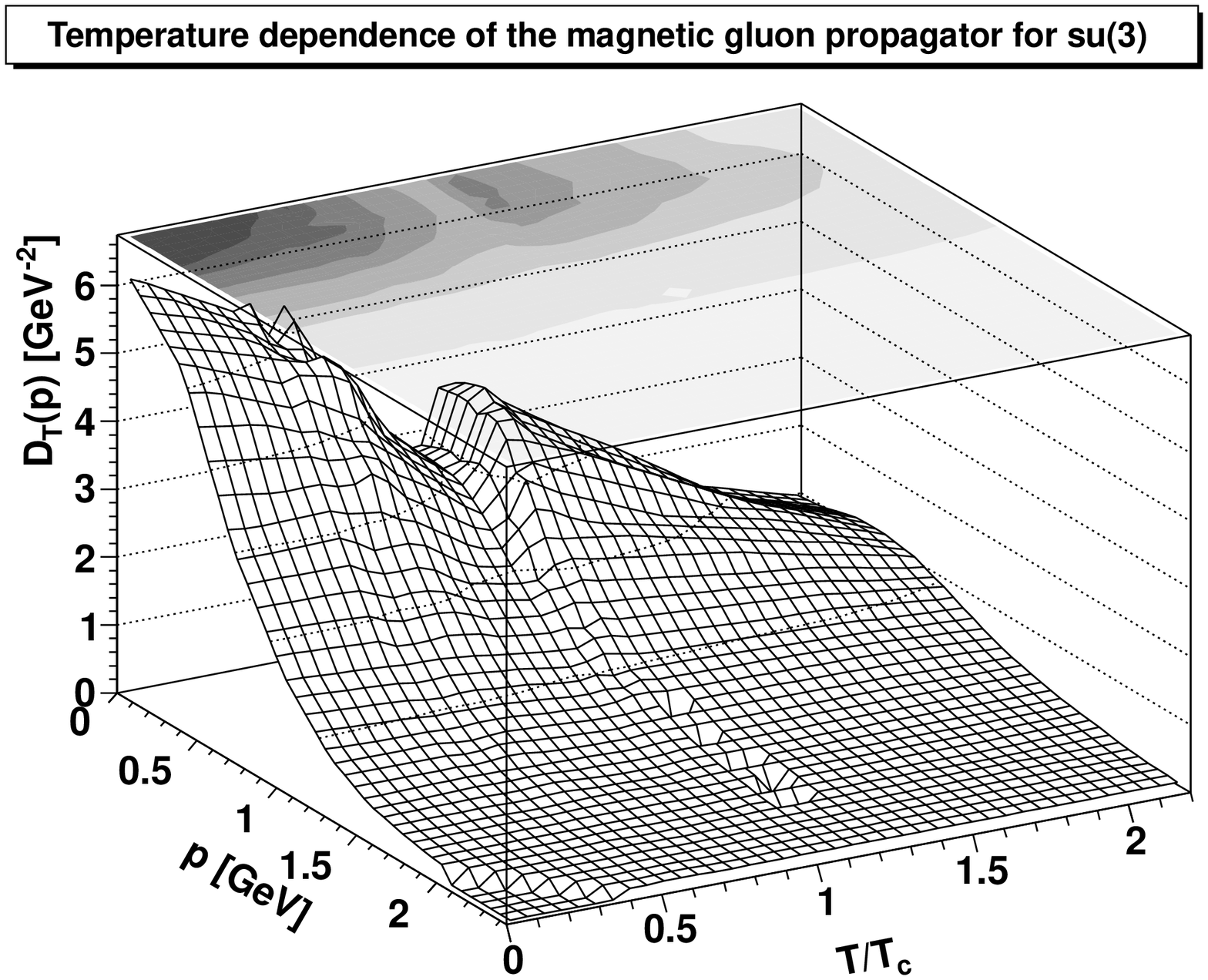}\\
\includegraphics[width=0.45\textwidth]{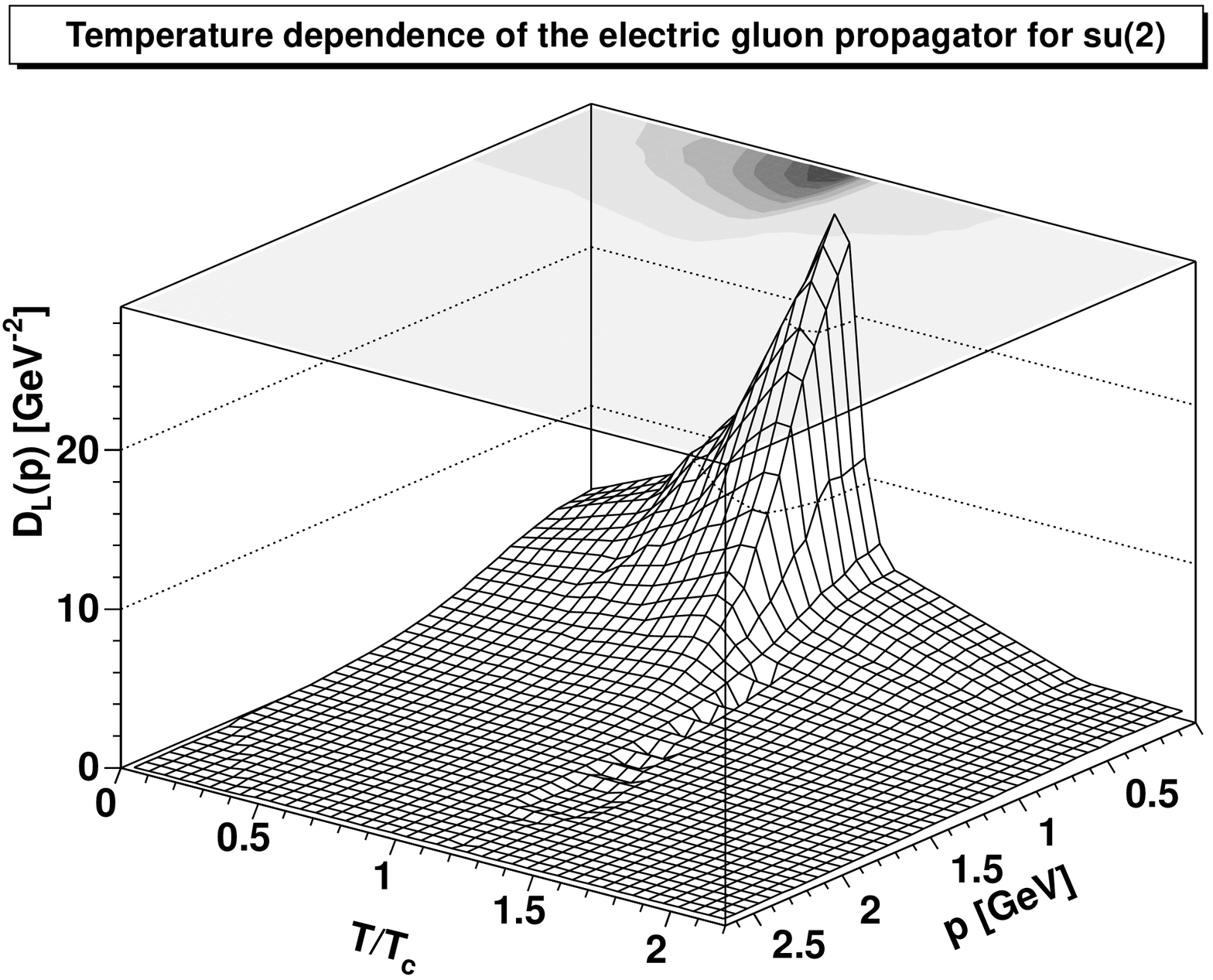}\includegraphics[width=0.45\textwidth]{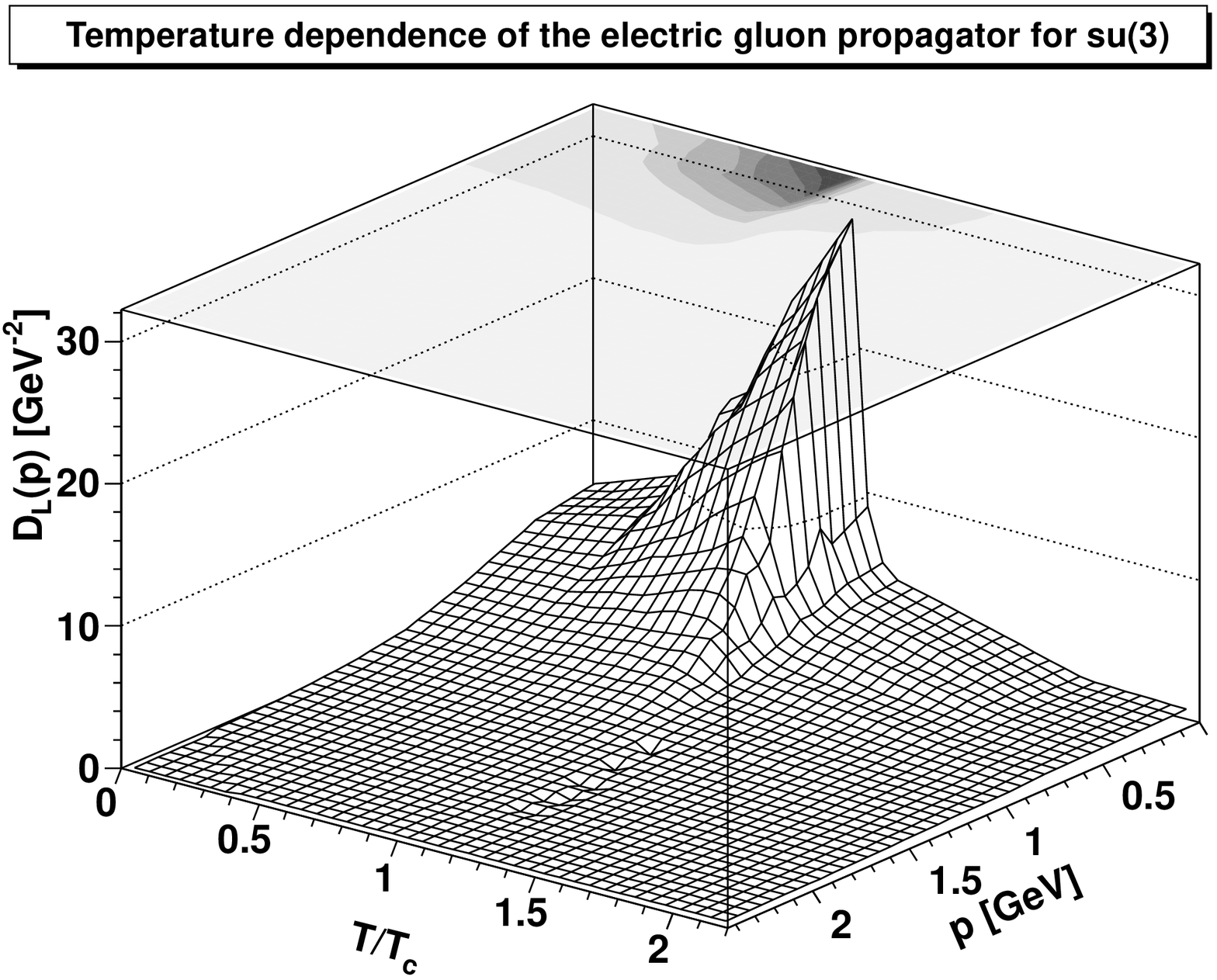}\\
\includegraphics[width=0.45\textwidth]{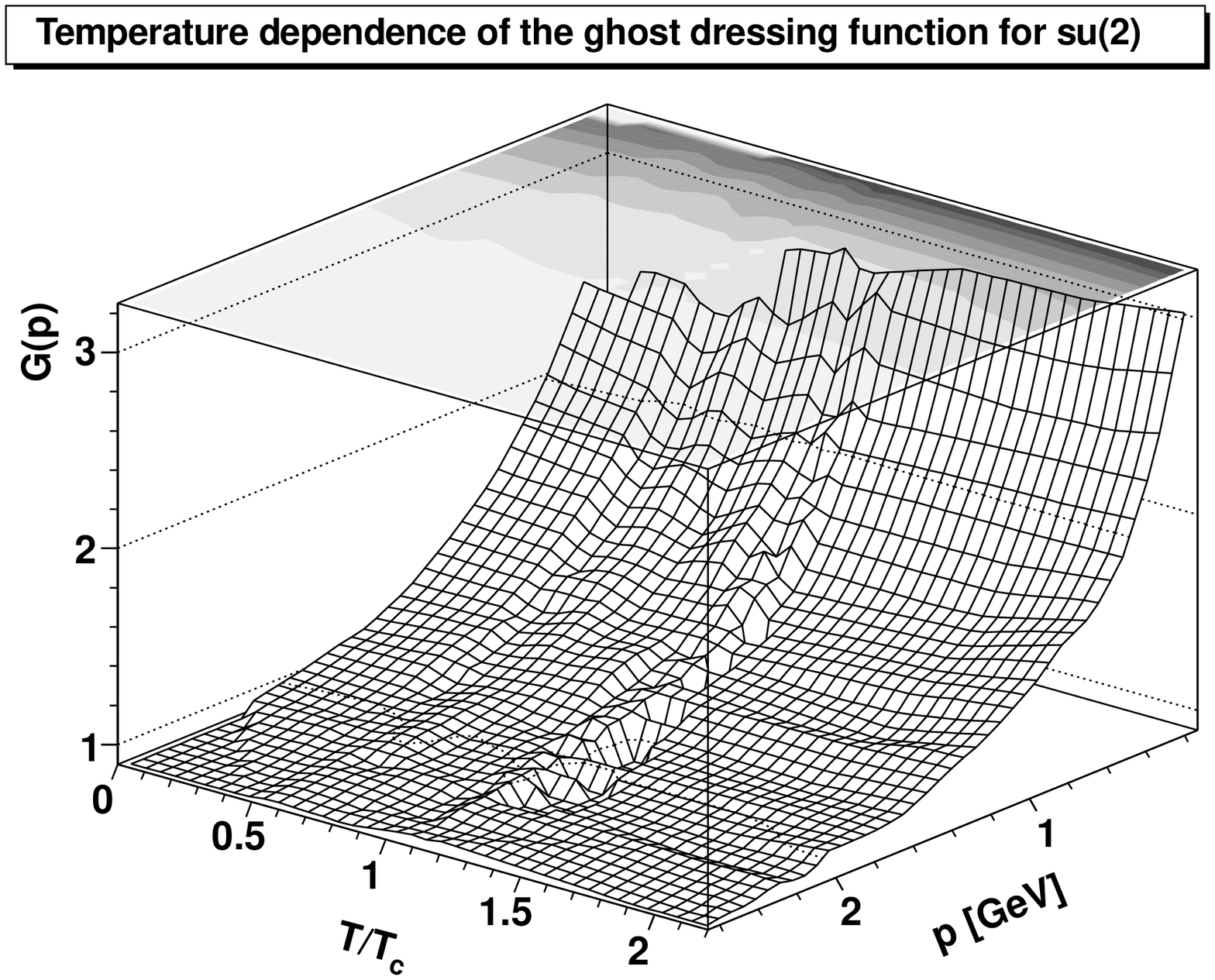}\includegraphics[width=0.45\textwidth]{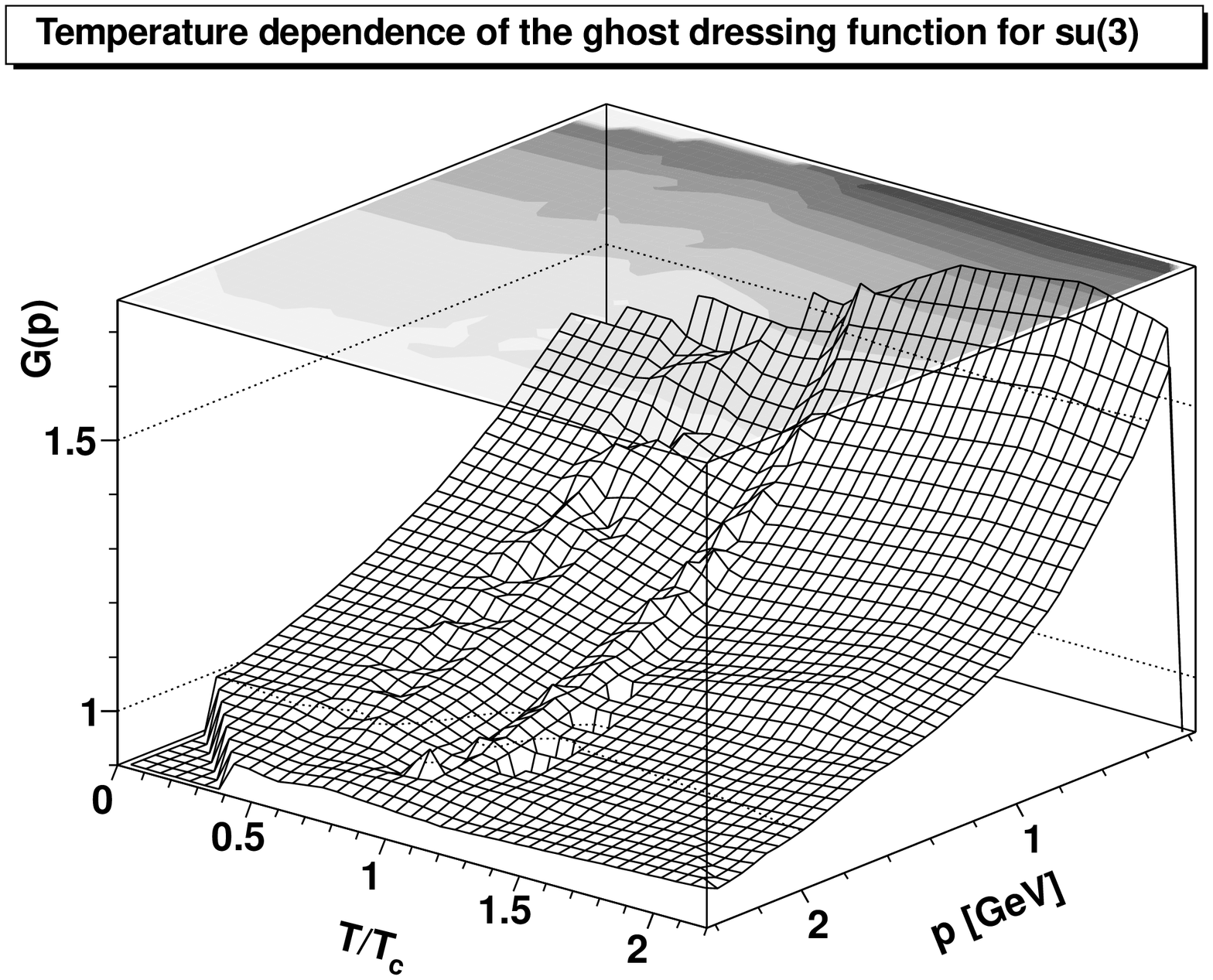}
\end{center}
\caption{\label{fig:finitet}The transverse (top panels) and longitudinal (middle panels) gluon propagator and the ghost dressing function (bottom panels) as a function of temperature and momentum \cite{Fischer:2010fx,Maas:unpublished2}. The left panels show results for the gauge algebra su(2) and the right panels for su(3). Independent determinations of the critical temperature $T_c$ can be found in \cite{Fingberg:1992ju,Lucini:2003zr}. Further results can be found for su(2) in \cite{Cucchieri:2007ta,Cucchieri:2001tw,Cucchieri:2000cy,Karsch:1994xh,Heller:1997nqa,Heller:1995qc,Bornyakov:2010nc,Bornyakov:2011jm} and for su(3) in \cite{Aouane:2011fv}.}
\end{figure}

\begin{figure}
\includegraphics[width=0.5\textwidth]{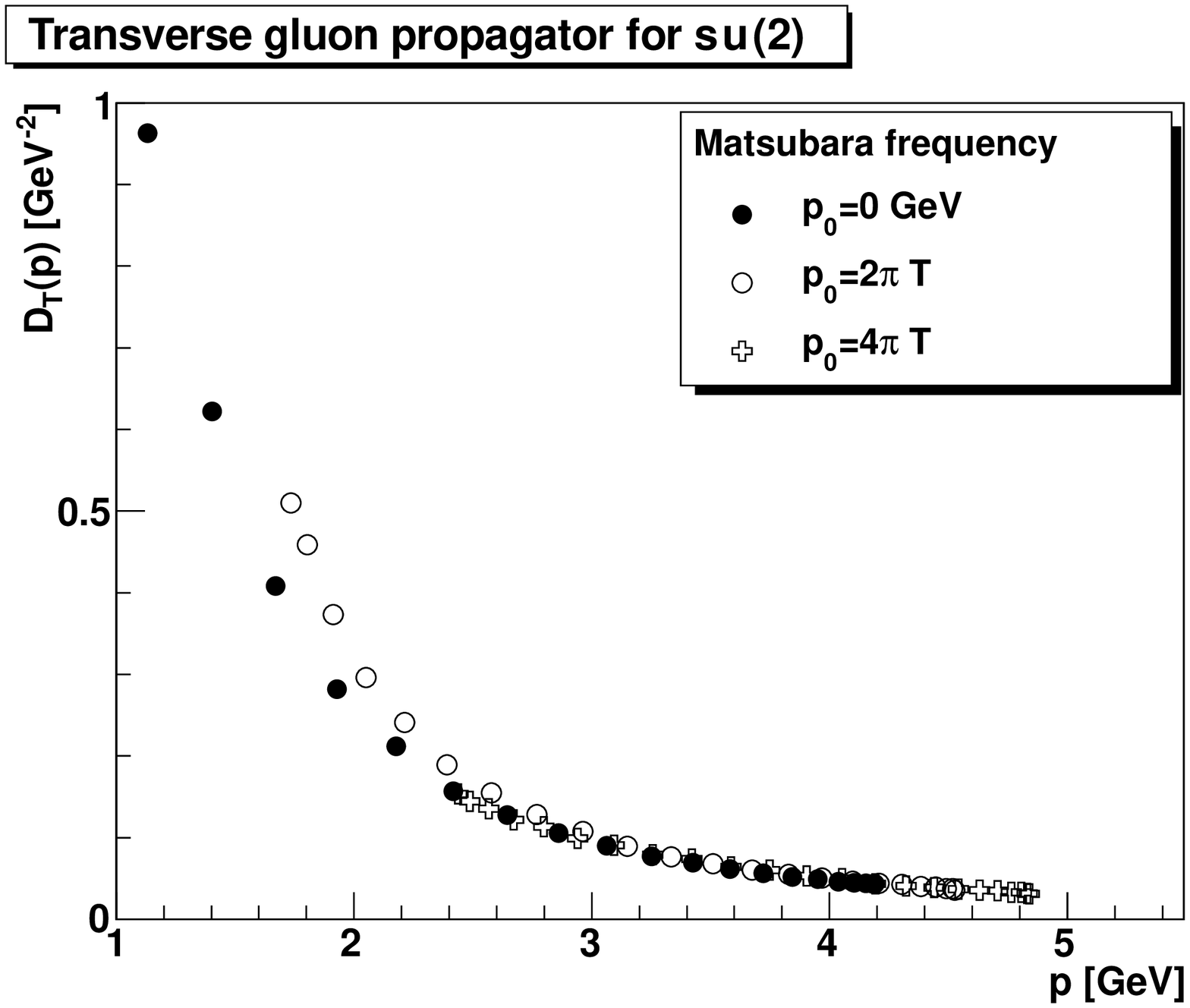}\includegraphics[width=0.5\textwidth]{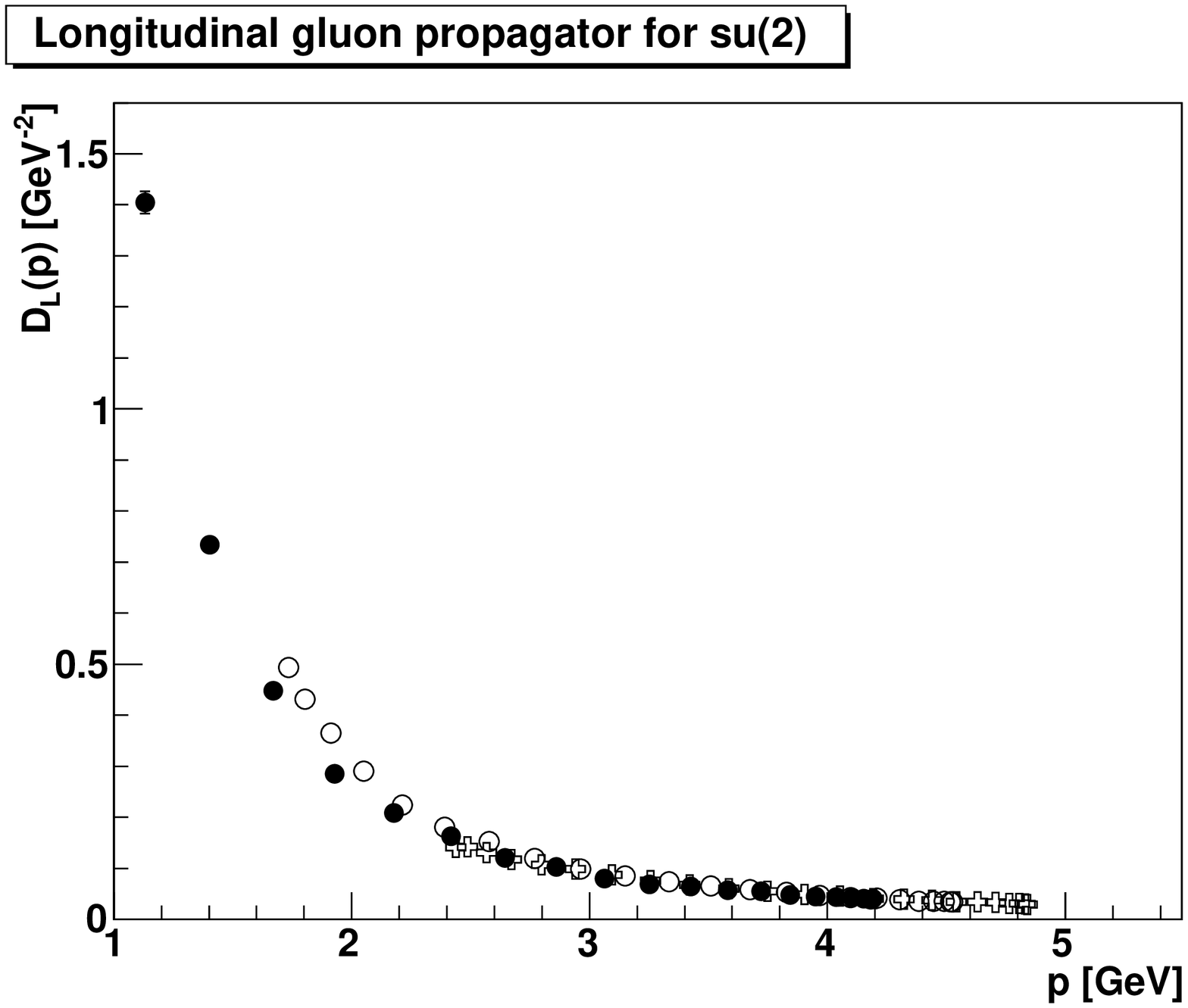}
\caption{\label{fig:hardmodes}Comparison of the soft mode to the first two higher Matsubara frequencies at $T=T_c$ for su(2) on a $4\times 46^{3}$ lattice. In the left panel the transverse gluon propagator is shown, and in the right panel the longitudinal one \cite{Fischer:2010fx}. All results are shown as a function of the four momentum $p^2=p_0^2+(\vec p)^2$.}
\end{figure}

Fortunately, lattice calculations do not have similar problems \cite{Rothe:2005nw,Gattringer:2010zz}, and the calculations performed at zero temperature can also be performed at finite temperature. The resulting zero-modes of the chromoelectric, chromomagnetic, and ghost propagators as a function of temperature are shown in figure \ref{fig:finitet}. To show the adequacy of the approximation \pref{ft:restore}, some examples for the hard modes are given in figure \ref{fig:hardmodes}. These results have been obtained in the minimal Landau gauge. The volumes yet accessed are rather small, so a significant gauge-dependence would naively only be expected for the ghost propagator \cite{Fischer:2010fx}. However, the comparison to absolute Landau gauge \cite{Bornyakov:2010nc,Bornyakov:2011jm} also shows an influence on the gluon propagators, though this may be intertwined with severe lattice artifacts observed at finite temperature \cite{Cucchieri:2007ta,Cucchieri:2010ft,Bornyakov:2011jm,Cucchieri:2011di,Maas:unpublished2}.

It is visible that the only significant temperature dependence is seen for the electric gluon propagator. The temperature dependence of the magnetic gluon propagator is rather weak, and in particular not qualitatively affected by the phase transition. For the ghost propagator, besides effects which can be easily due to the different spatial volumes at the different temperatures, no effect is seen \cite{Fischer:2010fx,Cucchieri:2007ta}. Since the relation \pref{ft:restore} is found to hold rather accurately even at the phase transition \cite{Fischer:2010fx}, none of the hard modes show a pronounced dependence on the phase transition \cite{Fischer:2010fx,Maas:unpublished}, not surprisingly given their effective energy of at least $2\pi T_c\approx 2$ GeV. Thus, the soft and hard magnetic sectors depend rather smoothly on temperature. Also, the hard electric sector does not show a significant dependence \cite{Fischer:2010fx,Cucchieri:2007ta}. This also implies that the electric-magnetic asymmetry observed \cite{Chernodub:2008kf,Vercauteren:2010rk} in dimension-two gluon condensates \cite{Dudal:2009tq} at finite temperature is driven just by the electric sector \cite{Fischer:2010fx}.

\begin{figure}
\begin{center}
\includegraphics[width=0.45\textwidth]{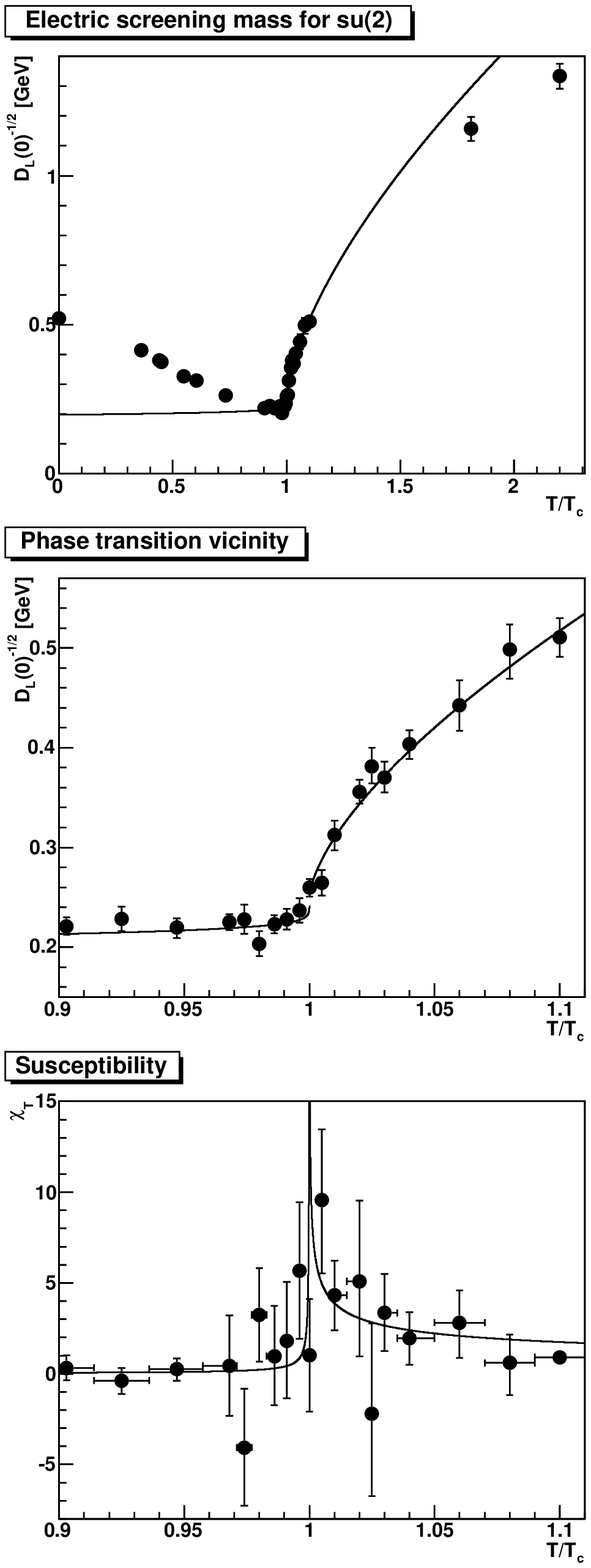}\includegraphics[width=0.45\textwidth]{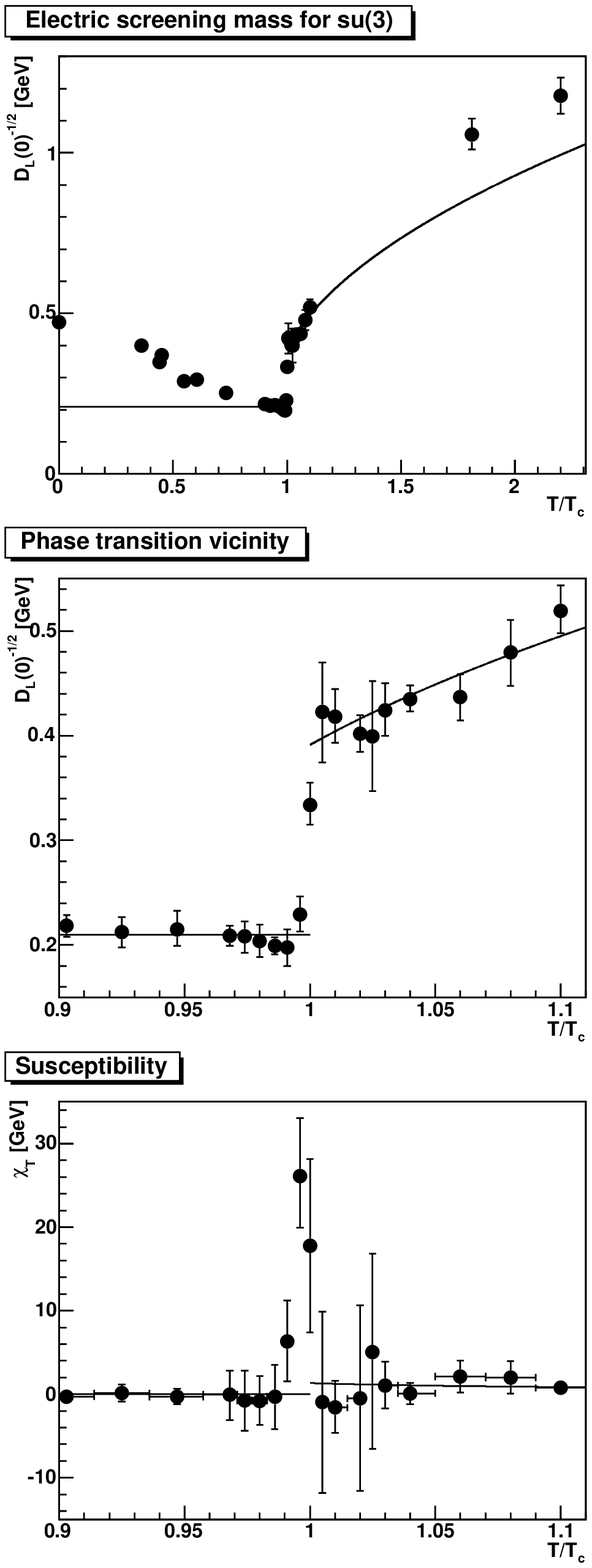}
\end{center}
\caption{\label{fig:t-em}The chromoelectric screening mass (top and middle panels) and the corresponding susceptibility \pref{finitet:susc} (bottom panel) for the gauge algebra su(2) (left panels) and for su(3) (right panels) \cite{Fischer:2010fx,Maas:unpublished2,Maas:unpublished}. The line for the su(2) case is a fit of type \pref{finitet:suscfit} and for the su(3) case of type \pref{finitet:suscfit1o}.}
\end{figure}

Only the soft electric part is then left. In agreement with the argumentation based on the infrared analysis in section \ref{sfinitet:ir}, there is a significant imprint of the phase transition visible for the chromoelectric propagator. In particular, its screening mass is severely affected, as is shown in figure \ref{fig:t-em}. The phase transition point can indeed be determined from the maximum of the susceptibility of the electric screening mass\footnote{The overall scale and many details of the curve shown in figure \ref{fig:t-em} are sensitive to lattice artifacts \cite{Cucchieri:2010ft,Maas:unpublished2}. However, only its temperature dependence is relevant, but this has to be studied further.}
\be
\chi_{m_T}=T_c\frac{\pd D_L(0)^{-\frac{1}{2}}}{\pd T}\label{finitet:susc}
\ee
\no for both su(2) and su(3). It is therefore possible to determine the gauge-invariant phase transition temperature from the propagators alone\footnote{One should be somewhat wary with this statement. It is known that gauge-dependent quantities can indicate a (quantum) phase transition, though there is no physical transition associated with it \cite{Caudy:2007sf,Greensite:2008ss}. However, in the present case the free energy and other physical observables confirm the presence of the phase transition \cite{Karsch:2001cy}.}. Note that the minimum of the screening mass, though close to the phase transition temperature, cannot be the location of the phase transition for a second order transition, since at this point the susceptibility would be zero instead of infinite. The location of the phase transition is thus rather indicated by the peak in the susceptibility \cite{Maas:unpublished2}. 

For a second order phase transition, as is the case for su(2) \cite{Karsch:2001cy}, it is somewhat surprising that the screening mass remains finite at the phase transition. This is even more peculiar since the temperature-dependence of the electric screening mass, after subtracting a zero-point mass, shows a critical behavior \cite{Maas:unpublished2}. This is best illustrated using a fit in the critical region close to $T_c$ of type 
\be
D_L(0)^{-\frac{1}{2}}(T)=m_0+a_l\theta(T_c-T)\left(1-\frac{T}{T_c}\right)^{\nu_<}+a_h\theta(T-T_c)\left(\frac{T}{T_c}-1\right)^{\nu_>}\label{finitet:suscfit},
\ee
\no where $\nu_{<>}$ is a critical exponent. This fit is shown also in figure \ref{fig:t-em}. It is a current question whether this critical behavior is related to the one expected for su(2) based on the universality class of the 3-dimensional Ising model. This would require a value of the critical exponent of $\nu_>=\nu_<\approx0.63$. The fitted values are $\nu_<=0.15(1)$ and $\nu_>=0.65(1)$ \cite{Maas:unpublished2}. These do not show the expected equality, though the exponent in the high-temperature phase is close to the expected one. In the low-temperature phase, the very small pre-factor $a_l$ makes the determination of the exponent rather complicated, given the statistical uncertainty. The constant offset $m_0$ may then be linked to the screening mass in the minimal Landau gauge, and thus obscuring the critical behavior. A more detailed analysis can be found in \cite{Maas:unpublished2}. Taking tall of his into account, the screening mass indeed shows a behavior consistent with the critical scaling of a second order phase transition.

In the su(3) case, the transition is of first order \cite{Karsch:2001cy}. This leads to the expectation of a discontinuity of the screening mass, which is reflected in the lattice results shown in figure \ref{fig:t-em}. Indeed, a behavior of type \cite{Maas:unpublished2}
\be 
D_L(0)^{-\frac{1}{2}}(T)=m_0+\theta(T-T_c)\, a \sqrt{\delta+t }\label{finitet:suscfit1o}
\ee
\no is describing the data acceptable close to $T_c$, as is visible in figure \ref{fig:t-em}. Thus, the order of the phase transition seems to be deducible from the longitudinal gluon propagator. Thus, all interesting properties of the phase transition - critical temperature, order, and critical behavior - seem to be accessible directly from the correlation functions, though systematic effects remain to be better understood \cite{Cucchieri:2007ta,Cucchieri:2010ft,Bornyakov:2011jm,Cucchieri:2011di,Fischer:2010fx,Maas:unpublished2}.

Of course, the overall scale of the figure \ref{fig:t-em}, and thus of the fit \pref{finitet:suscfit}, is renormalization-group-dependent, as is the screening mass in general. However, this is a rather generic feature of many order parameters \cite{Maas:unpublished2}, including such important quantities like the Polyakov loop and the chiral condensate discussed in section \ref{sorder}. Nonetheless, this implies that it cannot be measured directly, even if it would be gauge-invariant.

Another nice result is that, as in the discussion of the scaling case above, the resulting propagator here in the minimal Landau gauge also solves both aspects of the Linde problem immediately. One is the appearance of infrared divergent perturbative integrals as soon as a magnetic propagator is involved. This has been solved previously by a regularization procedure and the introduction of a phenomenological magnetic screening mass \cite{Blaizot:2001nr}. This screening mass is now recovered here from first principles. The second component is that the perturbative expansion breaks down already at order $g^6$, since from this order on all orders of perturbation theory equally contribute\footnote{To resolve this problem, resummation techniques have been introduced, like hard-thermal loops \cite{Blaizot:2001nr,Andersen:2011sf}. Though they provide an enormous improvement in the intermediate momentum regime, they eventually fail in the infrared, disagreeing with the results for the propagators presented here \cite{Maas:2005ym}.}. This problem is also alleviated by the present approach, since it provides a non-perturbative expression, which does not require any expansion in the coupling constant. It is worthwhile to note that using the Gribov-Zwanziger Lagrangian these problems are explicitly overcome, and permit even perturbative calculations in finite-temperature Yang-Mills theory \cite{Zwanziger:2006sc,Lichtenegger:2008mh}.

\subsubsection{Infinite-temperature limit}\label{sfinitet:inft}

At the current time there is quite a number of hints from both experiments \cite{Muller:2006ee,BraunMunzinger:2009zz,Jacobs:2007dw,Andronic:2009gj,Leupold:2011zz} and theoretical investigations  \cite{Muller:2006ee,Shuryak:2008eq,BraunMunzinger:2009zz} that the matter above the phase transition is strongly interacting up to temperatures as large or even larger than 2-3 times the critical temperature. It is thus far from the naively expected quark-gluon plasma. This expectation was based on the idea that the coupling evaluated at the temperature becomes small with increasing temperature. Of course, in a renormalizable and interacting quantum field theory, this simple argument is not fully adequate. However, it is imaginable that at very large temperatures all processes are essentially dominated by leading-order perturbative effects.

Indeed, when measuring thermodynamic observables using lattice gauge theory at very high temperatures, they show a behavior which could be in agreement with a logarithmic approach of a Stefan-Boltzmann behavior, and thus thermodynamic bulk quantities appear as being dominated by free gluons \cite{Karsch:2001cy,Borsanyi:2011zm}. However, there are a number of arguments against that this is a generic behavior. Already the infrared analysis of section \ref{sfinitet:ir} suggests that even at very high temperatures, at least in the soft magnetic sector, non-perturbative long-range interactions should remain \cite{Cucchieri:2007ta}.

Indeed, there is a very general argument supporting this line of thought \cite{Appelquist:1981vg,Maas:2005ym}: Take the limit of infinite temperature. This is equivalent to the static limit of the theory, as the extension of the time direction shrinks to zero \cite{Appelquist:1981vg}. Thus it becomes effectively three-dimensional. The corresponding three-dimensional Lagrangian can be obtained as an effective field theory \cite{Kajantie:1995dw,Maas:2004se}. In this course the original $A_0$ component of the gluon field becomes an additional adjoint Higgs field. This preserves the number of degrees of freedom, as a three-dimensional gluon field has only one transverse polarization. The constants that appear in the three-dimensional theory, like the Higgs mass, can be obtained by matching with the original theory. The most important quantity in this context is the (now dimensionful) three-dimensional gauge coupling, which to leading order is given by $g_3=g^2T$ with the four-dimensional gauge coupling $g$. The limit actually contains some subtleties, since the starting theory is a renormalizable theory, while the limiting theory is finite in the sense that all renormalization constants are finite, at least in Landau gauge\footnote{From the perturbative point of view it is superrenormalizable, and outside Landau gauge it is actually linear divergent. Both points are of no relevance here.} \cite{Maas:2005hs,Maas:2005rf}. The resulting effective Lagrangian is then given by
\bea
{\cal L}&=&-\frac{1}{4}F_{\mu\nu}^aF^{\mu\nu}_a+\frac{1}{2}(D_\mu^{ab}\phi_b D^\mu_{ac}\phi^c+m_h^2\phi^a\phi_a)+\frac{h}{4}\phi^a\phi_a\phi^b\phi_b\label{l3d}\\
D_\mu^{ab}&=&\delta^{ab}\pdm+g_3f^{ab}_{\;\; c}A_\mu^c\nonumber,
\eea
\no which still requires the usual gauge-fixing techniques. As noted, the coupling constants, $g_3$, $m_h$, and $h$, are not free, but tied to their finite-temperature expressions, and in particular functions of the four-dimensional gauge coupling and the temperature only.

Such a theory is actually confining, which could have been expected on the basis that a three-dimensional Yang-Mills theory is qualitatively not much different from a four-dimensional Yang-Mills theory \cite{Feynman:1981ss}. The adjoint scalar could actually change this \cite{Baier:1986ni}, but for the value of the parameters relevant this is not the case \cite{Karsch:1996aw,Cucchieri:2001tw}. In fact, the adjoint scalar is a residue of the electric sector of the theory, while the Yang-Mills sector corresponds to the magnetic sector. Thus, it should be expected that even the infinite-temperature limit is strongly interacting \cite{Maas:2005ym}.

Motivated by this it is possible to also find non-perturbative quantities in the finite temperature system which do not show a behavior compatible with a perturbatively dominated system, like the spatial string tension \cite{Bali:1993tz}. The system is therefore non-perturbative. Nonetheless, it remains that the thermodynamic bulk quantities are apparently perturbative. The resolution of this apparent contradiction will be discussed below in section \ref{sfinitet:bulk}.

\begin{figure}
\includegraphics[width=\textwidth]{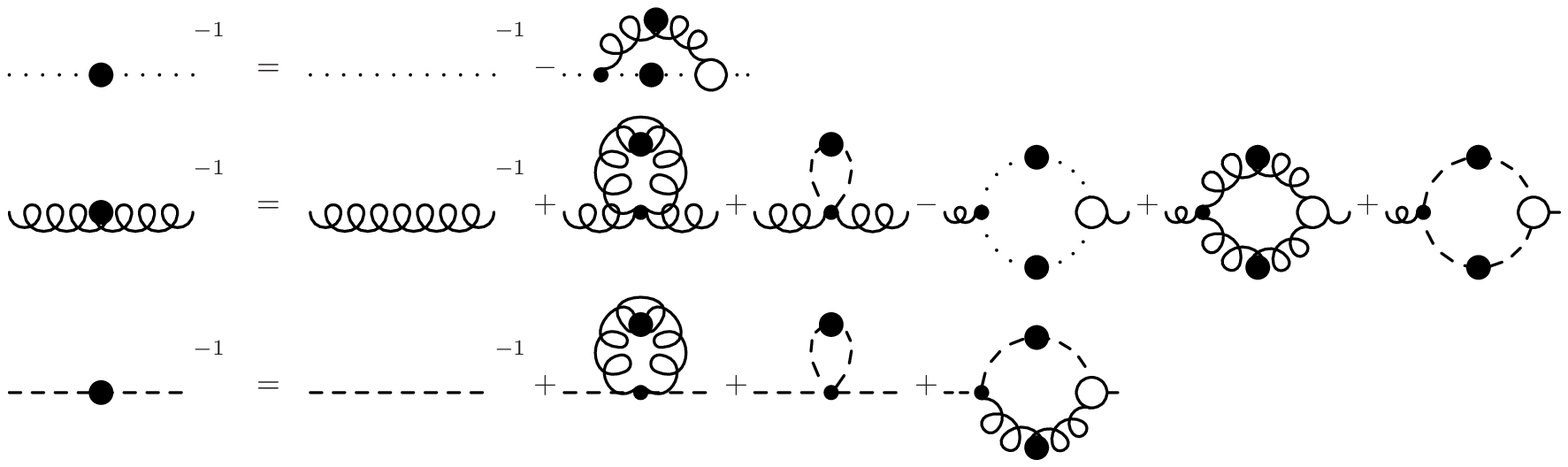}
\caption{\label{fig:3dsys}The graphical representation of the DSE truncation in the infinite-temperature limit used here \cite{Maas:2004se}. Curly lines are gluons, dotted lines are ghosts, and dashed lines are the adjoint scalars. Lines with a circle are full propagators, and open circles are full vertices, which are set bare in this truncation.}
\end{figure}

\begin{figure}
\includegraphics[width=\textwidth]{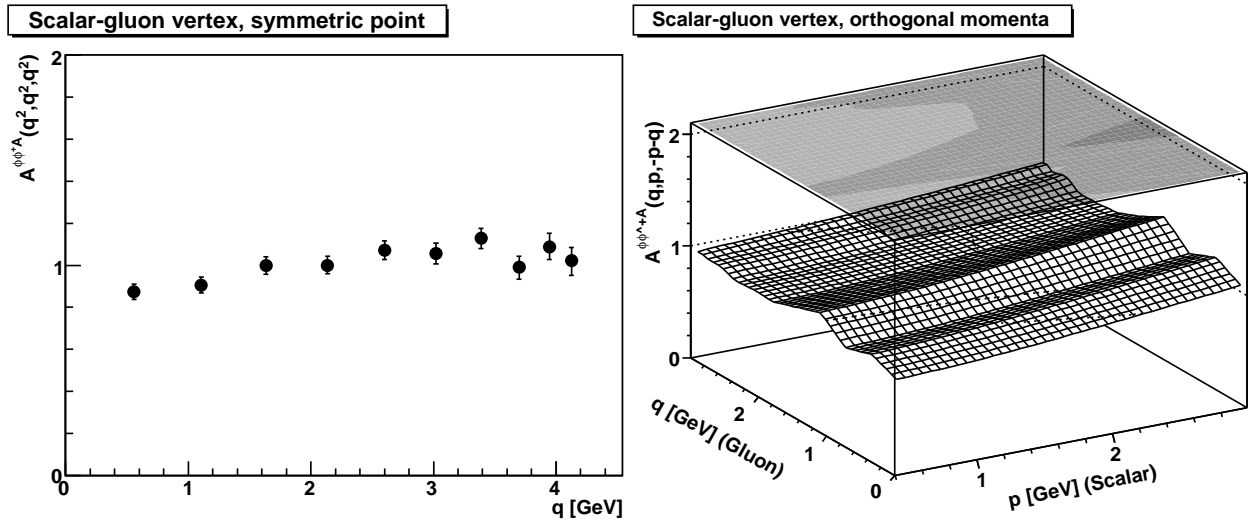}
\caption{\label{fig:agv}The quenched scalar-gluon vertex in three dimensions for su(2) in minimal Landau gauge \cite{Maas:2011yx,Maas:unpublished}. Note that the tensor structure of the scalar-gluon vertex is the same as for the ghost-gluon vertex, \pref{zerot:ggvtensors}. The lattice size is (3.1 fm)$^3$ with $a=0.13$ fm. The mass of the quenched scalar is 1 GeV.}
\end{figure}

Of course, such a system can be equally well solved using either lattice or functional methods. In case of DSEs, the truncation employed is shown in figure \ref{fig:3dsys}. Employing a bare scalar-gluon vertex can be justified at least by quenched lattice results, i.\ e.\ using only test scalars, but not with sea scalars. The result is shown in figure \ref{fig:agv}. As a consequence, the adjoint Higgs decouples in the infrared for any non-zero tree-level mass\footnote{This can radically change for a massless scalar \cite{Maas:2005rf,Macher:2010ad}.}. Since the tree-level mass is essentially the electric screening mass, and thus proportional to the temperature, this is an acceptable assumption.

Also the tadpoles have to be included in a very particular way in this approach, in contrast to the more simpler Yang-Mills case \cite{Maas:2004se,Maas:2005rf}. On the one hand, the finite Higgs mass yields a finite shift of its mass due to a Higgs tadpole. On the other hand, in principle any three-dimensional Higgs-Yang-Mills theory in Landau gauge features linear divergences. However, this particular theory is the infinite-temperature limit of a four-dimensional Yang-Mills theory in Landau gauge. Since temperature cannot introduce new divergences \cite{Das:1997gg}, and in the current truncation no new artificial ones appear \cite{Maas:2004se,Maas:2005hs,Cucchieri:2007ta}, the effective three-dimensional theory cannot have any divergent renormalization constants. This dictates that the additionally introduced coupling constants have to arrange themselves so that no such divergences arise, which has to be taken into account in the truncation, finally yielding the truncation used here and described in detail in \cite{Maas:2004se}.

\begin{figure}
\includegraphics[width=0.5\textwidth]{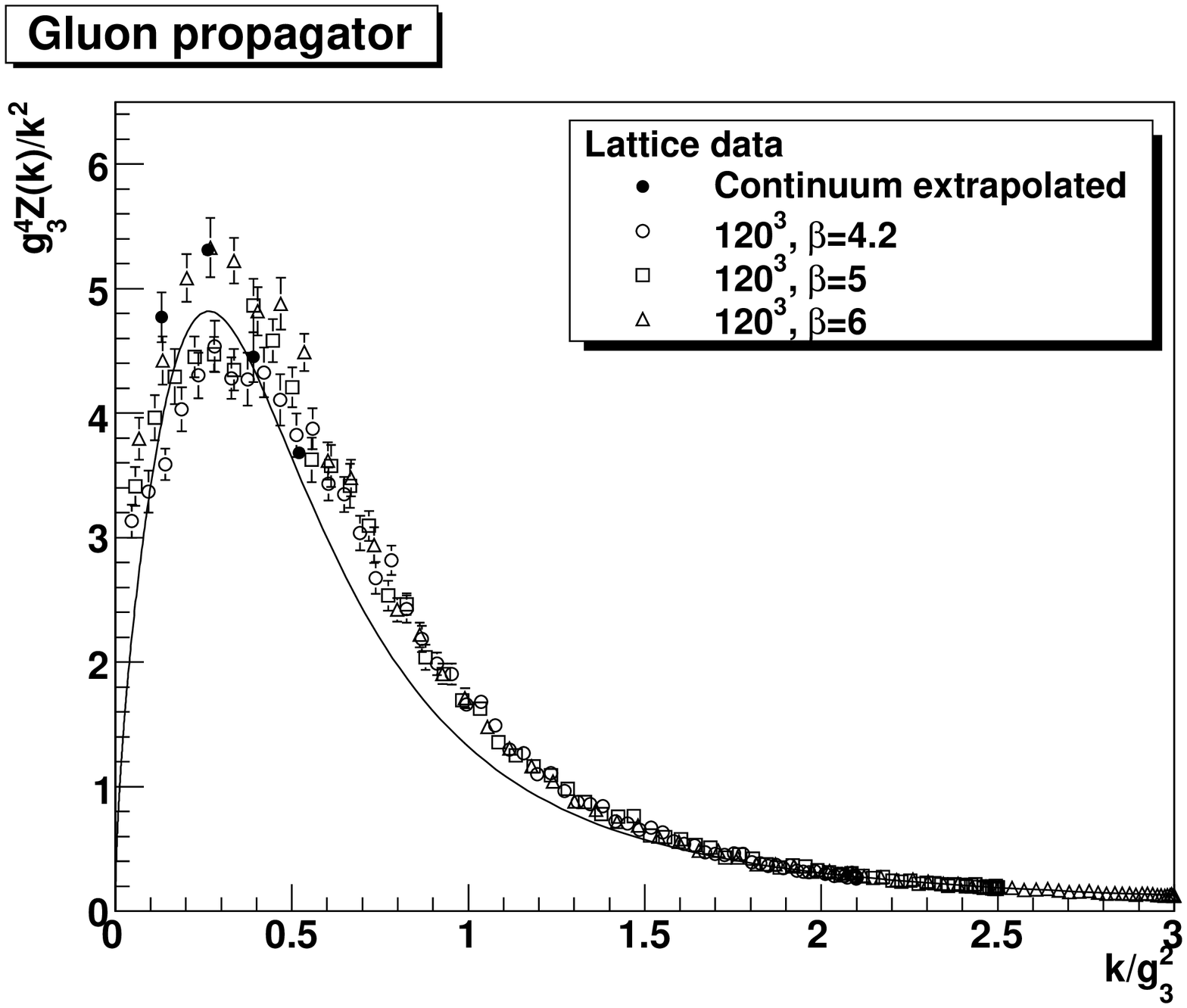}\includegraphics[width=0.5\textwidth]{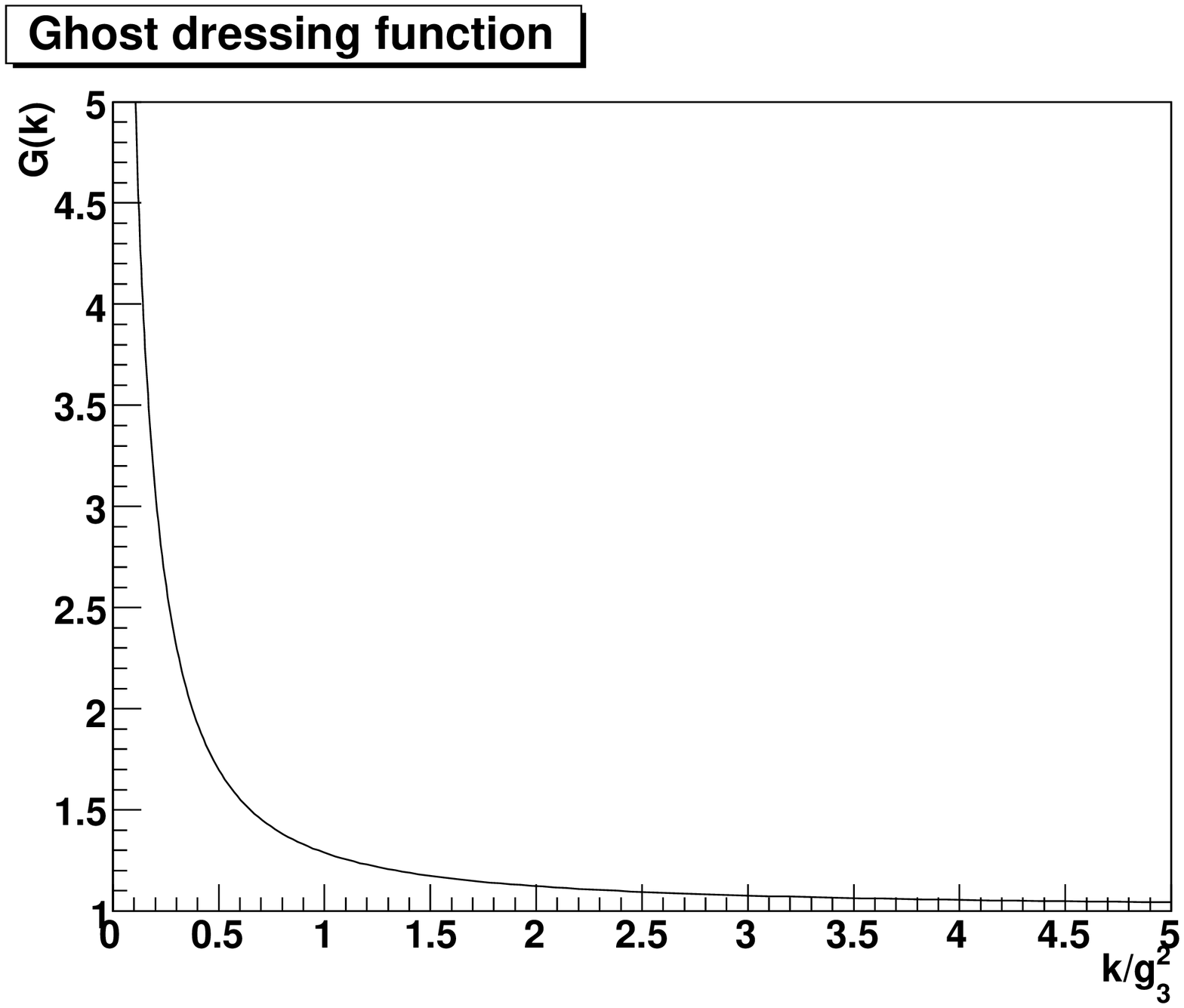}\\
\includegraphics[width=0.5\textwidth]{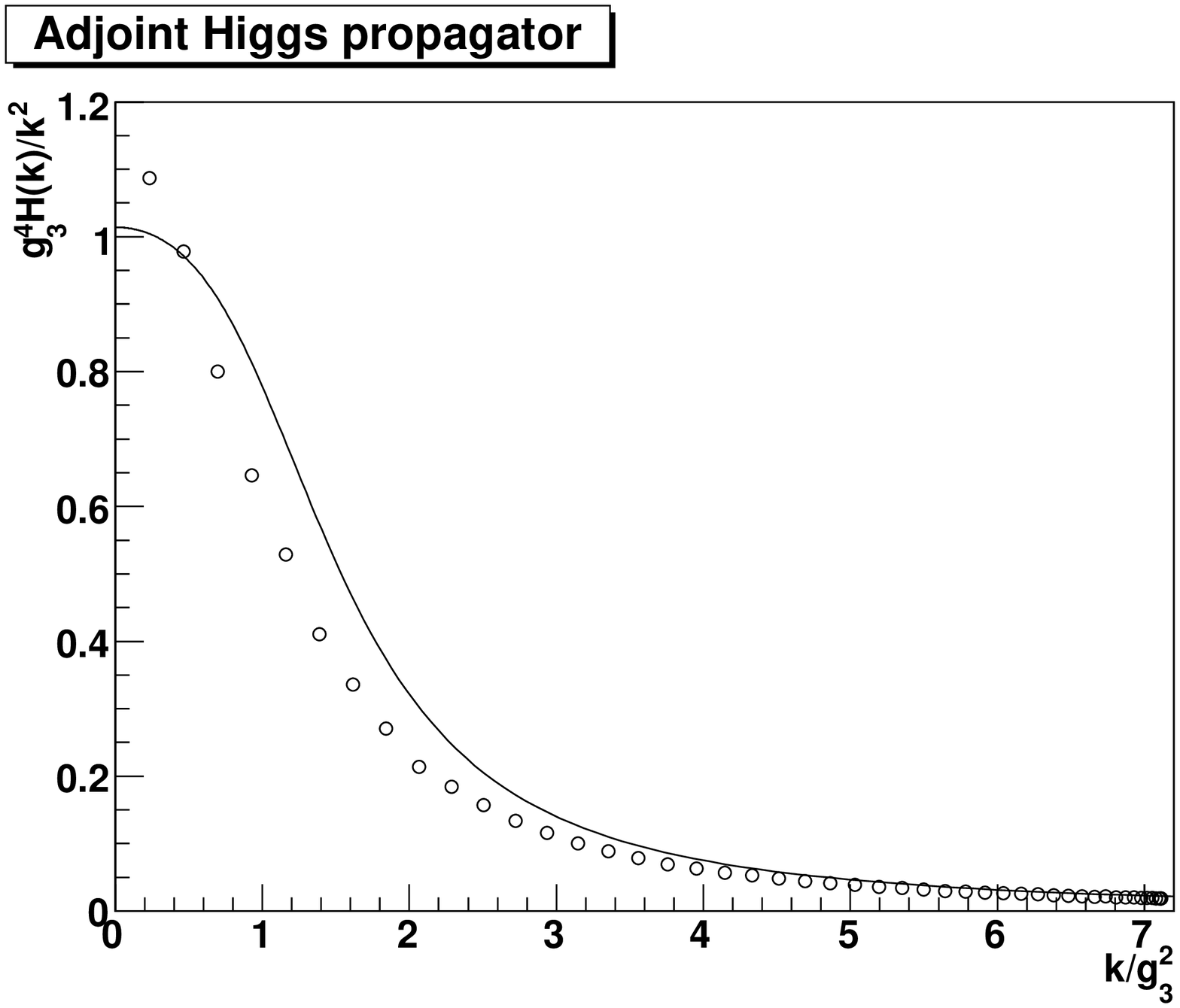}\includegraphics[width=0.5\textwidth]{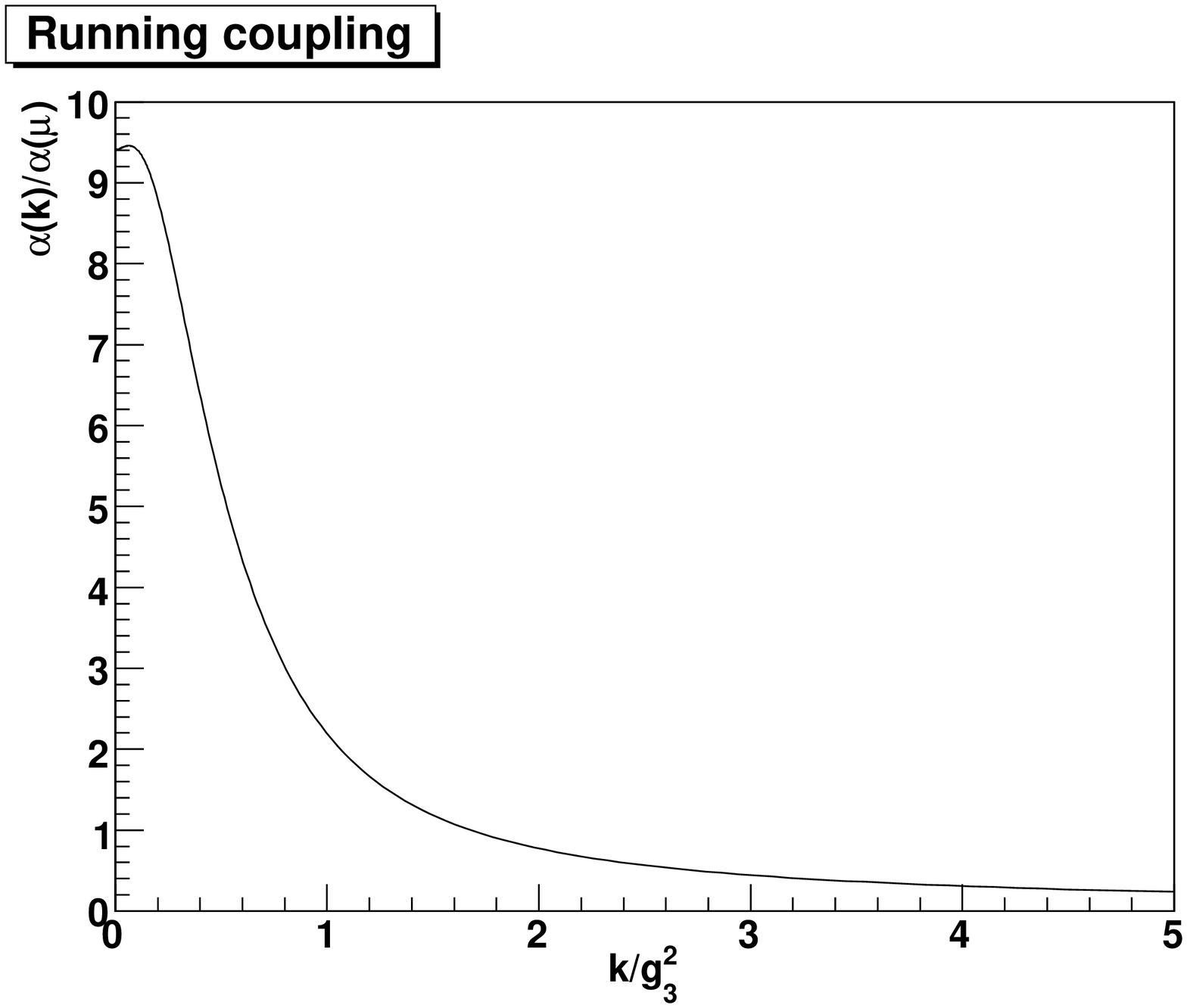}
\caption{\label{fig:infinitet}The results for the su(2) infinite temperature limit from DSEs (in the scaling case) \cite{Maas:2004se} and lattice calculations (in the minimal Landau gauge) \cite{Cucchieri:2001tw}. The top-left panel shows the gluon propagator (the lattice data are for the Yang-Mills case \cite{Cucchieri:2003di}, for comparison), the top-right panel the ghost dressing function, the bottom-left panel the adjoint Higgs propagator, and the bottom-right panel the running gauge coupling.}
\end{figure}

Similarly, the system can be simulated on the lattice \cite{Cucchieri:2001tw,Karsch:1996aw}. Again, it is necessary to tune the three independent constants in the system such that the theory is finite. This is a highly non-trivial task \cite{Cucchieri:2001tw}, and guidance can be taken from the perturbative equivalence \cite{Kajantie:1995dw}, in particular since this problem arises in the deep ultraviolet. It is then possible to determine the propagators of all involved fields in both lattice and DSE calculations. The results are shown in figure \ref{fig:infinitet}. It turns out that the presence of the Higgs is actually a negligible effect for the propagators of the gluon and the ghost, and hence the result for them is almost identical to the one of a three-dimensional Yang-Mills theory \cite{Cucchieri:2001tw,Maas:2004se}. This is in line with the argumentation above using functional methods that the chromoelectric sector essentially decouples at sufficiently high temperature, being dominated by its screening mass. The latter encodes the influence of the hard modes on the chromoelectric sector in the infinite temperature limit.

\subsubsection{Schwinger functions and confinement}\label{sfinitet:schwinger}

\begin{figure}
\begin{center}
\includegraphics[width=0.45\textwidth]{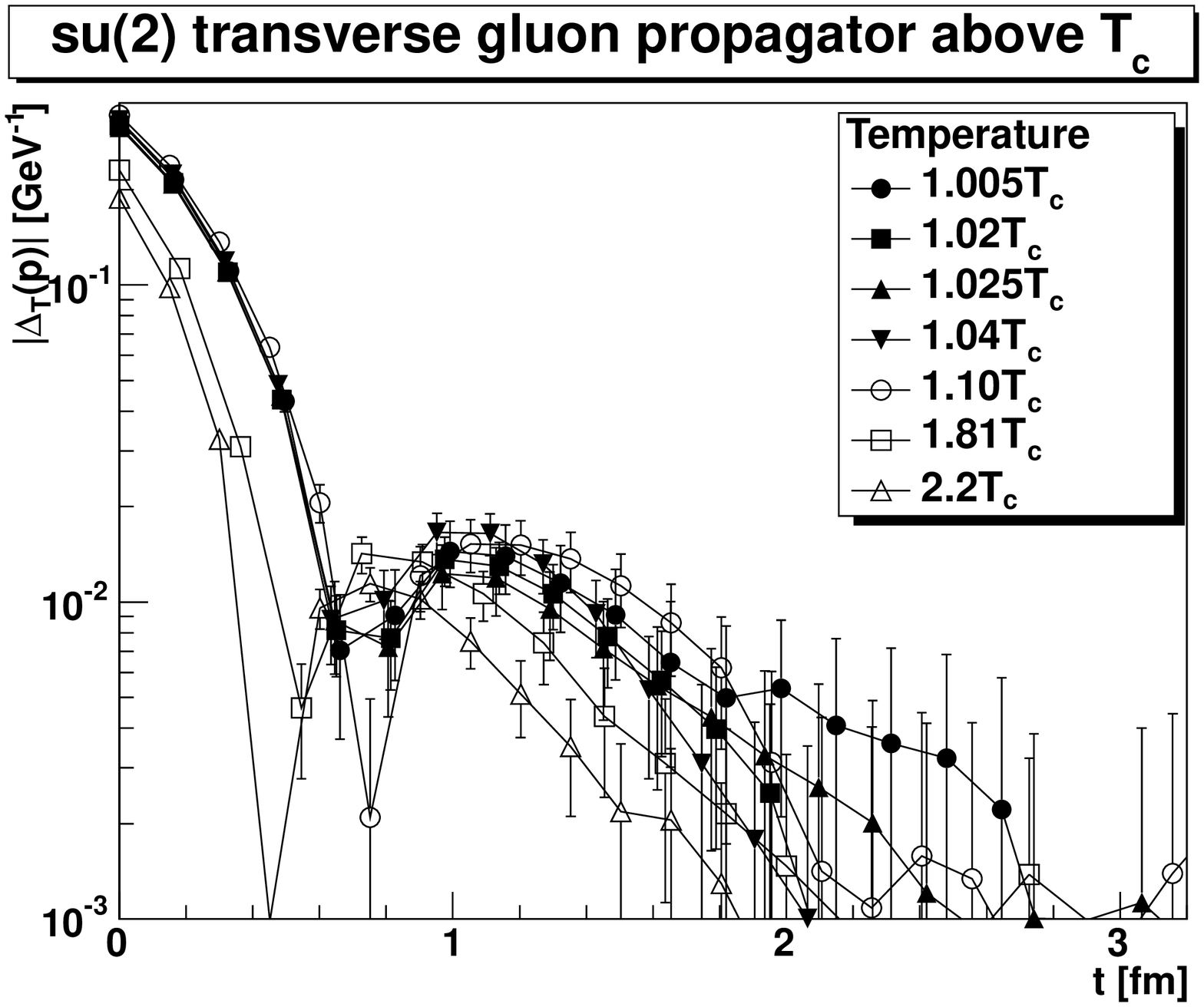}\includegraphics[width=0.45\textwidth]{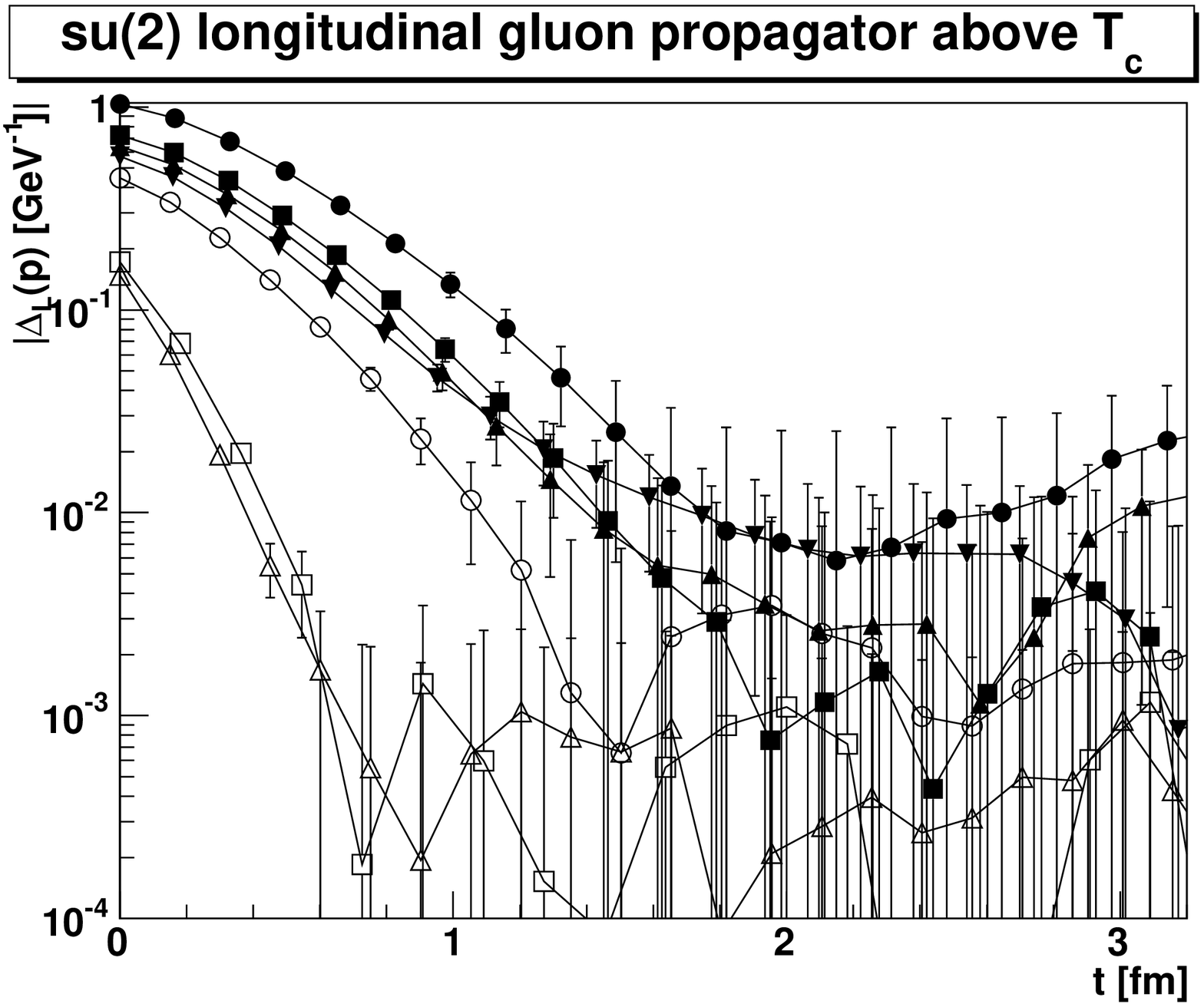}\\
\includegraphics[width=0.9\textwidth]{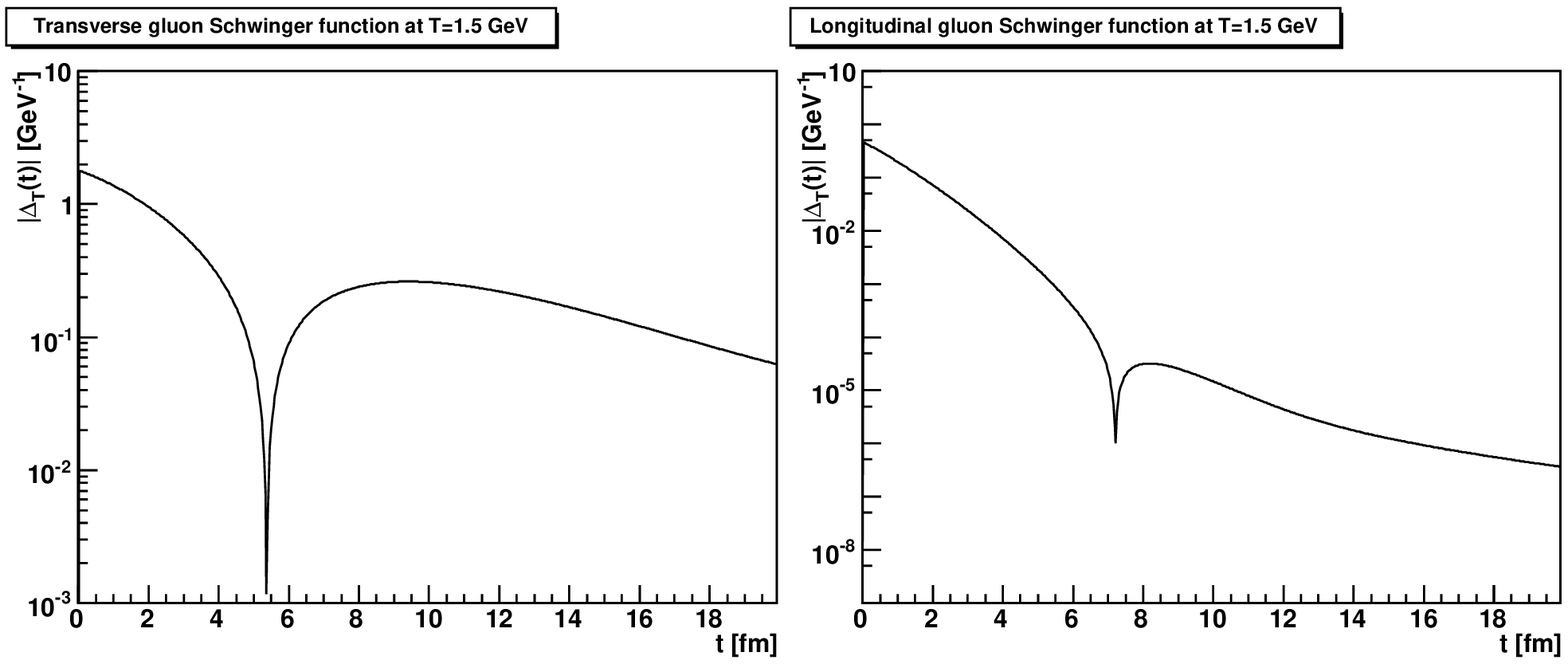}\\
\includegraphics[width=0.45\textwidth]{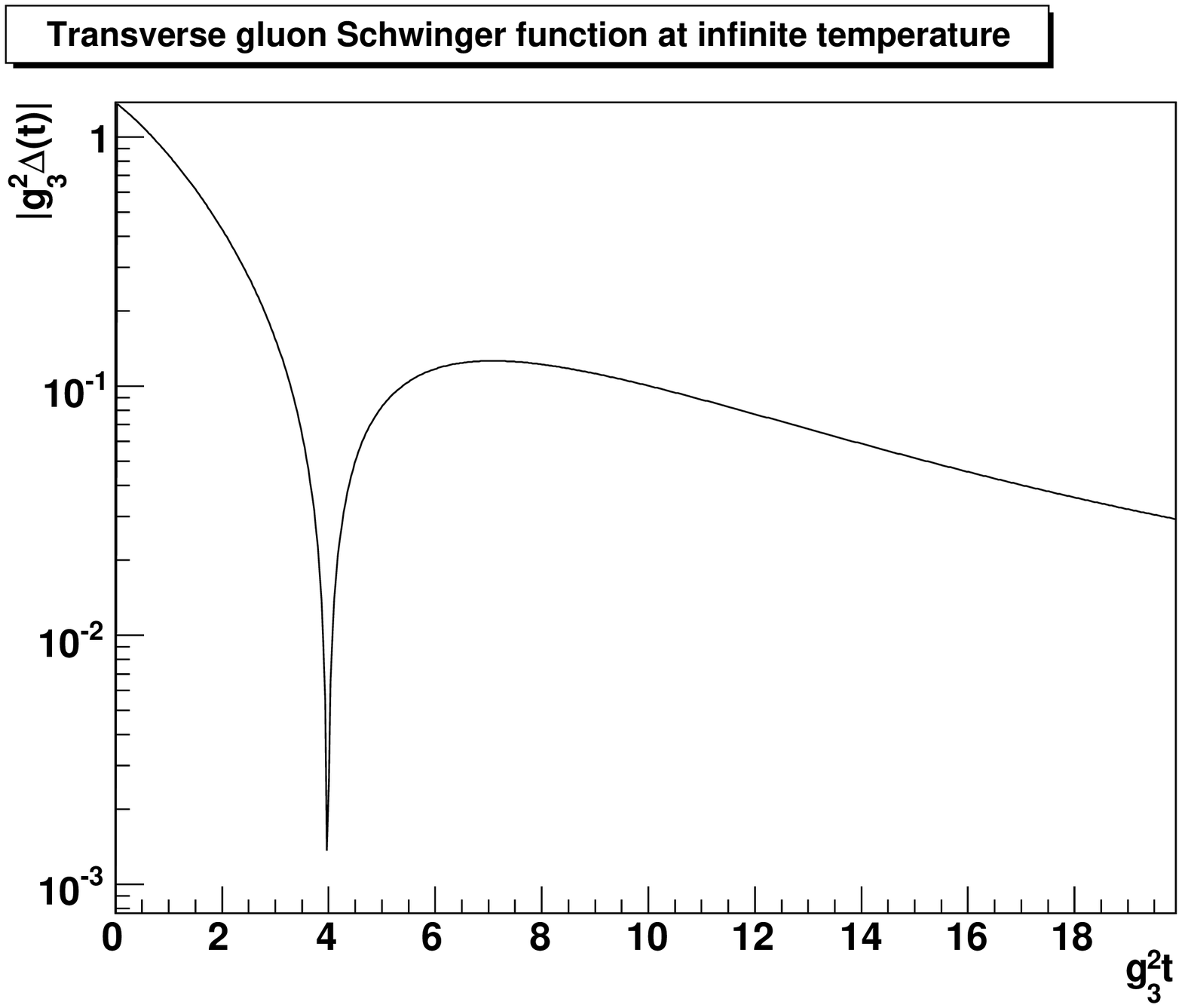}\includegraphics[width=0.45\textwidth]{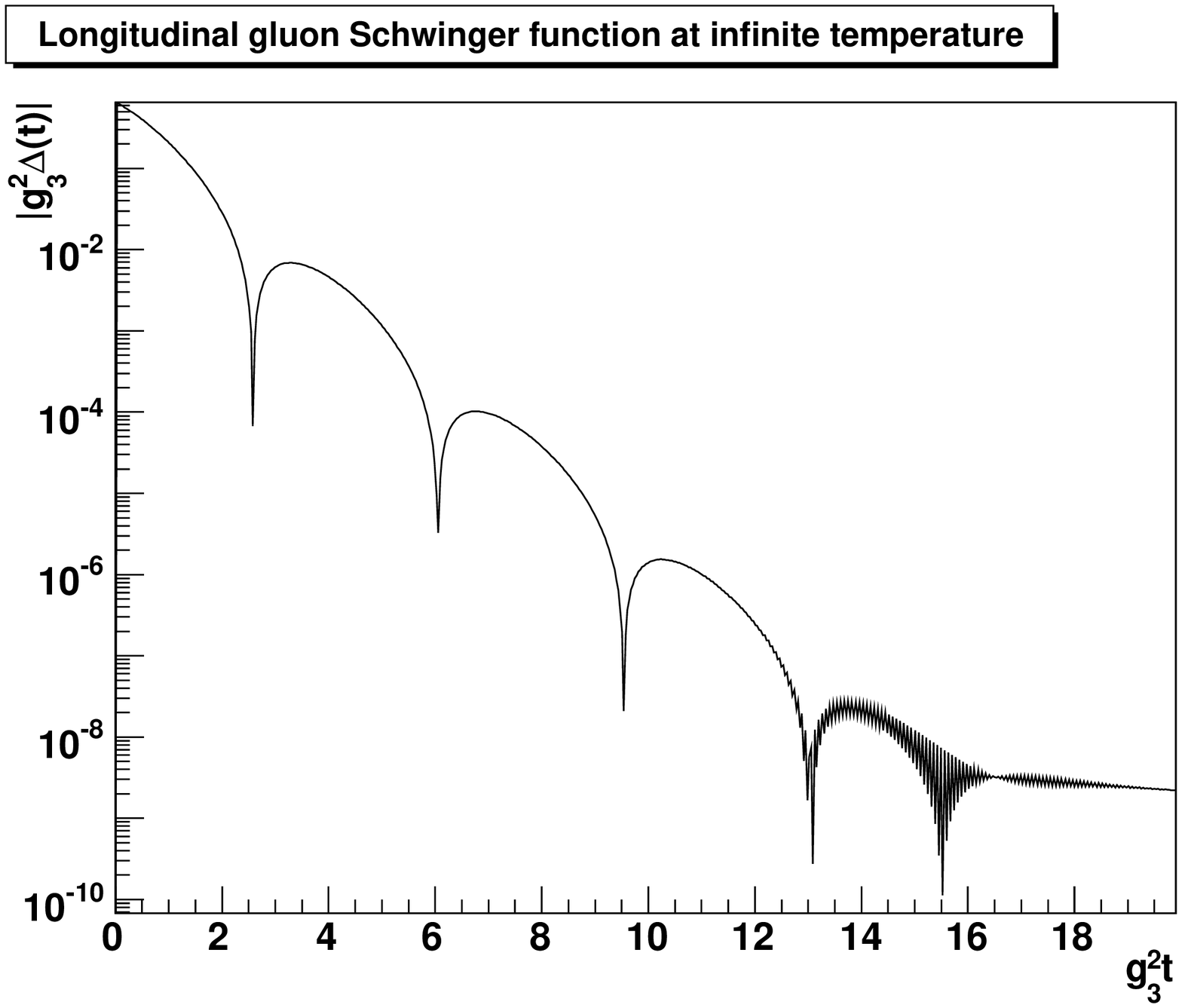}
\end{center}
\caption{\label{fig:ftschwinger}The soft transverse (left panel) and longitudinal (right panel) gluon Schwinger functions above the phase transition temperature. The top panels show the results from lattice calculation in the minimal Landau gauge \cite{Maas:unpublished}, the middle panels using DSEs in the scaling case at about 1.5 GeV temperature \cite{Maas:2005hs}, and the bottom panel in the infinite temperature limit using DSEs in the scaling case \cite{Maas:2004se}. Note that the 'time' in this equilibrium case is actually along a spatial direction, as discussed in section \ref{quant:schwinger}.}
\end{figure}

Once the propagators are available, it is natural to investigate once more what their analytic structure is. From the infrared analysis of section \ref{sfinitet:ir} it can already be expected that the transverse gluon propagator will not be an asymptotic state in very much the same way as in four dimensions. Indeed, the corresponding Schwinger function shows this behavior, as can be seen in figure \ref{fig:ftschwinger} for the finite temperature case, though the employed lattice volumes are not sufficient to make any final decision yet \cite{Cucchieri:2011di}. Only that the Schwinger function is not monotonously decaying at these volumes can be observed. Note that for the not shown hard modes their effective tree-level mass is so large that it is not yet possible to find any significant deviation from a massive particle-like behavior, nor any trace of their thermal width.

Thus, the magnetic sector again appears to be essentially unaffected by temperature. In particular, this implies that magnetic gluons are not asymptotic states at all temperatures, in contrast to the idea of a plasma of free gluons. However, given the problems encountered when trying to construct a gauge-invariant gluon state, this outcome is rather comforting.

The situation is somewhat different for the chromoelectric gluon. First of all \cite{Cucchieri:2007ta}, at the perturbative level the zero component of the gluon in Landau gauge is a member of an elementary (perturbative) BRST quartet \cite{Bohm:2001yx,Kugo:1979gm}. As such, it should not be observable, i.\ e., it is effectively confined. This is also not altered at finite temperature \cite{Das:1997gg}. However, pending a final understanding of a non-perturbative BRST construction, this is at the current time not fully satisfactory.

Therefore, again its Schwinger function is quite interesting, which is also shown in figure \ref{fig:ftschwinger}, for both the infinite and finite temperature case\footnote{It should be noted that the result is qualitatively independent of the interaction strength of the Higgs with the gluons in the infinite-temperature case \cite{Maas:2005rf}, a nice display of how even arbitrarily weak interactions can have an influence.}. At infinite temperature, both the propagator and the Schwinger function $\Delta_L$ can be well fitted by the expression \pref{zerot:unstable} for an unstable particle \cite{Maas:2004se}. Indeed, the propagator is in very good agreement with a double complex pole structure. It is even possible, at least in the infinite temperature case, to identify a subleading pole structure. The lattice data, however, are not yet sufficiently precise to make a final statement, except that the behavior is different from the transverse case: The Schwinger function, and a possible zero crossing of it, seems to depend stronger on temperature than in the transverse case. Thus, it is in principle possible that there could be a change of the analytic structure either when going to non-zero temperature or at the phase transition. In particular, the latter option is interesting, as this would be closer to the idea of a deconfining transition.

\subsection{Thermodynamic quantities}\label{sfinitet:bulk}

\subsubsection{Thermodynamic potential}

With all these non-perturbative interactions around, it is a non-trivial question, why various results \cite{Karsch:2001cy,Kajantie:2002wa,Borsanyi:2011zm} show that for asymptotically large temperatures a Stefan-Boltz\-mann-like behavior is observed for bulk thermodynamic quantities. It has been shown, employing the dimensional reduction as a tool, that the thermodynamic potential can be quite accurately reproduced using perturbation theory and only one further effective parameter to include sub-leading non-perturbative contributions of the three-dimensional Yang-Mills theory \cite{Kajantie:2002wa,Hietanen:2008tv}. These results show that the non-perturbative, effectively three-dimensional, contribution is sub-leading by at least one power of the temperature compared to the hard interactions, and thus the bulk thermodynamic quantities at sufficiently high temperatures are essentially given by the perturbative expression. Arguments based on DSEs \cite{Maas:2004se} and the Gribov-Zwanziger effective theory \cite{Zwanziger:2006sc} are in-line with this conclusion.

From the physical point of view this is just the statement that the average energy scale of interactions is given by the temperature, and this is then deep in the asymptotic domain. However, this is essentially an off-shell process, since, as discussed previously, the gluons themselves are likely confined. Therefore, a more appropriate picture may be a dense glueball soup, which undergoes rapidly hard collisions dominated by hard partonic processes \cite{Cucchieri:2007ta}.

Given the manifestation of the phase transition in the correlation functions, it is an interesting question whether also the other thermodynamic information can be obtained from the correlation functions. Indeed, it is, in principle, possible to determine the thermodynamic potential by a 2PI/Luttinger-Ward-Cornwall-Jackiw-Tomboulis construction \cite{Luttinger:1960ua,Cornwall:1974vz,Haeri:1993hi,Carrington:2003ut,Carrington:2004sn,Arrizabalaga:2002hn,Gruter:2004bb,Braun:2009gm,Nickel:2006vf}. The exact expression is given by \cite{Gruter:2004bb}
\bea
V(D_{\mu \nu},D_G)&=& V_0(D_{\mu \nu},D_G)+V_2(D_{\mu \nu},D_G) \label{cjtfull}\\
V_0(D_{\mu \nu},D_G)&=& \int \frac{d^4p}{(2\pi)^4} {\rm tr}\left\{ \frac12 \left[{D^{\mathrm{tl}}_{\mu \alpha}}^{-1}(p)D_{\alpha \nu}(p) -\right. \right.\left. g_{\mu \nu} \right] -\frac12 \ln \left({D^{\mathrm{tl}}_{\mu \alpha}}^{-1}(p)D_{\alpha \nu}(p) \right) \nonumber \\
&&-\left[D_G(^{\mathrm{tl}}p)^{-1}D_G(p)-1 \right]+\ln \left(D_G^{\mathrm{tl}}(p)^{-1}D_G(p) \right) \Big\} \nn,
\eea
\no where tl denotes the tree-level propagators. The contribution $V_2$ contains contributions from the vertices, which are dropped in the truncation at propagator level. In the infinite-temperature limit, this expression reduces to \cite{Maas:2004se}
\bea
\Omega&=&\frac 1 2 N_A T\sum\int\frac{d^3q}{(2\pi)^3}\left(-\ln\left(\frac{Z(q)}{Z_0(q)}\right)+\left(\frac{Z(q)}{Z_0(q)}-1\right)\right.\nonumber\\
&&\left.+\ln\left(\frac{G(q)}{G_0(q)}\right)-\left(\frac{G(q)}{G_0(q)}-1\right)-\frac{1}{2}\ln\left(\frac{H(q)}{H_0(q)}\right)+\frac{1}{2}\left(\frac{H(q)}{H_0(q)}-1\right)\right)\label{cjt},
\eea
where $Z_0$, $G_0$, and $H_0$ are the tree-level dressing functions of the dimensionally-reduced theory \pref{l3d}
\bea
Z_0(p)=G_0(p)=1\nonumber\\
H_0(p)=\frac{p^2}{p^2+m_h^2}.\nonumber
\eea
\no By dropping the expression $V_2$, although consistent with the truncation scheme of the functional equations, this approximation has the serious drawback of not closing thermodynamically \cite{Gruter:2004bb}. In particular, this yields spurious divergences in the necessary cut-off, which are very hard to determine at finite temperature. The spurious divergences can be better isolated and subtracted in the infinite-temperature limit \cite{Maas:2004se,Maas:2005rf}, yielding an expression proportional to the temperature cubed, and therefore as expected sub-leading compared to the Stefan-Boltzmann behavior generated by the hard interactions \cite{Kapusta:2006pm}.

\begin{figure}
\includegraphics[width=\textwidth]{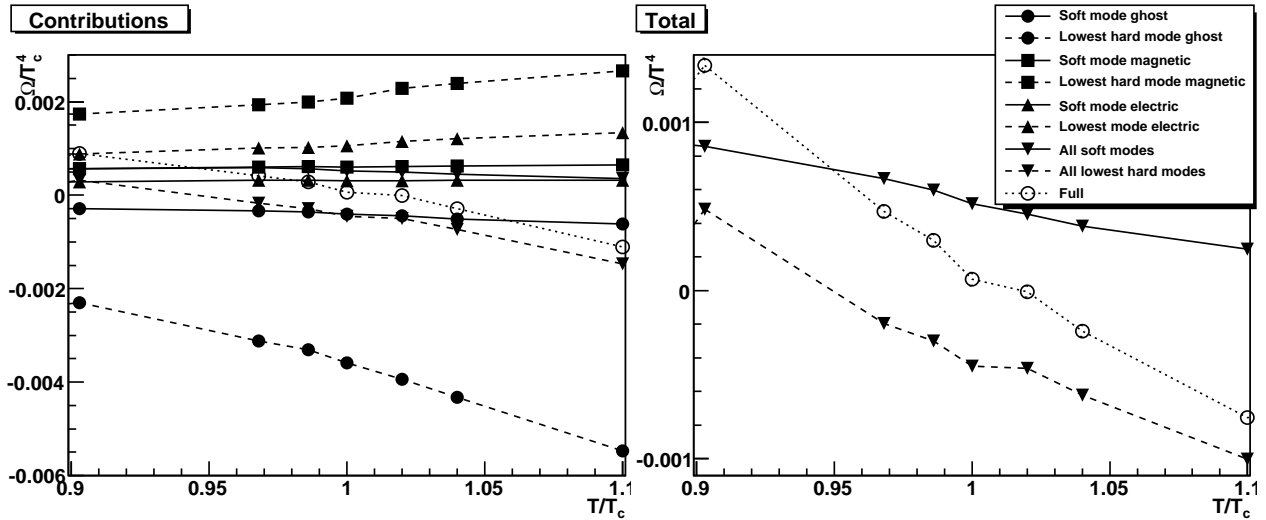}
\caption{\label{fig:ftia}Results for the approximation \pref{cjtfull} of the thermodynamic action using the soft mode and the lowest hard mode and a three-dimensional cut-off of 1.5 GeV for the lattice data \cite{Maas:unpublished}.}
\end{figure}

A possibility to obtain a comparison at finite temperature is to use the lattice result in the expression \pref{cjtfull} at the same truncation level\footnote{I am grateful to Jan M.\ Pawlowski for motivating this investigation.}. However, it suffers from discretization artifacts, which require to include an artificially low cut-off of about 1 to 2 GeV with the available lattice data. The corresponding temperature-dependence of this approximation is shown in figure \ref{fig:ftia}. It is essentially dominated by the truncation of the Matsubara sum, and has little resemblance to the expected behavior. This indicates that possibly contributions from higher energy and higher Matsubara frequencies are important. This is to be expected,  since the cut-off imposed is below $2\pi T_c\approx 1.8$ GeV, and thus below the natural scale of the hard interactions at the phase transition. Thus interesting in principle, this approach requires much better input.

Of course, other bulk quantities, like the entropy, can in principle be determined as soon as this problem is under control \cite{Kapusta:2006pm}.

\subsubsection{Order parameters}\label{sorder}

As already noted, the critical temperature and the order of the phase transition can be accessed rather directly from the propagators. It is therefore interesting whether it is possible to also determine other order parameters. Of course, this is in principle not of direct relevance, since QCD likely does not exhibit a genuine phase transition at finite temperature at all \cite{Philipsen:2010gj}, though this may change for the standard model \cite{vonSmekal:2010la}. Moreover, for most gauge algebras the finite temperature phase transition of Yang-Mills theories is first order \cite{Lucini:2005vg,Holland:2003kg,Braun:2010cy,Holland:2003jy,Pepe:2006er,Cossu:2007dk}. Therefore no order parameters need to be associated with them \cite{Negele:1988vy}. However, it has been quite a striking observation that the order parameters employed for su(2) Yang-Mills theory, in particular the Polyakov loop \cite{Greensite:2003bk}, remain to be rather good indicators for the first order phase transitions \cite{Lucini:2005vg,Holland:2003kg,Braun:2010cy,Holland:2003jy,Pepe:2006er,Cossu:2007dk}, or even for the cross-overs observed in theories with matter fields \cite{Philipsen:2010gj,Cheng:2009be,Borsanyi:2010cj,Langguth:1985dr,Evertz:1985fc,Bonati:2009pf}. It is thus worthwhile to determine them using the present setup.

This will be done here for the one central order parameter in Yang-Mills theory, the temporal Polyakov loop \cite{Gattringer:2010zz}
\be
P=\left\langle\tr\exp\left(\int_{\cal C} dt A_0\right)\right\rangle\label{sfinitet:polyakov},
\ee
\no where the closed curve ${\cal C}$ wraps once around the compactified time direction. It is associated with the center symmetry Z$_N$ of SU($N$) theory\footnote{Strictly speaking, this symmetry has to be factored out in the standard model for anomaly reasons \cite{O'Raifeartaigh:1986vq}. However, even for theories which do not possess a non-trivial center, the Polyakov loop is an acceptable order parameter \cite{Holland:2003jy,Pepe:2006er,Cossu:2007dk,Braun:2010cy}, and can therefore be expected to remain so also for the standard model. It is then, however, no longer related to a symmetry-breaking transition, and is also not strictly zero in either phase \cite{Danzer:2008bk}, though to a very good approximation it turns out to be so \cite{Danzer:2008bk,Pepe:2006er,Cossu:2007dk}. There are further arguments concerning the product-group structure of the standard model suggesting its usefulness \cite{vonSmekal:2010la}.}, being non-zero if the symmetry is broken \cite{Greensite:2003bk}.

Unfortunately, the expression for the Polyakov loop \pref{sfinitet:polyakov} is an exponential in the fields. Therefore, as for any Wilson line \cite{Montvay:1994cy}, it is an infinite series of contracted correlation functions of arbitrary high order. Therefore, an exact evaluation using correlation functions is in general not possible.

As a consequence, a number of approximate evaluation schemes have been developed, aside from the direct evaluation using lattice gauge theory \cite{Gattringer:2010zz}. Two of them are closely tied to the correlation functions.

One is based on the Weisz-potential to determine an upper bound of the Polyakov loop \cite{Marhauser:2008fz,Braun:2010cy,Braun:2009gm,Braun:2007bx}. This approach involves only the ghost and gluon propagators, and even with the approximation of using their zero-temperature behavior provides rather good qualitative and even quantitative results. This indicates that the dynamics of the Polyakov-loop at the phase transition is not tied strongly to the finite-temperature behavior of the propagators, but is dominated by the interaction strength mediated by the gluon propagator essentially at the scale of the temperature.

Another possibility is to use dual observables, which denote a class of observables obtained by performing Fourier transformations of quantities with respect to the (fermionic) boundary condition of a test particle. I.\ e., the boundary condition is taken to be not just (anti-)periodic, but periodic up to a phase $\exp(i\phi)$. The simplest example is using a test quark\footnote{An alteration to a sea quark corresponds to the introduction of an imaginary chemical potential \cite{Roberge:1986mm}, as discussed in \cite{Braun:2009gm}.}. The momentum integral of the trace of the quark propagator $S$ yields then the gauge-invariant dual chiral condensate \cite{Gattringer:2006ci,Fischer:2009wc}
\be
 \langle\bar{\psi}\psi \rangle_{\varphi} =  Z_2\, C_A\, T\sum_{n}\int\frac{d^3\vec p}{(2\pi)^3}\tr S\left(\vec{p},2\pi T\left(n+\frac{\varphi}{2\pi}\right)\right), \label{trace}
\ee
\no where $Z_2$ is the wave-function renormalization constant of the quark, and the trace is over both Dirac and color space. Of course, at $\varphi=\pi$ this is just the usual, renormalization-group-dependent quark condensate.

\begin{figure}
\includegraphics[width=0.5\textwidth]{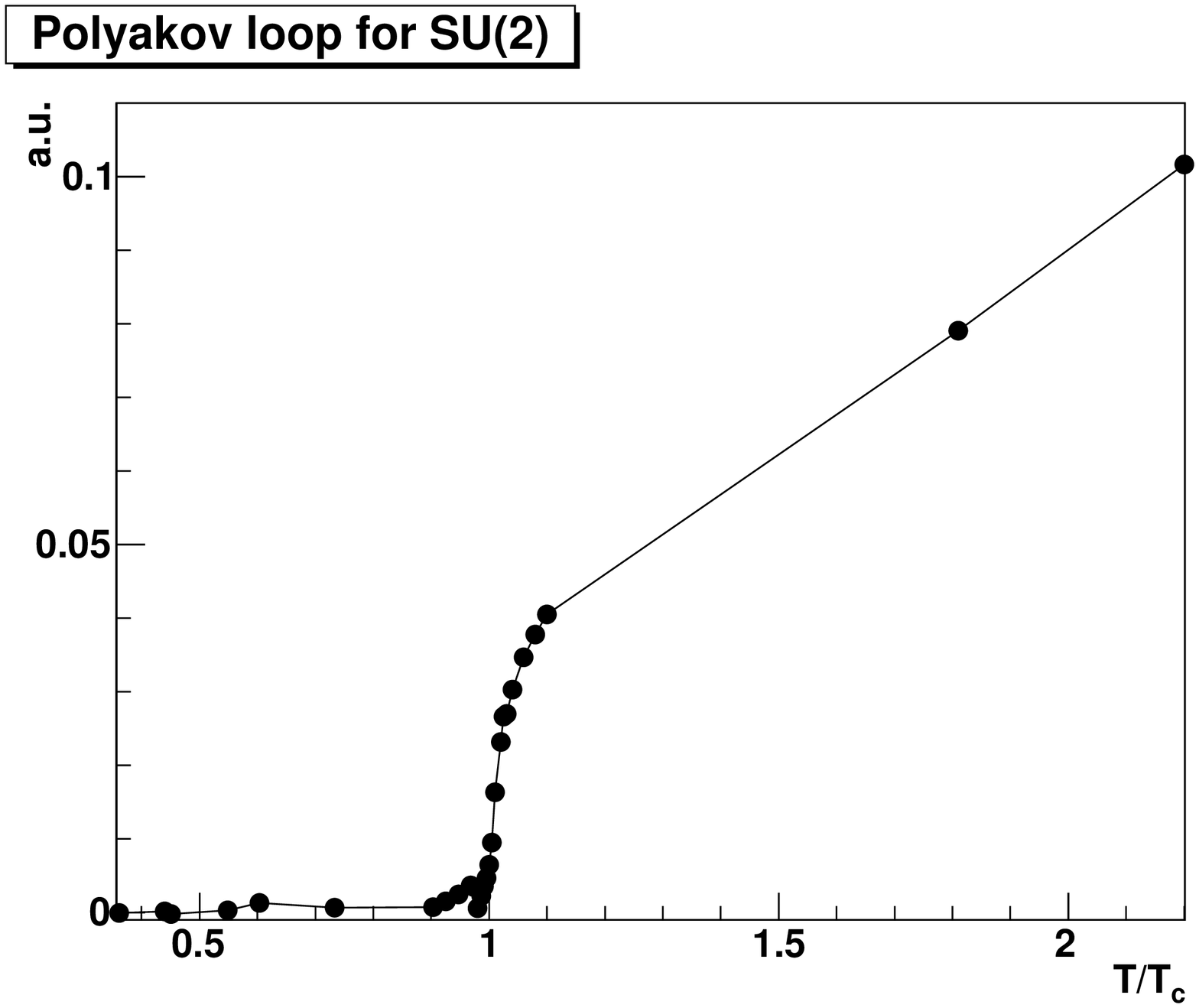}\includegraphics[width=0.5\textwidth]{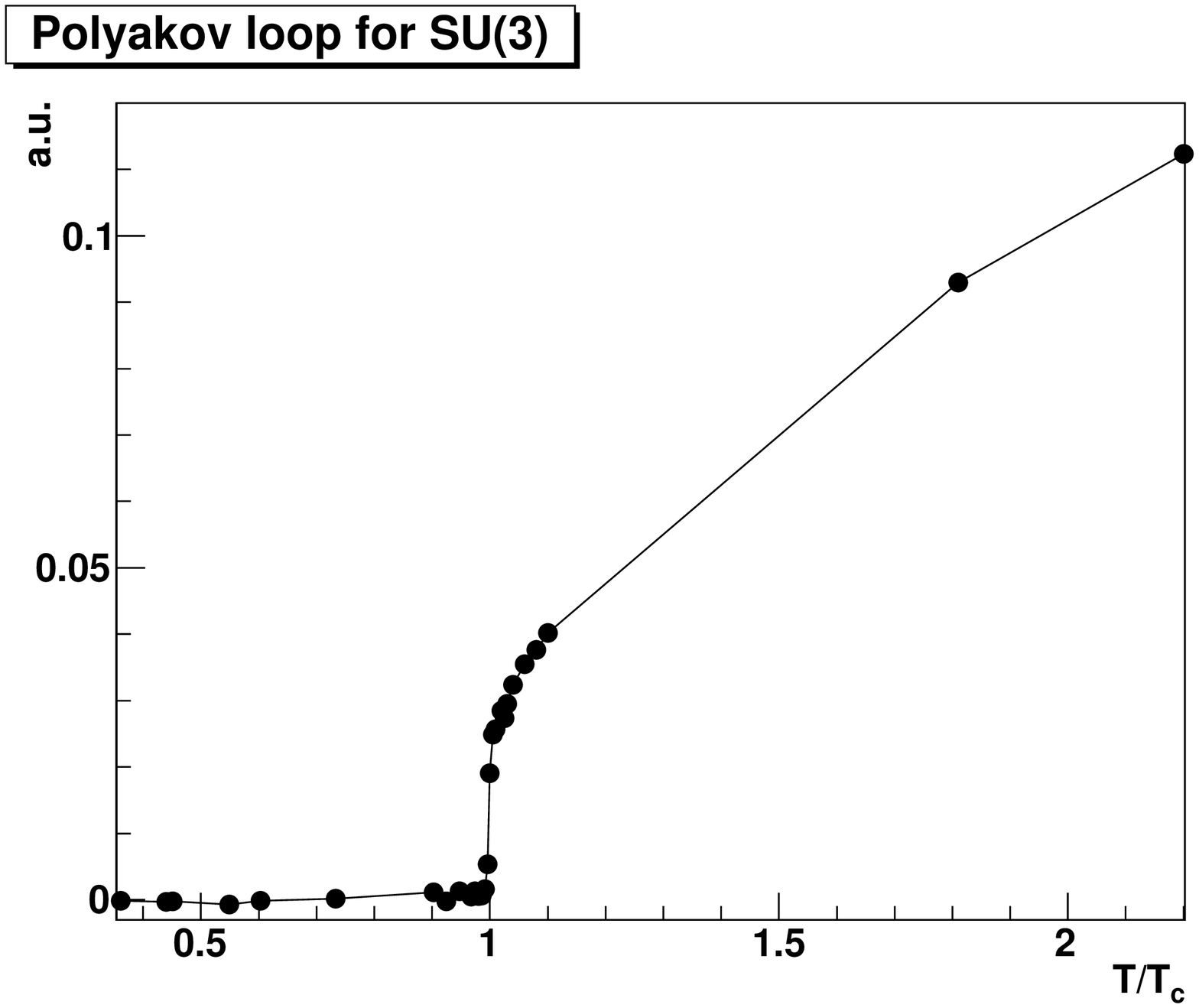}\\
\includegraphics[width=0.5\textwidth]{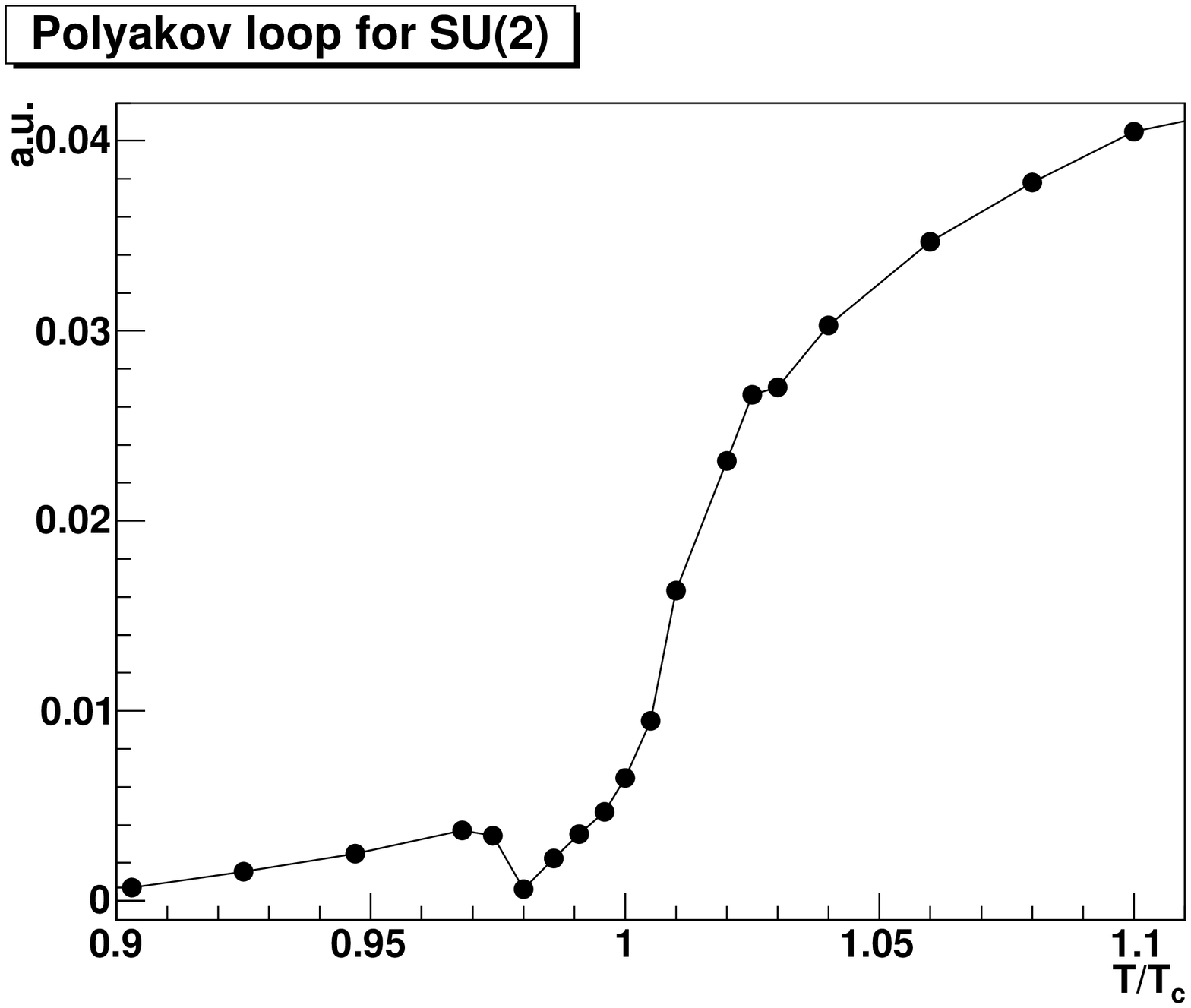}\includegraphics[width=0.5\textwidth]{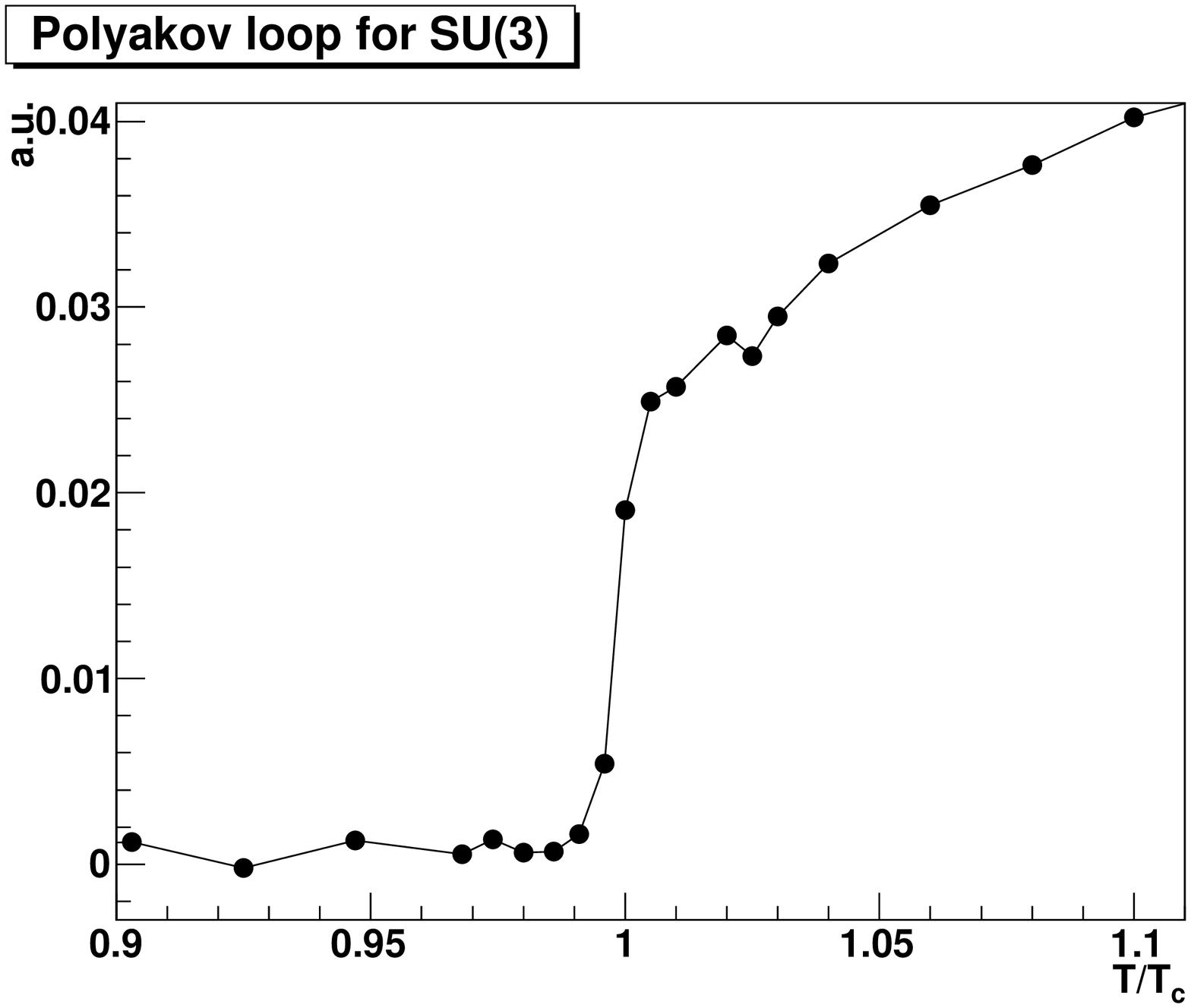}
\caption{\label{fig:ft-polyakov}The Polyakov loop in the quenched approximation obtained from the dual chiral condensate \pref{trace} by Fourier-transformation \pref{dual} for su(2) (left panel) and su(3) (right panel) \cite{Fischer:2010fx,Luecker:2011dq}. The lower panels show a magnification around the critical temperature.}
\end{figure}

It can now be shown that the corresponding Fourier transform
\be
\Sigma = \int_0^{2\pi} \frac{d \varphi}{2\pi} e^{-i\varphi}\langle \overline{\psi} \psi \rangle_\varphi \label{dual}
\ee
\no equals the Polyakov-loop in the quenched case \cite{Gattringer:2006ci}. Thus, it is only necessary to determine the trace of the quenched quark propagator \cite{Fischer:2009wc}. This can be done either by lattice calculations \cite{Zhang:2004gv} or using functional equations\footnote{Where currently actually lattice results for the gluon propagator are used as input, and an assumption has to be made on the quark-gluon vertex \cite{Fischer:2009wc,Fischer:2009gk,Fischer:2010fx}.}. The results obtained with the latter method are shown in figure \ref{fig:ft-polyakov} for the quenched su(2) and su(3) case. Though the involved approximations make the Polyakov loop not exactly zero in the low temperature phase, the phase transition is clearly discernible. However, since the available input data, as shown in figure \ref{fig:t-em}, do not clearly signal the properties of the phase transition yet, the distinction of su(2) and su(3) is also not completely evident in the Polyakov loop yet. Thus, again better data for the gluon propagator would be necessary for this purpose. Nonetheless, this demonstrates how it is possible to access thermodynamic order parameters using correlation functions.

\subsection{Vertices}

As remarked, vertices become tremendously complex at finite temperature due to the large number of possible vertex structures once Lorentz/Euclidean symmetry is no longer manifest \cite{Das:1997gg,Blaizot:2001nr}. As a consequence, at non-zero temperature there have been only few investigations of the vertices beyond perturbative evaluations yet. The ghost-gluon vertex has been investigated in \cite{Fister:2011ym}, and found to be moderately affected by the temperature, but in a way intricately linked to the propagators.

In the infinite-temperature limit it is possible to also make some statements about the vertices, since the tensor structure becomes simple once more. However, even in this case, reliable investigations without quenching the adjoint Higgs are not yet available. The quenched case has been shown in figures \ref{fig:ggv}, \ref{fig:g3v}, and \ref{fig:agv} \cite{Cucchieri:2008qm,Maas:unpublished,Schleifenbaum:2004id,Maas:2011yx} for the ghost-gluon, the three-gluon, and the two-adjoint-Higgs-gluon vertices, respectively. The former two are identical to the ones of pure Yang-Mills theory, and show a very similar behavior to the zero-temperature case. For the latter it is found that the vertex is essentially flat and close to tree-level, supporting the approximations made in the DSE treatment of the infinite-temperature limit in section \ref{sfinitet:inft}, though this requires confirmation in the unquenched case.

\section{Summary and discussion}\label{ssum}

\subsection{The state of the art, pitfalls, and odds and ends}

Summarizing, quite a lot of progress has been made in the determination of (gauge-dependent) correlation functions since the first investigations using functional \cite{Mandelstam:1979xd} or lattice methods \cite{Mandula:1987rh}. At the present time the technical tools for investigations using lattice methods in Landau gauge are quite well under control. In functional methods, a wealth of understanding of different approximations and truncations has been obtained. As a consequence, sophisticated truncation schemes have been developed, which on the one hand permit to reach rather good agreement with lattice results for the propagators \cite{Fischer:2008uz}, while at the same time already give access to more complex quantities like three-point vertices \cite{Cucchieri:2008qm,Alkofer:2008tt,Alkofer:2008dt,Schleifenbaum:2004id}, and to some extent four-point vertices \cite{Kellermann:2008iw,Huber:2007kc,Fischer:2009tn}, as well as genuine bound-states \cite{Alkofer:2000wg,Fischer:2006ub,Blank:2010pa} and thermodynamic properties \cite{Fischer:2010fx,Braun:2007bx,Roberts:2000aa}. In particular the latter two points bring the functional approach, though quite less intensively developed than lattice methods, to the same general level: Being a tool for investigating physical processes.

This progress has been made possible by a synergistic development of lattice and continuum methods, which also includes non-functional approaches, like effective theories. Only this combination has made it possible to have as good a control over and understanding of the correlation functions as has been presently achieved.

With this a turning-point has been reached. The investigation of correlation functions has so far mainly been conducted for its own sake, and for the exploratory investigation of how to access physical observables\footnote{Of course, in particular concerning the hadronic spectrum these have already been quite successful \cite{Alkofer:2000wg,Fischer:2006ub,Roberts:2007jh}, but so far with somewhat ad-hoc approximations, mainly guided only by chiral symmetry and neglecting most of the Yang-Mills sector. The reasons why these drastic approximations can deliver a rather good description of physical phenomena can only with the full solutions now be fully understood \cite{Blank:2010pa,Fischer:2005en}.} and to understand more about fundamental non-perturbative phenomena like confinement, chiral symmetry breaking and topology directly on the level of correlation functions \cite{Alkofer:2000wg,Fischer:2006ub,Alkofer:2006fu}. Now the technology developed is sufficiently advanced to start accessing completely new fields. And there are two avenues of investigations, which are currently followed. One is the application of the combined framework of lattice and functional methods to new systems. First successes have been achieved in this program \cite{Maas:2010nc,Maas:unpublished,Maas:2004se,Cucchieri:2001tw,Karsch:1996aw,Maas:2011jf,Aguilar:2010ad,Braun:2010qs,Braun:2011pp}, but a lot of work remains to bring to bear the full possibilities of the synergistic effects. The other is to actually use the functional methods to extrapolate the lattice results into domains where the lattice cannot venture currently for practical reasons, like cold, dense matter \cite{Roberts:2000aa,Nickel:2006vf,Nickel:2006kc,Nickel:2008ef,Braun:2008pi,Braun:2011pp}, non-equilibrium \cite{Berges:2008zt}, and scattering processes \cite{Bicudo:2001jq}. It can therefore be expected that the combination of methods will provide access to a number of unresolved problems in the future.

This said, there are still topics to be understood, even on a fundamental level. Though quite a number of hints have been presented here for the idea of how the various solutions available fit together, this is far from a proof. Being skeptical, it could easily turn out that indeed all possible ways to select Gribov copies inside the first (or any fixed) Gribov region could coincide in the thermodynamic limit. In particular, the discussion given in section \ref{sec:kugo} implies that it is very well possible that the scaling case of functional calculations can never be connected with a feasible numerical simulation, though this does not imply that it does not exist.

However, as has been repeatedly emphasized, this is actually not a limitation, since the physical results turn out to be independent of the choice of solution \cite{Blank:2010pa,Luecker:2009bs,Fischer:2009gk,Fischer:2009jm,Costa:2010pp}, including questions like confinement \cite{Fischer:2008uz,Braun:2007bx}. Thus, in principle, it is possible to sit back relaxed, and wait what is realized, and thus available, in Yang-Mills theory, and all the while keep on working with the already confirmed set of correlation functions. Their properties are anyhow equally well under control.

One has to be a bit less relaxed when thinking about their interpretation. Though by now a non-perturbative realization of BRST symmetry \cite{Neuberger:1986xz,vonSmekal:2007ns,vonSmekal:2008es,vonSmekal:2008ws}, and with it an algebraic construction of the state space \cite{Kugo:1979gm,Fischer:2008uz}, is available, it is by far not clear how to connect it to any of the proposed gauges on the lattice, as emphasized above. On the other hand, in the finite-ghost cases, it is still not entirely clear how a Hilbert space has to be constructed \cite{Fischer:2008uz,Baulieu:2008fy,Baulieu:2009ha,Sorella:2009vt,Sorella:2010fs,Sorella:2010it}, and in particular this may involve explicit non-localities \cite{Sorella:2010it,Sorella:2010fs,Baulieu:2009ha,Sorella:2009vt,Baulieu:2008fy,Kondo:2009qz,Capri:2010hb}. Though this is not something with direct relevance to the determination of observables, it is a pain in the back of the head, and conceptually not yet satisfactory. In particular, though manifest quark and possibly gluon confinement has been obtained from correlation functions \cite{Fischer:2008uz,Braun:2007bx,Cucchieri:2004mf,Bowman:2007du,Dudal:2008sp,Cornwall:2009ud}, its full understanding remains a challenge.

Another unclear question is yet how the string-like properties of Yang-Mills theory are realized. This entails two questions. One is, whether it is in principle possible to encode the string-like features in low-order correlation functions at all \cite{Alkofer:2006fu,Greensite:2003bk,Alkofer:2008tt,Schwenzer:2008vt}. Since they are already manifest in, e.\ g., the Regge trajectories \cite{Greensite:2003bk} of mesons encoding the string tension, the answer is an affirmative yes, since the meson properties can be determined from the four-point functions \cite{DeGrand:2006zz}. But this implies that possibly full self-consistency at the level of four-point functions may be required, and this is definitely not yet achieved. It is also not clear whether and how a corresponding truncation can be determined. The situation is even worse for string-breaking, which requires at least six-point functions to observe.

Even more complicated is the Wilson potential. Determining it may require rather different gauges, like the Coulomb gauge \cite{Zwanziger:2002sh,Popovici:2010mb}, though some approximate results have been obtained also in Landau gauge \cite{Alkofer:2008tt,Zayakin:2009jz}. But even with the approximate results in place, the understanding of how to realize the difference between Yang-Mills theory with different gauge groups or for Yang-Mills theory coupled to matter in different representations remains a subject of research \cite{Macher:2010ad,Maas:2005ym,Maas:2007af,Fister:2010yw,Macher:2011ad}.

Finally, though again exploratory results are available, the connection of the correlation function approach to other views of low-energy Yang-Mills theory and QCD still requires a better understanding. In particular, only how to relate correlation functions to topological scenarios  \cite{Maas:2008uz,Maas:2005qt,Maas:2006ss,Boucaud:2003xi,Gattnar:2004bf,Langfeld:2001cz,Quandt:2010yq,Reinhardt:2006fq,Langfeld:2002dd,Greensite:2009zz} as well as to stochastic quantization ideas \cite{Zwanziger:2001kw,Zwanziger:2002ia,Zwanziger:2003cf,Pawlowski:2009iv} has been investigated so far. Still a direct relation to many other views, like the stochastic vacuum, analytic perturbation theory, and many others remains unknown.

\subsection{Summary and outlook}

In total, in this manuscript an overview has been given on the determination of (gauge-dependent) correlation functions with both lattice and functional methods at zero and finite temperature in gauge theories. Particular attention has been paid to the problem introduced by the Gribov-Singer ambiguity. The emphasis has been put on Yang-Mills theory, with a simple Lie group, though connections to other theories and/or setups have been indicated. The results here have concentrated on gauges fulfilling the perturbative Landau gauge condition. Therefore, many results in Coulomb gauge \cite{Popovici:2010mb,Reinhardt:2008ij,Zwanziger:2002sh,Gribov:1977wm,Campagnari:2010wc,Leder:2010ji,Watson:2010cn,Burgio:2009xp,Pak:2009em,Reinhardt:2008pr,Burgio:2008jr,Reinhardt:2008ek,Epple:2007ut,Alkofer:2009dm,Watson:2007fm,Feuchter:2007mq,Epple:2006hv,Schleifenbaum:2006bq,Lichtenegger:2009dw,Cucchieri:2000gu,Langfeld:2004qs,Greensite:2010hn,Greensite:2009eb,Greensite:2004ke,Popovici:2010ph,Szczepaniak:2001rg,Nakagawa:2011ar,Reinhardt:2011hq,Greensite:2010tm,Campagnari:2011bk,Watson:2011kv,Burgio:2012rm}, interpolating gauges \cite{Cucchieri:2007uj,Fischer:2005qe,Baulieu:1998kx,Cucchieri:2001tw,Iritani:2011zg,Maas:2011ej}, linear covariant gauges \cite{Alkofer:2000wg,Cucchieri:2009kk,Cucchieri:2008zx,Mendes:2008ux,Sobreiro:2005vn}, maximal Abelian gauges \cite{Mendes:2008ux,Huber:2009wh,Bornyakov:2003ee,Mihara:2008zz,Capri:2010an,Capri:2008vk,Capri:2008ak,Capri:2007ph,Capri:2007hw,Capri:2005zj}, direct Laplacian gauges \cite{Alexandrou:2002gs,Alexandrou:2001fh,Alexandrou:2000ja,vanBaal:1994ai}, and other gauges \cite{Alkofer:2000wg,Buttner:1995hg,Fischer:2004uk,Gies:2002af,Marhauser:2008fz,vonSmekal:2008es,Kalloniatis:2005if,Alkofer:2003jr,Laufer:2012tg,Cucchieri:2012tm} have not been covered, though results at a similar level have been obtained in these cases as well. In particular, a similar situation with the finite-ghost case and scaling appears to be realized in Coulomb gauge \cite{Watson:2010cn,Epple:2007ut,Epple:2006hv,Szczepaniak:2001rg}, though this has to be investigated further.

Of course, the current important topic is the full inclusion of matter fields to investigate QCD, the standard model, and beyond-the-standard-model theories. In the end, however, all this will rest on the concepts presented here, and will be achieved with similar methods. It is thus the creation of this technological framework and the understanding of its results within the last fifteen years, which is now paving the way for a complete new generation of applications. Thus, with the next decade it will be decided whether the promise of the possibilities these methods and approaches hold will be realizable, and perhaps even open up the way to new frontiers in non-perturbative physics.

\subsection*{Acknowledgments}

I am grateful for many discussions and/or collaborations on the topic of this review and on related subjects to Reinhard Alkofer, Daniel August, Martina Blank, Jens Braun, Falk Bruckmann, Michael Buballa, Giuseppe Burgio, Attilio Cucchieri, Dennis Dietrich, Luigi Del Debbio, David Dudal, Leonard Fister, Christof Gattringer, Jeff Greensite, Holger Gies, Lisa Haas, Markus Huber, Ernst-Michael Ilgenfritz, Christian Kellermann, Andreas Krassnigg, Christian Lang, Daniel Litim, Veronika Macher, Tereza Mendes, Daniel Mohler, Tajdar Mufti, Jens M\"uller, Michael M\"uller-Preussker, {\v S}tefan Olejn\`ik, Joannis Papavassiliou, Agostino Patella, Claudio Pica, Willibald Plessas,  Antonio Rago, Hugo Reinhardt, Francesco Sannino, Wolfgang Schleifenbaum, Bernd-Jochen Schaefer, Wolfgang Schweiger, Kai Schwenzer, Daniel Spielmann, Silvio Sorella, Andr\'e Sternbeck, Adam Szczepaniak, Nele Vandersickel, Jochen Wambach, Peter Watson, Bj\"orn Wellegehausen, Richard Williams, Andreas Wipf, Daniel Zwanziger, and especially my long-term collaborators Christian S.\ Fischer, Jan M.\ Pawlowski, and  Lorenz von Smekal. I am indebted to Michael M\"uller-Preussker, Wolfgang Schweiger, and Hartmut Wittig for comments on my habilitation thesis at the Karl-Franzens-University of Graz, on which this review is based.

This research has been financially supported by the DFG (grant numbers MA 3935/1-1, MA 3935/1-2, MA 3935/2-1, and MA 3935/5-1) and the FWF (grant numbers P20330 and M1099-N16). The ROOT framework \cite{Brun:1997pa} has been used in this project.

I would also like to thank my wife Renate Knobloch-Maas for being patient with me while writing this, and for her love.

I dedicate this work to my parents, Rainer Maas and Karina Maas-Margraf.

\bibliographystyle{bibstyle}
\bibliography{bib}

\end{document}